\documentclass{aa}

\usepackage{graphicx}
\usepackage{placeins}
\usepackage{array}
\usepackage{caption}
\usepackage{subcaption}
\usepackage{gensymb, footnote}
\usepackage{tikz}
\usepackage{sidecap}
\usepackage{changepage}
\usepackage[toc,page]{appendix}
\usepackage{multirow}
\usepackage[hidelinks]{hyperref}
\usepackage{bm}
\usepackage{float}
\usepackage{lscape}

\usepackage{tablefootnote}
\usepackage{threeparttable}
\usepackage{upgreek}
\usepackage[hidelinks]{hyperref}
\hypersetup{
 colorlinks = true,
 citecolor ={blue}
}

\bibpunct{(}{)}{;}{a}{}{,}   

\newcommand{\Msun}[0]{{M_{\odot}}}
\newcommand{\Lsun}[0]{{L_{\odot}}}
\newcommand{\Rsun}[0]{{R_{\odot}}}
\newcommand{\kms}{km\,${\rm s}^{-1}$}
\newcommand{\teff}[0]{T_{\rm eff}}
\newcommand{\logg}[0]{{\log\,g}}

\newcommand{\vsini}[0]{v\,\sin\,i}
\newcommand{\vmac}[0]{v_{\mathrm{mac}}}

\renewcommand*{\thesubfigure}{(\arabic{subfigure})}
\begin{document}

\title{Identifying quiescent compact objects in massive Galactic single-lined spectroscopic binaries}

\author{L. Mahy\inst{1} \and  
    H. Sana\inst{2} \and 
    T. Shenar\inst{3} \and  
    K. Sen\inst{4,5} \and  
    N. Langer\inst{4,5} \and  
    P. Marchant\inst{2} \and
    M. Abdul-Masih\inst{6} \and 
    G. Banyard\inst{2} \and 
    J.\,Bodensteiner\inst{7} \and 
    D. M. Bowman\inst{2} \and 
    K. Dsilva\inst{2} \and
    M. Fabry\inst{2} \and 
    C. Hawcroft\inst{2}  \and  
    S. Janssens\inst{2} \and  
    T. Van Reeth\inst{2} \and 
    C. Eldridge\inst{8}
}
\institute{Royal Observatory of Belgium, Avenue Circulaire/Ringlaan 3, B-1180 Brussels, Belgium \\ \email{laurent.mahy@oma.be}
\and 
Institute of Astronomy, KU Leuven, Celestijnenlaan 200D, 3001, Leuven, Belgium 
\and
Anton Pannekoek Institute for Astronomy, University of Amsterdam, Postbus 94249, 1090 GE Amsterdam, The Netherlands
\and
Argelander-Institut f{\"u}r Astronomie, Universit{\"a}t Bonn, Auf dem H{\"u}gel 71, 53121 Bonn, Germany
\and
Max-Planck-Institut f{\"u}r Radioastronomie, Auf dem H{\"u}gel 69, 53121 Bonn, Germany
\and
ESO – European Organisation for Astronomical Research in the Southern Hemisphere, Alonso de Cordova 3107, Vitacura, Santiago de Chile, Chile
\and
ESO – European Organisation for Astronomical Research in the Southern Hemisphere, Karl-Schwarzschild-Strasse 2, 85748
Garching, Germany
\and
Amateur astronomer
}

   \date{Received 2022; accepted 2022}

 
  \abstract
   {The quest to detect dormant stellar-mass black holes (BHs) in massive binaries (i.e. OB+BH systems) is challenging; only a few candidates have been claimed to date, all of which must still be confirmed.}
   {To search for these rare objects, we study 32 Galactic O-type stars that were reported as single-lined spectroscopic binaries (SB1s) in the literature. In our sample we include Cyg~X-1, which is known to host an accreting stellar-mass BH, and HD~74194, a supergiant fast X-ray transient, in order to validate our methodology. The final goal is to characterise the nature of the unseen companions to determine if they are main-sequence (MS) stars, stripped helium stars, triples, or compact objects such as neutron stars (NSs) or stellar-mass BHs.}
    {After measuring radial velocities and deriving orbital solutions for all the systems in our sample, we performed spectral disentangling to extract putative signatures of faint secondary companions from the composite spectra. We derived stellar parameters for the visible stars and estimated the mass ranges of the secondary stars using the binary mass function. Variability observed in the photometric TESS light curves was also searched for indications of the presence of putative companions, degenerate or not. }
   {In 17 of the 32 systems reported as SB1s, we extract secondary signatures, down to mass ratios of $\sim 0.15$. For the 17 newly detected double-lined spectroscopic binaries (SB2s), we derive physical properties of the individual components and discuss why they have not been detected as such before. Among the remaining systems, we identify nine systems with possible NS or low-mass MS companions. For Cyg~X-1 and HD~130298, we are not able to extract any signatures for the companions, and the minimum masses of their companions are estimated to be about $7\,\Msun$. Our simulations show that secondaries with such a mass should be detectable from our dataset, no matter their nature: MS stars, stripped helium stars or even triples. While this is expected for Cyg~X-1, confirming our methodology, our simulations also strongly suggest that HD~130298 could be another candidate to host a stellar-mass BH. }
   {The quest to detect dormant stellar-mass BHs in massive binaries is far from over, and many more systems need to be scrutinised. Our analysis allows us to detect good candidates, but confirming the BH nature of their companions will require further dedicated monitorings, sophisticated analysis techniques, and multi-wavelength observations.}

\maketitle

\section{Introduction}
Massive stars tend to end their short but energetic lives as core-collapse supernovae \citep{heger03}, producing compact remnants such as neutron stars (NSs) or stellar-mass black holes (BHs). With their final supernova outflows and their powerful stellar winds, they are one of the most important cosmic engines that drive the evolution of galaxies, by providing chemical enrichment, ionising radiation, and mechanical feedback \citep[e.g.][]{maclow04,hopkins14,crowther16}. 

One of the most striking properties of massive stars is their high degree of multiplicity \citep{sana12,sana14,moe17,offner22}. As a consequence, the presence of a companion severely impacts the evolution of these stars \citep{Pod92}. The strong binary interactions make the understanding of their evolution more complex such that many aspects are not yet completely understood. This has been confirmed with the detections of gravitational waves (GWs) by LIGO (Laser Interferometer Gravitational-Wave Observatory) and Virgo \citep[][and subsequent papers]{abbott16}. These GW events have shown that tight pairs of compact objects exist and occasionally merge. To explain how massive stars in binary systems evolve to produce these GW events, different scenarios have been proposed. They include (i) chemically homogeneous evolution \citep[CHE,][]{maeder87, Langer92, martins13} in very massive short-period stellar binaries, which prevents mass transfer and allows compact MS binaries to directly evolve into compact BH binaries \citep{mandel16, Marchant16, demink16, abdulmasih19, dubuisson20, riley21, abdulmasih21, menon21}, (ii) evolution through a common-envelope phase \citep[e.g.][]{Paczynski76,vandenheuvel76, tutukov93, belczynski02, belczynski16, Giacobbo18}, even though current theoretical predictions are highly uncertain and observational constraints of these specific stages are missing, (iii) stable mass transfer \citep{vandenheuvel17, neijssel19, bavera20, marchant21,menon21,Sen2021b}, and (iv) Population III stars \citep{belczynski04,kinugawa14,inayoshi17}. 

It is commonly accepted that NSs are the remnants of stars with initial masses between $8$ and $20\,\Msun$. Stars with initial masses between $\sim 20$ and $40\,\Msun$ have stellar-mass BHs as their end-of-life remnants, which are formed by fallback of mass after an initial supernova shock has been launched. Stars with initial masses between $\sim 40$ and $150\,\Msun$ are thought to experience direct collapse, forming stellar-mass BHs without spectacular explosions. Stars initially more massive than $\sim 150~\Msun$ are expected to explode in pair-instability supernovae (PISNe; \citealt{fryer01,woosley07,Sukhbold2016}) without leaving a remnant behind. The above mass ranges are model-dependent and also depend on the metallicity and initial rotation rate (e.g. \citealt{Sukhbold2016}). 

Given the star formation history, over 100 million stellar-mass BHs are predicted to lurk in the Milky Way \citep[][]{brown94, mashian17,breivik17, lamberts18, Yalinewich18, Yamaguchi18, janssens22}. So far, about 100 compact objects have been detected in X-ray binaries \citep{walter15,corral16}, accreting material from their stellar companions through Roche-lobe overflow or wind accretion \citep{postnov14}. However, most of the known X-ray binaries involve a NS, and only a few are believed to harbour a stellar-mass BH. In addition, and in the vast majority of these cases, the BH accretes material from a low-mass companion, leaving only Cyg~X-1 \citep[][and, possibly, Cyg~X-3 ,\citealt{Zdziarski13,Koljonen17}]{orosz11,Miller-Jones21} in our Galaxy as the prototypical and widely accepted example of a BH accreting from a massive companion, that is massive enough to end its life as yet another compact object. However, such X-ray emission only arises in tight binary systems or when the secondary star starts filling its Roche lobe, hence where substantial accretion can occur \citep[e.g.][]{Shapiro1976, Sen2021,hirai21}. In wider binaries or binaries with largely unevolved stellar companions, it is natural to expect a stellar-mass BH in a quiescent stage, that is without X-ray emission. The fact that the majority of OB$+$ BH binaries are expected to be in wide orbits was notably highlighted by \citet{langer20}.

Over 90\% of GW detections come from BH$+$BH binary systems, leading to the discovery of almost 100 additional stellar-mass BHs. Finding and characterising binary systems that host a dormant BH in the Milky Way would not only help test the validity of the binary evolution channel to produce GW events, but would also provide a critical anchor point to test and validate the physical assumptions made regarding BH formation (e.g. the presence of a kick) as well as the prediction of binary interaction theories \citep{langer12}.

From the above discussion, OB+BH systems have so far predominantly been found when the BH is accreting material from its companion, producing X-ray emission. Several exceptions exist, such as MWC~656 \citep{casares14} or \object{HD~96670} \citep{Gomez21}, where the stellar-mass BH was not found to be X-ray bright. The BHs in these systems are therefore referred to as quiescent. Other reports of quiescent OB+BH systems (e.g. \object{LB-1}, \citealt{Liu2019}; \object{HR\,6819}, \citealt{Rivinius2020}; \object{NGC~1850} BH1, \citealt{saracino21}; \object{NGC~2004 \#115}, \citealt{lennon22}) and quiescent stripped giants+BH exist in the literature (e.g. V723~Mon, \citealt{Jayasinghe21}), but all of these reports were challenged in subsequent studies \citep[e.g.][]{AbdulMasih2020LB1, Shenar2020LB1, bodensteiner20, Gies2020, ElBadry2021, elbadry21b, elbadry22,frost22,elbadry22b}. 

Recent theoretical computations, however, predict that about 3\% of massive O or early-B stars in binary systems should have a dormant BH as companion \citep{shao19,langer20}. If the theoretical predictions are correct, these systems should hide in plain sight. A number of Galactic and extra-galactic young open clusters and OB associations have been probed to derive the binary status of massive stars and to investigate their orbital properties through dedicated long-term spectroscopic and interferometric campaigns \citep[e.g.][]{sana08,mahy09,sana09,sana11,sana12,kobulnicky12,mahy13,sana14,barba17,trigueros21,banyard22}. One way to look for these OB$+$BH systems is to probe the population of single-lined spectroscopic binaries (SB1s), which are systems where only one component is visible, either because the companion is a low-mass star or because it is a compact companion. One must, however, be careful because some of these systems might be hidden double-lined spectroscopic binaries (SB2s), where the secondary is very diluted, or rotates very rapidly. Some could also simply be pulsating single stars, in which the line profile variability due to pulsations mimics a binary signature \citep[see e.g.][]{aerts93}.

Many attempts have been made to find compact objects in binary systems \citep[e.g.][]{guseinov66, Gu19}. The masses of the unseen components are deduced from the binary mass function and the spectroscopic mass of its counterpart (obtained from the stellar radius and surface gravity). When this mass exceeds the critical mass of $3 - 5\,\Msun$, the unseen object can be considered as a candidate stellar-mass BH \citep[see reviews, e.g. ][and references therein]{cowley92,remillard06,casaresJonker14}. With the developments of new instruments, photometric \citep{zucker07, masuda19, gomel21b}, asteroseismic \citep{Shibahashi20}, and astrometric \citep{breivik17,Yamaguchi18, andrews19, janssens22} methods have also been developed to find hidden BHs, but no conclusive discovery has been achieved so far. 

In the present paper we combine high-resolution spectroscopy, high-precision space-based photometry, and state-of-the-art spectral disentangling to constrain the nature of unseen companions in systems classified as SB1s in the literature that host an O- or an early-B-type star. In addition to searching for stellar-mass BHs, we use the detected low-mass MS companions to characterise the low-mass end of the companion mass function. The paper is organised as follows. Section~\ref{sec:sample} describes the sample, the observations, and the data reduction procedures. Section~\ref{sec:methodology} details the methodology we used to constrain the nature of the unseen objects and provides the results. Section~\ref{sec:discussion} discusses the results, and Sect.~\ref{sec:conclusion} summarises our conclusions.

\section{Sample, observations, and data reduction}
\label{sec:sample}

\subsection{Sample selection}
Our initial sample is based on the list of systems reported as SB1s by the Galactic O-Stars Spectroscopic Survey catalogues \citep[GOSSS,][and references therein]{sota11,sota14,maiz16}, dedicated monitorings of young open clusters \citep[][among others]{sana08,mahy09,sana09,sana11,mahy13} and by the Southern Galactic O- and WN-type Stars (OWN) Survey \citep{barba17}. We selected objects for which archival and/or new observed spectra exist to uniformly cover their orbital cycles, and to compute the orbital parameters with uncertainties close to 10\% without further selection criteria. Our final sample contains 32 stars split over the northern and southern hemispheres and includes Cyg~X-1, known to host a stellar-mass BH \citep{orosz11,Miller-Jones21}, and HD~74194, a supergiant fast X-ray transient (i.e. a sub-class of high-mass X-ray binaries showing sporadic and bright X-ray flares) that hosts a NS \citep{gamen15}. 

An overview of the sample stars as well as the details of the observations (number of spectra, instruments, etc) can be found in Table\,\ref{table:database}.

\subsection{Observations and data reduction}
Objects with declinations higher than $-25^{\circ}$ were mainly observed with the High-Efficiency and high-Resolution Mercator Echelle Spectrograph (HERMES) mounted on the 1.2m Flemish Mercator telescope \citep{raskin11} at the Spanish Observatorio del Roque de los Muchachos in La Palma (Spain). The data were taken in the high-resolution fibre mode, which has a spectral resolving power of $R = 85\,000$. The spectra cover the 4000–9000\,\AA\ wavelength domain. All the stars were randomly observed over one or several semesters. The raw exposures were reduced using the dedicated HERMES pipeline and we worked with the extracted cosmic-removed, merged and wavelength-calibrated spectra afterwards.

We also retrieved spectra taken with the spectrographs ELODIE and SOPHIE, both mounted on the 1.93-m telescope at the Observatoire de Haute-Provence (France). ELODIE \citep{baranne96} was operational from 1993 to 2006. This instrument covered the spectral range from 3850 to 6800~\AA\ and has $R \sim 42\,000$. SOPHIE \citep{Bouchy06,Perruchot08} covers the wavelength range 3870–6940 Å with $R \sim 40~000$. The data were processed by the SOPHIE fully automatic data reduction pipeline. 

We also retrieved data collected with the Echelle SpectroPolarimetric Device for the Observation of Stars \citep[ESPaDOnS][]{donati06} mounted on the Canada-France-Hawaii Telescope on Mauna Kea (Hawaii, USA). Spectra were retrieved from the Polarbase website\footnote{\url{http://polarbase.irap.omp.eu}} and cover the 3700–10500~\AA\ wavelength range with $R \sim 80~000$.

For the stars with declinations lower than $-25^{\circ}$, we retrieved optical spectra from the ESO archives observed with the Fibre-fed Extended Range Optical Spectrograph (FEROS), the UV and Visible Echelle Spectrograph (UVES) and XShooter. FEROS is mounted on the MPG/ESO 2.2-m telescope at La Silla (Chile). FEROS \citep{Kaufer1997b,Kaufer1999b} provides a resolving power of $R = 48\,000$ and covers the entire optical range from $3700$ to $9200$\,\AA. The data were reduced following the procedure described in \citet{mahy10, mahy17}.

UVES \citep{dekker00} is mounted on the VLT/UT2 at Paranal (Chile), has a resolving power of $R = 80\,000$ and covers different wavelength ranges in the near-UV and optical domains depending on the setup. The data were reduced with the UVES pipeline.

X-Shooter \citep{vernet11} is an intermediate-resolution ($R \sim 4~000-17~000$) slit spectrograph covering a wavelength range from 3000 to $25~000$\,\AA, divided over three arms: UV-Blue, visible, and near-infrared. The data were reduced with the XShooter pipeline.

We also collected one spectrum of HD~130298 with the High Resolution Spectrograph on the Southern African Large Telescope \citep{bramall10,bramall12,crause14} under programme 2021-1-SCI-014 (PI: Manick). The data were taken in medium-resolution mode and reduced with the \textsc{midas} pipeline \citep{kniazev2016} based on the {\'e}chelle \citep{ballester92} and FEROS \citep{stahl99} packages. We applied heliocentric corrections to the data and confirmed the wavelength calibrations by using the diffuse interstellar bands that are present within the wavelength coverage.

\section{Methodology and results}
\label{sec:methodology}
\subsection{Orbital solution}
\label{subsec:orbitalsolution}
SB1s are characterised as binary systems with a visible star that shows periodic radial velocity (RV) variations that implies the presence of a binary companion. This companion can be either a low-mass main-sequence (MS) object, a stripped helium star, or a degenerate object (white dwarf, NS, or stellar-mass BH). Other factors, for example the rotation of the companion or the brightness ratio between the two stars, can also lead to the non-detection of a companion \citep[see e.g. ][]{Shenar2020LB1}, and therefore to their classification as SB1. The term SB1 does not, however, automatically involve binary systems. Some objects, classified as SB1s in the literature, are in fact single stars where their RVs show a periodic motion, reminiscent of the orbital motion, but with lower RV semi-amplitudes (i.e. these are false positive). These variations might be intrinsic to a single star, and produced by pulsations \citep[see e.g.][]{DeCat2000b} or inhomogeneities in their stellar winds \citep[see e.g.][]{eversberg98,bouret03}. It is therefore useful to complement the spectroscopy with photometry to detect intrinsic variability or specific signals related to their orbital motions/pulsation patterns (Sect.\,\ref{subsec:tess}).

\begin{figure*}
    \centering
    \begin{subfigure}{0.33\linewidth}
    \includegraphics[width = \textwidth]{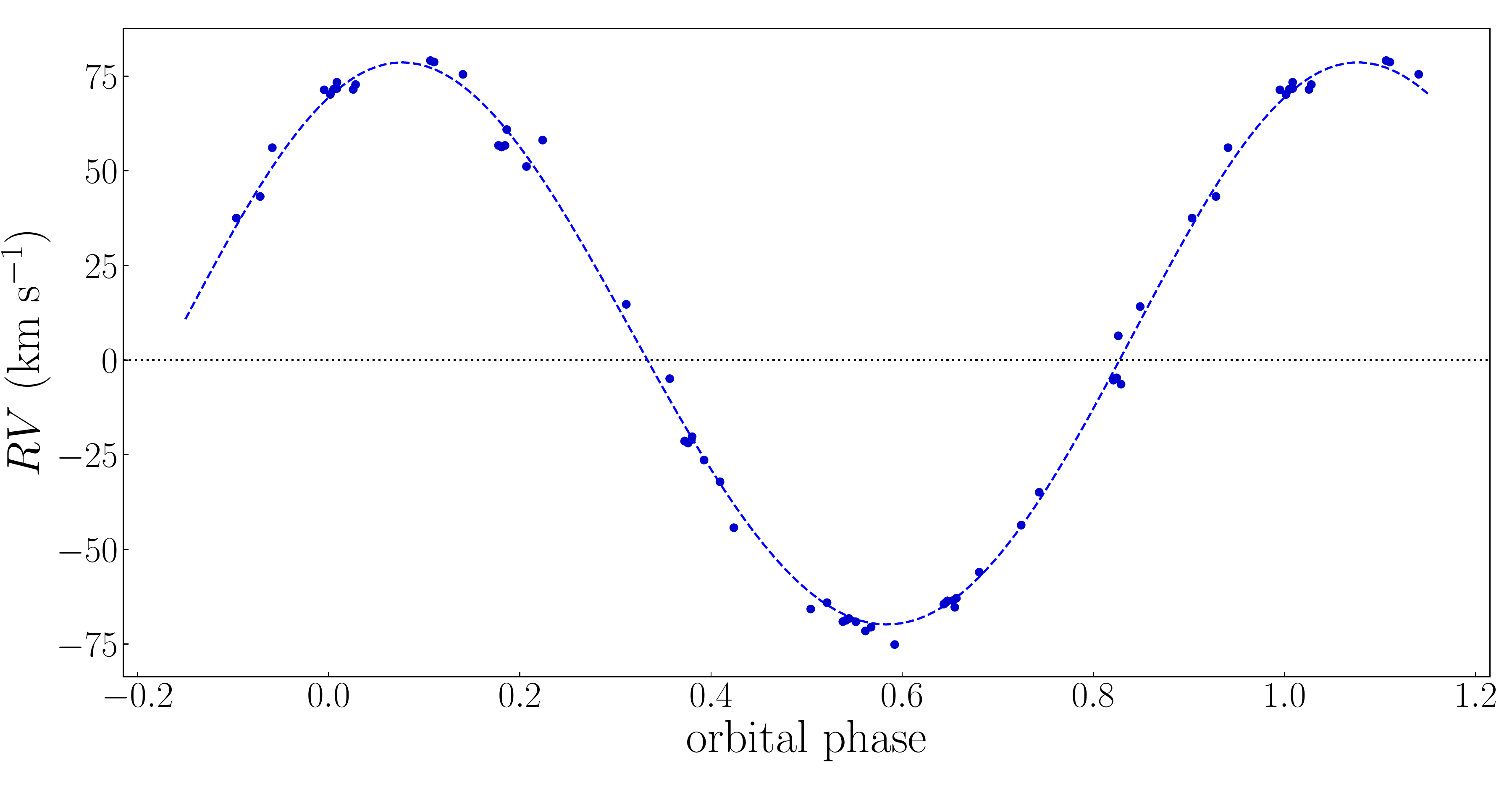}
    \caption{Cyg~X-1}
    \end{subfigure}
    \begin{subfigure}{0.33\linewidth}
    \includegraphics[width = \textwidth]{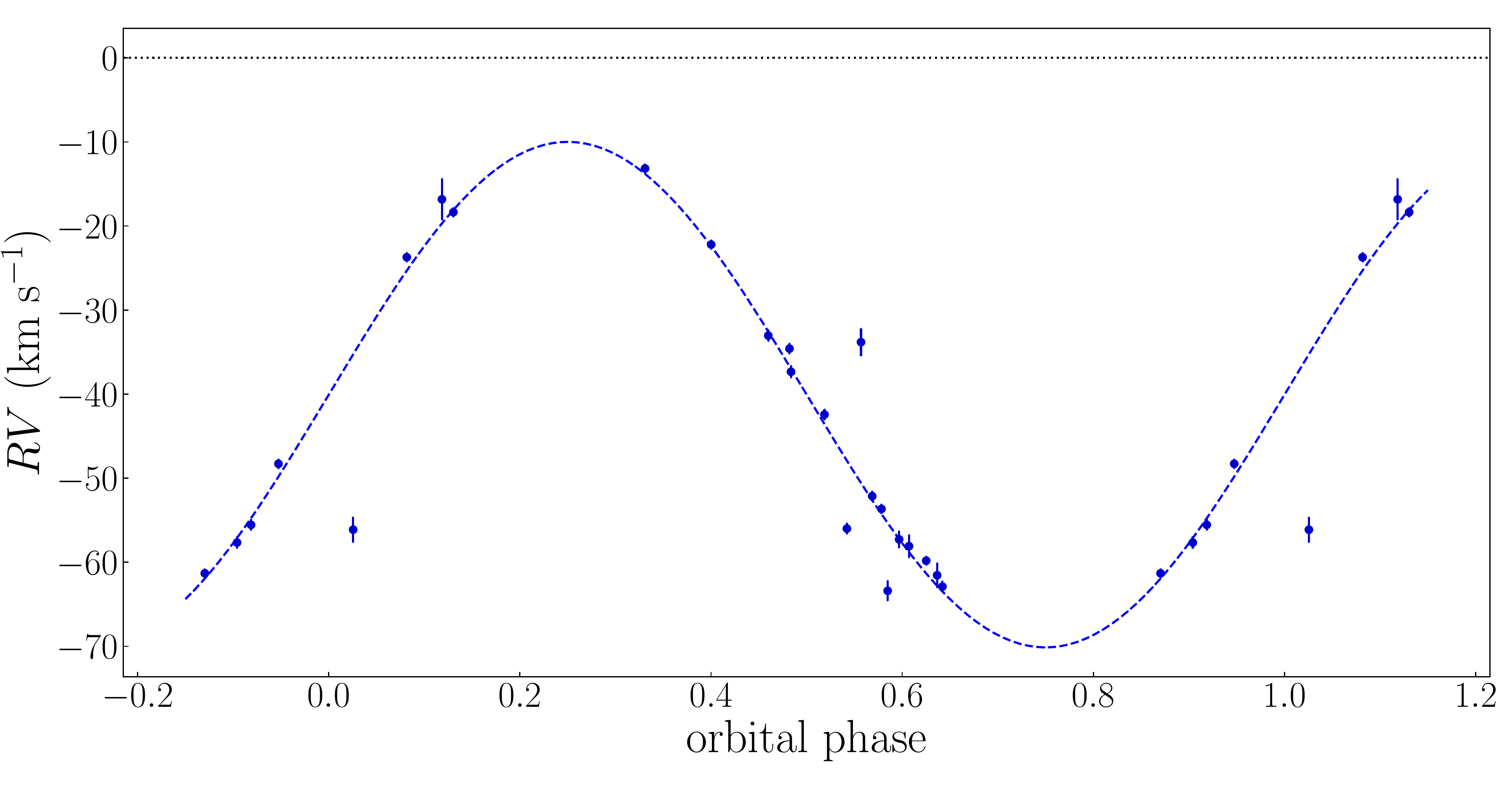}
    \caption{HD~12323}
    \end{subfigure}
    \begin{subfigure}{0.33\linewidth}
    \includegraphics[width = \textwidth]{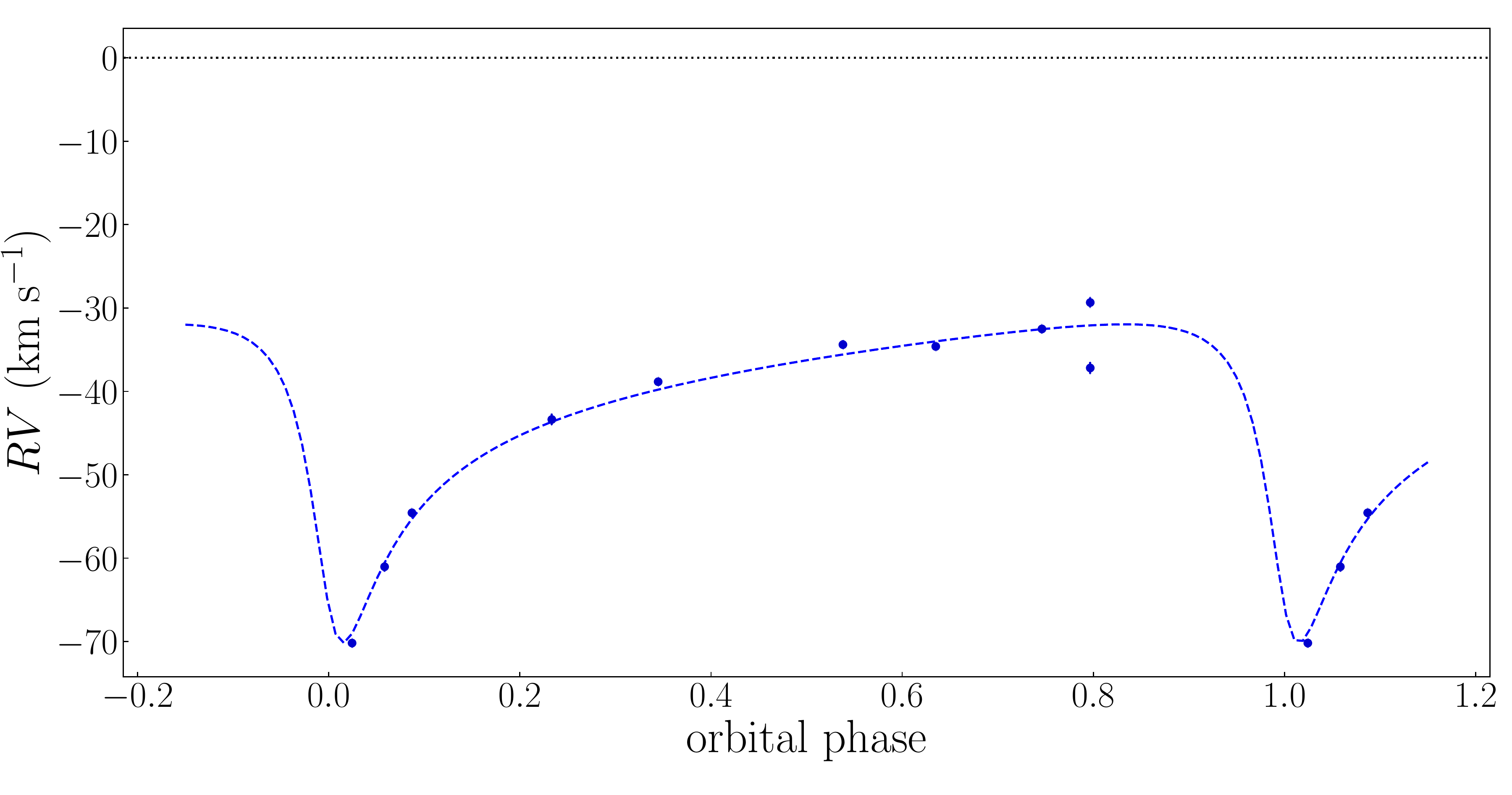}
    \caption{HD~14633}
    \end{subfigure}
    \begin{subfigure}{0.33\linewidth}
    \includegraphics[width = \textwidth]{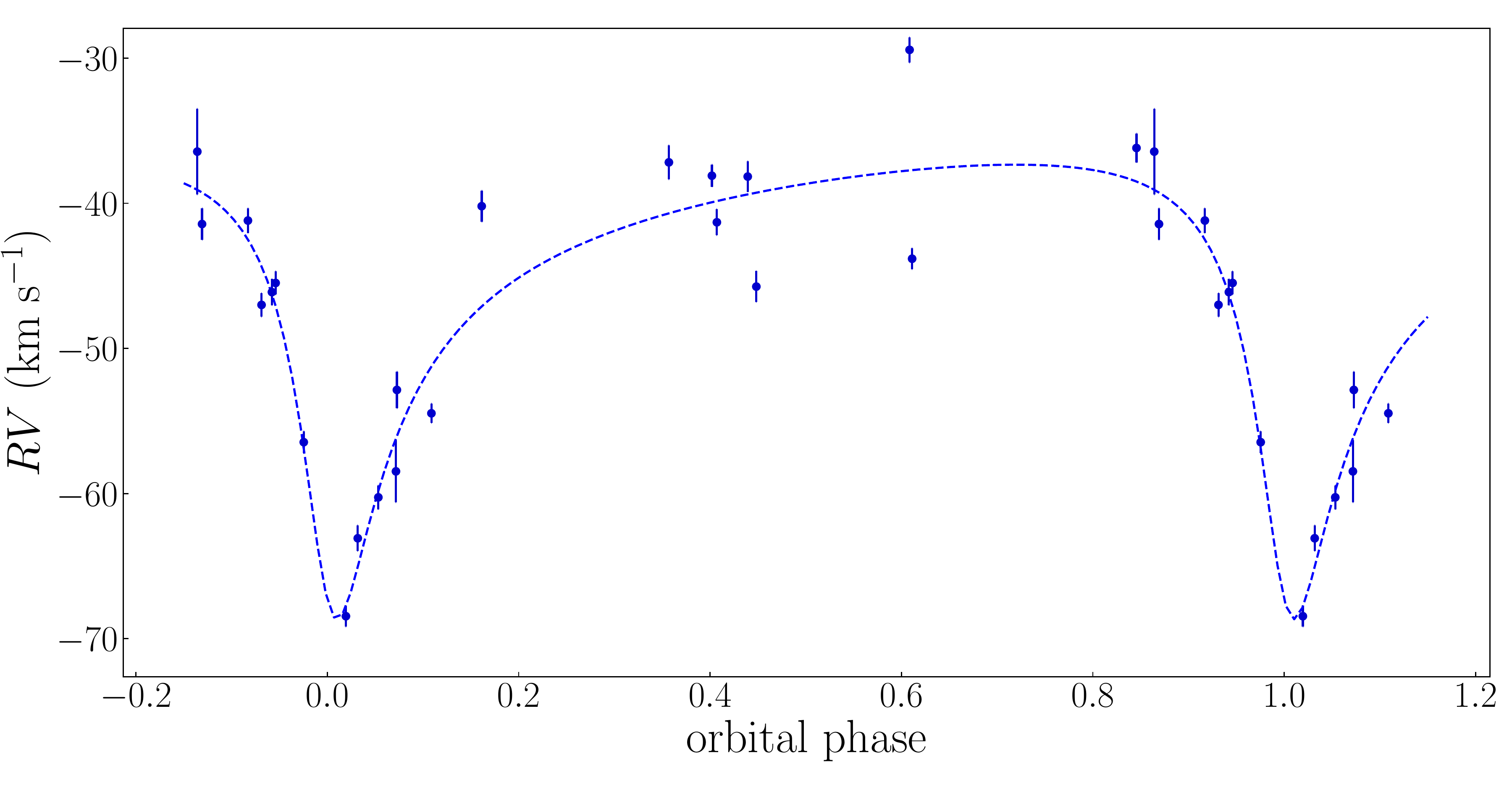}
    \caption{HD~15137}
    \end{subfigure}
    \begin{subfigure}{0.33\linewidth}
    \includegraphics[width = \textwidth]{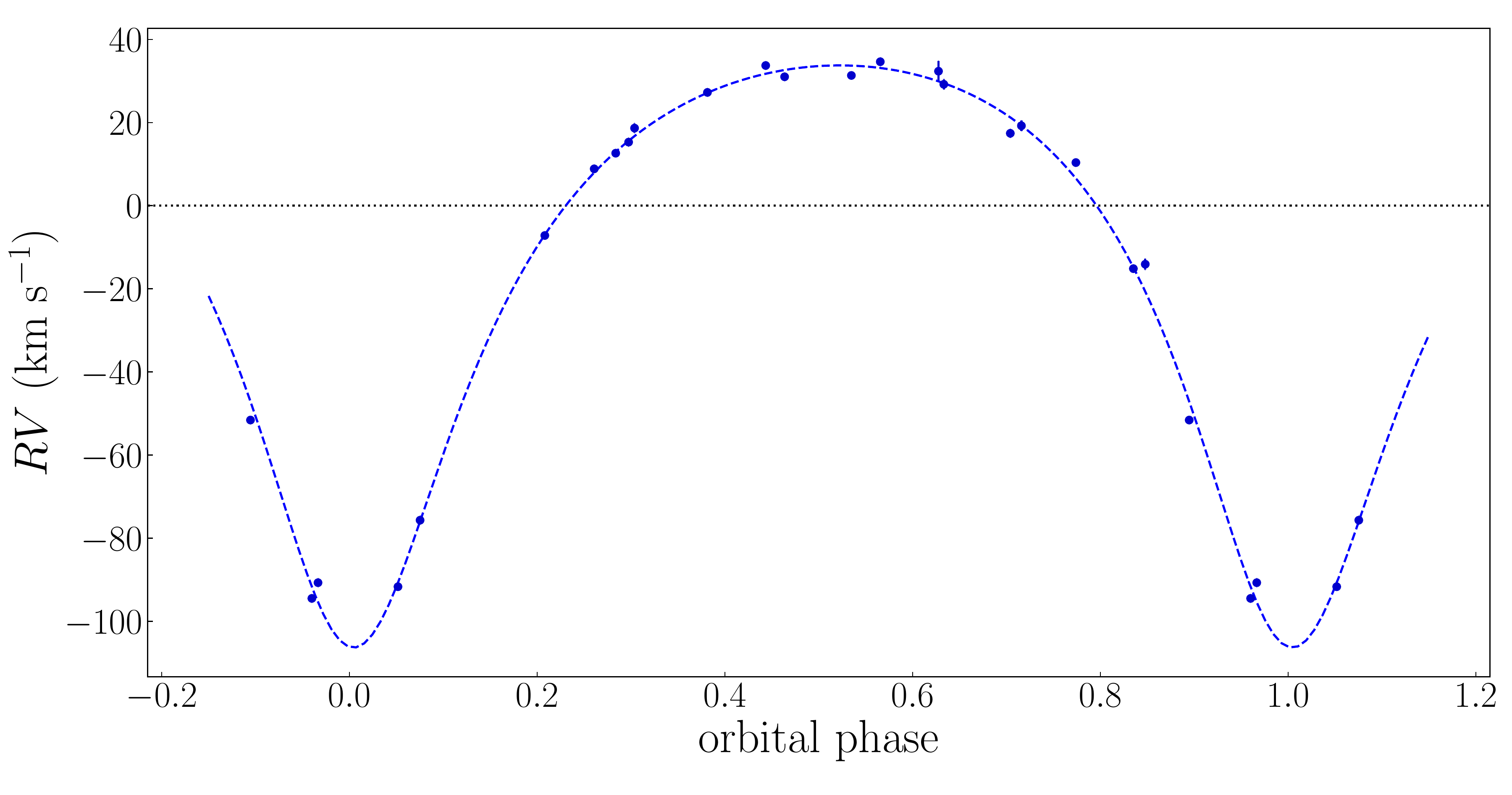}
    \caption{HD~37737}
    \end{subfigure}
    \begin{subfigure}{0.33\linewidth}
    \includegraphics[width = \textwidth]{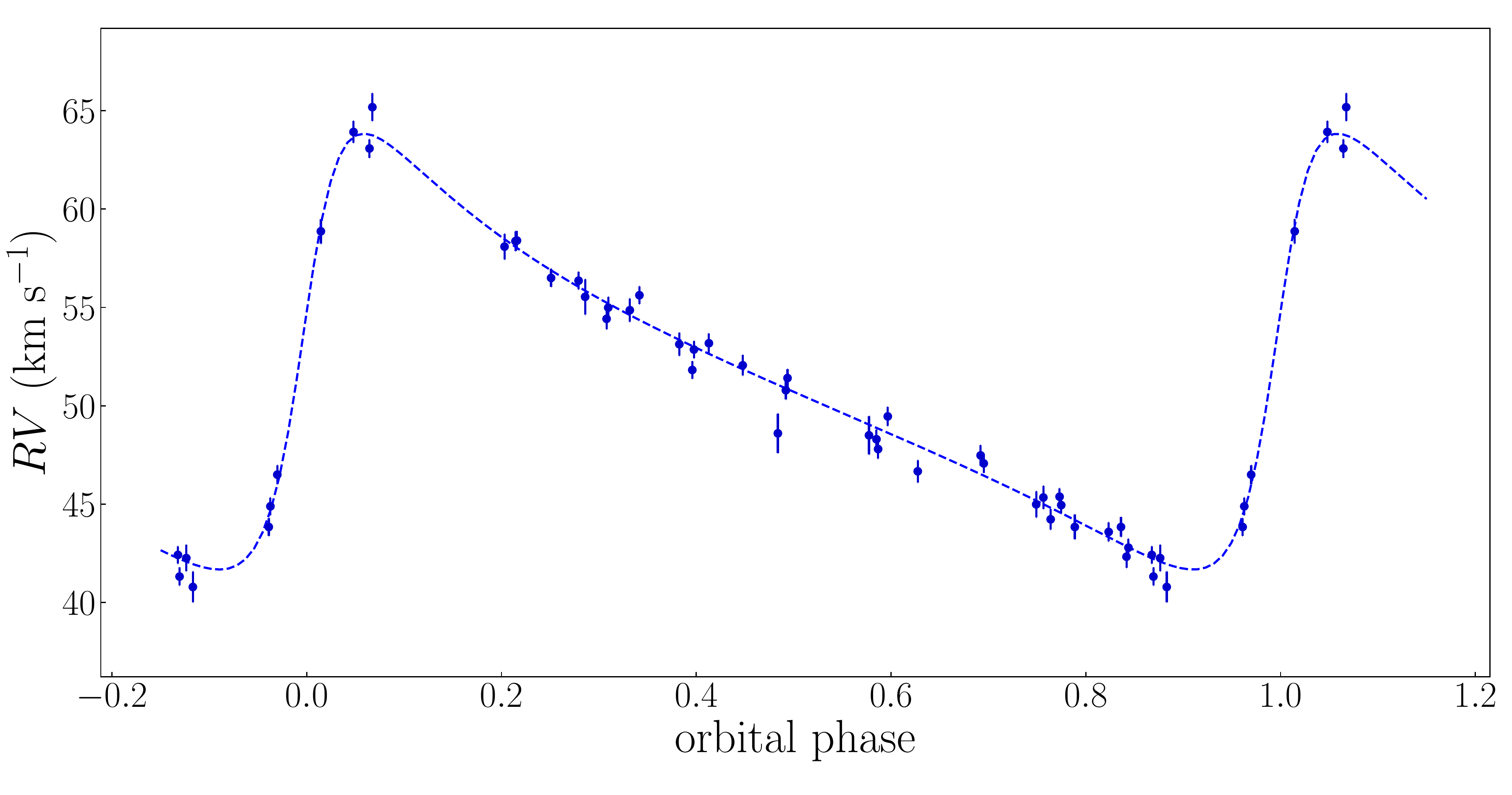}
    \caption{HD~46573}
    \end{subfigure}
    \begin{subfigure}{0.33\linewidth}
    \includegraphics[width = \textwidth]{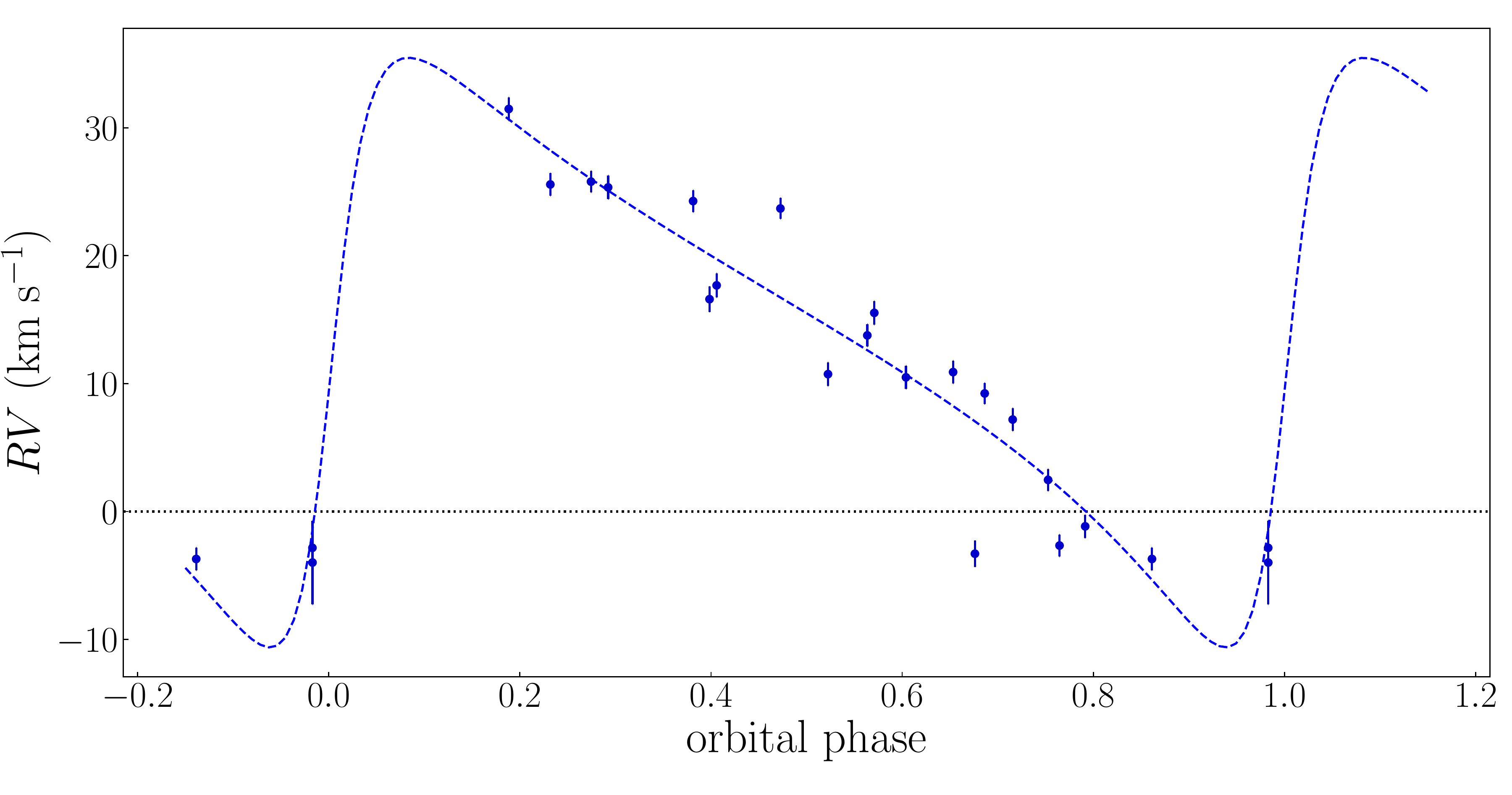}
    \caption{HD~74194}
    \end{subfigure}
    \begin{subfigure}{0.33\linewidth}
    \includegraphics[width = \textwidth]{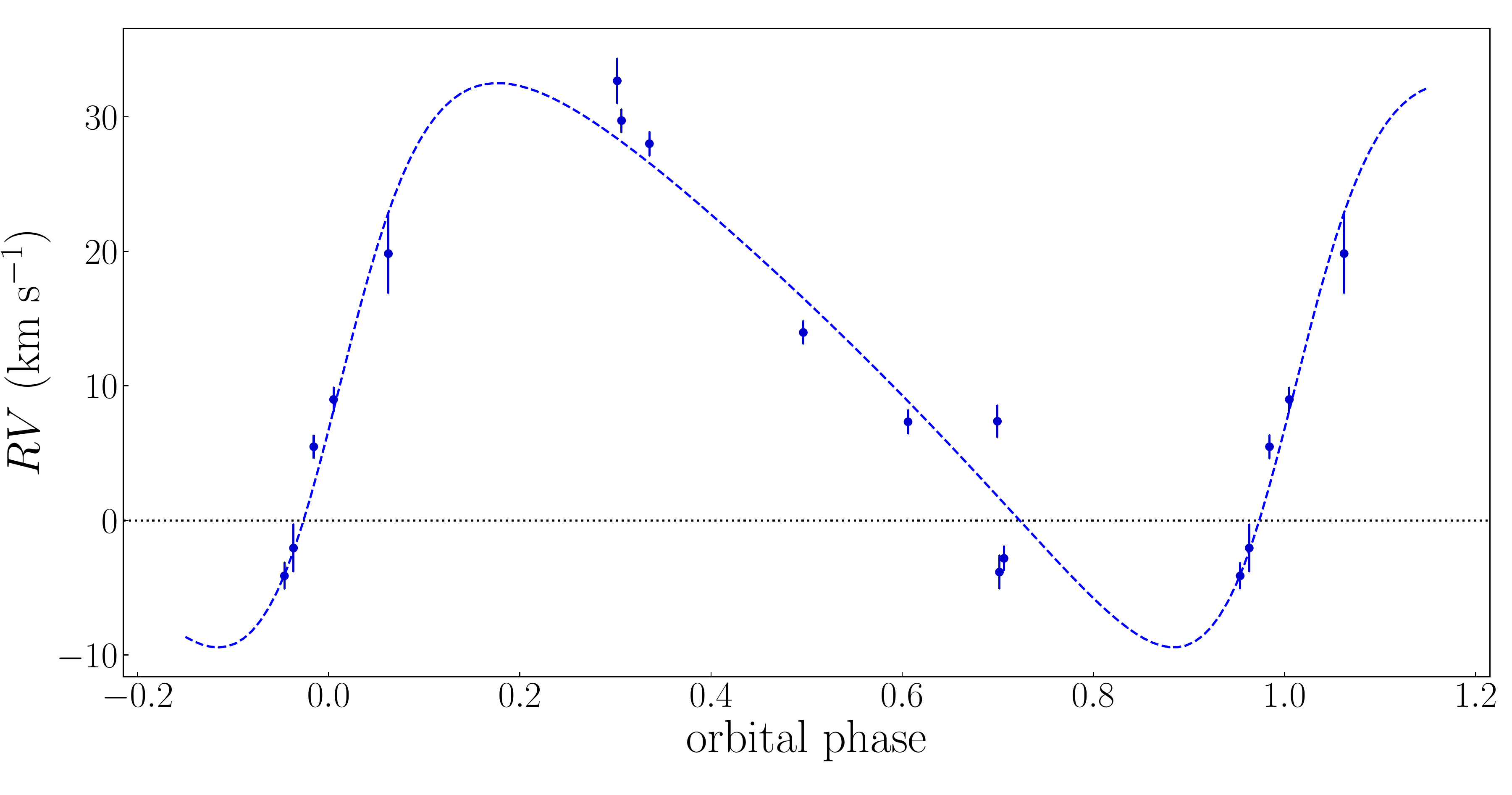}        \caption{HD~75211}
    \end{subfigure}
    \begin{subfigure}{0.33\linewidth}
    \includegraphics[width = \textwidth]{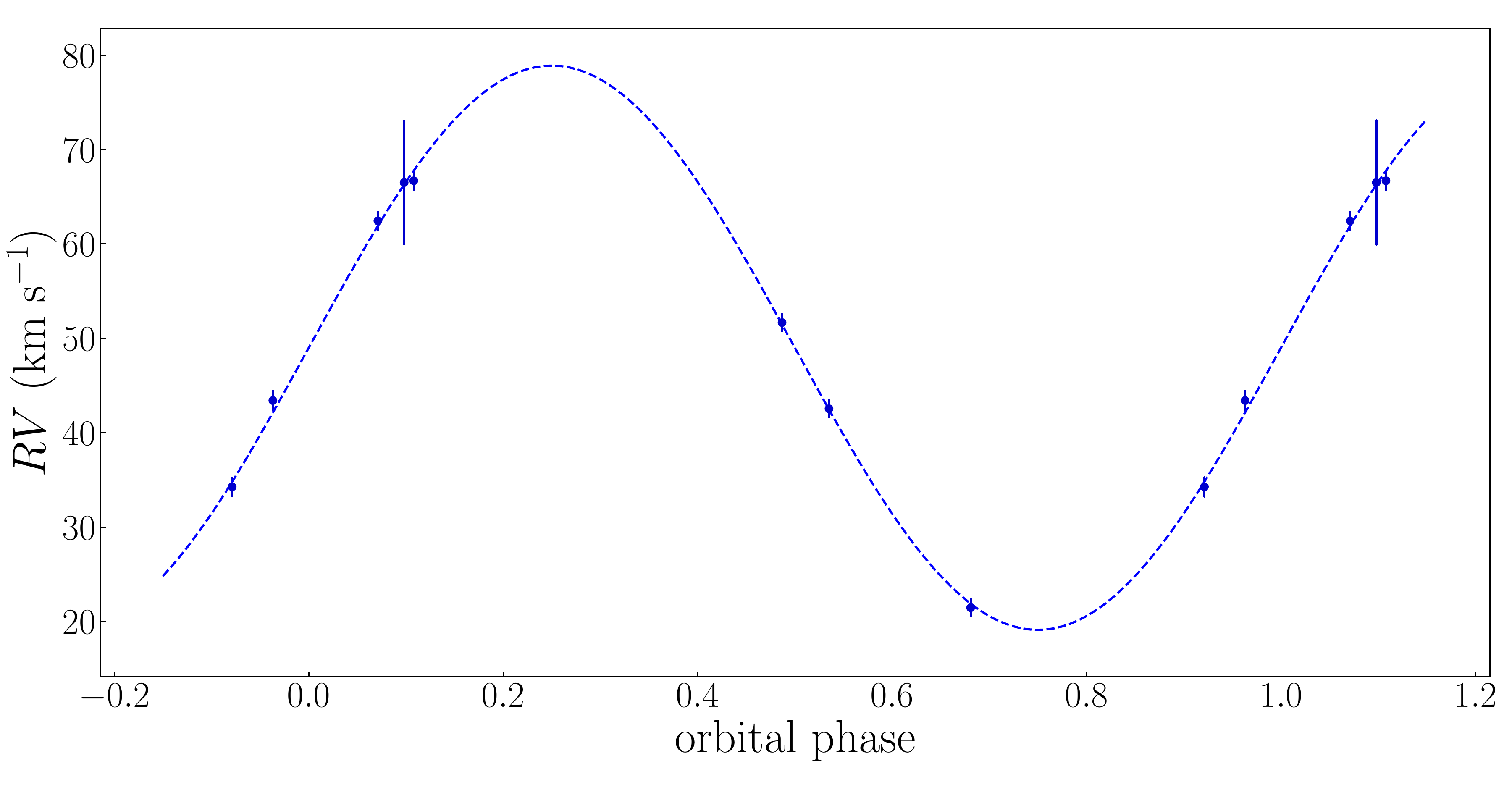}
    \caption{HD~94024}
    \end{subfigure}
    \begin{subfigure}{0.33\linewidth}
    \includegraphics[width = \textwidth]{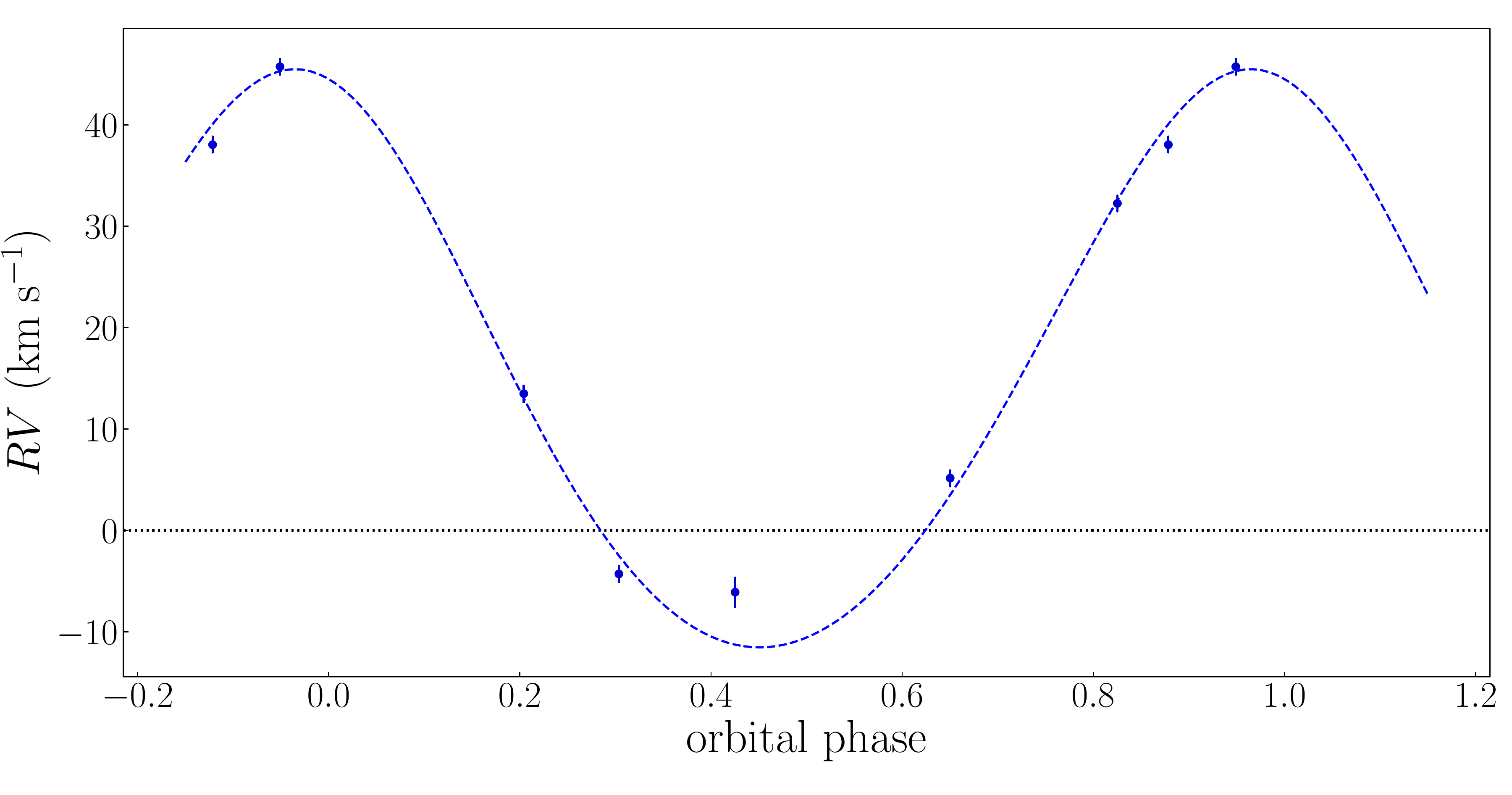}
    \caption{HD~105627}
    \end{subfigure}
    \begin{subfigure}{0.33\linewidth}
    \includegraphics[width = \textwidth]{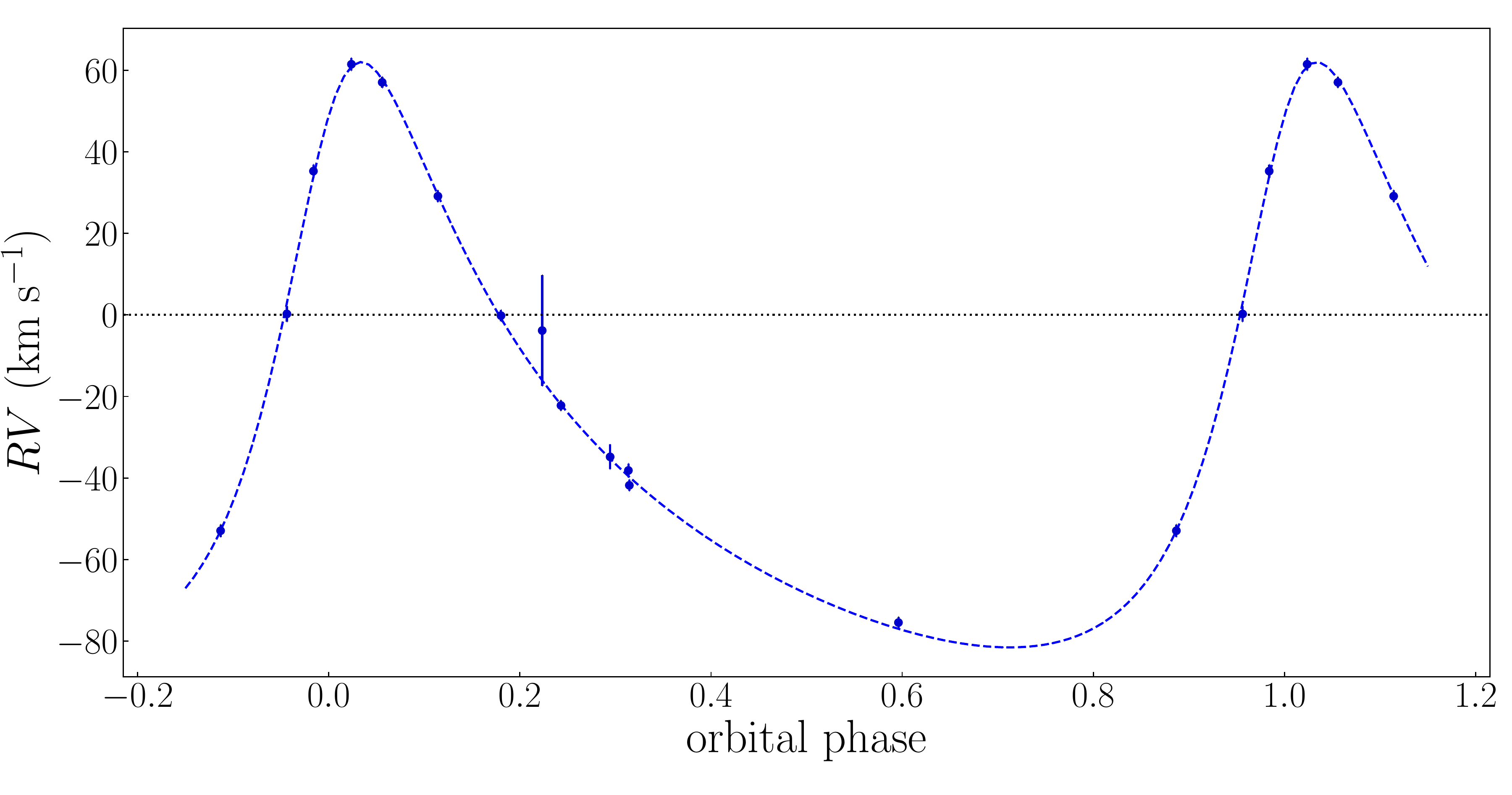}
    \caption{HD~130298}
    \end{subfigure}
    \begin{subfigure}{0.33\linewidth}
    \includegraphics[width = \textwidth]{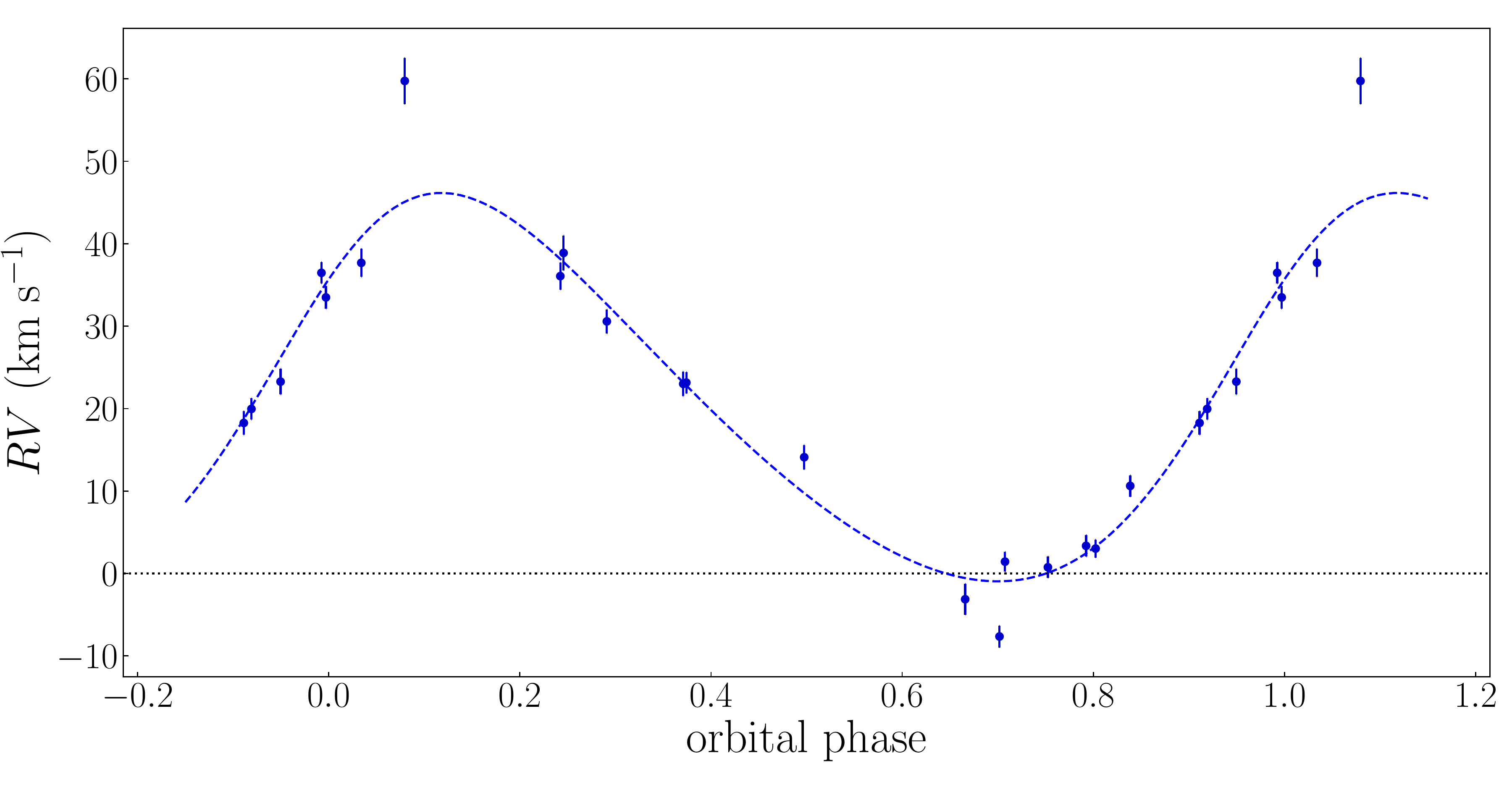}  \caption{HD~165174}
    \end{subfigure}
    \begin{subfigure}{0.33\linewidth}
    \includegraphics[width = \textwidth]{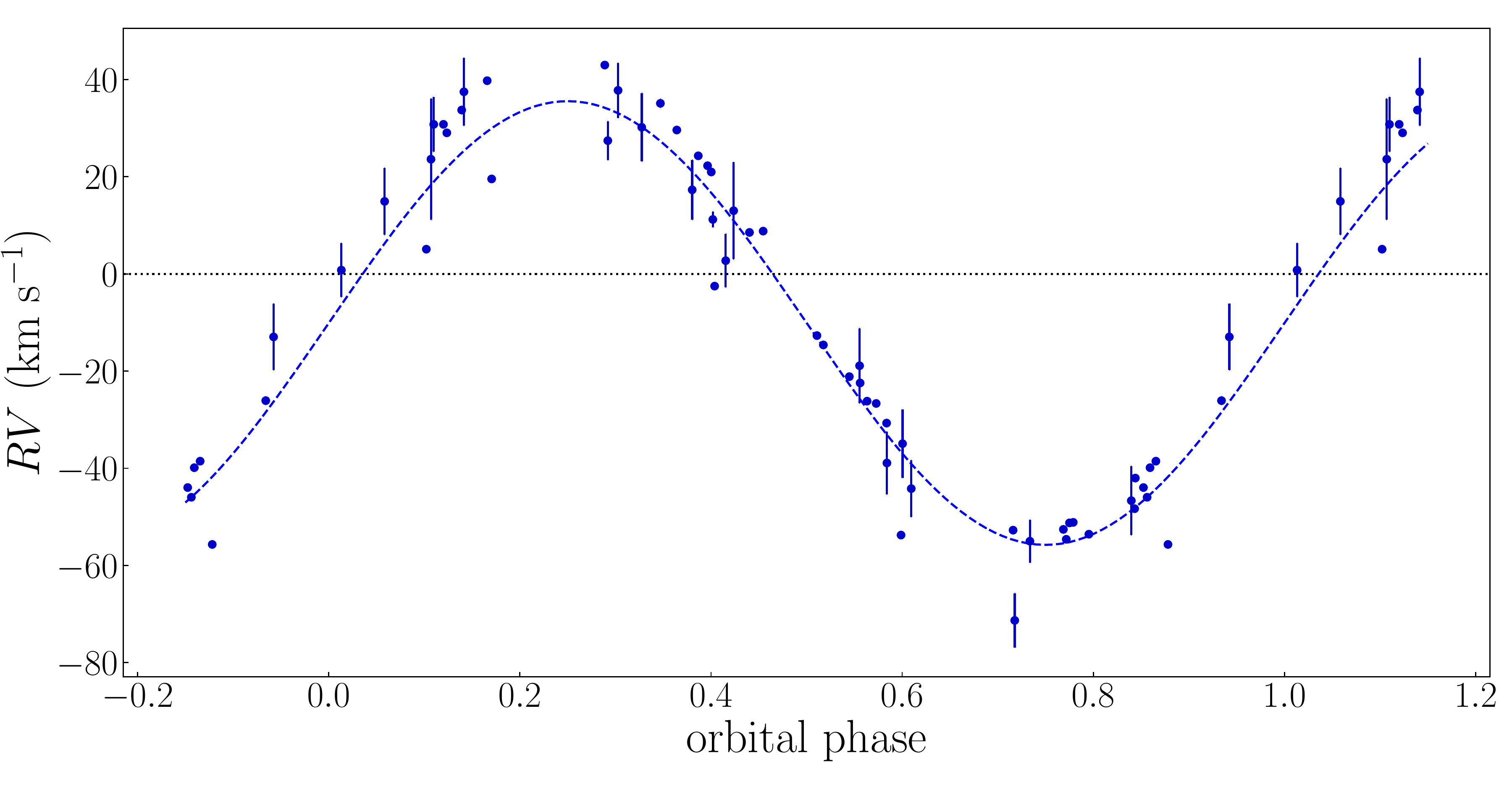}
    \caption{HD~229234}
    \end{subfigure}
    \begin{subfigure}{0.33\linewidth}
    \includegraphics[width = \textwidth]{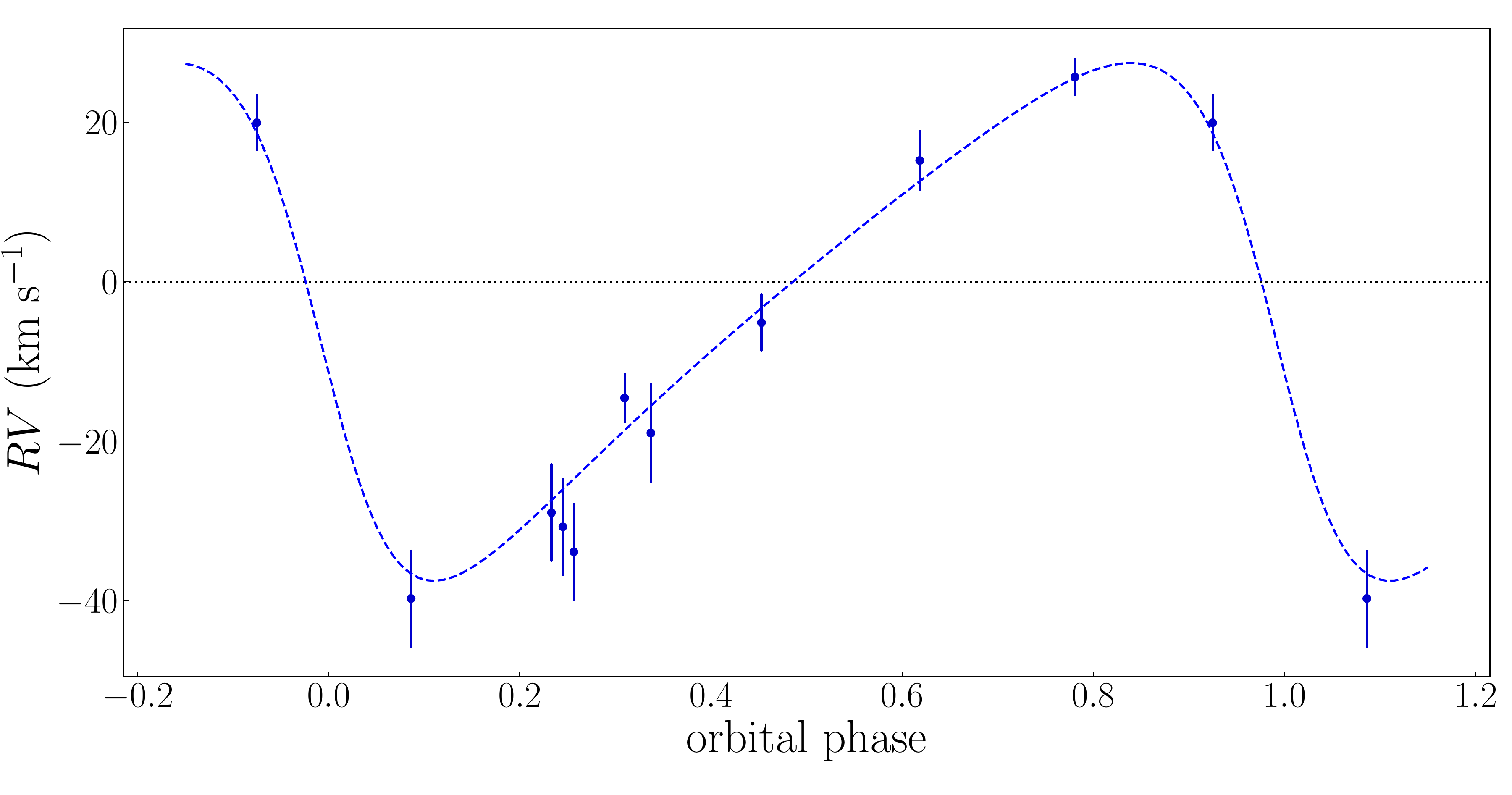}
    \caption{HD~308813}
    \end{subfigure}
    \hspace{-0.1in}
    \begin{subfigure}{0.33\linewidth}
    \includegraphics[width = \textwidth]{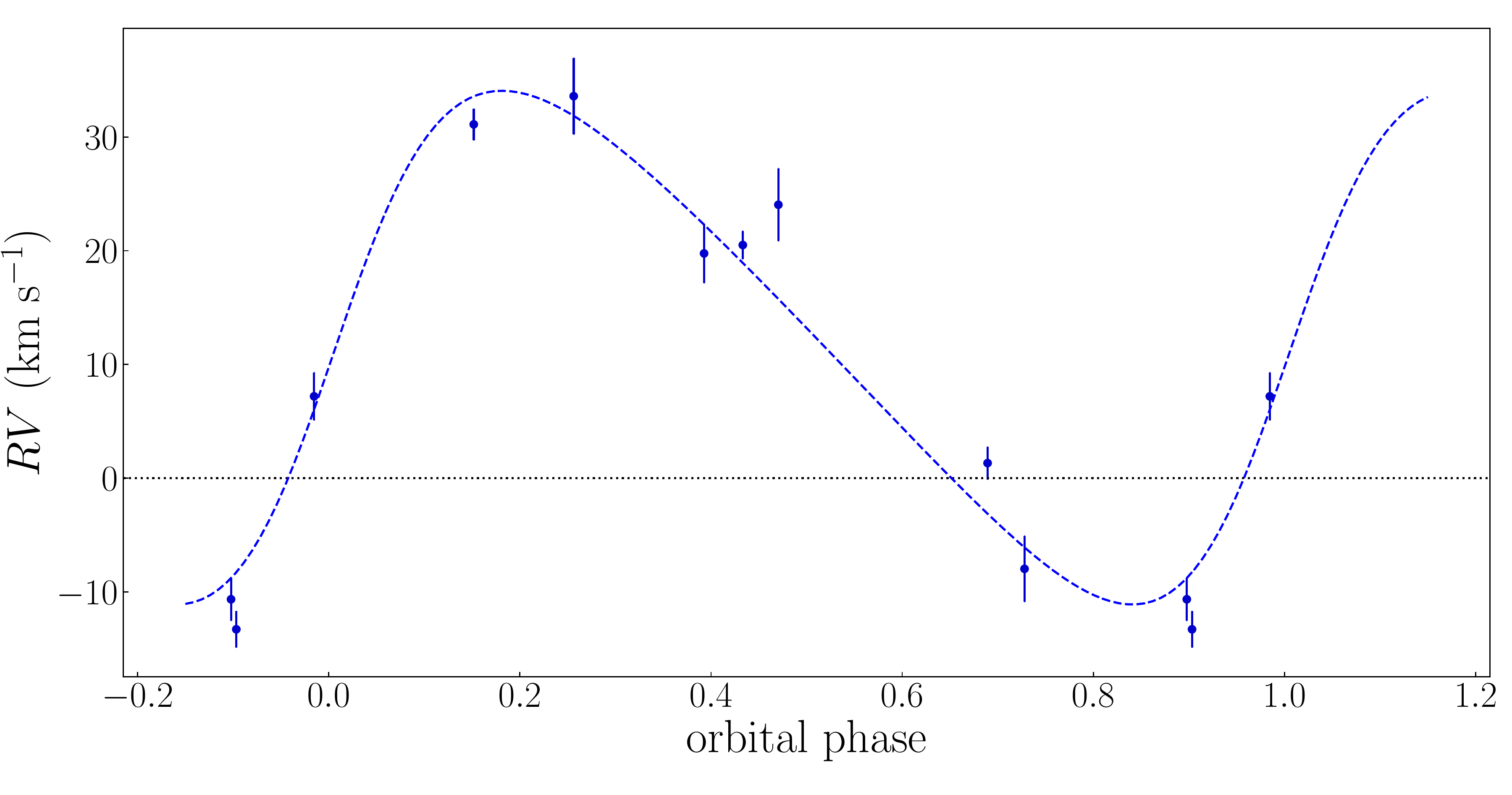}
    \caption{LS~5039}
    \end{subfigure}
    \hspace{-0.1in}
    \caption{Orbital solutions of all the SB1s in our sample, for which the spectral disentangling did not provide us with any spectral signatures for the secondary companions.}
    \label{fig:SB1RVcurves}
\end{figure*}

\begin{figure*}
    \centering
    \begin{subfigure}{0.33\linewidth}
    \includegraphics[width = \textwidth]{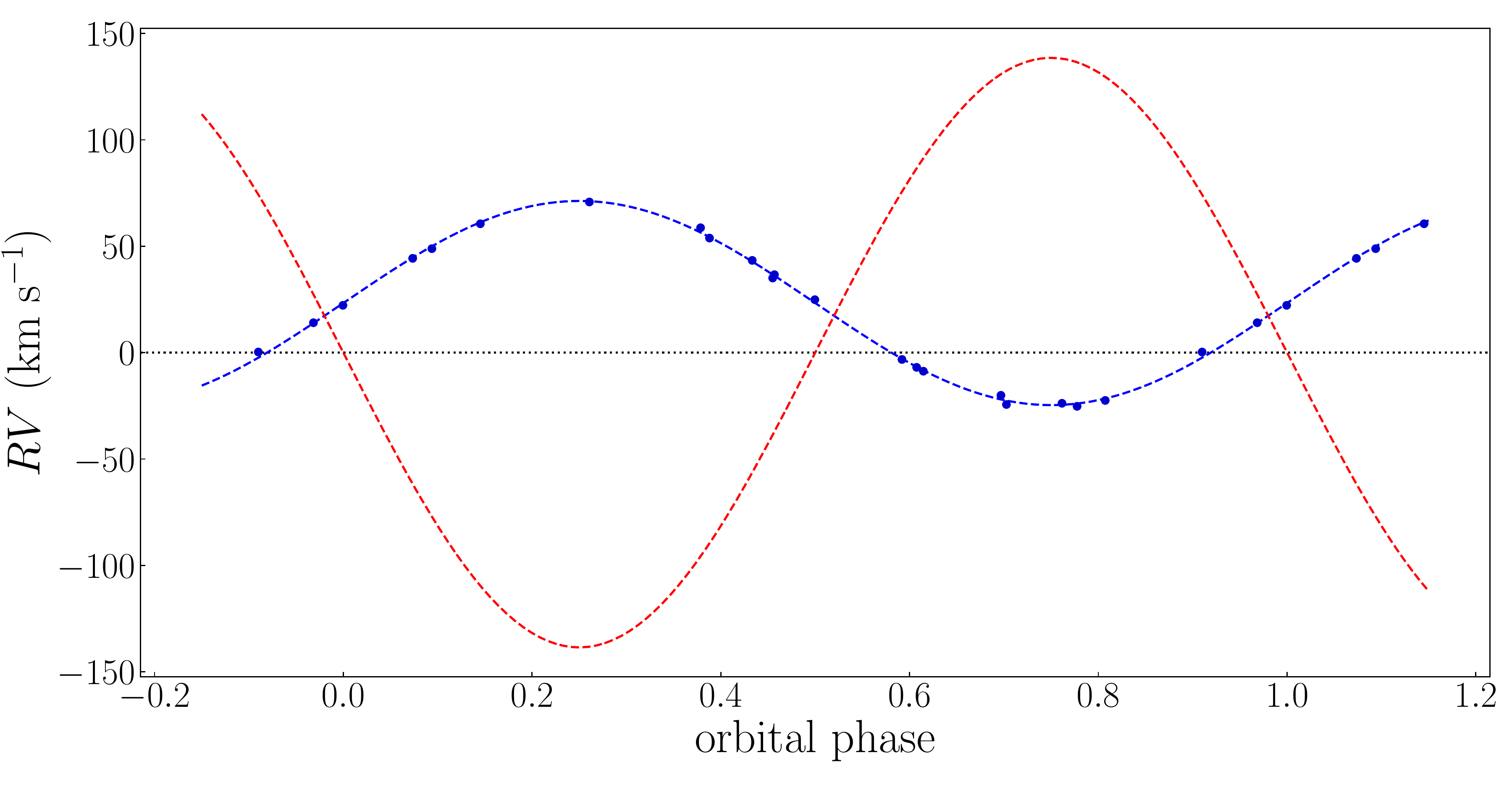}
    \caption{HD~29763}
    \end{subfigure}
    \begin{subfigure}{0.33\linewidth}
    \includegraphics[width = \textwidth]{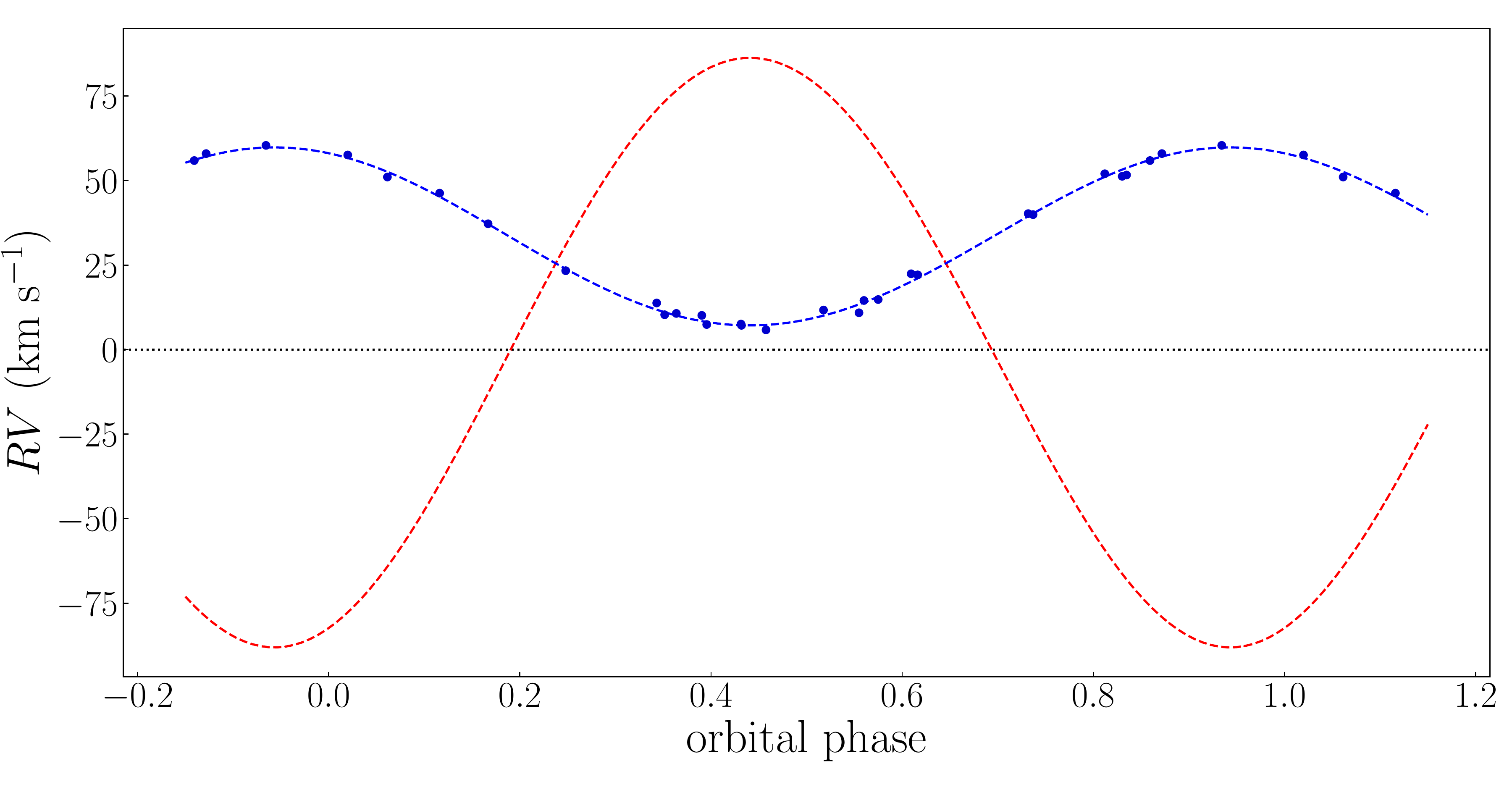}
    \caption{HD~30836}
    \end{subfigure}
    \begin{subfigure}{0.33\linewidth}
    \includegraphics[width = \textwidth]{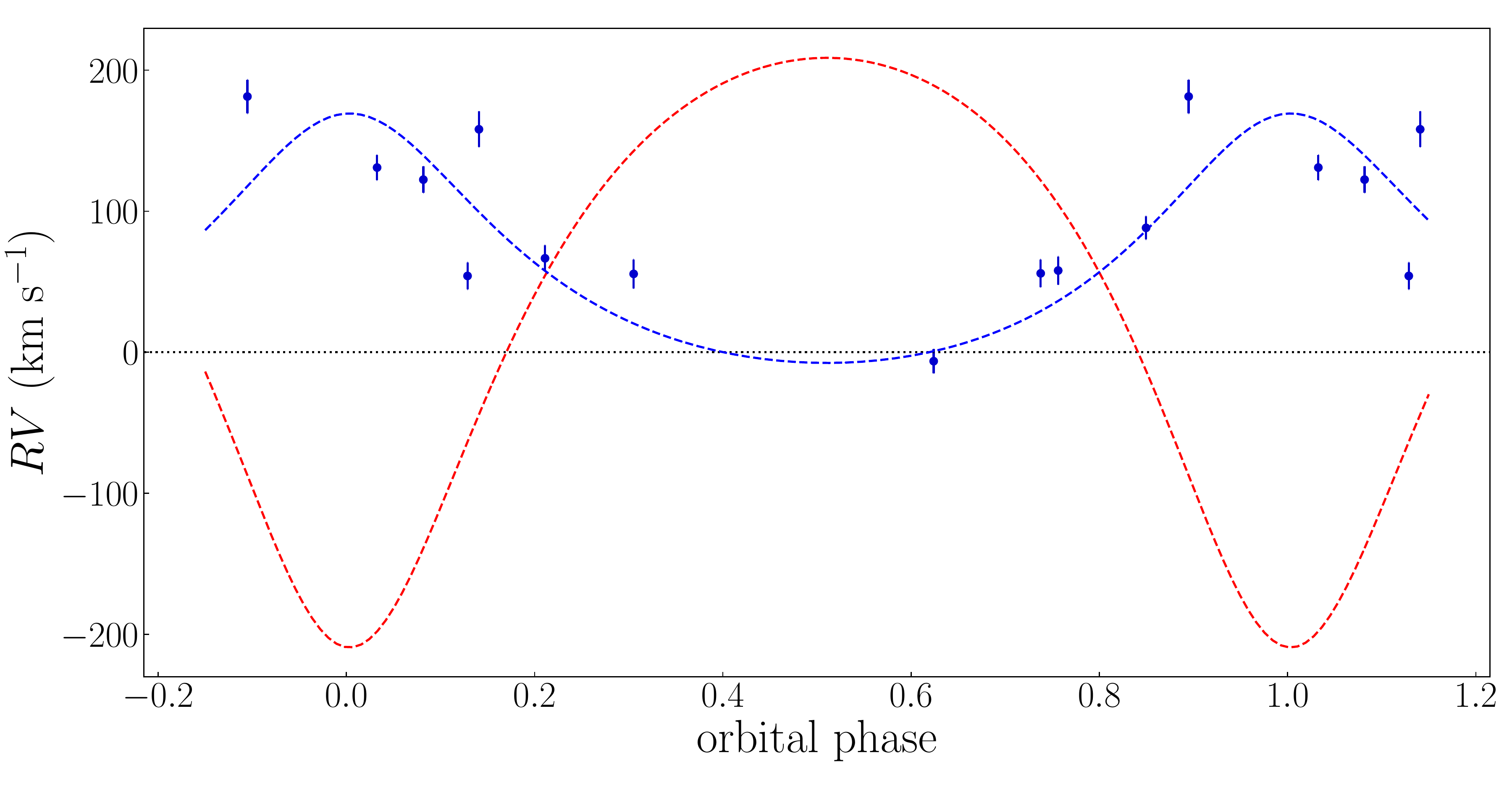}
    \caption{HD~52533}
    \end{subfigure}
    \begin{subfigure}{0.33\linewidth}
    \includegraphics[width = \textwidth]{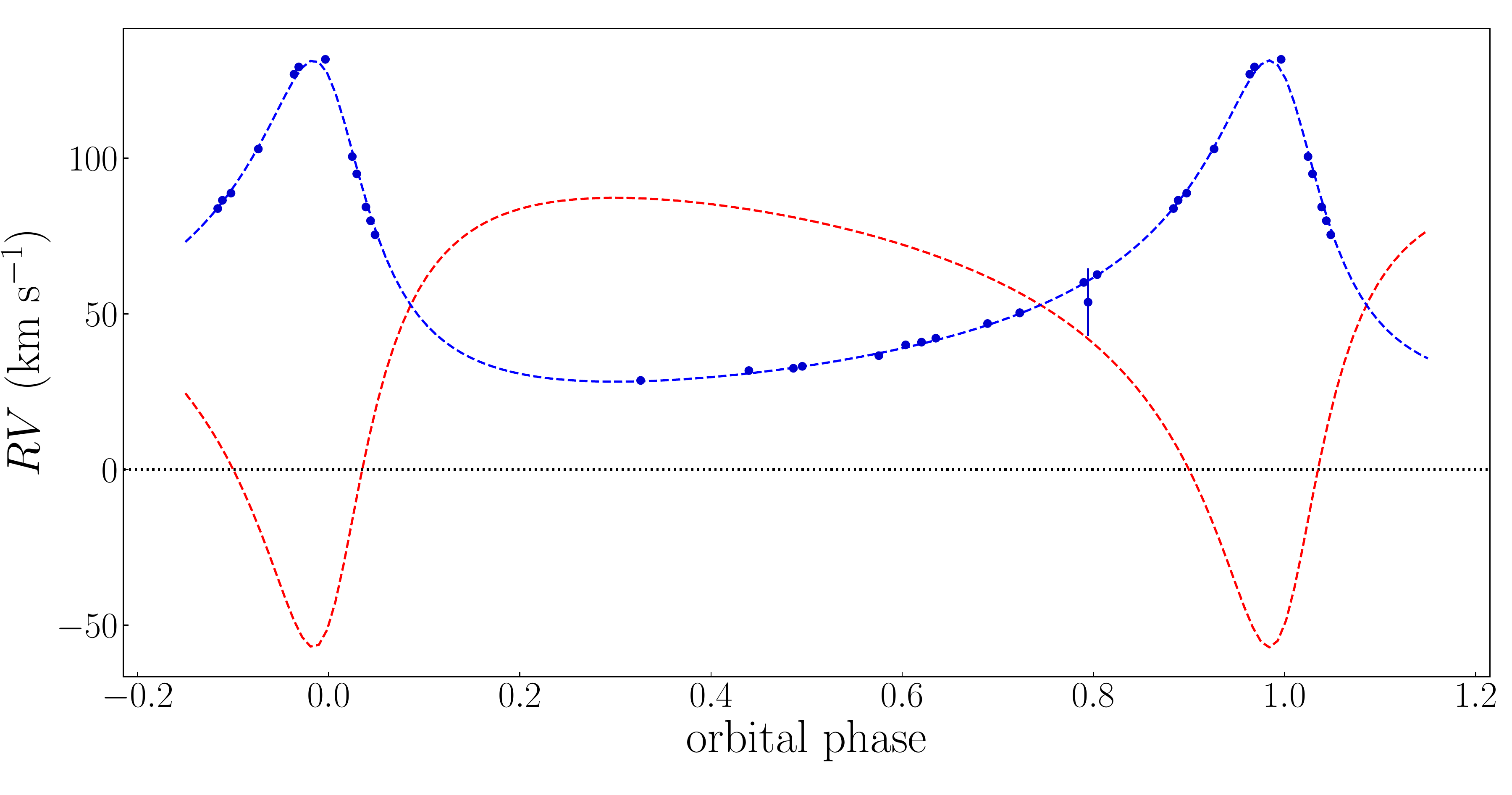}
    \caption{HD~57236}
    \end{subfigure}
    \begin{subfigure}{0.33\linewidth}
    \includegraphics[width = \textwidth]{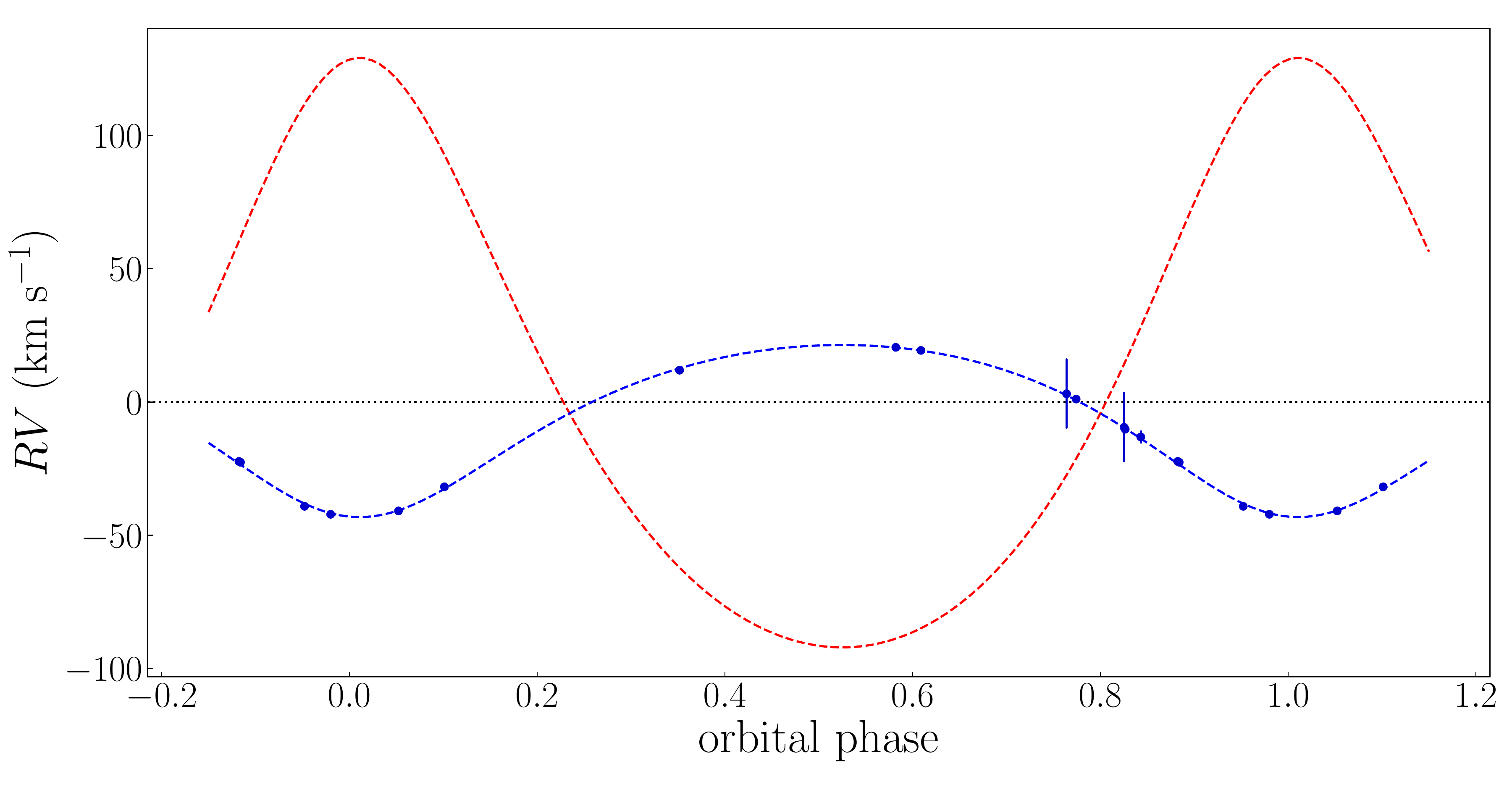}
    \caption{HD~91824}
    \end{subfigure}
    \begin{subfigure}{0.33\linewidth}
    \includegraphics[width = \textwidth]{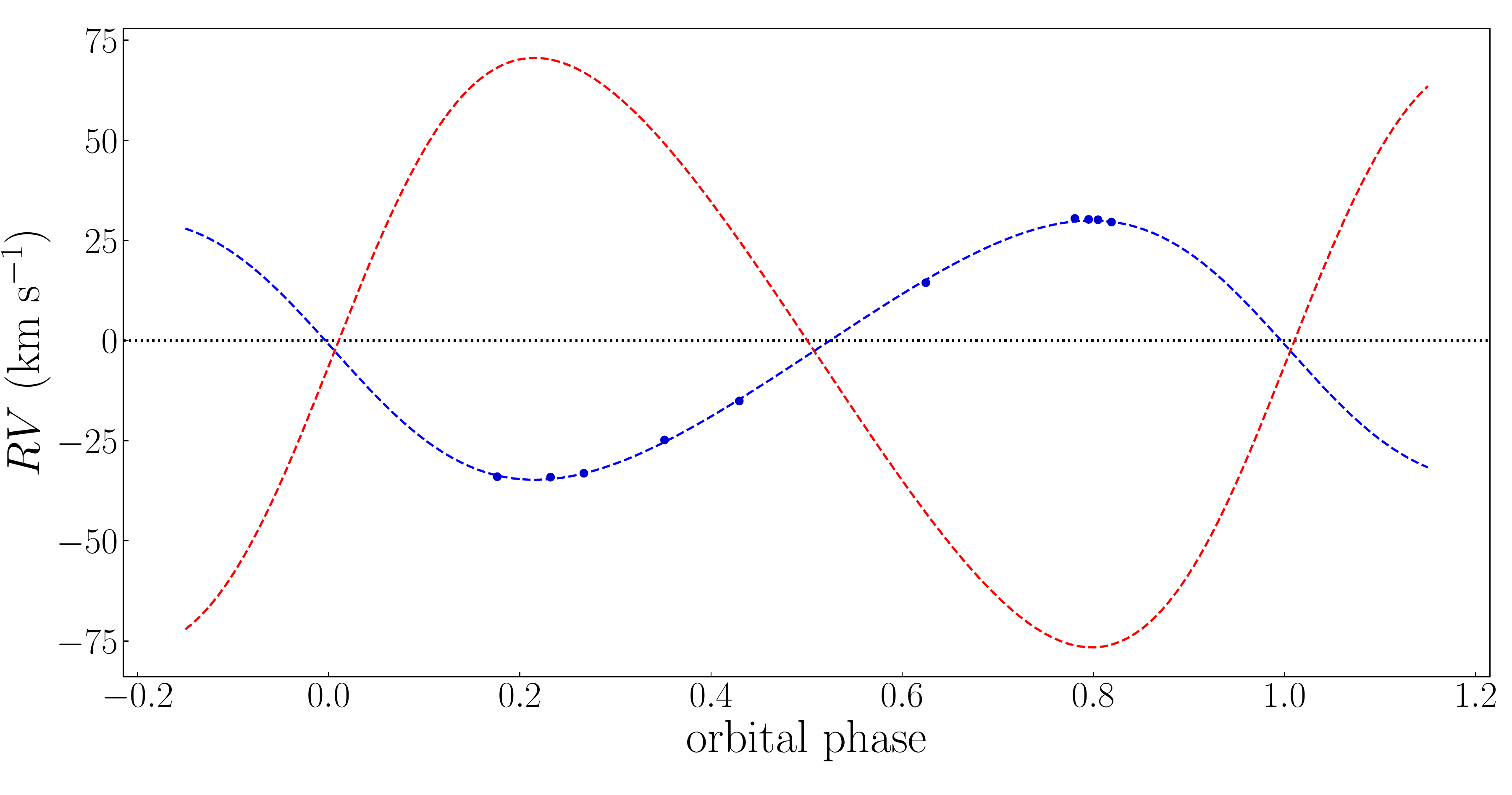}
    \caption{HD~93028}
    \end{subfigure}
    \begin{subfigure}{0.33\linewidth}
    \includegraphics[width = \textwidth]{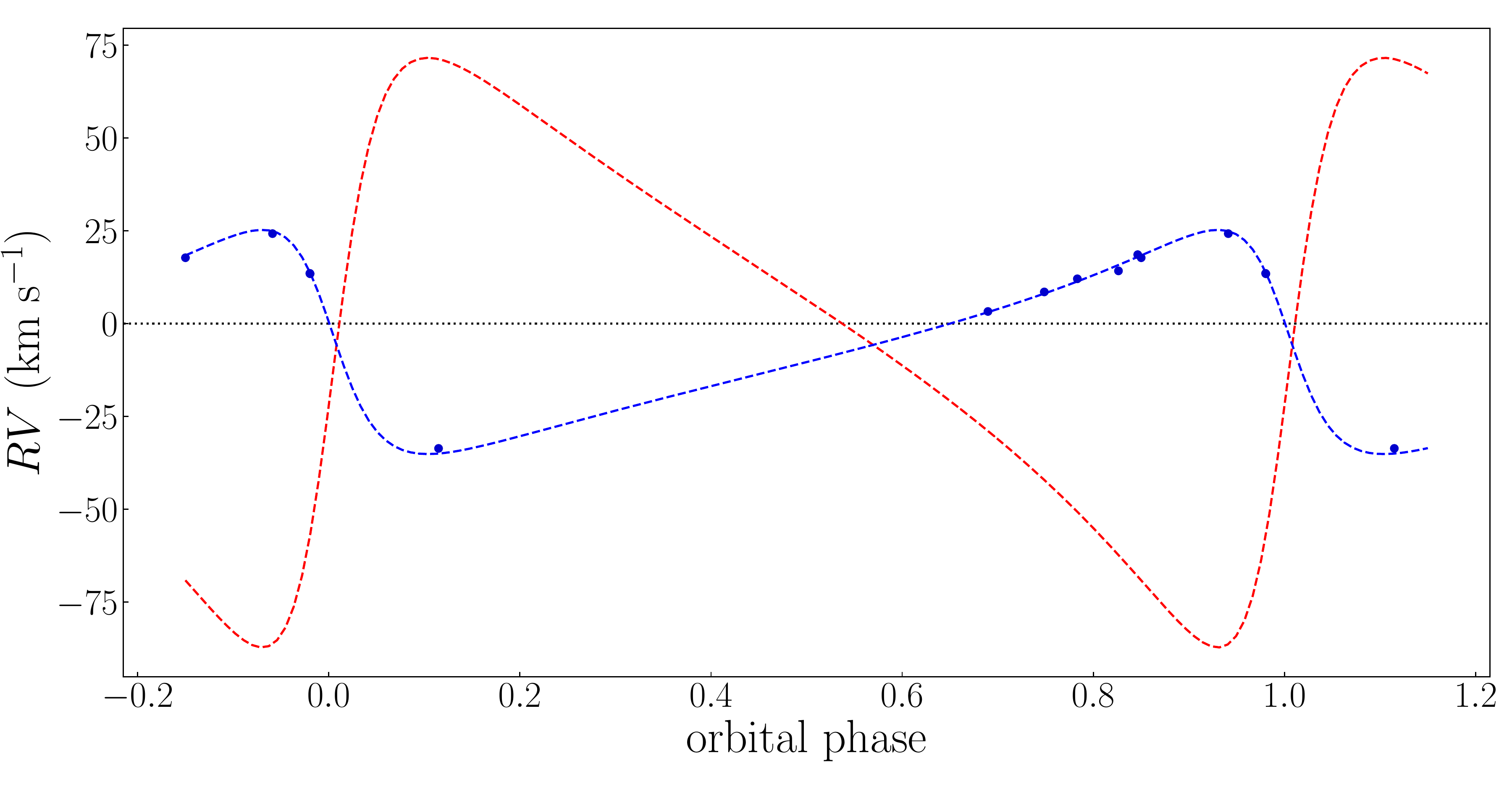}
    \caption{HD~152405}
    \end{subfigure}
    \begin{subfigure}{0.33\linewidth}
    \includegraphics[width = \textwidth]{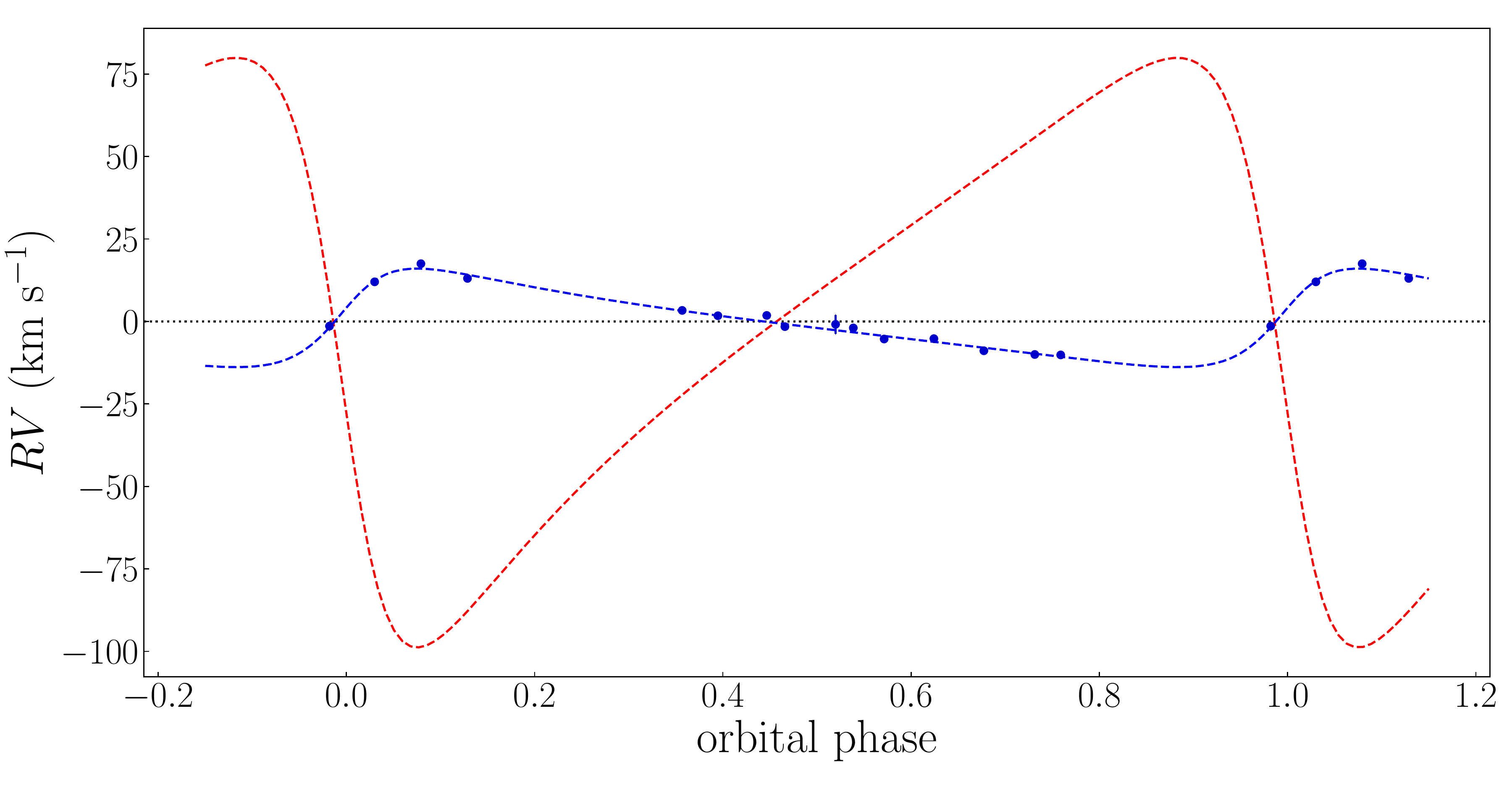}        \caption{HD~152723}
    \end{subfigure}
    \begin{subfigure}{0.33\linewidth}
    \includegraphics[width = \textwidth]{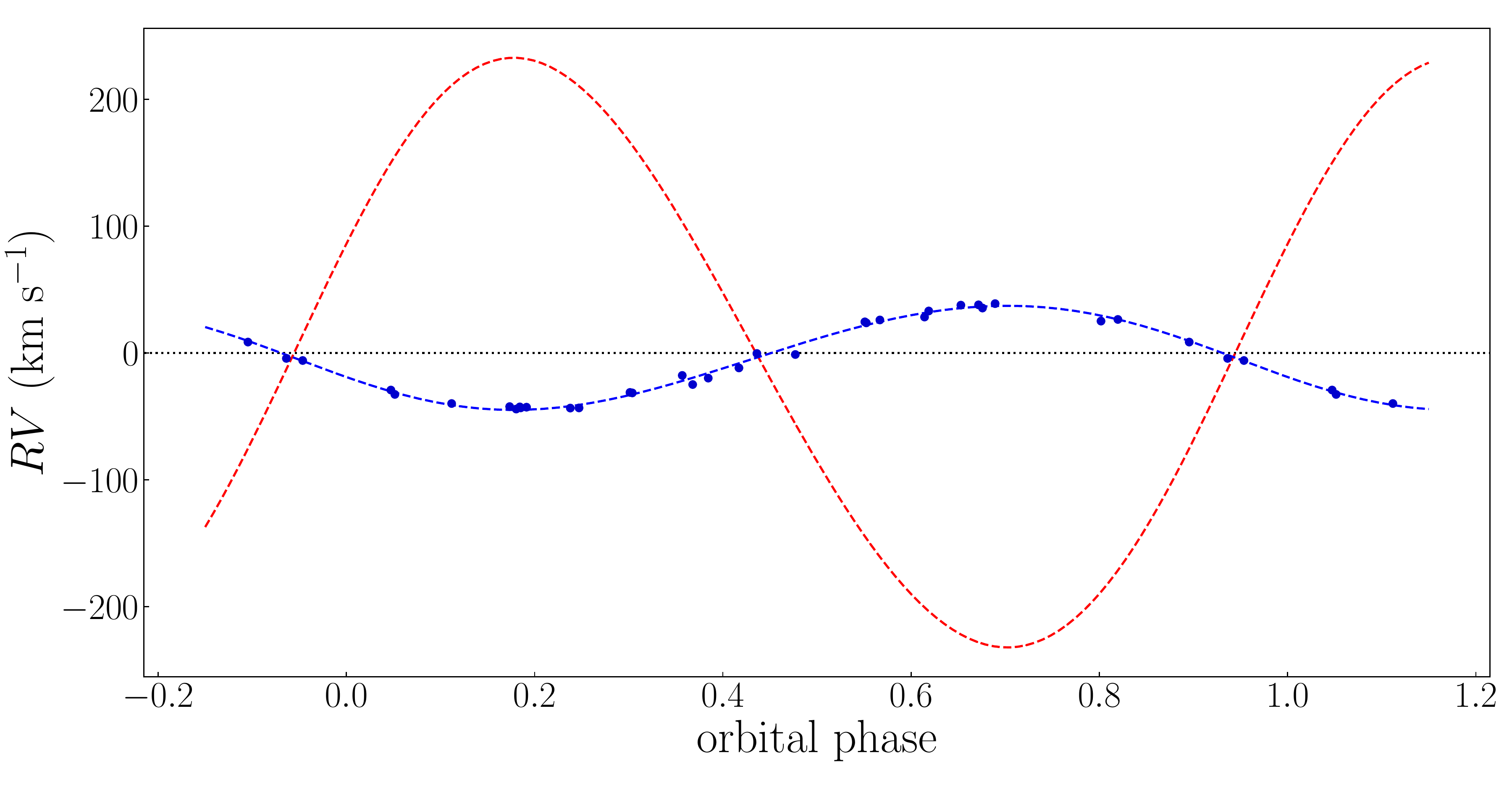}
    \caption{HD~163892}
    \end{subfigure}
    \begin{subfigure}{0.33\linewidth}
    \includegraphics[width = \textwidth]{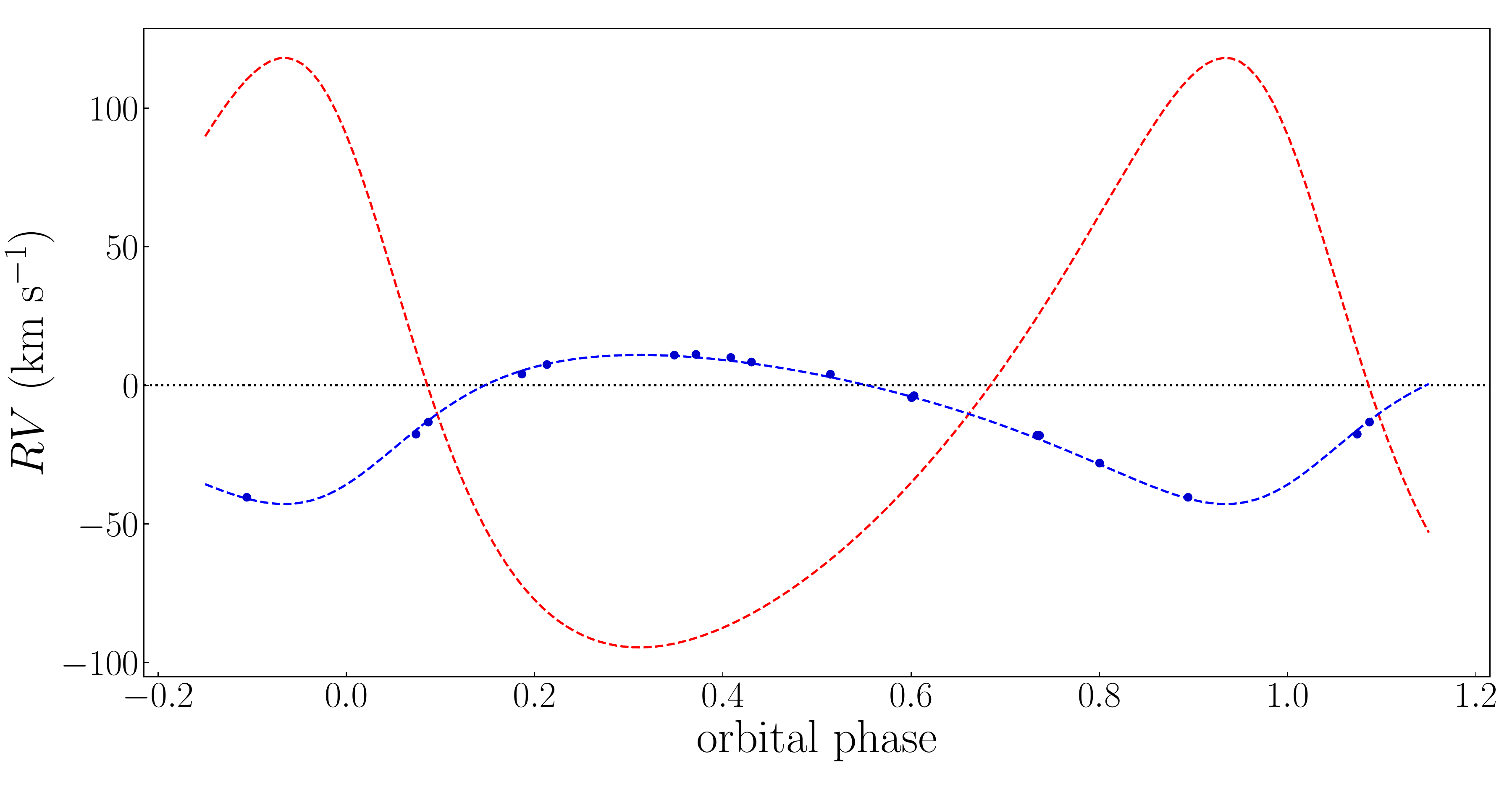}
    \caption{HD~164438}
    \end{subfigure}
    \begin{subfigure}{0.33\linewidth}
    \includegraphics[width = \textwidth]{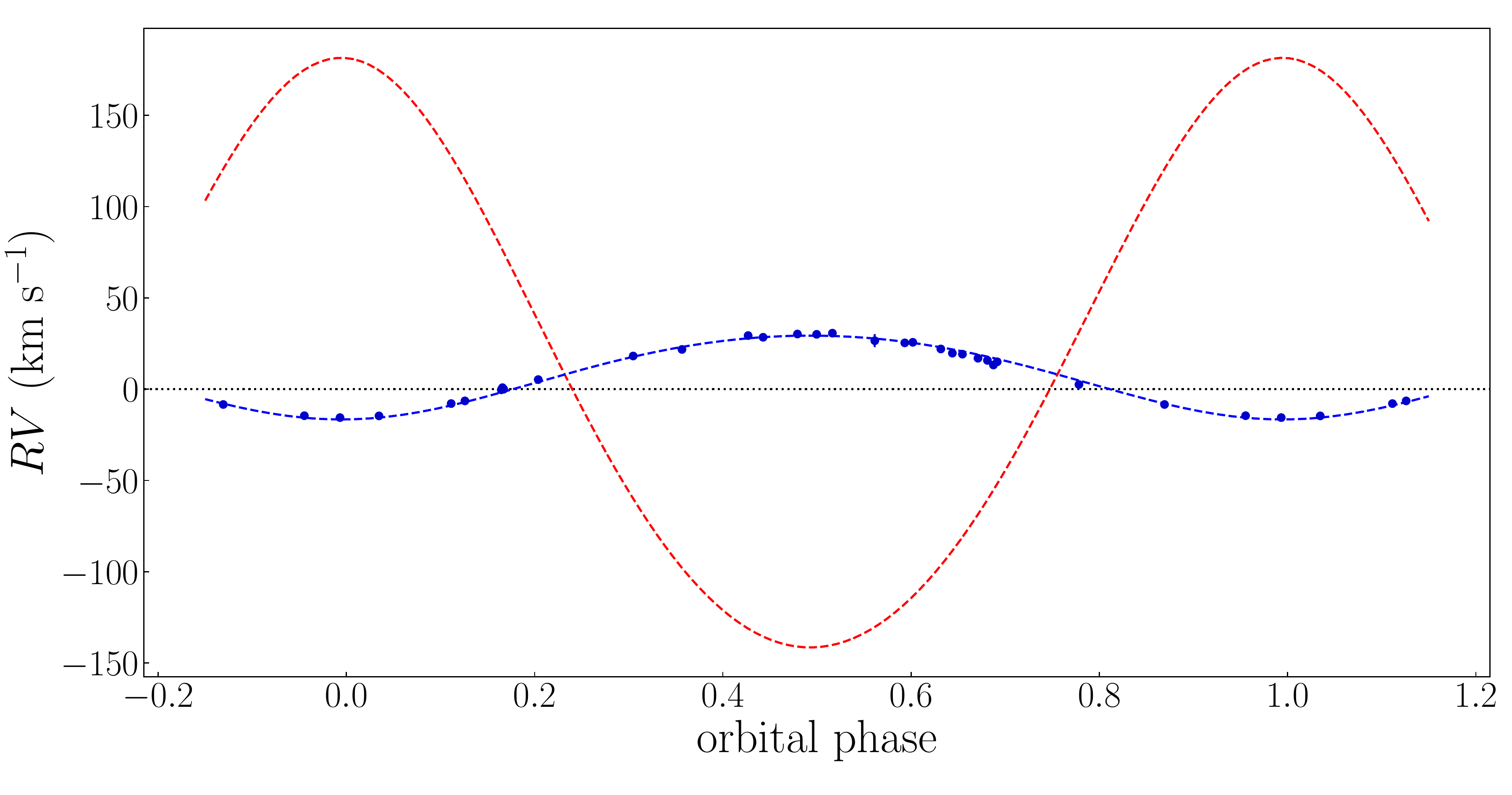}
    \caption{HD~164536}
    \end{subfigure}
    \begin{subfigure}{0.33\linewidth}
    \includegraphics[width = \textwidth]{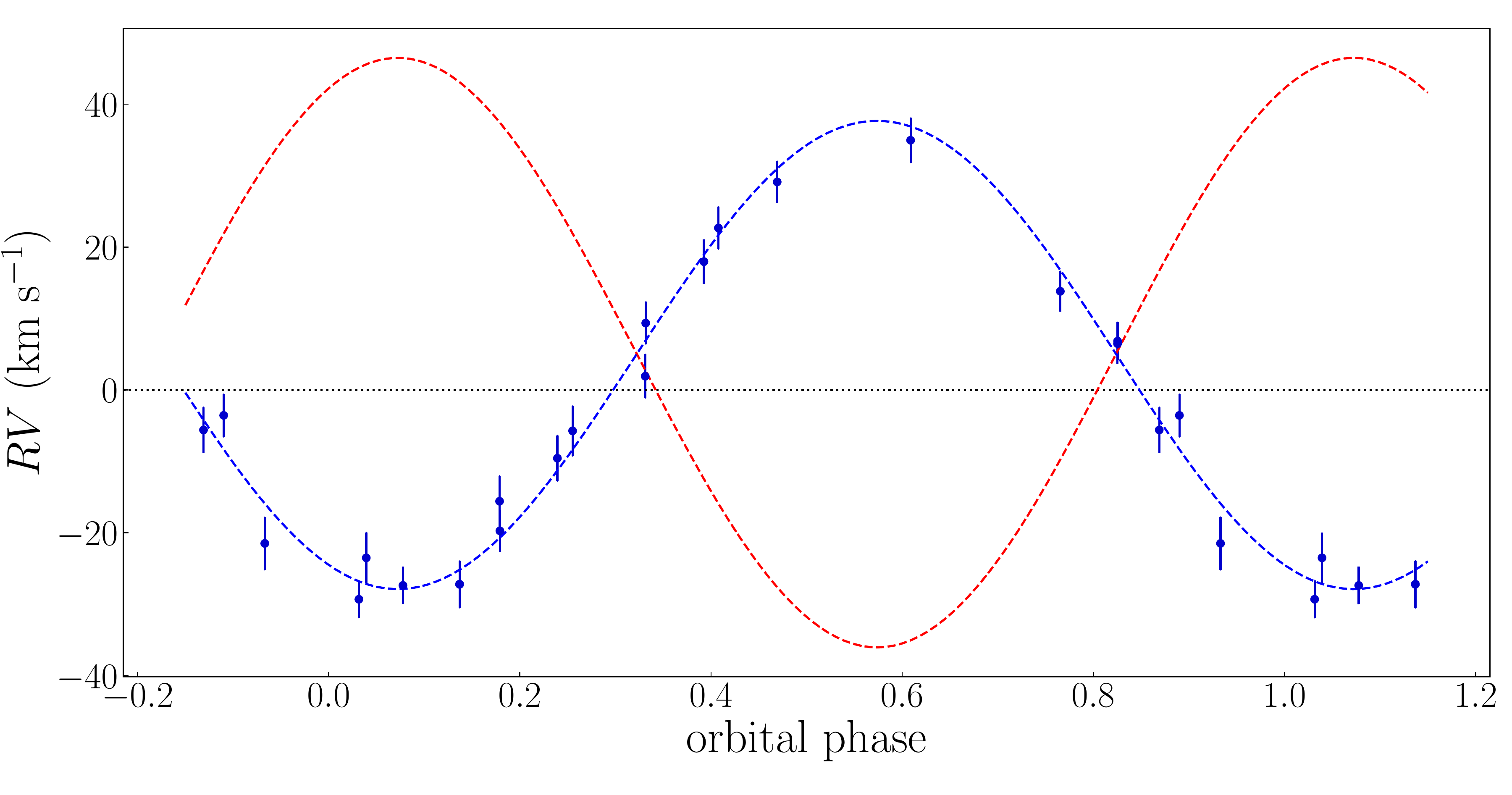}
    \caption{HD~167263}
    \end{subfigure}
    \begin{subfigure}{0.33\linewidth}
    \includegraphics[width = \textwidth]{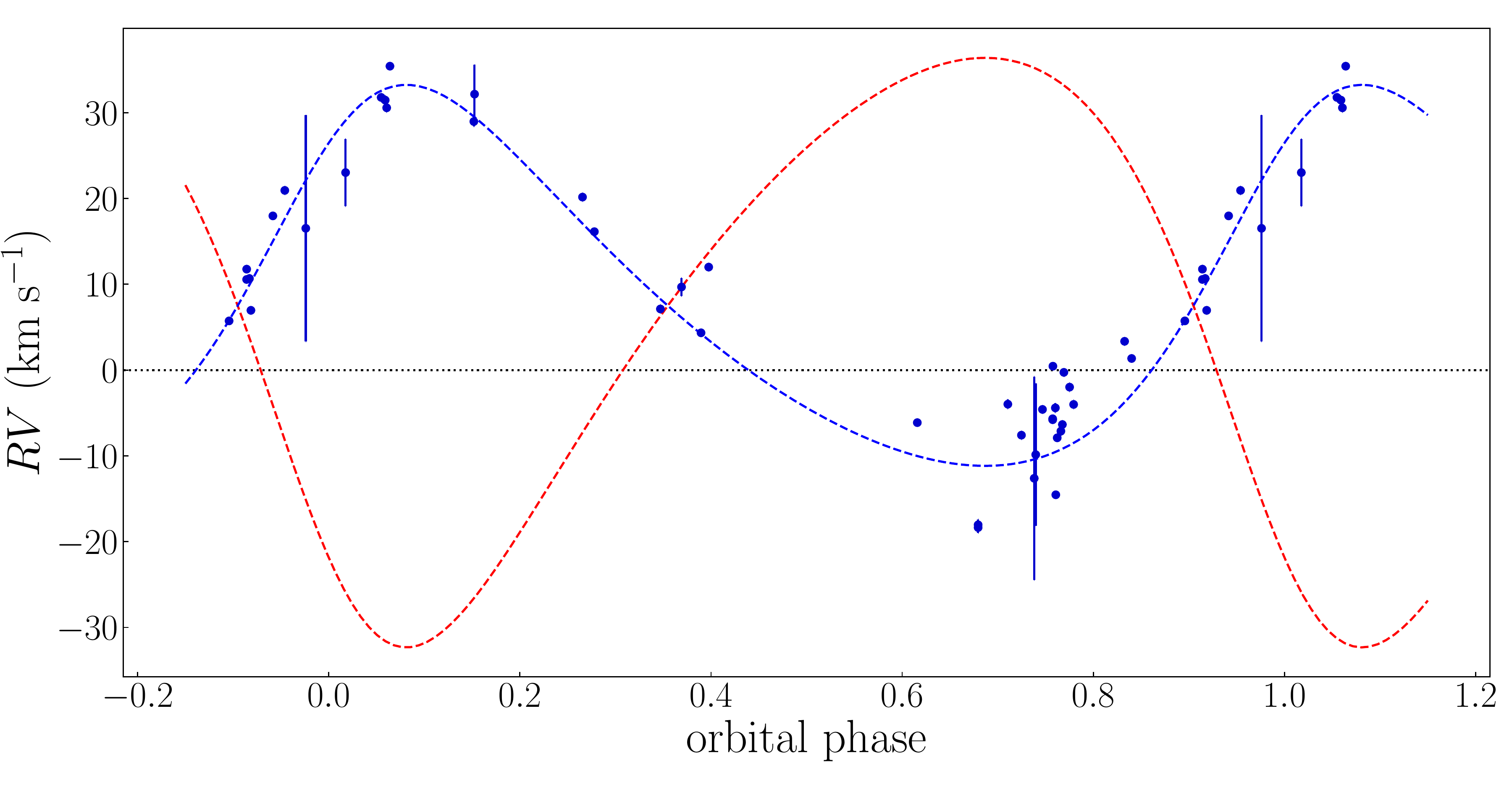}  \caption{HD~167264}
    \end{subfigure}
    \begin{subfigure}{0.33\linewidth}
    \includegraphics[width = \textwidth]{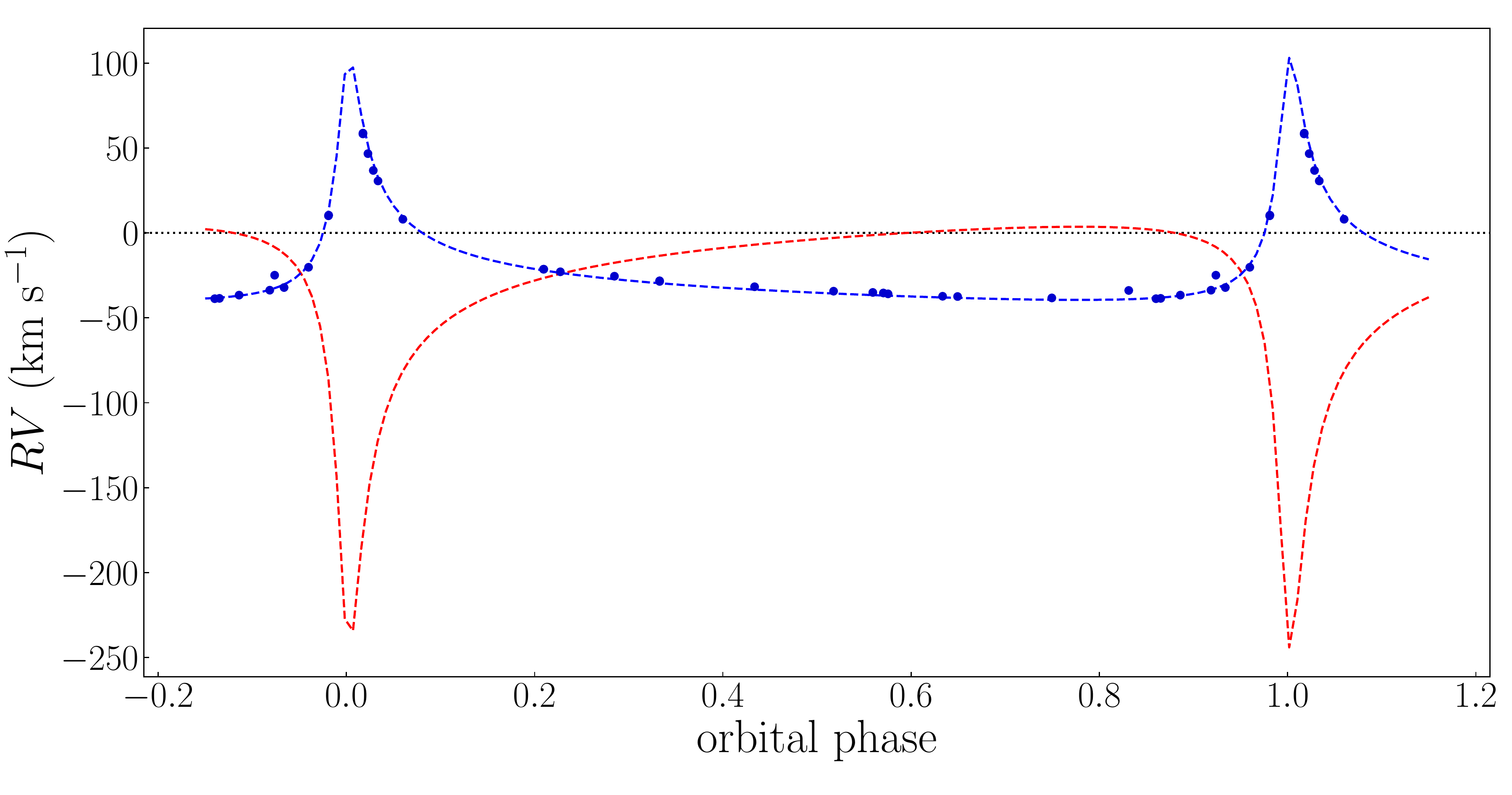}
    \caption{HD~192001}
    \end{subfigure}
    \begin{subfigure}{0.33\linewidth}
    \includegraphics[width = \textwidth]{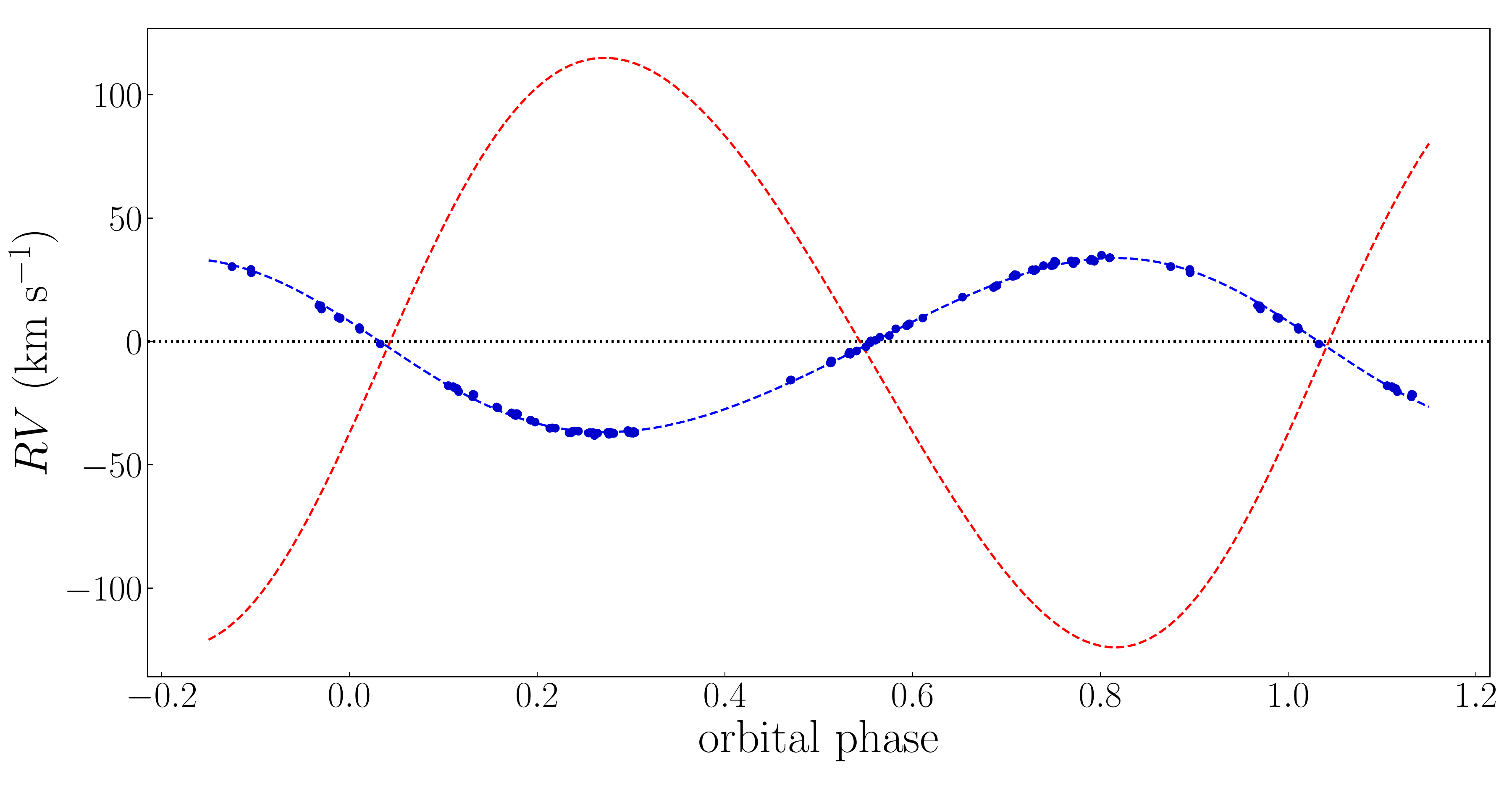}
    \caption{HD~199579}
    \end{subfigure}
    \hspace{-0.1in}
     \begin{flushleft}    
    \begin{subfigure}{0.33\linewidth}
    \includegraphics[width = \textwidth]{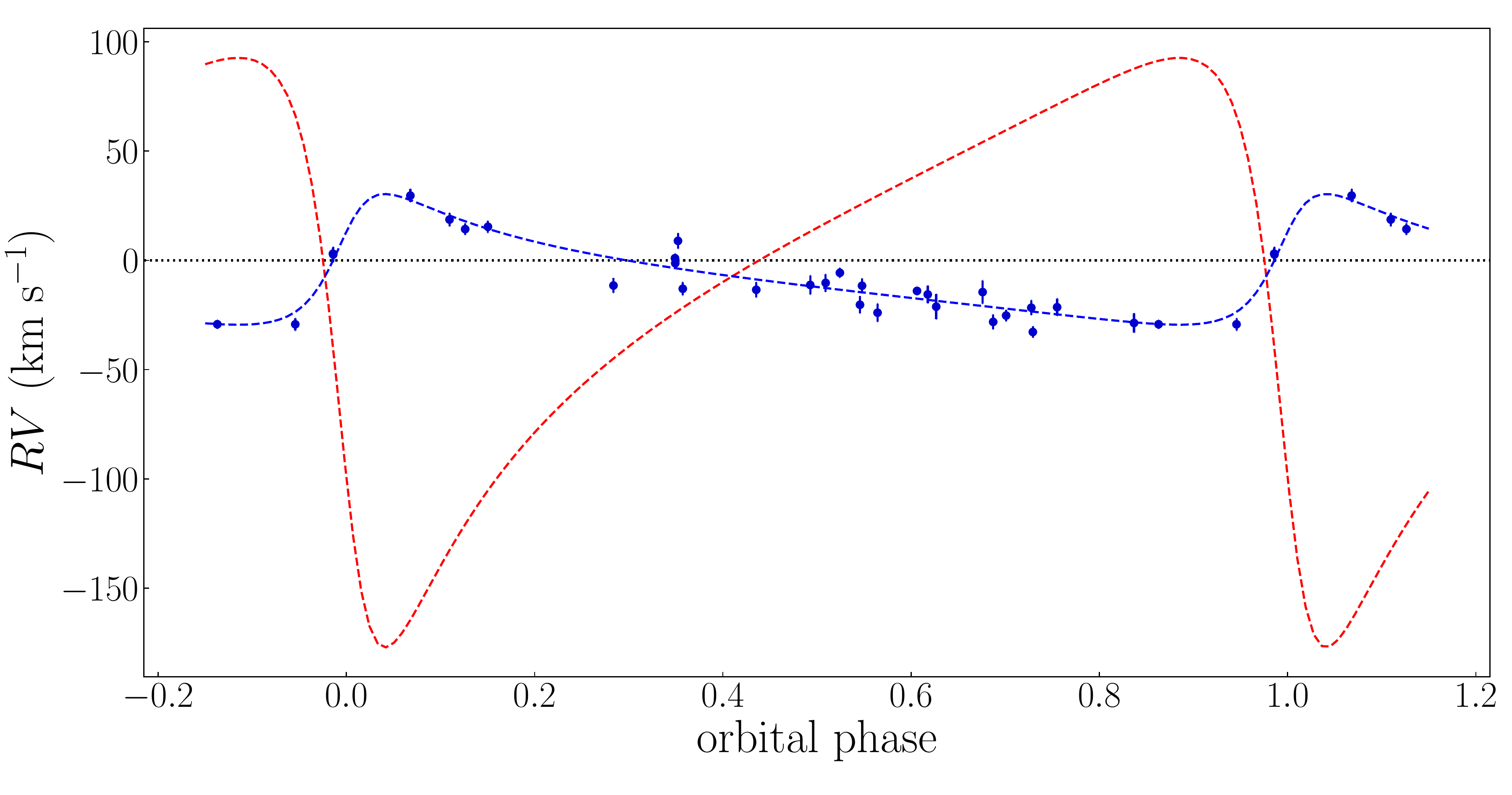}
    \caption{Schulte~11}
    \end{subfigure}
    \begin{subfigure}{0.33\linewidth}
    \includegraphics[width = \textwidth]{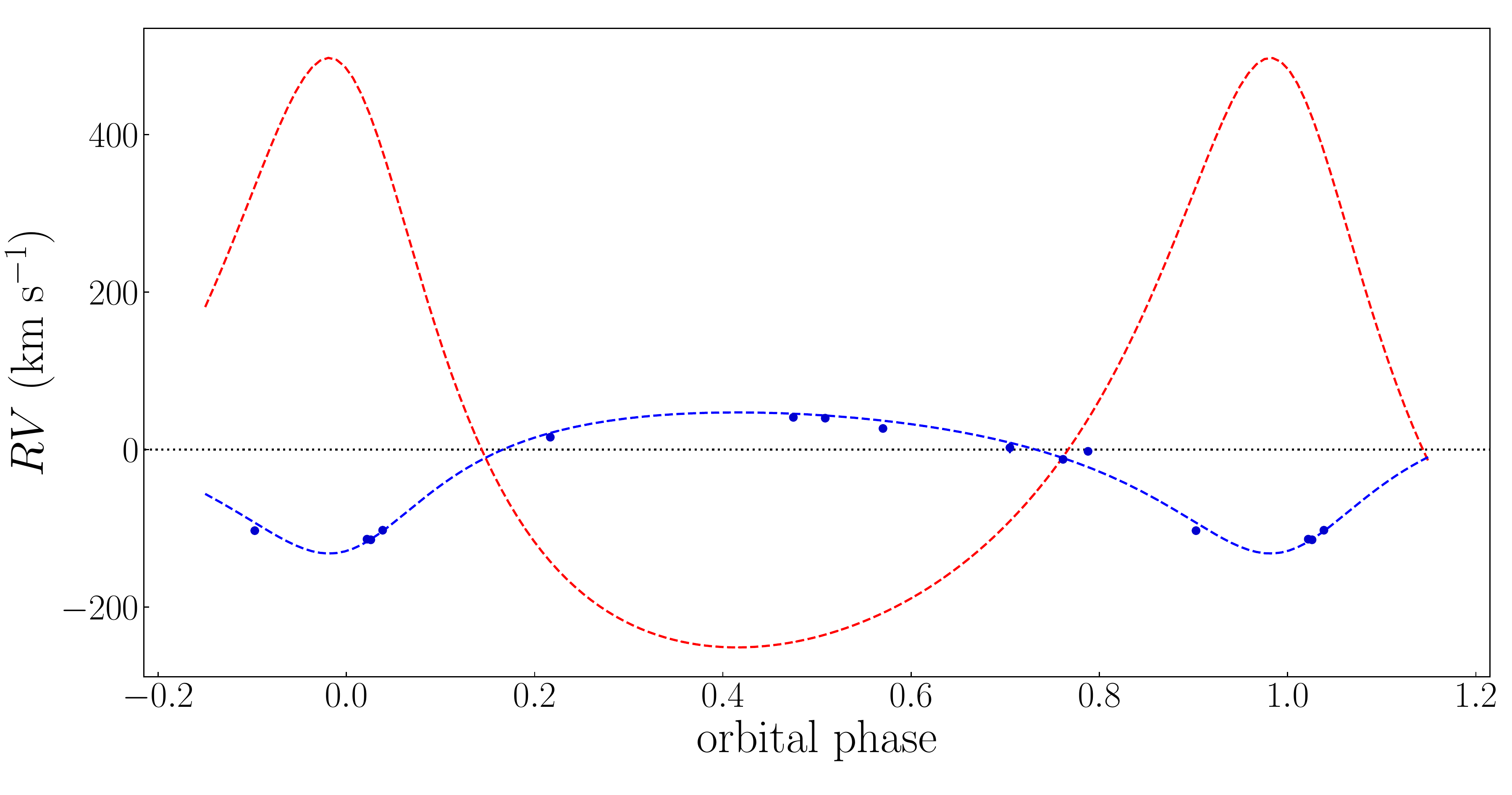}
    \caption{V747~Cep}
    \end{subfigure}
    \hspace{-0.1in}
     \end{flushleft}
    \caption{Orbital solutions of all the newly classified SB2s in our sample. The dashed red line represents the RV curve of the secondary star. We stress that the RVs of the secondaries are not fitted individually from the spectra, but are forced to vary in anti-phase with the primary and with the semi-amplitude derived through spectral disentangling (Sect.~\ref{subsec:disentangling}).}
    \label{fig:SB2RVcurves}
\end{figure*}

As we deal with objects where only one star is visible in the spectra, we measured the RVs of the visible stars by performing a 1D cross-correlation technique \citep{zucker03} to different wavelength domains. This technique, described by \cite{shenar17}, uses a master-spectrum built from the observations themselves. This method adopts, as a reference frame, the RVs computed at the first epoch, so that all the RVs are shifted accordingly. The absolute RVs were then obtained by cross-correlating the high S/N template with a suitable atmosphere model (Sect.~\ref{subsec:atmosphere}). The RVs for all stars at all epochs are given electronically at the CDS (Centre de Donn{\'e}es astronomiques de Strasbourg). 

\begin{sidewaystable*}
\caption{Orbital parameters of the SB1 systems. The errors correspond to $1\sigma$. }\label{table:SB1orbitalparameters}
\centering
\begin{tabular}{lrrrrrrrr}
\hline \hline
 Star & $P_{\mathrm{orb}}$ & $e$ & $\omega$ & K & $T_0$ & $\gamma$  & $a\,\sin\,i$ & $f_{\mathrm{mass}}$  \\
  & [d] &  & [$^{\circ}$] & [\kms] & $- 2\,450\,000$ & [\kms] & [$\Rsun$] & [$\Msun$]  \\
  \hline
 Cyg\,X-1 & $5.599711 \pm 0.000015$ & $0.023 \pm 0.003$ & $331.12 \pm 6.36$ & $74.25 \pm 0.20$ & $5847.3261 \pm 0.1008$ & $2.90 \pm 0.25$  & $8.22 \pm 0.02$ & $0.2374 \pm 0.0019$\\
 HD\,12323 & $1.925124 \pm 0.000011 $ & $ 0.000$ (fixed) & $ 270.00$ (fixed) & $ 30.07 \pm 1.43$ & $ 9099.7222 \pm 0.0099$ & $-40.06 \pm 0.70$  &$ 1.14 \pm 0.05$&$ 0.0054 \pm 0.0008$ \\
 HD\,14633 & $15.409090 \pm 0.000164$ & $0.698 \pm 0.007$ & $140.74 \pm 1.46$ & $19.11 \pm 0.27$ & $3718.7827 \pm 0.0329$ & $-40.74 \pm 0.56$ & $4.17 \pm 0.07$ & $0.0041 \pm 0.0002$ \\
 HD\,15137 & $55.336938 \pm 0.008895 $ & $ 0.663 \pm 0.038 $ & $ 156.32 \pm 4.66 $ & $15.66 \pm 1.47$ & $ 9016.2268 \pm 0.4854 $ & $-43.49 \pm 0.44$ &$ 12.81 \pm 1.32	$&$ 0.0092 \pm 0.0029$ \\
 HD\,37737 & $7.846907 \pm 0.000470 $ & $ 0.383 \pm 0.016 $ & $ 176.14 \pm 1.95 $ & $ 70.03 \pm 1.16$ & $ 8783.2099 \pm 0.0445 $ & $-9.51 \pm 0.54$ & $ 10.04 \pm 0.18	$&$ 0.2224	\pm 0.0127$ \\
 HD\,46573 & $10.6549205 \pm 0.0001422$ & $0.595 \pm 0.014$ & $280.64 \pm 2.27$ & $11.08 \pm  1.64$ & $8996.1876 \pm 0.0400$ & $51.54 \pm 0.10$  & $1.87 \pm 0.04$ & $0.0008 \pm 0.0001$ \\
 HD\,74194 & $9.544240 \pm 0.000160$ & $0.599 \pm 0.014$ & $262.31 \pm 3.41$ & $23.05 \pm 0.60$ & $4700.0447 \pm 0.0518$ & $14.27 \pm 0.75$ & $3.48 \pm 0.36$ & $0.0062 \pm 0.0019$ \\
 HD\,75211 & $20.447972 \pm 0.000365 $ & $ 0.340 \pm 0.009 $ & $ 256.90 \pm 3.40 $ & $ 20.97 \pm 0.28 $ & $ 8677.4749 \pm 0.1715 $ & $13.15 \pm 0.82$  &$ 7.97 \pm 0.11	$&$ 0.0162 \pm 0.0007$ \\
 HD\,94024 & $2.463962 \pm 0.000054$ & $0.000$ (fixed) & $270.00$ (fixed) & $29.89 \pm 0.98$& $9110.8068 \pm 0.0950$ & $49.00 \pm 0.26$ & $1.46 \pm 0.05$ & $0.0068 \pm 0.0007$ \\
 HD\,105627 & $4.340874 \pm 0.000095$ & $0.084 \pm 0.020$ & $15.03 \pm 17.33$ & $28.51 \pm 0.67$ & $3718.4721 \pm 0.2278$ & $14.66 \pm 0.71$ & $2.44 \pm 0.06$ & $0.0103 \pm 0.0007$ \\
 HD\,130298 & $14.62959 \pm 0.000854$& $0.457 \pm 0.007$ & $324.75 \pm 1.28$ & $71.78 \pm 0.68$ & $9001.5647 \pm 0.0332$ & $-36.54 \pm 0.36$ & $17.38 \pm 0.12$ & $0.3292 \pm 0.0073$ \\
 HD\,165174 & $23.876059 \pm 0.007301 $ & $ 0.156	\pm 0.058 $ & $ 303.76 \pm 23.68 $ & $ 23.56 \pm 1.85 $ & $ 9298.9069 \pm 1.4667 $& $20.56 \pm 0.70$ &$ 11.02 \pm 0.97	$&$ 0.0313 \pm 0.0071$ \\
 HD\,229234 & $3.510361 \pm 0.000107$ & $ 0.000$ (fixed) & $ 270.00$ (fixed) & $ 45.66 \pm 1.99 $ & $ 9000.5922 \pm 0.0262$ & $-10.11 \pm 1.06$  &$ 3.17 \pm 0.14	$&$ 0.0351 \pm 0.0057$ \\
 HD\,308813 & $6.346190 \pm 0.000380$& $0.375 \pm 0.012$ & $101.44 \pm 2.12$& $32.49 \pm 0.44$ & $4543.5312 \pm 0.0659$ & $-2.65 \pm 1.03$ & $3.90 \pm 0.20$ & $0.0198 \pm 0.0030$ \\
 LS\,5039 & $3.906080 \pm 0.000084$ & $ 0.254 \pm 0.040 $ & $ 265.63 \pm 9.01 $ & $ 22.59 \pm 1.47$ & $ 9001.7403 \pm 0.1557 $ & $11.91 \pm 1.12$ &$ 1.69 \pm 0.11$&$ 0.0042 \pm 0.0008$ \\
 \hline
\end{tabular}
\end{sidewaystable*}

\begin{sidewaystable*}
\tiny
\caption{Orbital parameters of the SB2 systems (the primary star - 1 - being the most massive object of the system). The errors correspond to $1\sigma$. }\label{table:SB2orbitalparameters}
\centering
\begin{tabular}{lrrrrrrrrrrrrr}
\hline \hline
 Star & $P_{\mathrm{orb}}$ & $e$ & $\omega$ & $q$ & K$_1$ & K$_2$ & $T_0$ & $\gamma$ & $M_1\,\sin^3\,i$ & $M_2\,\sin^3\,i$ & $a_1\,\sin\,i$ & $a_2\,\sin\,i$ \\
  & [d] &  & [$^{\circ}$] & $[M_2/M_1]$ & [\kms] & [\kms] & $-2\,450\,000$ & [\kms]& [$\Msun$] & [$\Msun$] & [$\Rsun$] & [$\Rsun$]  \\
  \hline
 HD\,29763 & $2.956526 \pm 0.000010$ & $0.000$ (fixed) & $270.00$ (fixed) & $0.39 \pm 0.02$ &$53.28 \pm 0.44$ & $138.53 \pm 4.33$ & $7267.1839 \pm 0.0041$ & $23.31 \pm 0.31$ & $1.56 \pm 0.12$ & $0.60 \pm 0.03$ & $3.11 \pm 0.06$ & $8.10 \pm 0.25$ \\
 HD\,30836 & $9.519999 \pm 0.000409$ & $0.011 \pm 0.002$ & $20.89 \pm 107.59$ & $0.30 \pm 0.03$ & $26.33 \pm 0.41$ & $87.21 \pm 5.86$ & $9002.9824 \pm 2.8569$ & $33.22 \pm 0.24$ & $1.11 \pm 0.19$ & $0.34 \pm 0.04$ & $4.96 \pm 0.24$ & $16.41 \pm 1.09$ \\
 HD\,52533 & $21.969943 \pm 0.014958$ & $0.273 \pm 0.082$ & $357.78 \pm 30.12$ & $0.40 \pm 0.11$ & $88.42 \pm 14.79$ & $208.98 \pm 41.44$ & $6060.2366 \pm 1.7721$ & $56.76 \pm 2.27$ & $32.96 \pm 16.57$ & $13.18 \pm 5.40$ & $33.79 \pm 6.09$ & $84.53 \pm 17.01$ \\
 HD\,57236 & $212.497879 \pm 0.035930$ & $0.580 \pm 0.002$ & $25.63 \pm 0.55$ & $0.83 \pm 0.16$ & $59.81 \pm 0.22$ & $72.22 \pm 13.74$ & $8434.1662 \pm 0.2304$ & $52.84 \pm 0.29$  & $15.02 \pm 6.00$ & $12.44 \pm 2.68$ & $204.71 \pm 5.97$ & $247.17 \pm 47.03$ \\
 HD\,91824 & $112.397158 \pm 0.005960$ & $0.207 \pm 0.006$ & $173.94 \pm 1.66$ & $0.51 \pm 0.22$ & $36.19 \pm 0.20$ & $110.59 \pm 4.34$ & $5069.2193 \pm 0.4258$ & $-4.25 \pm 0.22$ & $26.05 \pm 2.60$ & $8.52 \pm 0.64$ & $78.68 \pm 2.48$ & $240.45 \pm 9.44$ \\
 HD\,93028 & $204.942537 \pm 0.810187$ & $0.131 \pm 0.027$ & $87.46 \pm 23.92$ & $0.48 \pm 0.13$ & $35.58 \pm 0.58$ & $73.60 \pm 9.73$ & $4545.8786 \pm 13.3840$ & $-2.54 \pm 0.15$ & $18.19 \pm 11.47$ & $8.80 \pm 3.24$ & $142.94 \pm 6.27$ & $295.64 \pm 79.27$ \\
 HD\,152405 & $25.489568 \pm 0.000519$ & $0.547 \pm 0.010$ & $79.64 \pm 0.97$ & $0.38 \pm 0.13$ & $30.18 \pm 0.15$ & $79.38 \pm 15.44$ & $8676.6017 \pm 0.0615$ & $-7.90 \pm 0.27$ & $1.48 \pm 1.16$ & $0.56 \pm 0.27$ & $12.73 \pm 1.03$ & $33.49 \pm 10.74$ \\
 HD\,152723 & $18.898193 \pm 0.003882$ & $0.514 \pm 0.030$ & $281.87 \pm 6.46$ & $0.21 \pm 0.07$ & $18.37 \pm 2.71$ & $89.37 \pm 25.44$ & $5015.0280 \pm 0.1614$ & $-0.48 \pm 0.58$ & $1.28 \pm 1.09$ & $0.26 \pm 0.15$ & $5.88 \pm 0.89$ & $28.64 \pm 9.10$ \\
 HD\,163892 & $7.835566 \pm 0.000062$ & $0.041 \pm 0.008$& $111.64 \pm 10.96$ & $0.18 \pm 0.02$ & $41.05 \pm 2.50$ & $232.46 \pm 14.93$& $3972.8985 \pm 0.24812$ & $-3.09 \pm 0.48$ & $14.11 \pm 2.46$ & $2.49 \pm 0.34$ & $6.35 \pm 0.39$ & $35.98 \pm 2.31$ \\
 HD~164438 & $10.249635 \pm 0.000140$ & $0.282 \pm 0.013$ & $222.41 \pm 2.18$ & $0.27 \pm 0.04$ & $28.68 \pm 0.20$ & $106.34 \pm 14.44$ & $8494.0208 \pm 0.0589$ & $-9.99 \pm 0.25$ & $1.82 \pm 0.64$ & $0.49 \pm 0.12$ & $5.58 \pm 0.51$ & $20.68 \pm 2.81$  \\
 HD\,164536 & $11.682917 \pm 0.000350$ & $0.074 \pm 0.006$ & $182.27 \pm 10.18$& $0.14 \pm 0.03$ & $22.95 \pm 0.16$ & $161.48 \pm 11.53$ & $4017.2911 \pm 0.3454$ & $8.10 \pm 0.29$ & $6.61 \pm 1.34$ & $0.94 \pm 0.26$ & $5.29 \pm 1.06$ & $37.20 \pm 2.66$  \\
 HD\,167263 & $64.835288 \pm 0.005845$ & $0.005 \pm 0.002$ & $153.72 \pm 37.33$& $0.58 \pm 0.25$ & $32.77 \pm 0.60$ & $41.26 \pm 19.51$ & $6003.5353 \pm 6.6274$ & $5.04 \pm 0.63$& $1.52 \pm 0.74$ & $1.21 \pm 0.33$ & $41.99 \pm 2.05$ & $52.88 \pm 12.18$ \\
 HD\,167264 & $674.416352 \pm 1.643820$ & $0.229 \pm 0.053$ & $314.06 \pm 15.21$& $0.76 \pm 0.07$ & $26.28 \pm 1.18$ & $34.36 \pm 2.53$ & $6138.2722 \pm 27.4503$& $7.50 \pm 0.66$ & $8.16 \pm 1.37$ & $6.24 \pm 0.85$ & $341.12 \pm 18.97$ & $446.05 \pm 33.36$ \\
 HD\,192001 & $189.443477 \pm 0.028975$ & $0.829 \pm 0.009$ & $340.25 \pm 0.74$& $0.58 \pm 0.10$ & $71.64 \pm 3.80$ & $124.50 \pm 21.44$ & $9126.9959 \pm 0.0801$ & $-23.71 \pm 0.51$& $16.48 \pm 6.57$ & $9.48 \pm 2.28$ & $150.05 \pm 6.96$ & $260.75 \pm 45.34$ \\
 HD\,199579 & $48.518999 \pm 0.002043$ & $0.072 \pm 0.003$ & $74.14 \pm 2.67$ & $0.33 \pm 0.03$ & $39.37 \pm 0.08$ & $119.48 \pm 11.27$ & $8702.3770 \pm 0.4055$ & $-2.29 \pm 0.07$ & $15.07 \pm 3.57$ & $4.97 \pm 0.73$ & $37.67 \pm 0.99$ & $114.32 \pm 10.78$ \\
 Schulte~11 & $72.620500 \pm 0.013936$ & $0.612 \pm 0.034$ & $295.04 \pm 7.33$ & $0.22 \pm 0.07$ & $29.91 \pm 2.22$ & $134.92 \pm 44.13$ & $8996.9718 \pm 0.4754$ & $-7.29 \pm 0.98$ & $13.68 \pm 11.89$ & $3.03 \pm 1.68$ & $33.97 \pm 2.76$ & $153.24 \pm 50.37$ \\
 V747\,Cep & $5.336561 \pm 0.000364$ & $0.370 \pm 0.086$ & $195.36 \pm 4.85$ & $0.24 \pm 0.05$ & $89.61 \pm 15.19$ & $374.44 \pm 32.44$ & $9003.1844 \pm 0.3003$ & $-11.02 \pm 2.82$ & $35.81 \pm 2.58$ & $8.57 \pm 1.25$ & $8.78 \pm 1.54$ & $36.68 \pm 3.25$ \\
 \hline
\end{tabular}
\end{sidewaystable*}

After measuring the RVs, we first used the Heck-Manfroid-Mersch (HMM) method \citep[][revised by \citealt{gosset01}]{heck85} to derive an initial guess for the orbital periods of the systems. The HMM method has the advantage of giving a better expression for the power spectrum than, for example, the one of \citet{scargle82}. These periods were then used as input for the SPectroscopic and INterferometric Orbital Solution finder code (\textsc{spinOS}\footnote{\url{https://github.com/matthiasfabry/spinOS}}, \citealt{fabry21}). This code allows us to compute the orbital parameters of the different systems in our sample from a set of RV measurements. The orbital parameters are derived by minimising a $\chi^2$ metric using a Levenberg-Marquardt optimisation algorithm, and the uncertainties are subsequently estimated using Markov Chain Monte Carlo sampling. \textsc{spinOS} was built to model astrometric and spectroscopic orbits simultaneously, so that the longitude of the periastron passage is shifted by $180^{\circ}$ with respect to the spectroscopic value of the primary star ($\omega_{\textsc{spinOS}} = \omega_{\rm spec} + 180^{\circ}$). We adopt the spectroscopic definition of $\omega$ in the rest of the paper. The orbital parameters of the SB1 and newly detected SB2 systems (Sect.\ref{subsec:disentangling}) are listed in Tables~\ref{table:SB1orbitalparameters}, and \ref{table:SB2orbitalparameters}, respectively. The SB1 and SB2 RV curves are displayed in Figs.~\ref{fig:SB1RVcurves} and ~\ref{fig:SB2RVcurves}, respectively.

\subsection{Spectral disentangling and detection limit}
\label{subsec:disentangling}

To search for non-degenerate companions and characterise their spectral features, we performed spectral disentangling. This technique aims at providing individual spectra of each component in a binary or multiple system, and it allows the orbital solution of the system to be directly refined by finding a self-consistent solution from a time series of composite spectra. To extract the spectral signatures of putative faint companions from the spectra, we applied Fourier spectral disentangling \citep{hadrava95}. This technique takes as inputs the orbital parameters derived in Sect.~\ref{subsec:orbitalsolution} and optimises them, using the Nelder \& Mead simplex \citep{NeldMead65}, to find the best solution in a multi-dimensional (6D) space (i.e. $P_{\rm orb}$, $e$, $\omega$, $K_1$, $K_2$, and $T_0$).

\begin{figure*}[htbp]
\centering
    \hspace{0cm}
    \begin{subfigure}{0.45\linewidth}
    \includegraphics[width=6cm,trim=10 20 10 10,clip,angle=270]{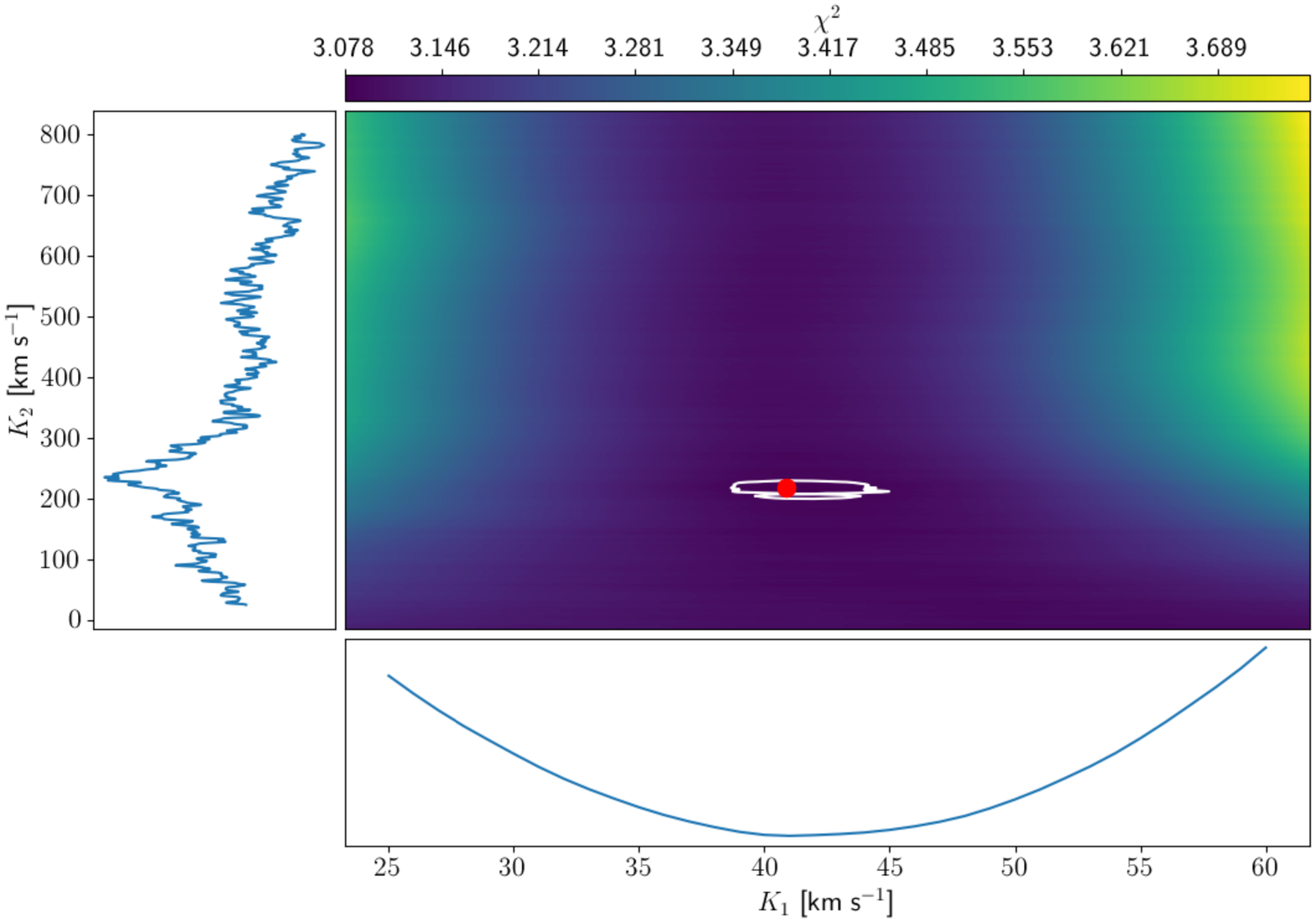}
    \end{subfigure}
    \qquad
    \begin{subfigure}{0.45\linewidth}
    \includegraphics[width=8.5cm,trim=0 0 35 35,clip]{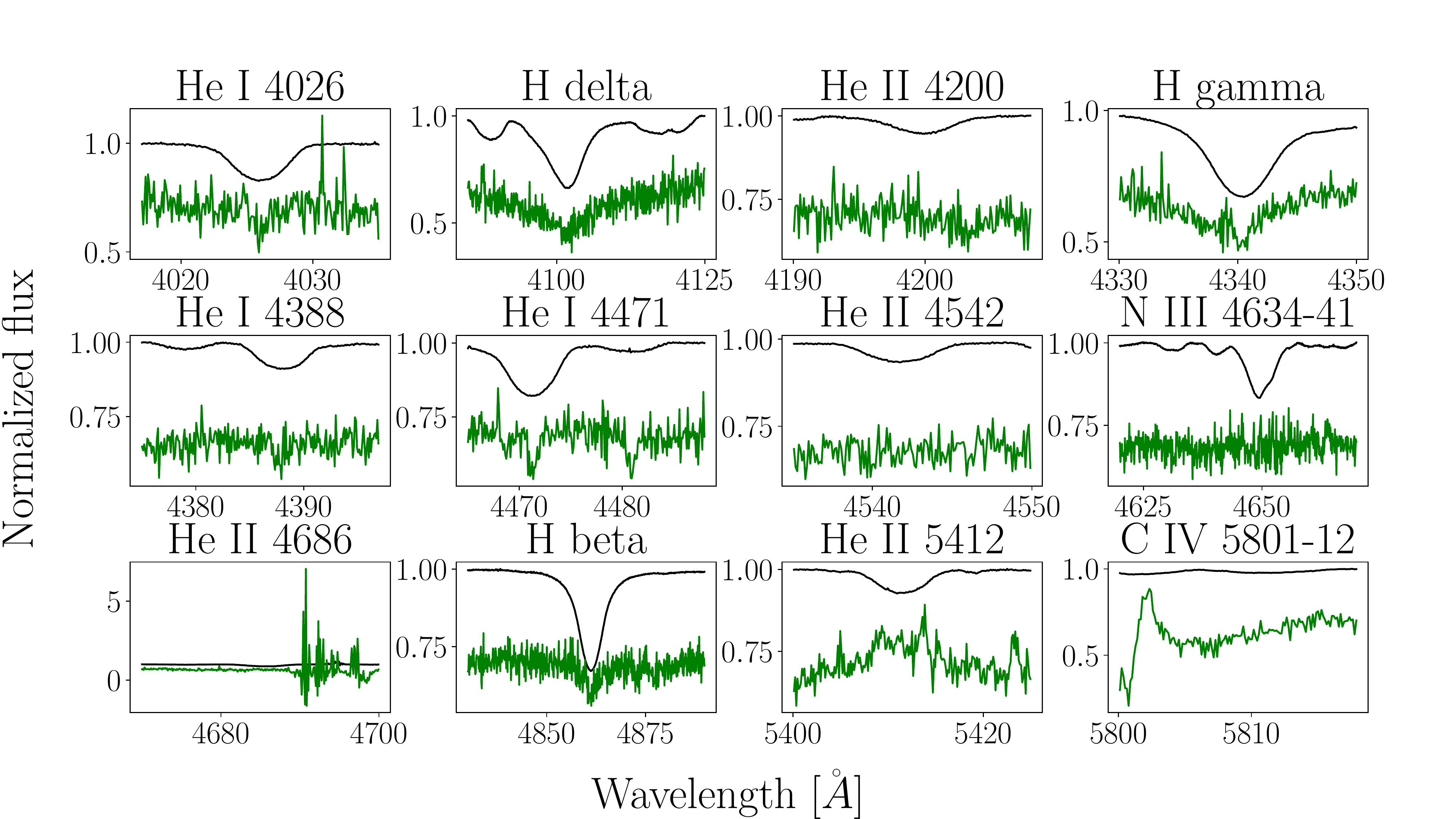}
    \end{subfigure}
\caption{\label{fig:disent} Left: Reduced $\chi^2$ map given by the grid disentangling for HD~163892 by combining the \ion{He}{i}~4471, \ion{Mg}{ii}~4481 and H$\gamma$ lines. The minimal reduced $\chi^2$, at $K_1 = 41.05 \pm 2.50$\,\kms, $K_2 = 232.5 \pm 14.9$\,\kms\ is shown with a red dot. The solid white lines represent the $1\sigma$ contours. The 1D cuts in both directions are given as indications. Right: Disentangled spectra of the primary (black) and secondary (green) of HD~163892. The latter has been shifted vertically by $-0.25$ for clarity.}
\end{figure*}

The efficiency of extracting faint companions depends on the number of spectra, their resolution, their signal-to-noise ratios (S/Ns), their distribution over the orbital cycle, and on the brightness ratios between the two components forming the binary systems. The simplex optimisation requires initial parameters that are close to the real solutions to avoid possible local minima. When the secondaries are bright enough but diluted due to their high rotation and when the number of observed spectra meets all the conditions mentioned above, the Nelder \& Mead simplex is very efficient to extract the spectral signatures of the companions. However, when the companion is very diluted in the composite spectra, its extraction is more complicated. We therefore decided to also apply a grid approach to limit the number of free parameters fitted simultaneously. In our analysis, the light ratios need to be derived from models (unless there are eclipses). This approach was successfully used by \citet{bodensteiner20} and \citet{Shenar2020LB1} to discard the presence of stellar-mass BHs orbiting around stripped B-type stars. Using this technique, the authors disclaimed the presence of stellar-mass BHs as secondaries and were able to extract two non-degenerate components (a stripped primary and a rapidly rotating secondary) from their spectra. 

In our 2D grid approach, we fixed the orbital period, eccentricity, longitude of the periastron passage and the time of reference, and only let the RV semi-amplitudes of the primary and secondary ($K_1$ and $K_2$) vary. We recorded the reduced $\chi^2$ for each point in our grid. We extracted the spectra of the faint secondary companions down to a mass ratio of about 0.15. By construction, the two-component spectral disentangling produces a spectrum for the primary and the secondary components. If the secondary is 'dark', one would ideally expect a flat spectrum. In practice, the disentangled secondary spectrum of a dark companion will contain noise and possible artefacts due to, for example, the normalisation uncertainties or non-Keplerian variations, etc. For each result, one must therefore decide whether the resulting secondary spectrum is compatible with that of a stellar object or not. For that purpose, we visually inspect the disentangled spectrum of each component and compare it with typical stellar spectra of \citet{gray09}\footnote{see also \url{https://ned.ipac.caltech.edu/level5/Gray/frames.html}}. These steps allow us to detect and extract the individual spectra of 17 non-degenerate stellar companions, turning these SB1 systems into newly classified SB2 systems. For all these systems, there is no doubt about the non-degenerate nature of the stellar companions. However, for the secondary component in Schulte~11, while its Balmer lines are clearly detected, other spectral features such as the \ion{He}{i} lines are too noisy to allow a spectral classification. An example of grid disentangling, for HD~163892, is shown in Fig.~\ref{fig:disent}. The disentangled spectra for the other systems are given in Fig.~\ref{fig:disent_all}.

\begin{figure*}[htbp]
    \centering
    \begin{subfigure}{0.49\linewidth}
    \includegraphics[width = \textwidth, trim=0 0 0 0,clip]{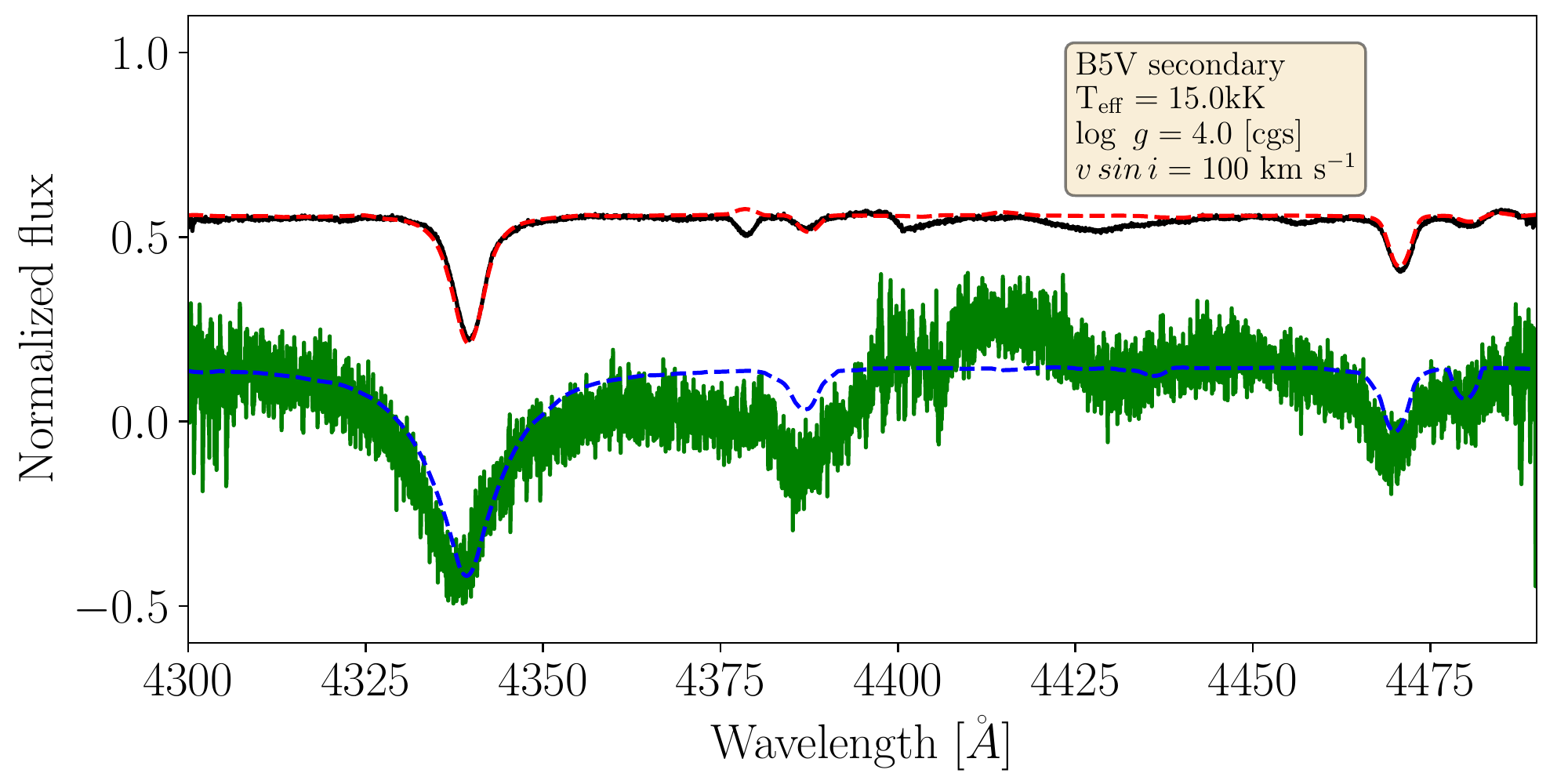}
    \caption{B5~V MS secondary with $\teff = 15.0$~kK, $\logg = 4.0$~[cgs] and $f2/f1=0.04$}
    \end{subfigure}
    \hfill
    \begin{subfigure}{0.49\linewidth}
    \includegraphics[width = \textwidth, trim=0 0 0 0,clip]{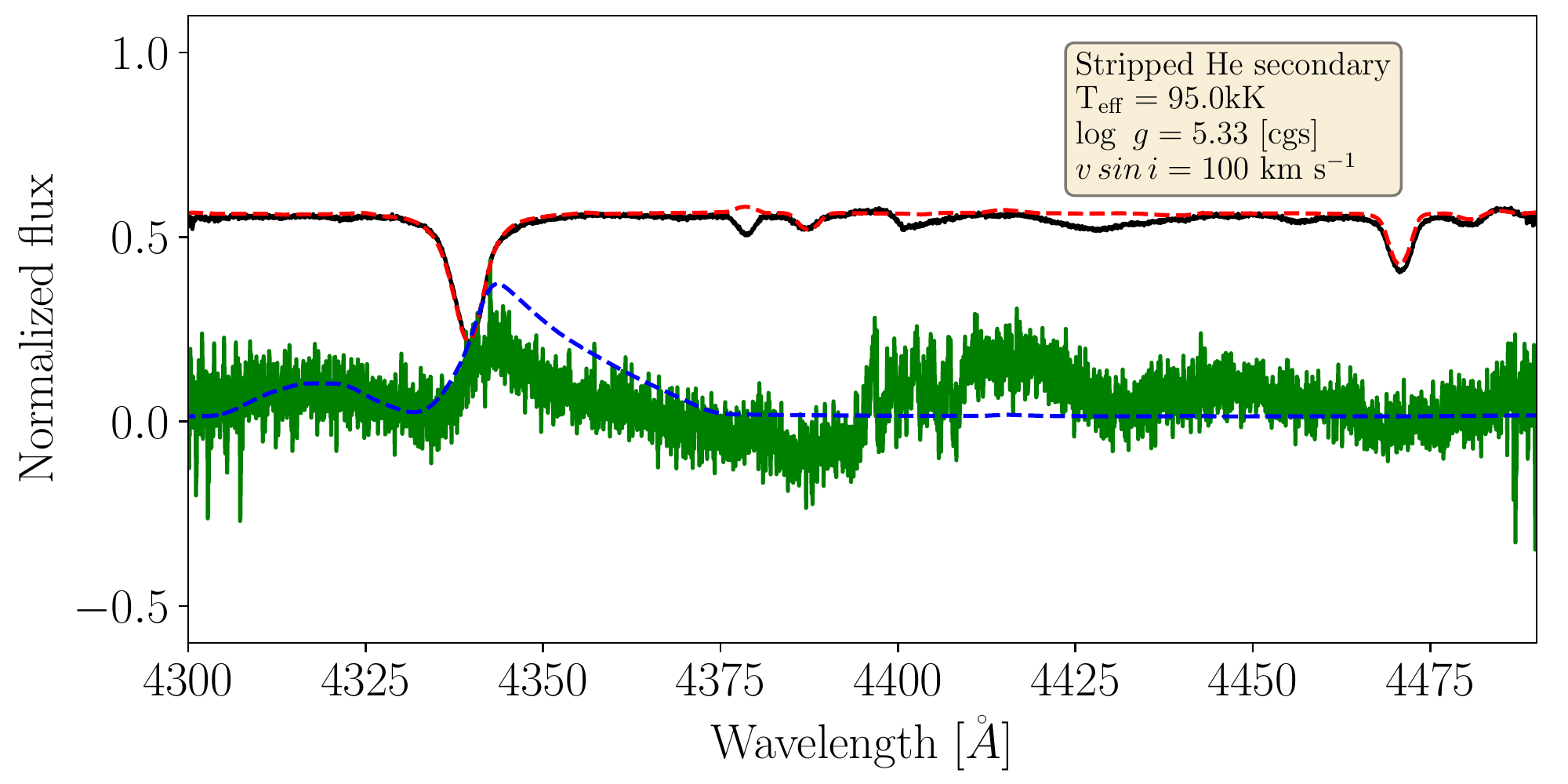}
    \caption{Stripped He secondary with $\teff = 95.0$~kK, $\logg = 5.33$~[cgs] and $f2/f1=0.02$}
    \end{subfigure} 
    \hfill
    \begin{subfigure}{0.49\linewidth}
    \includegraphics[width = \textwidth, trim=0 0 0 0,clip]{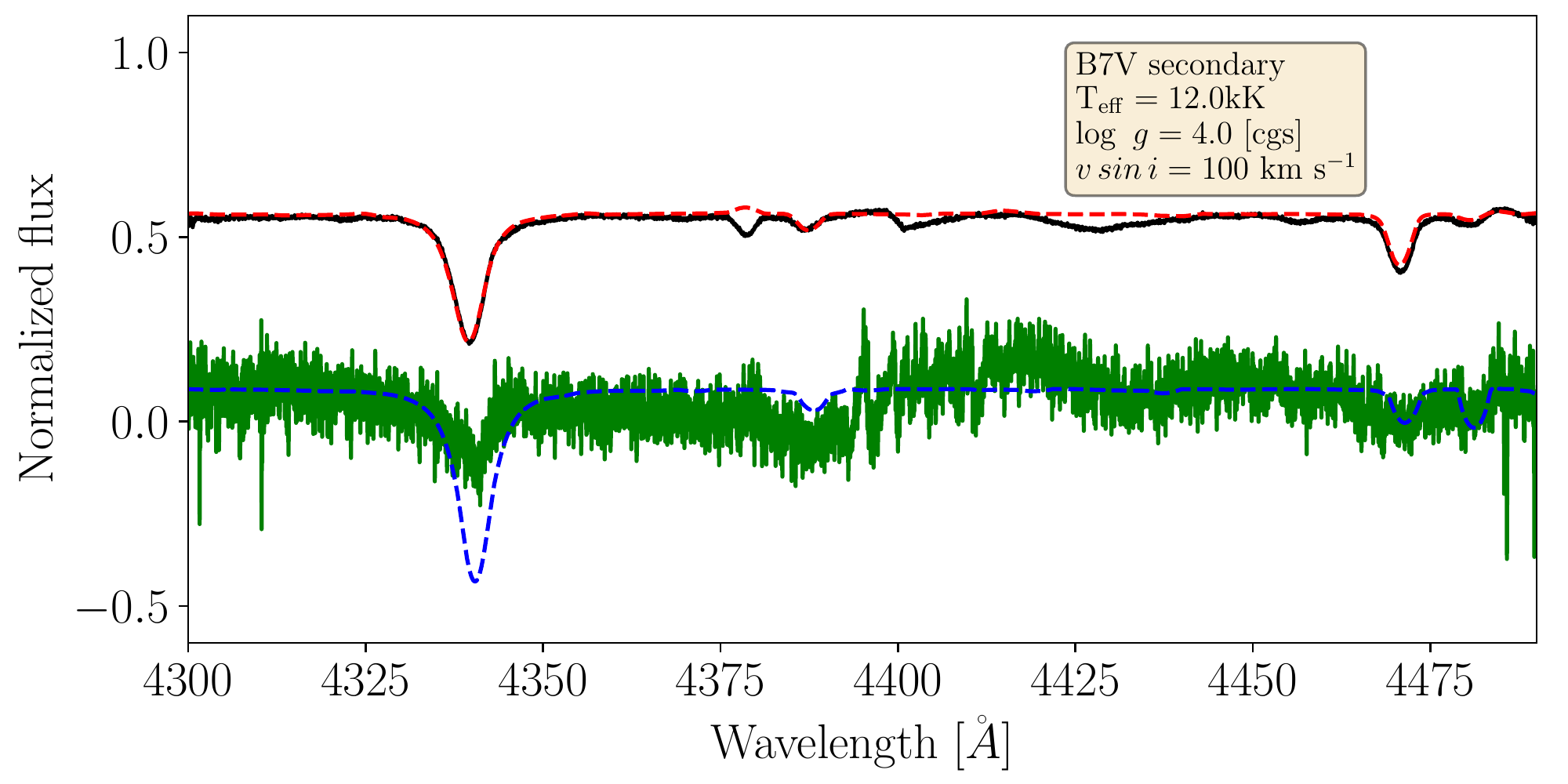}
    \caption{B7~V MS secondary with $\teff = 12.0$~kK, $\logg = 4.0$~[cgs] and $f2/f1=0.02$}
    \end{subfigure}
    \hfill
    \begin{subfigure}{0.49\linewidth}
    \includegraphics[width = \textwidth, trim=0 0 0 0,clip]{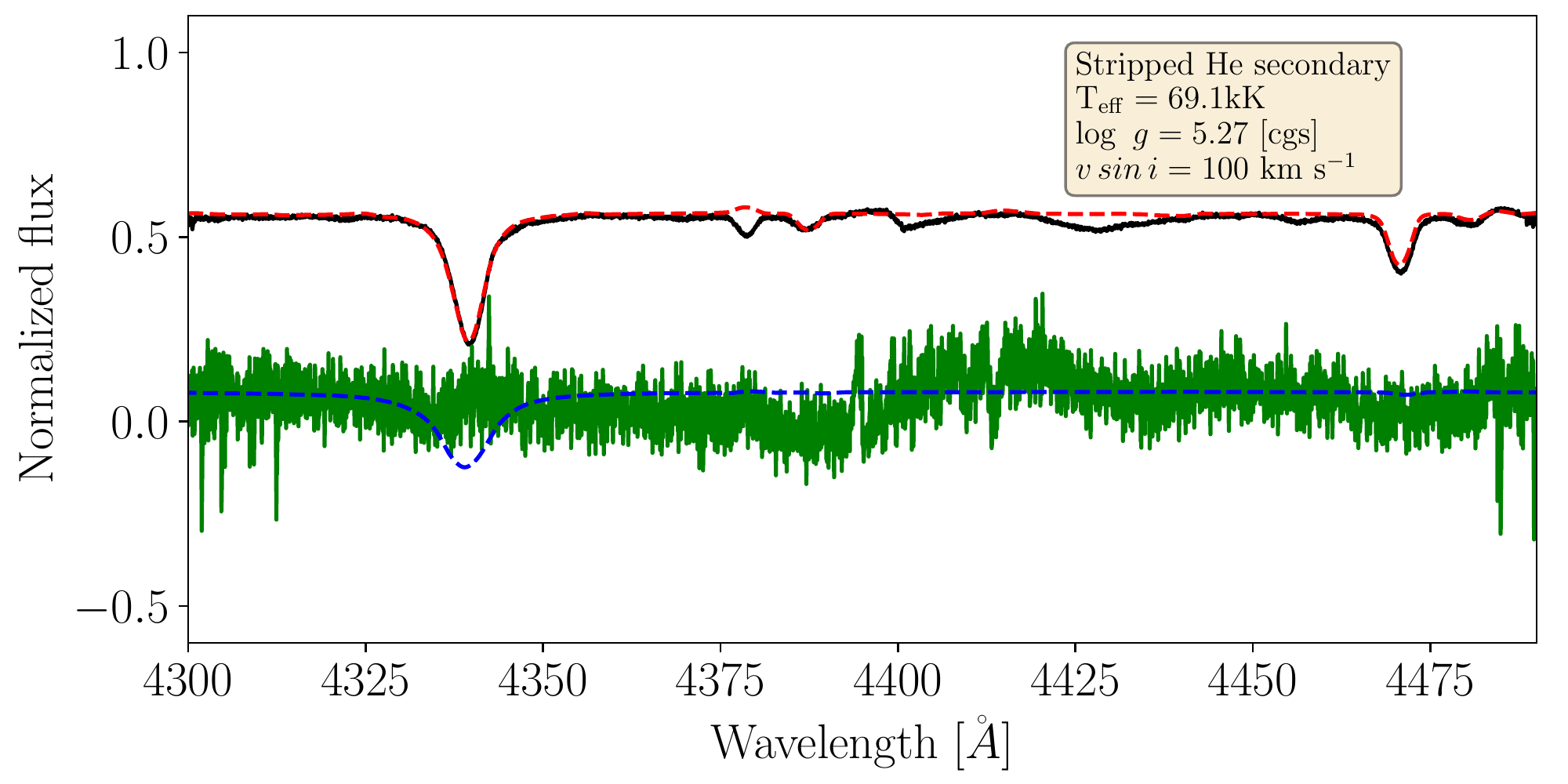}
    \caption{Stripped He secondary with $\teff = 69.1$~kK, $\logg = 5.27$~[cgs] and $f2/f1=0.008$}
    \end{subfigure}
    \hfill
    \begin{flushleft}
    \begin{subfigure}{0.49\linewidth}
    \includegraphics[width = \textwidth, trim=0 0 0 0,clip]{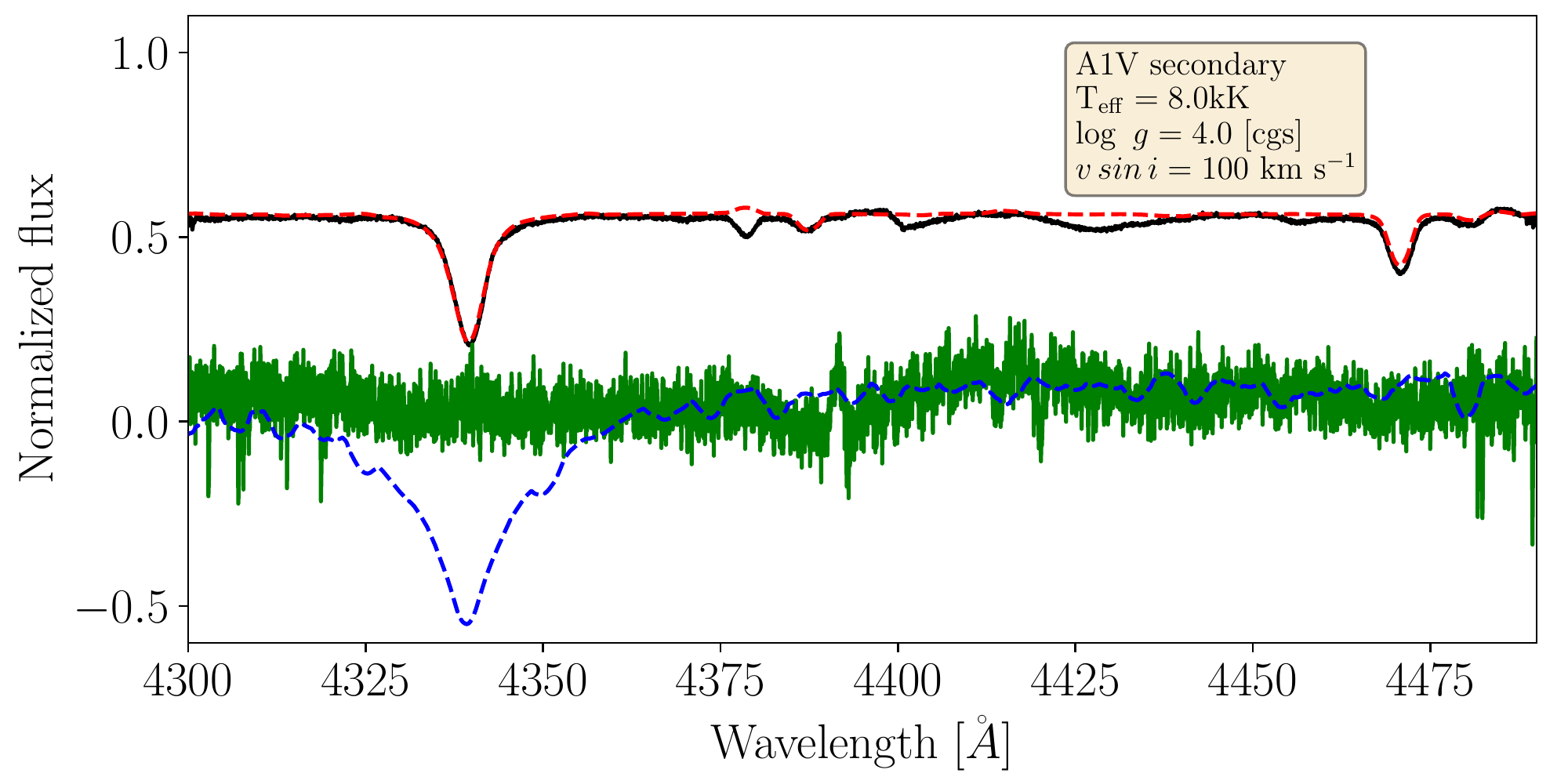}
    \caption{A1~V MS secondary with $\teff = 8.0$~kK, $\logg = 4.0$~[cgs] and $f2/f1=0.006$}
    \end{subfigure}
    \end{flushleft}
    
    \caption{Simulations of the H$\gamma$, \ion{He}{i}~4388, 4471, and \ion{Mg}{ii}~4481 spectral lines in the disentangled spectra of a mock system composed of an O6.5~III primary. Red and blue spectra are the synthetic models. We entangle 12 observed spectra with synthetic models of a MS secondary (left panels), and with synthetic models of a stripped helium star \citep[right panels][]{gotberg18}. The secondary spectra were convolved by a rotational profile of 100~\kms. Black and green spectra are the disentangled spectra. Companions with flux ratios lower than 0.02 would not have been detected from our data. The spectra are shifted vertically for clarity. We correct for the dilution by taking adopted values of 95\% for the primary and 5\% for the secondary for all the simulations.}
    \label{fig:simulation}
\end{figure*}

Without the presence of eclipses in the light curves, the disentangled spectra can be extracted but the strengths of the spectral lines strongly depend on the brightness (or scaling factor) that we adopt. The brightness factor for each component is a fraction of the total flux of the system and is given by $l_1 = f_1 / (f_1 + f_2)$ and $l_2 = 1-l_1 = f_2 / (f_1 + f_2)$. They were estimated through an iterative process, that ensured that the strengths of the hydrogen and helium lines of the disentangled spectra can be fitted with synthetic models, as was done in \citet{mahy20a}. We give the flux contributions for each object in Table~\ref{tab:parameters_SB2} with the individual parameters of each component derived in Sects.~\ref{subsec:atmosphere} and~\ref{sec:finalparameters}. The spectral disentangling process gives us the RV semi-amplitudes for both components, allowing us to compute the mass ratios. Knowing the mass of the primaries, we can compute the masses of the secondaries and have additional constraints on how the spectrum of the secondary must look like. In case no contribution of the secondary star is obtained through the disentangling process, the output spectra appear featureless, as shown in Figs.~\ref{fig:simulation}.

For systems still classified as SB1s, we must understand the mass limit up to which the spectral disentangling allows us to extract and characterise the nature of non-degenerate stellar companions. Indeed, since we cannot directly confirm the presence of a compact object, one must rule out all other possibilities before accepting the presence of such an object. 

For this purpose, we ran simulations to determine the detection limits for the systems in our sample. We considered three different cases: 1) a binary system with a MS secondary, 2) a binary system with a stripped helium star as secondary, and 3) a triple system where the visible OB star is the outer object and where the inner close system is composed of two lower-mass stars.

\subsubsection{Detection limit for MS companions}
\begin{figure*}[htbp]\centering
    \includegraphics[trim=10 0 60 0,clip,width=18cm]{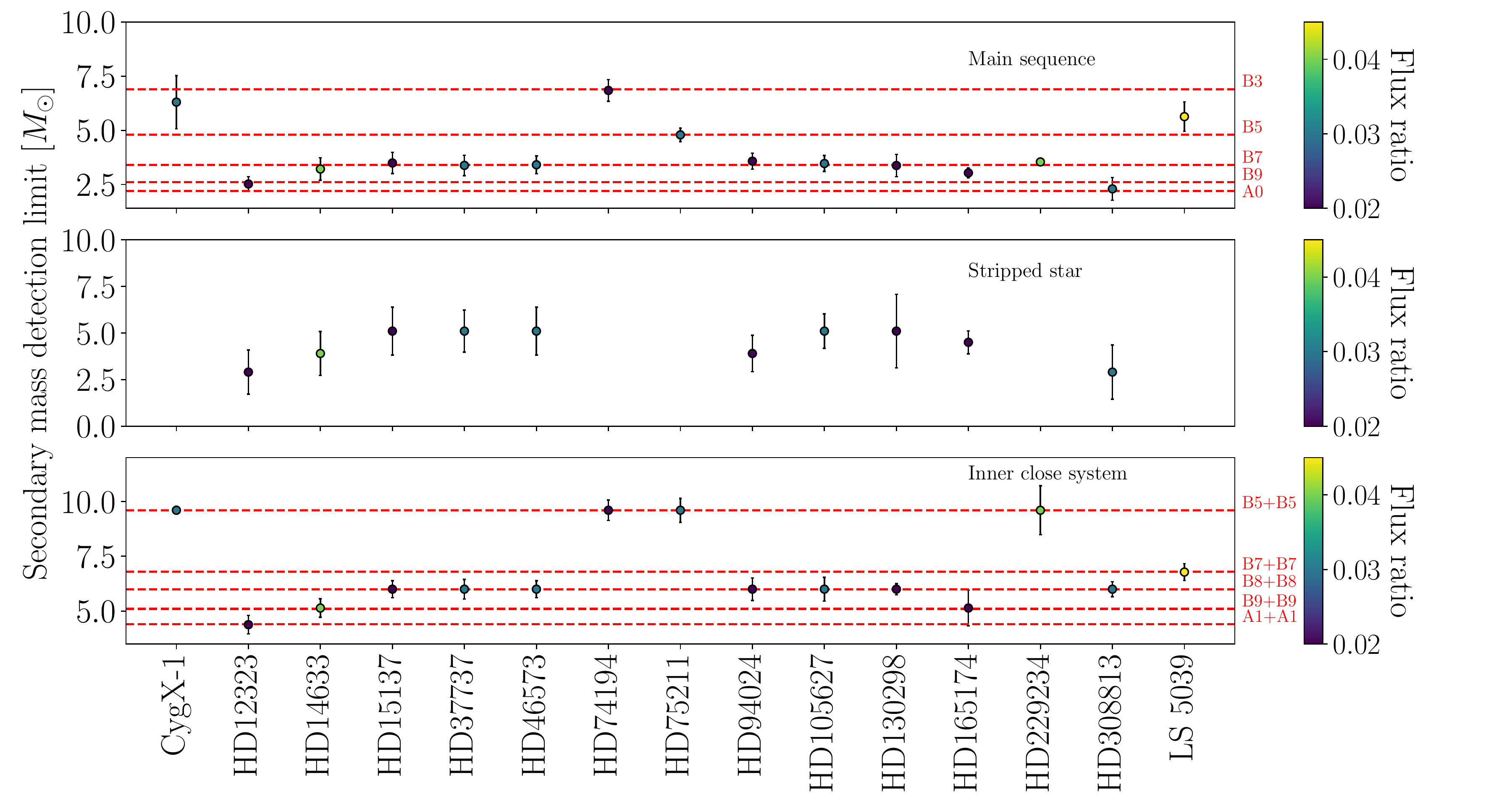}
    \caption{\label{fig:DetectionLimit} Maximum mass of a secondary object that can elude detection in our disentangling approach. The simulations have been run with $\vsini = 100$~\kms. These detection limits are given when the secondary is considered as a MS star (top), a stripped helium star (middle), or an inner close binary in a hierarchical triple system (bottom). In the middle panel, no markers have been indicated for four systems because we were not able to detect the stripped companion. In the bottom panel, the y axis corresponds to the total mass of the inner system.} \label{fig:detectionlimit}
\end{figure*}

To quantify the lower limit of detectability that spectral disentangling can reach for each of our datasets, we constructed mock composite spectra using the disentangled spectrum of the primary star, and we included a secondary companion, using synthetic spectra that mimic the stellar properties of the companion (effective temperatures, surface gravities, fluxes, and mass ratios). We used TLUSTY \citep[][for stars with effective temperatures higher than 15~kK]{lanz07} or MARCS models \citep[][for stars with effective temperatures lower than 15~kK]{delaverny12} from the Pollux database\footnote{\url{http://npollux.lupm.univ-montp2.fr/DBPollux/PolluxAccesDB/}} \citep[][]{palacios10}. We adopted for the synthetic spectra a surface gravity of $\logg = 4.0$ [cgs]. The properties of the mock stars (spectral types, effective temperatures, masses, and absolute visual magnitudes) were taken from the tables provided by \citet{Schmidt-Kaler1982}\footnote{\url{https://xoomer.virgilio.it/hrtrace/Calibr.htm}}. We tested the spectral disentangling by allowing i) all the orbital parameters to vary through the use of a multi-dimensional simplex and ii) only the RV semi-amplitudes of both components ($K_1$, $K_2$) to vary. All these simulations have been performed assuming a projected rotational velocity of $\vsini = 100$~\kms\ for the secondary (Fig.~\ref{fig:simulation}). A summary of the detection limit is given in Fig.~\ref{fig:detectionlimit}.

Despite the low flux contribution for the secondaries, a MS companion is readily retrieved down to mass and flux ratios of 0.13 and 0.02, respectively. Those ratios correspond to MS companions earlier than B3-B5 in most of the cases. Only for a few objects with later spectral types are the data of sufficient quality to extract companions down to early A-type stars.

\subsubsection{Detection limit for stripped helium companions}
In a second set of simulations we consider the secondaries to be stripped helium stars. The likelihood of occurrence for such systems is roughly ten times lower than for a stellar-mass BH companion \citep{Shao21}. For those simulations, we used the models computed by \citet{gotberg18}. Using their parameters, the detection limits for these objects are given in Fig.~\ref{fig:detectionlimit} and the results from the mock spectra are displayed in the bottom panels of Fig.\,\ref{fig:simulation}. 

\subsubsection{Detection limit considering triple systems}
Our last simulations focus on possible triple systems, composed of an outer O-type star (the visible star in the observed SB1s) orbiting around an inner close binary system composed of two lower-mass stars. We considered the conditions of stabilisation of hierarchical triple systems for our simulations as provided by \citet{toonen20}. All our simulations were done under the assumption that the inner close binaries are composed of two equal-mass intermediate-mass MS stars, which is the worst case scenario. We performed these simulations for all our systems, but given their periods (less than 55 days), only the longest-period systems are suitable for this scenario. To remain stable, the two low-mass inner stars are expected to orbit around each other with periods shorter than 1-2 days. We assume that the stars are on a 1.5-day circular orbit, moving anti-phase, and we assume a projected rotation of 50~\kms\ for each. Under these conditions, a double-lined system may appear single if the RV separation between the two profiles is not sufficient for them to be clearly de-blended \citep{bodensteiner21,banyard22}. Our simulations only provides us with a mean spectrum of each inner close system, and not of individual intermediate-mass stars in the inner systems. We show in Fig.~\ref{fig:detectionlimit} a summary of our results. 

\subsection{Minimum masses of the unseen companions}
\label{minMass}
\begin{figure*}[h!]
    \centering
    \begin{subfigure}{0.33\linewidth}
    \includegraphics[width = \textwidth, trim=0 0 0 0,clip]{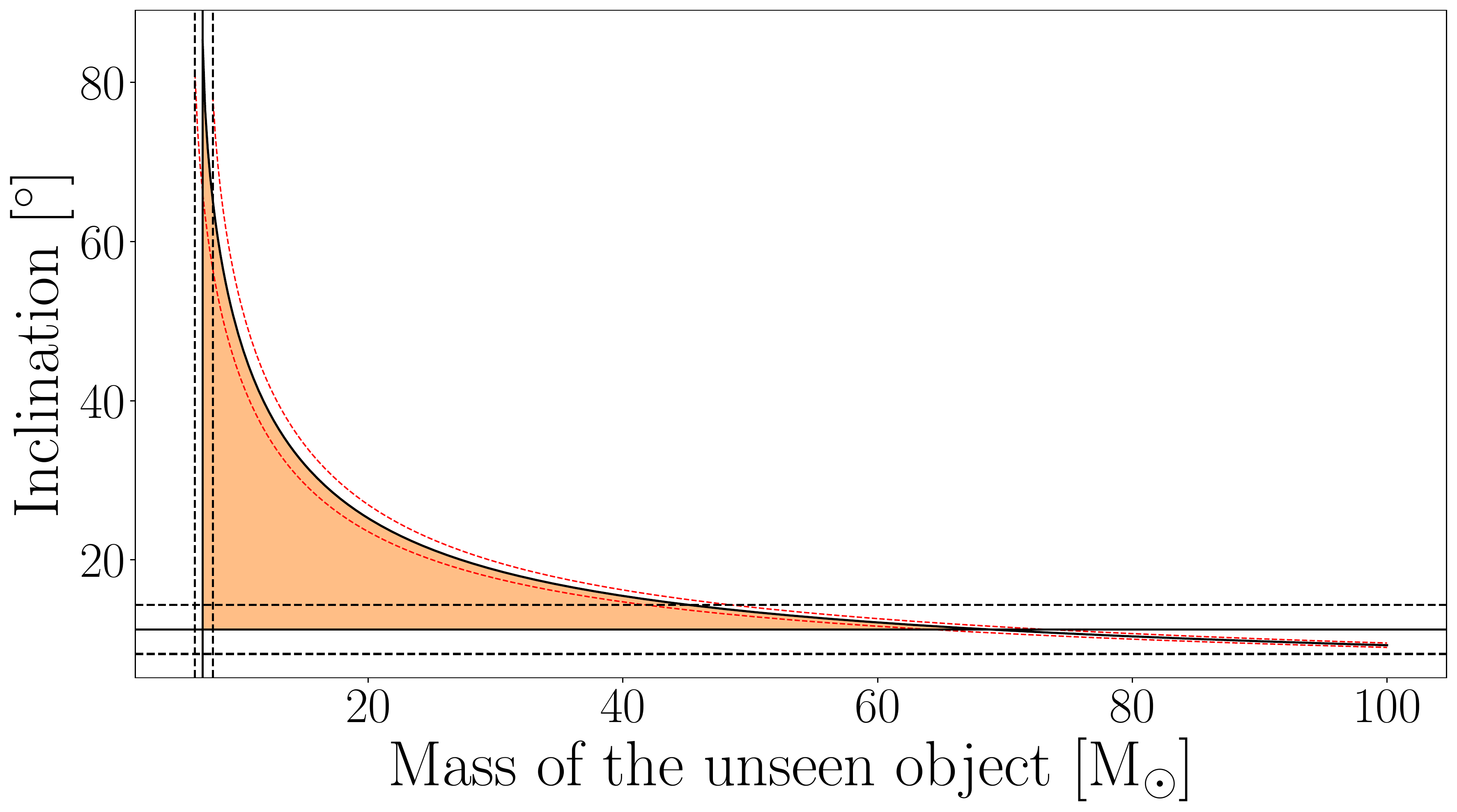}
    \caption{Cyg~X-1}
    \end{subfigure}
    \hfill
    \begin{subfigure}{0.33\linewidth}
    \includegraphics[width = \textwidth, trim=0 0 0 0,clip]{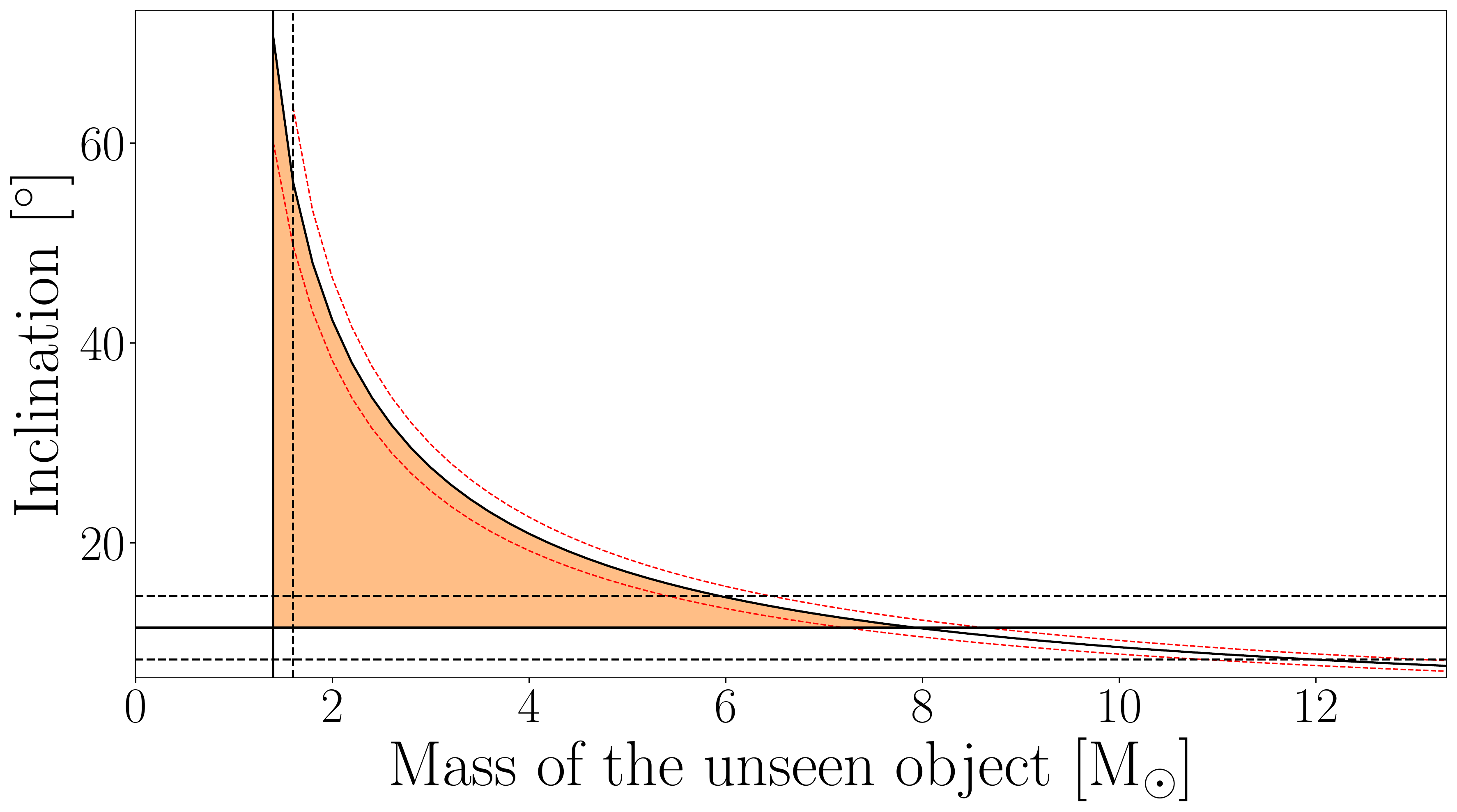}
    \caption{HD~12323}
    \end{subfigure}
    \hfill
    \begin{subfigure}{0.33\linewidth}
    \includegraphics[width = \textwidth, trim=0 0 0 0,clip]{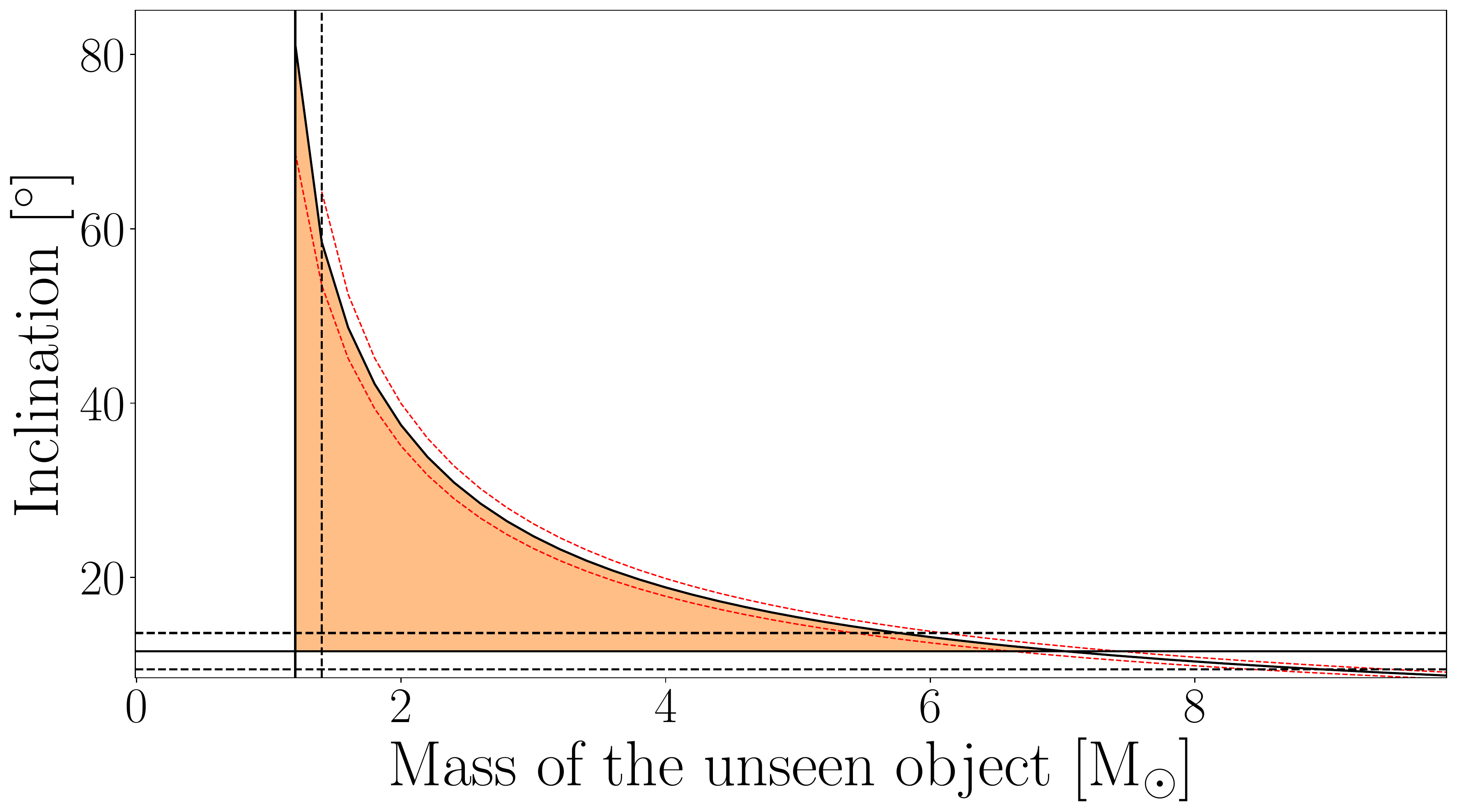}
    \caption{HD~14633}
    \end{subfigure} 
    \hfill
    \begin{subfigure}{0.33\linewidth}
    \includegraphics[width = \textwidth, trim=0 0 0 0,clip]{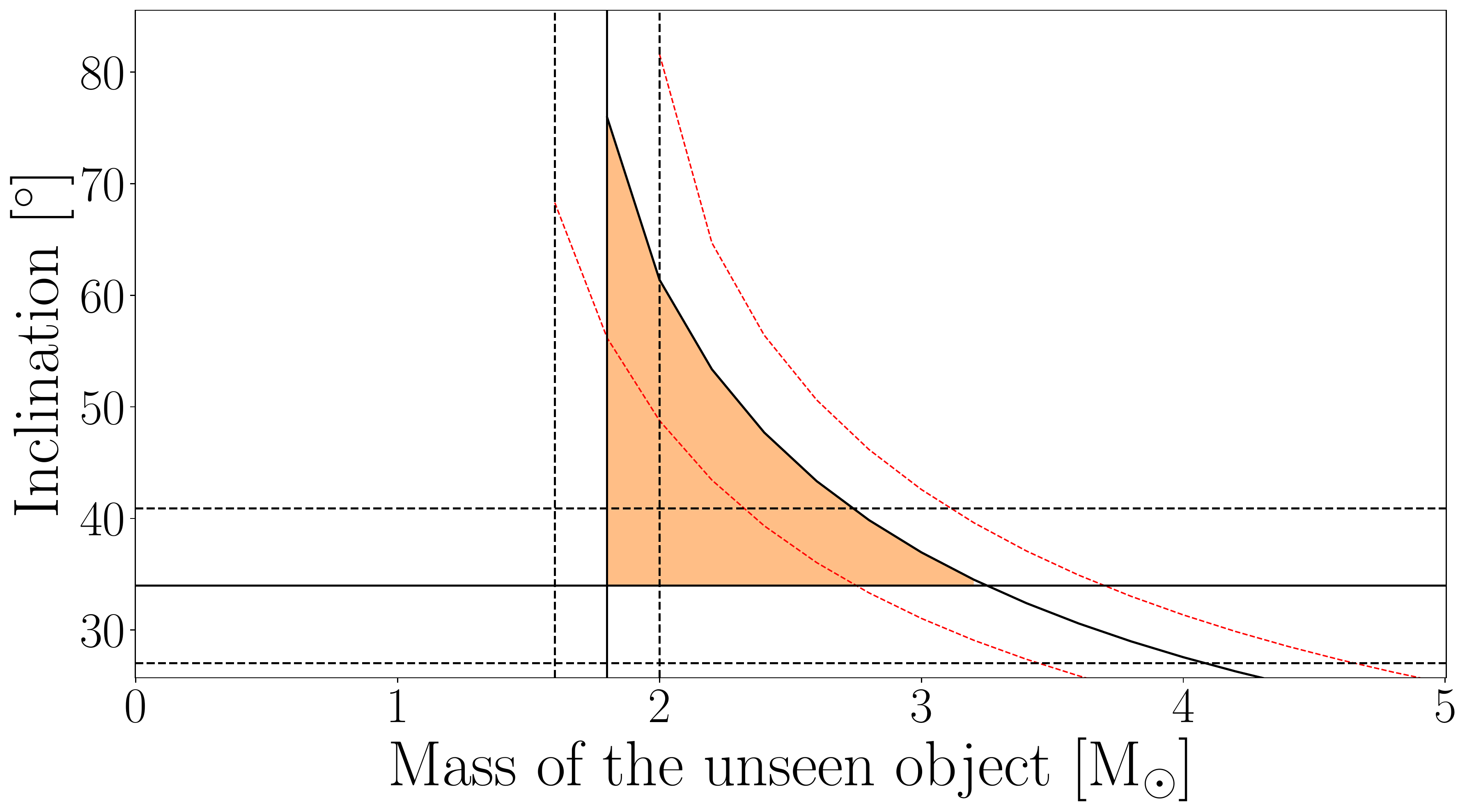}
    \caption{HD~15137}
    \end{subfigure}
    \hfill
    \begin{subfigure}{0.33\linewidth}
    \includegraphics[width = \textwidth, trim=0 0 0 0,clip]{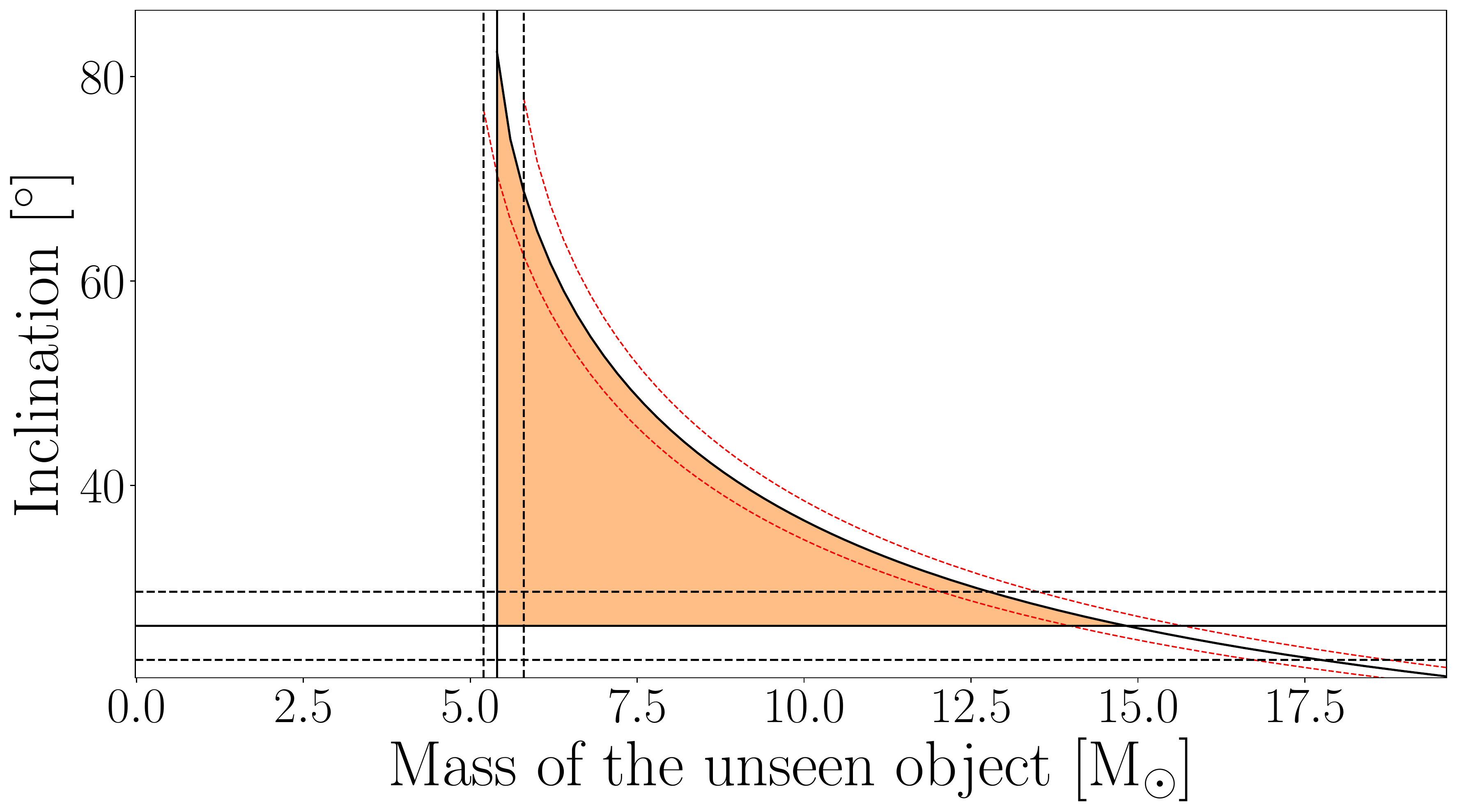}
    \caption{HD~37737}
    \end{subfigure}
    \hfill
    \begin{subfigure}{0.33\linewidth}
    \includegraphics[width = \textwidth, trim=0 0 0 0,clip]{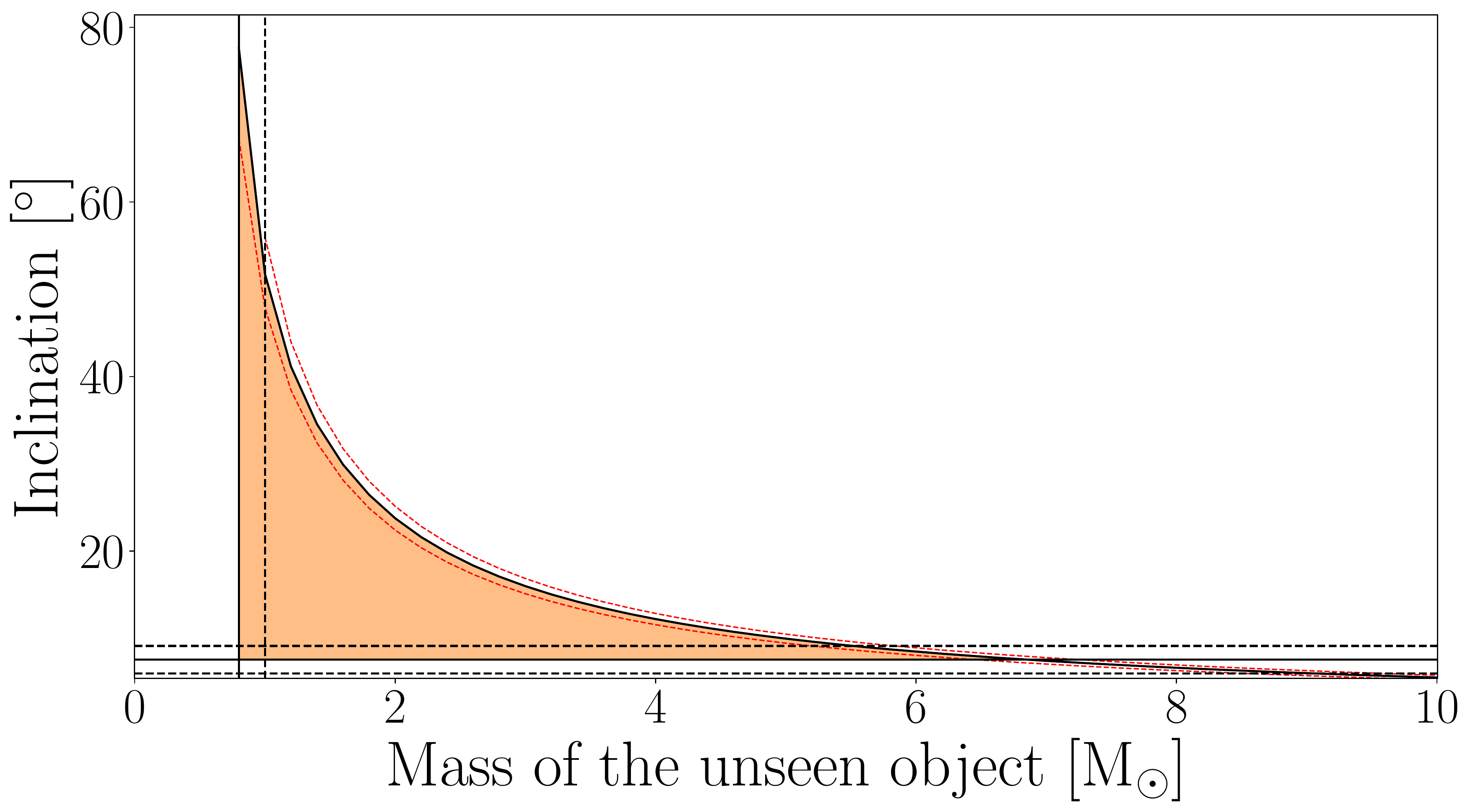}
    \caption{HD~46573}
    \end{subfigure} 
    \hfill
    \begin{subfigure}{0.33\linewidth}
    \includegraphics[width = \textwidth, trim=0 0 0 0,clip]{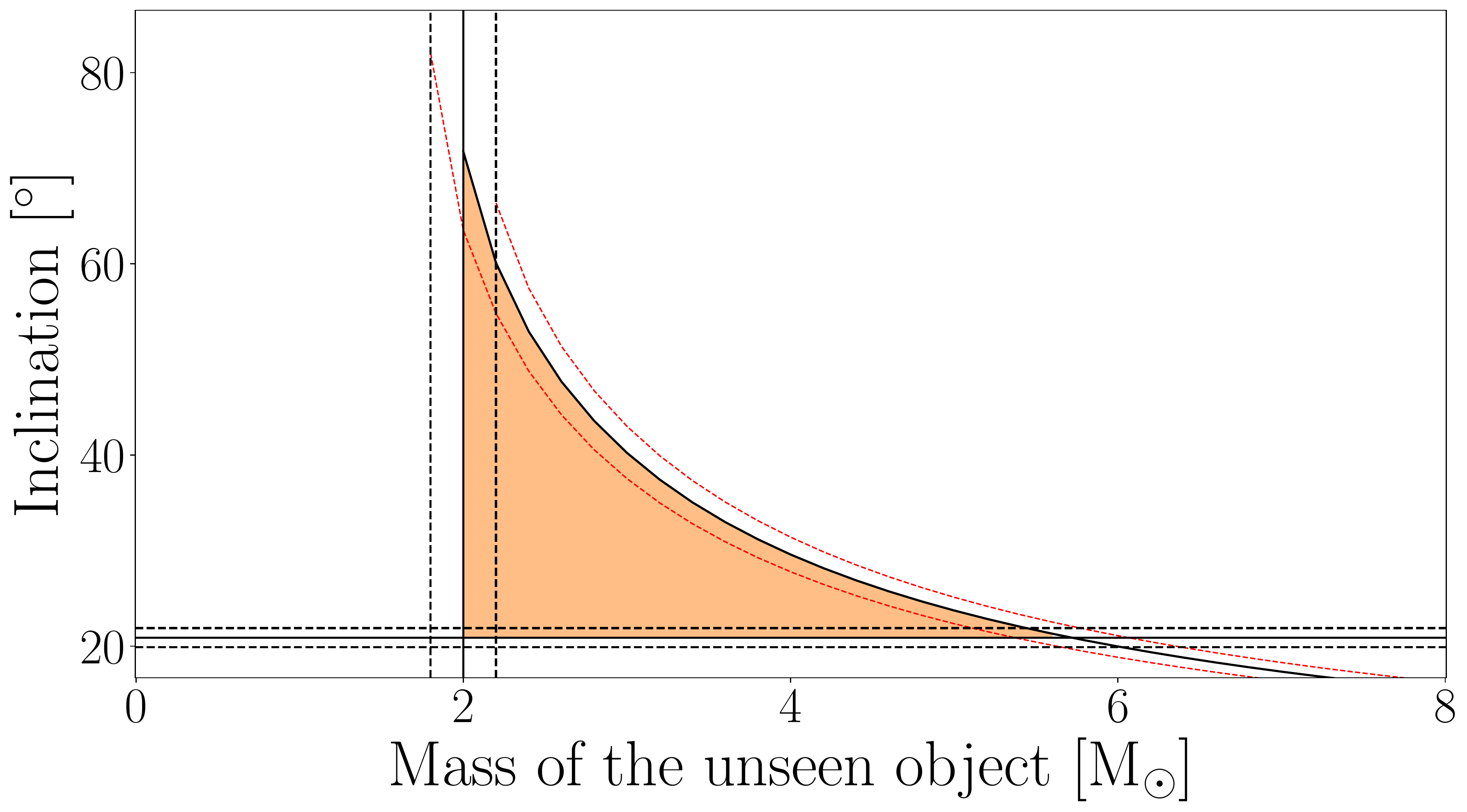}
    \caption{HD~74194}
    \end{subfigure} 
    \hfill
    \begin{subfigure}{0.33\linewidth}
    \includegraphics[width = \textwidth, trim=0 0 0 0,clip]{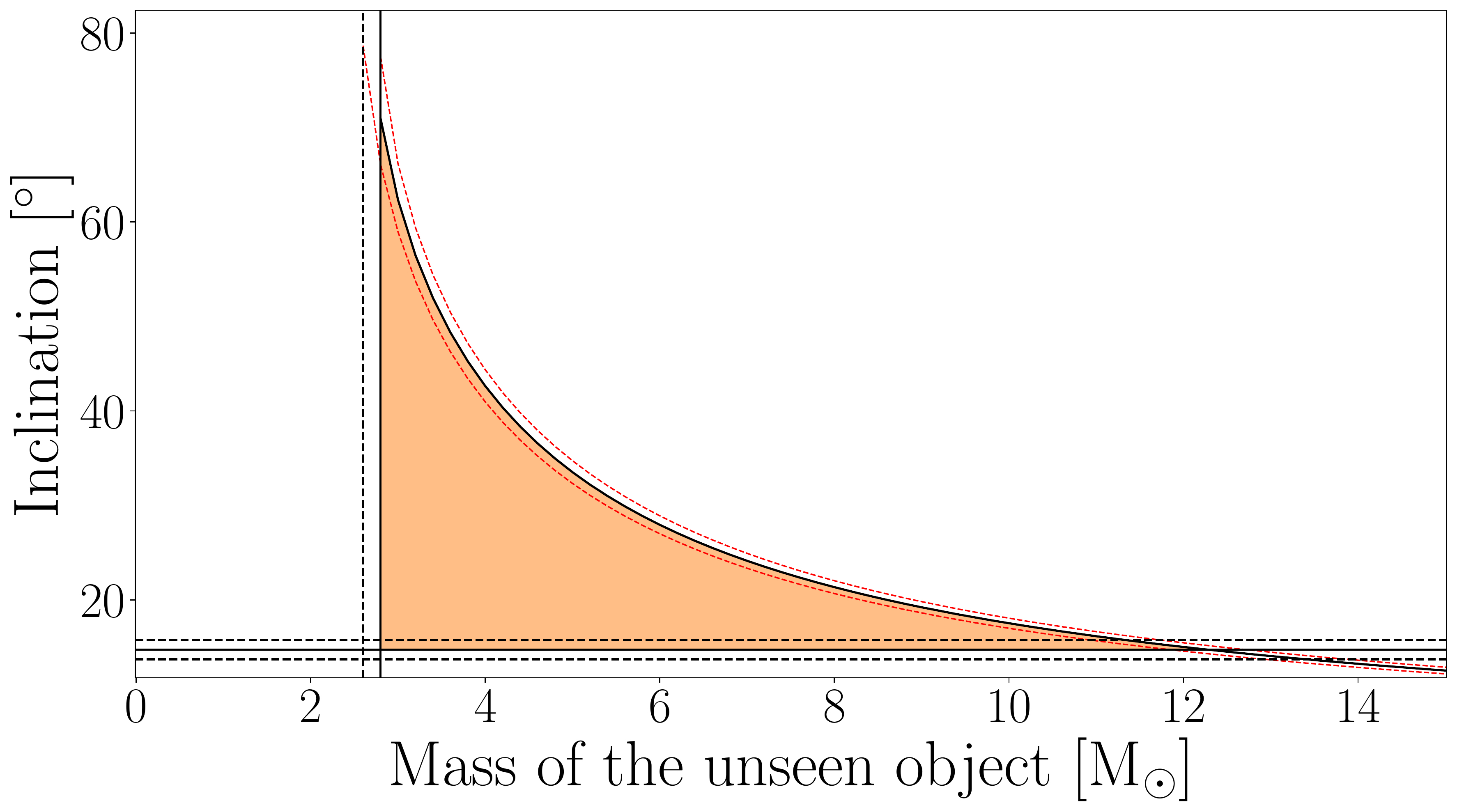} \caption{HD~75211}
    \end{subfigure}
    \hfill
    \begin{subfigure}{0.33\linewidth}
    \includegraphics[width = \textwidth, trim=0 0 0 0,clip]{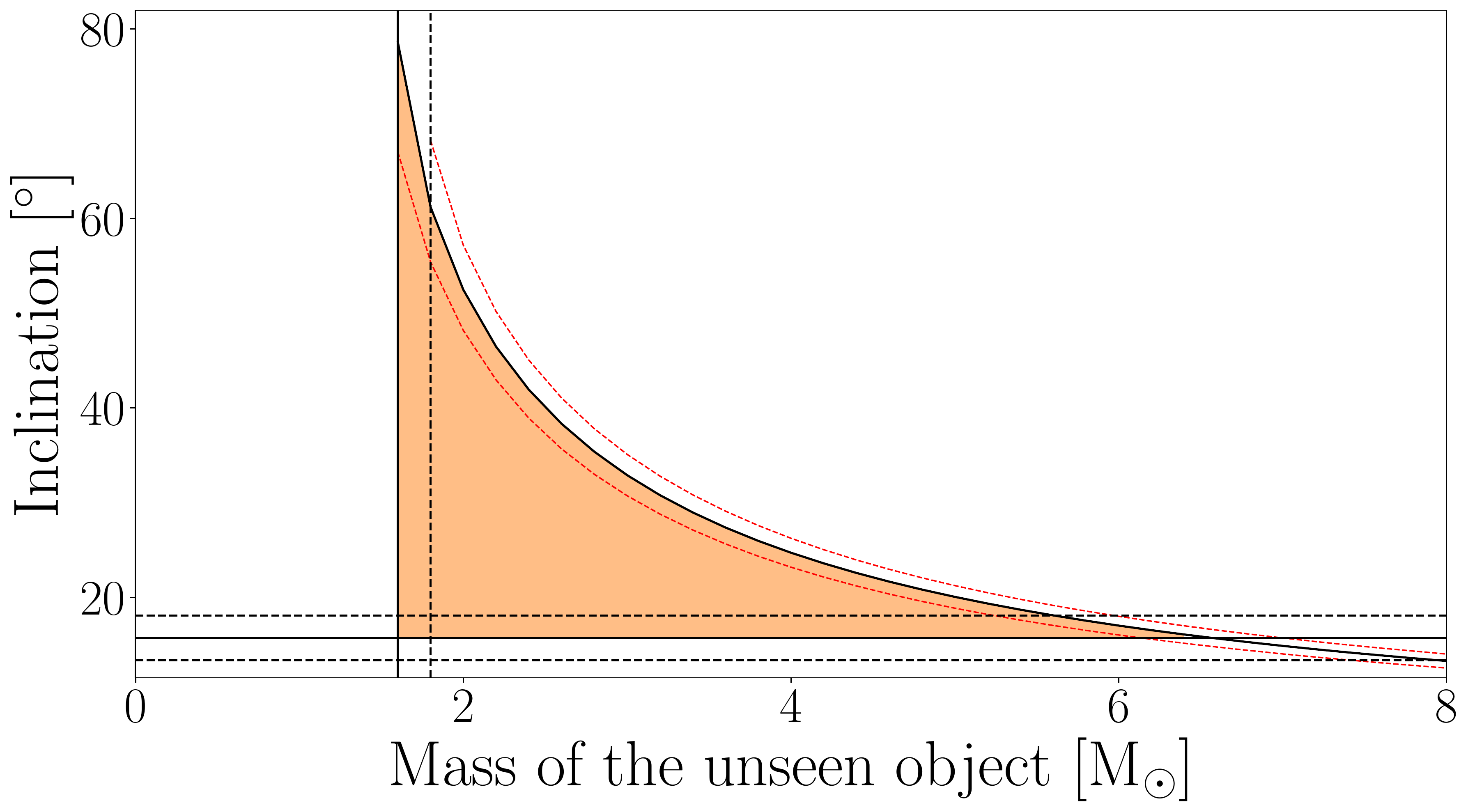}
    \caption{HD~94024}
    \end{subfigure} 
    \hfill
    \begin{subfigure}{0.33\linewidth}
    \includegraphics[width = \textwidth, trim=0 0 0 0,clip]{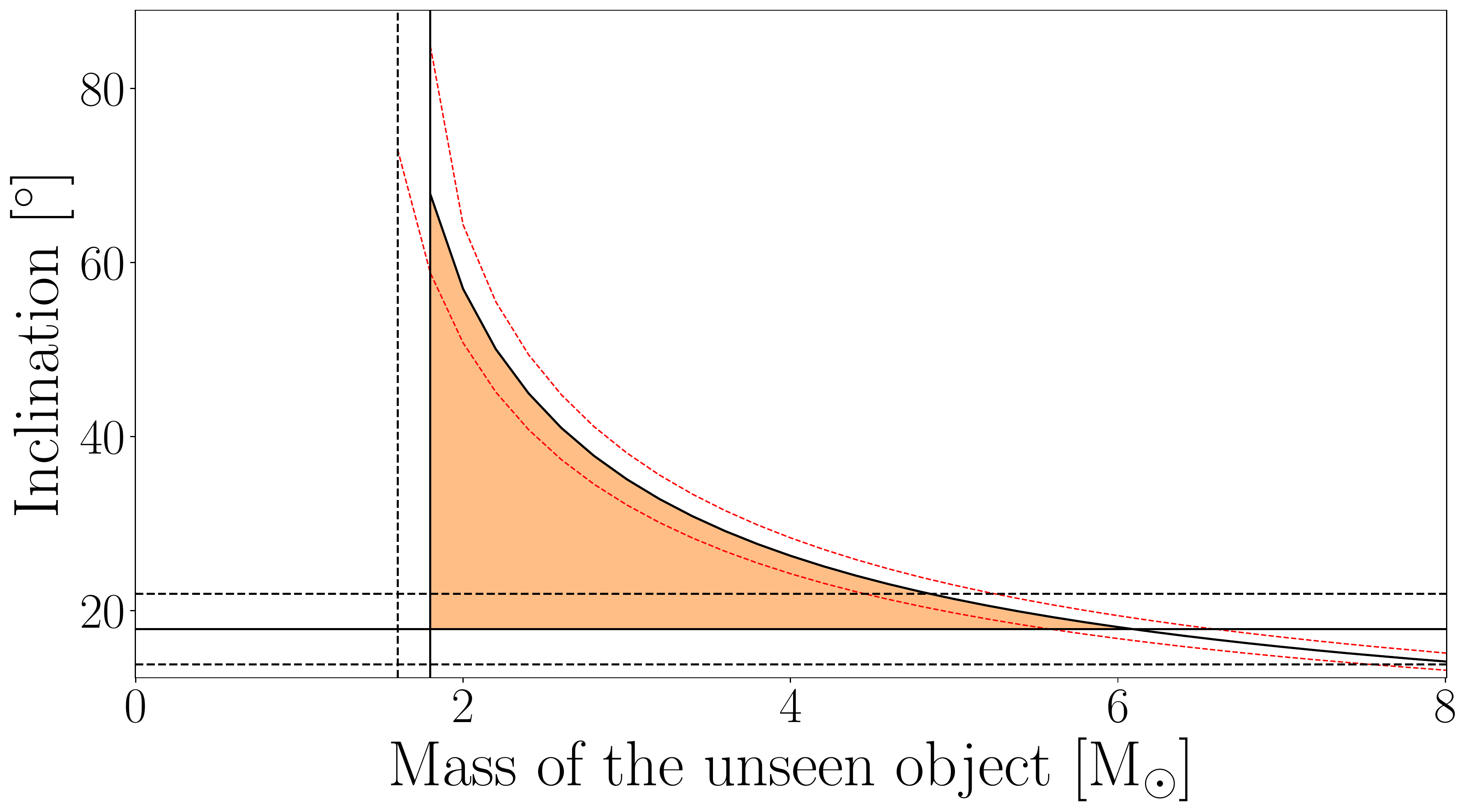}
    \caption{HD~105627}
    \end{subfigure}
    \hfill
    \begin{subfigure}{0.33\linewidth}
    \includegraphics[width = \textwidth, trim=0 0 0 0,clip]{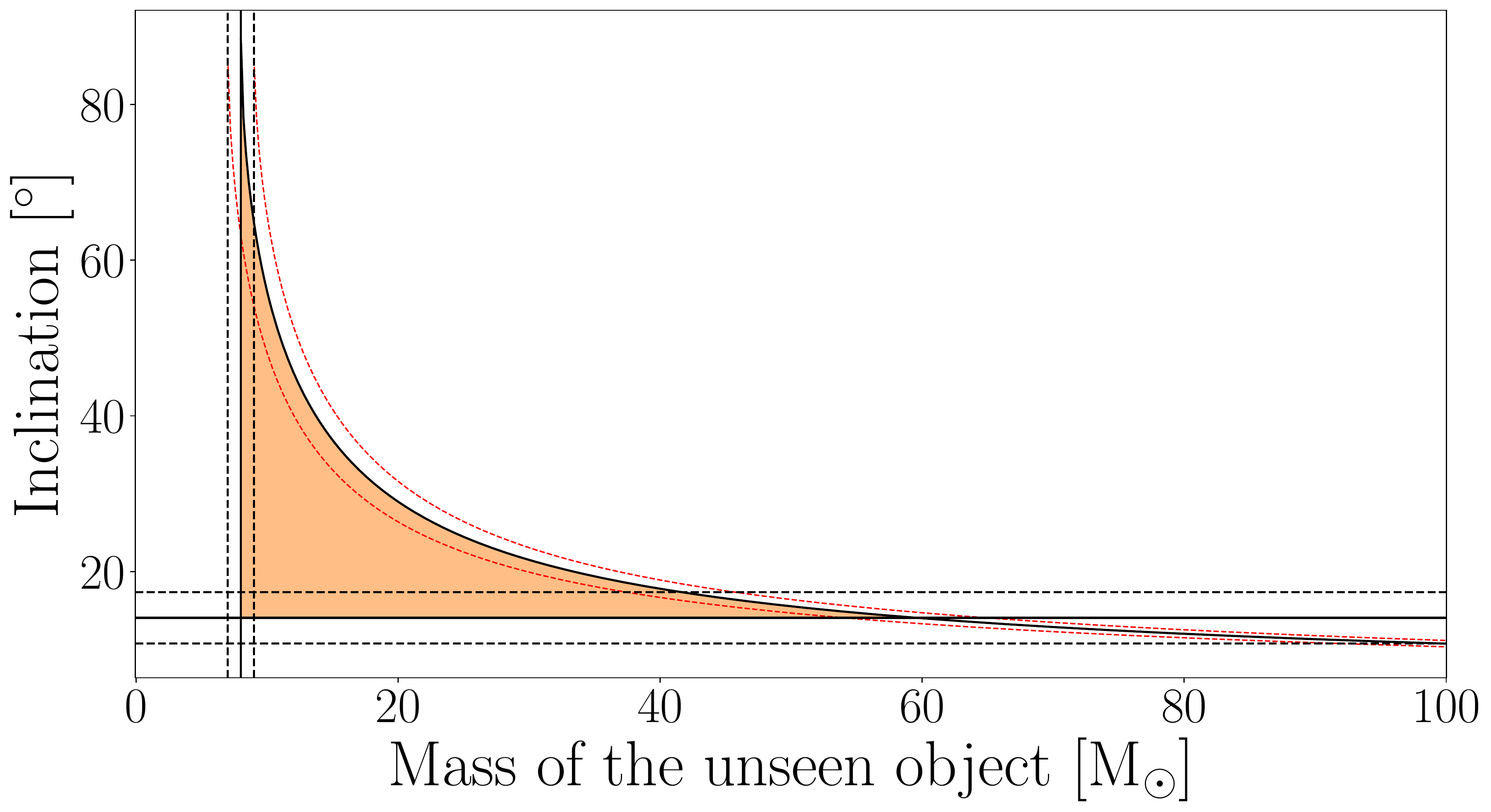}
    \caption{HD~130298}
    \end{subfigure}
    \hfill
    \begin{subfigure}{0.33\linewidth}
    \includegraphics[width = \textwidth, trim=0 0 0 0,clip]{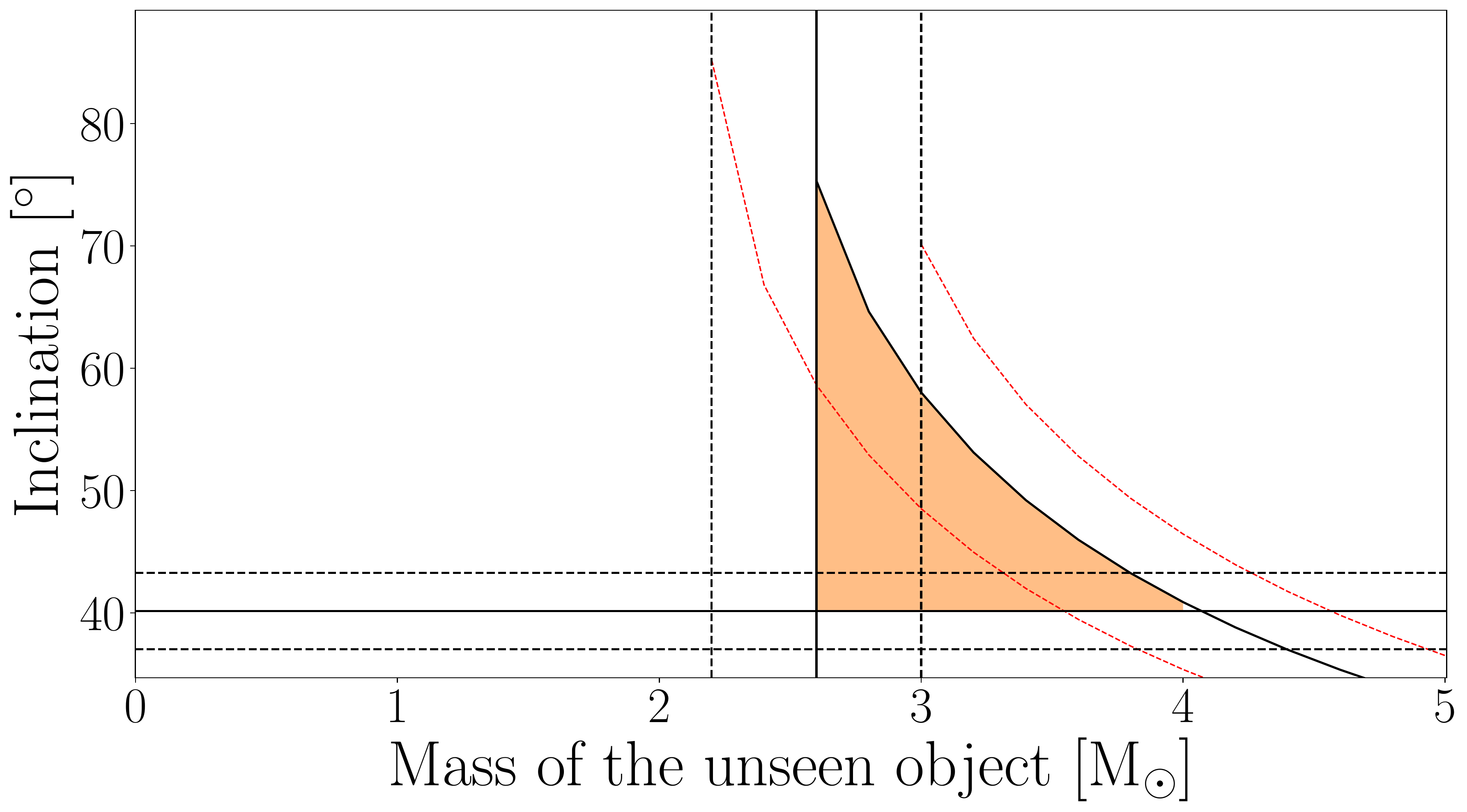}   \caption{HD~165174}
    \end{subfigure}
    \hfill
    \begin{subfigure}{0.33\linewidth}
    \includegraphics[width = \textwidth, trim=0 0 0 0,clip]{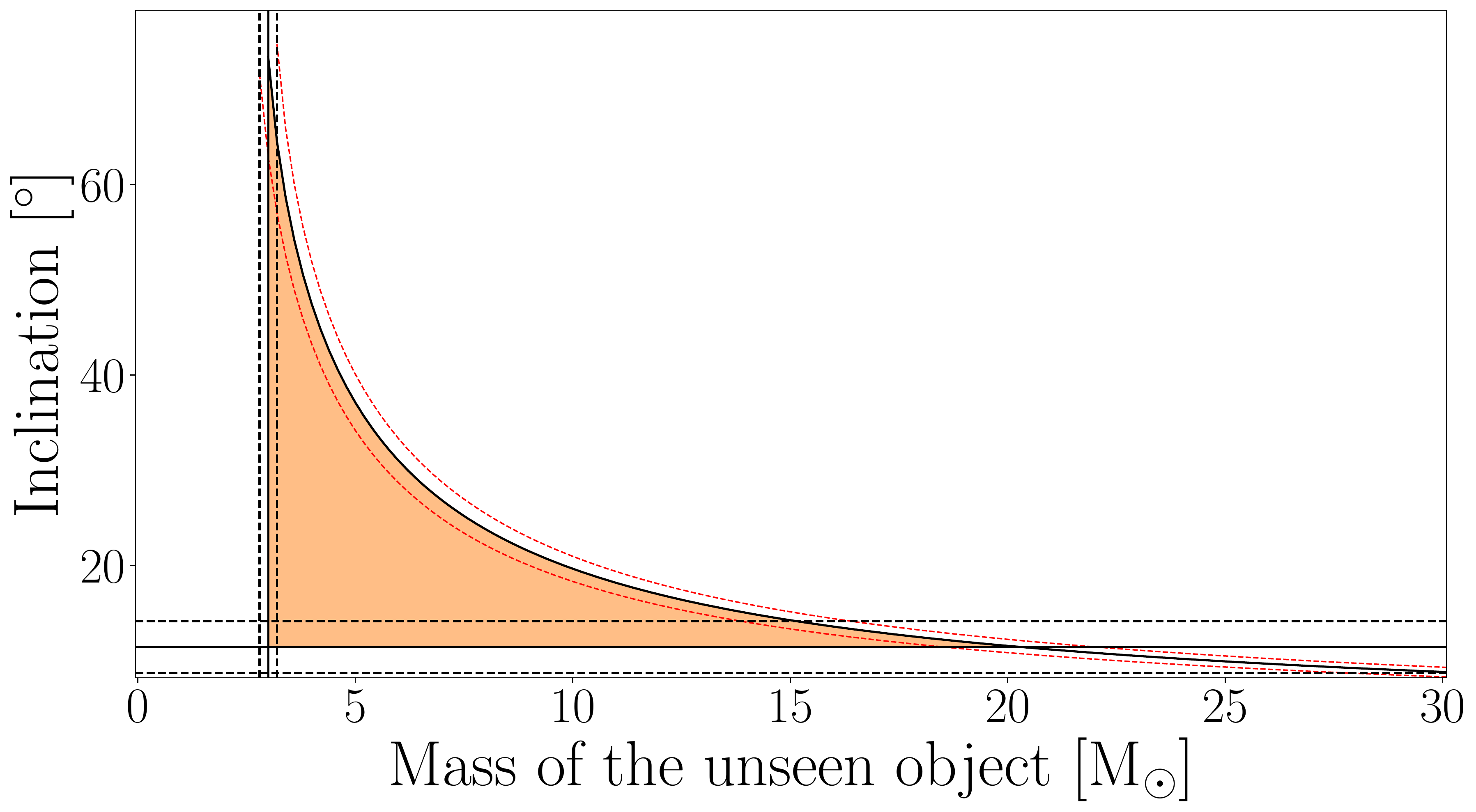}
    \caption{HD~229234}
    \end{subfigure} 
    \hfill
    \begin{subfigure}{0.33\linewidth}
    \includegraphics[width = \textwidth, trim=0 0 0 0,clip]{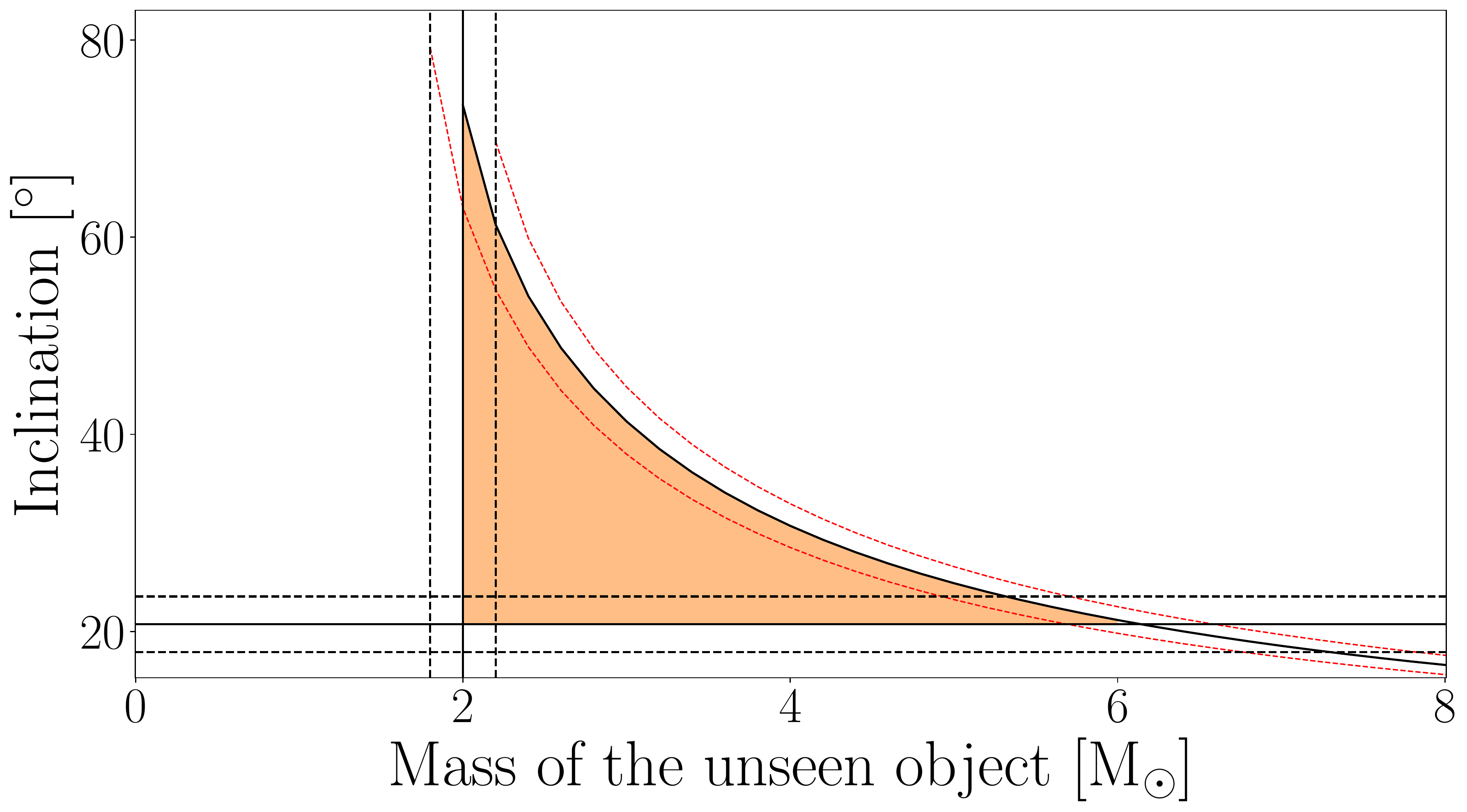}
    \caption{HD~308813}
    \end{subfigure}
    \hfill 
    \begin{subfigure}{0.33\linewidth}
    \includegraphics[width = \textwidth, trim=0 0 0 0,clip]{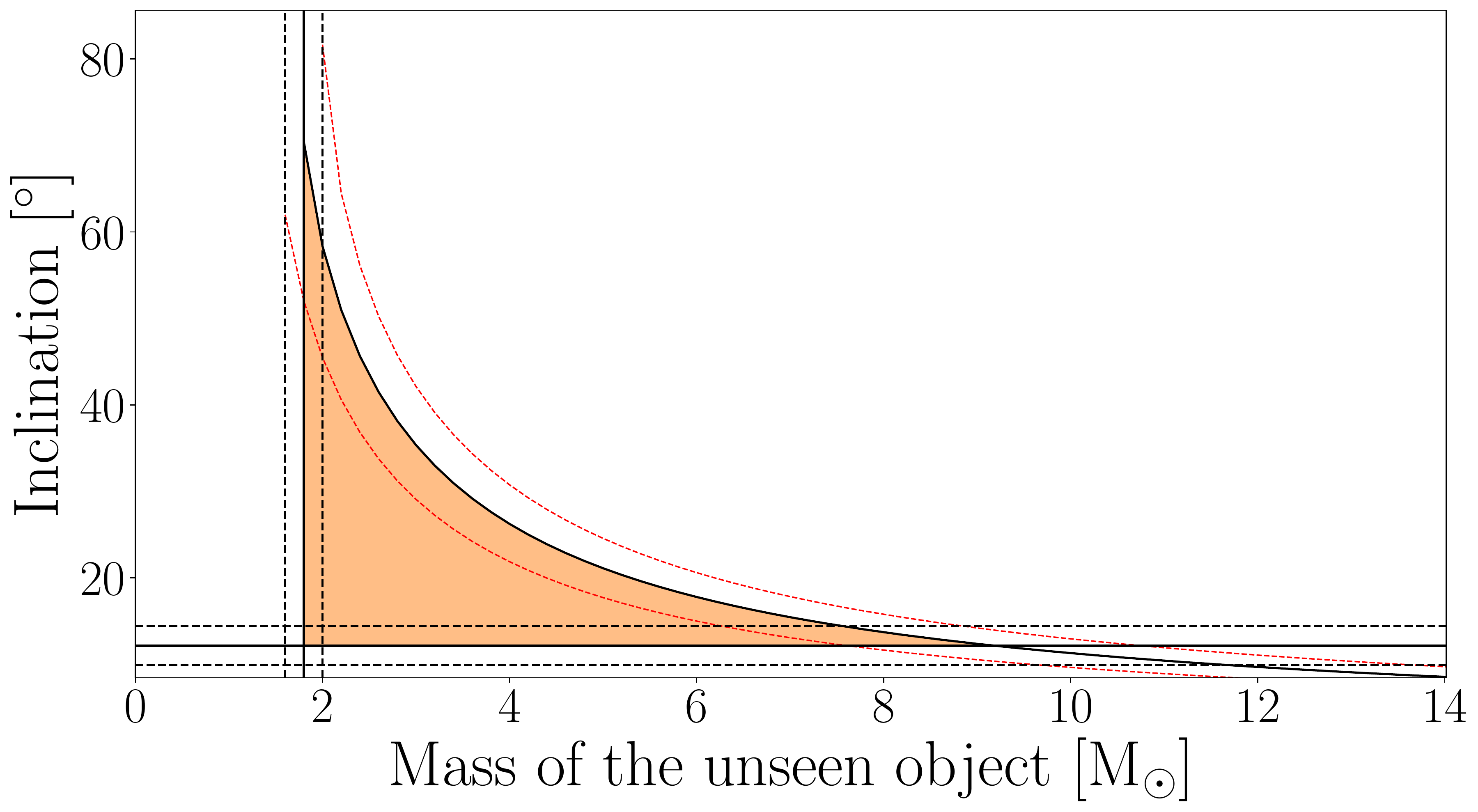}
    \caption{LS~5039}
    \end{subfigure} 
    \caption{Inclinations as a function of the secondary mass for all the SB1 systems, where no spectroscopic signatures of the secondaries were found. These diagrams were computed using the evolutionary masses and the radii estimated with BONNSAI. The vertical solid line indicates the minimum masses of the unseen companions, and the dashed lines show the error bars on those values. The horizontal solid line indicates the minimum inclination of the systems, and the horizontal dashed lines show the error bars on the minimum inclinations. The orange shaded regions correspond to the possible values for the system inclinations and masses of the unseen objects. The dashed red lines indicate the error bars on the binary mass function and are computed by propagating the $1\sigma$ errors on the other parameters.}
    \label{fig:binarymassfunction}
\end{figure*}

\begin{figure*}[h!]
    \centering
    \begin{subfigure}{0.33\linewidth}
    \includegraphics[width = \textwidth, trim=0 0 0 0,clip]{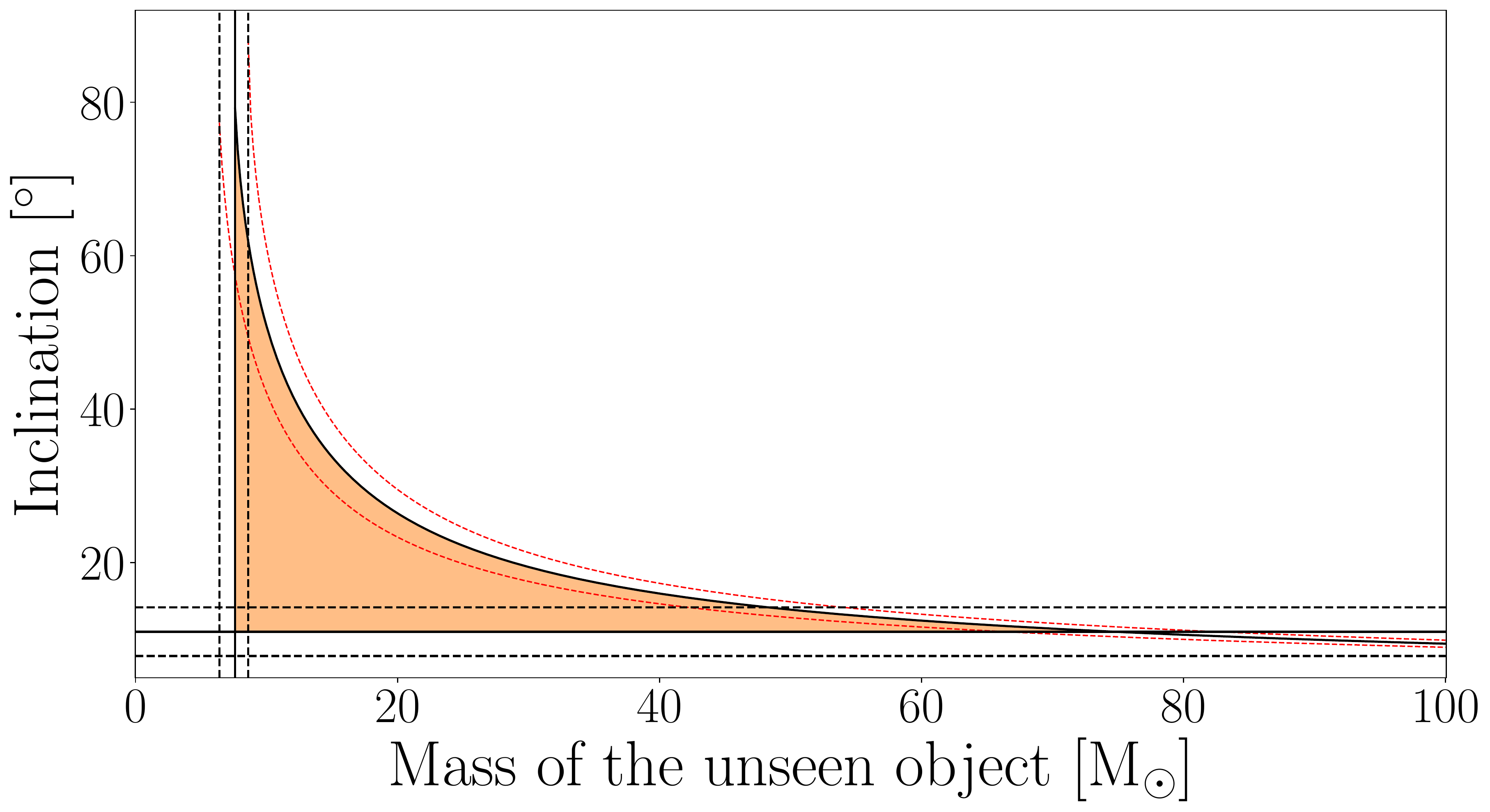}
    \caption{Cyg~X-1}
    \end{subfigure}
    \hfill
    \begin{subfigure}{0.33\linewidth}
    \includegraphics[width = \textwidth, trim=0 0 0 0,clip]{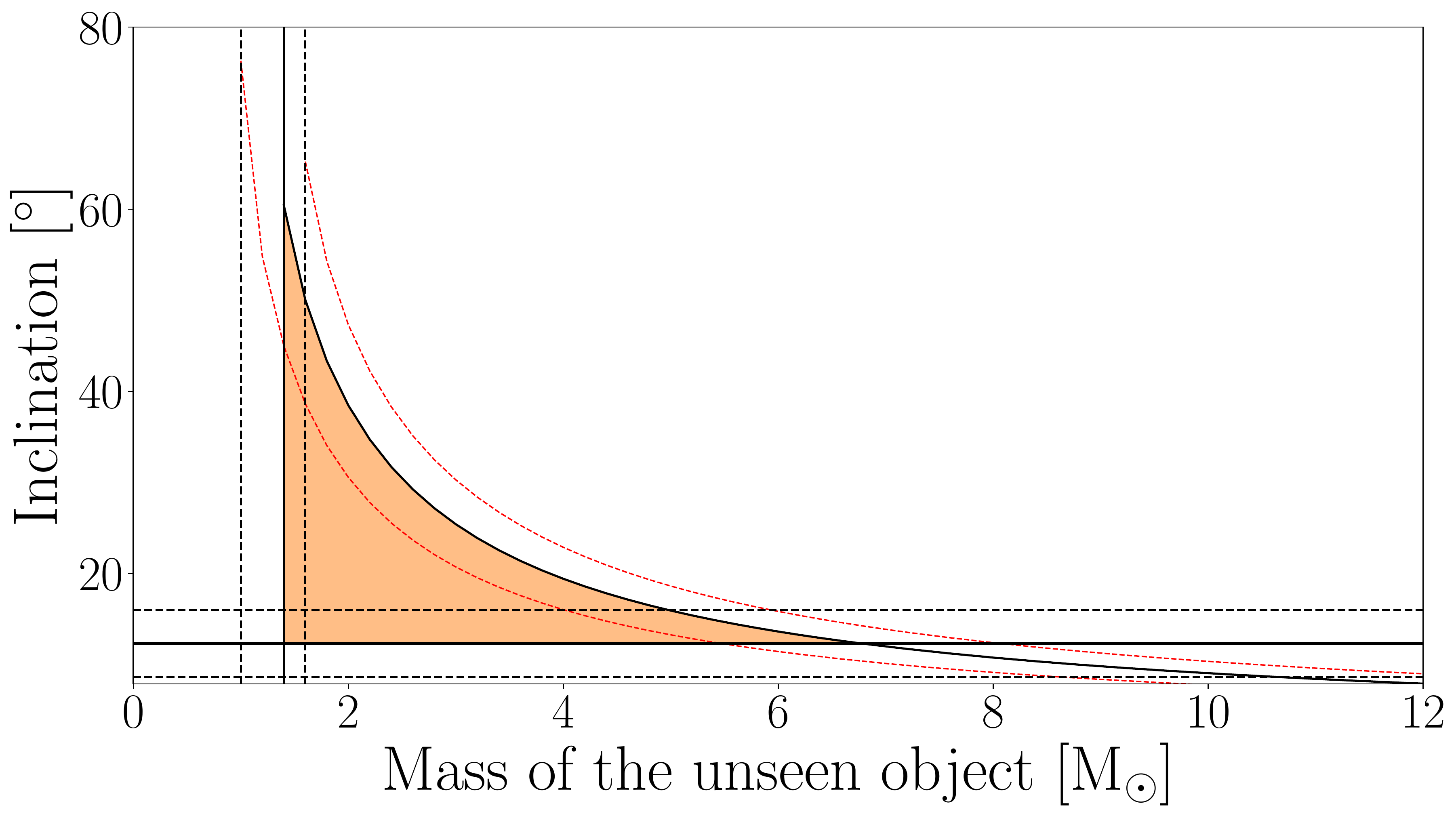}
    \caption{HD~12323}
    \end{subfigure}
    \hfill
    \begin{subfigure}{0.33\linewidth}
    \includegraphics[width = \textwidth, trim=0 0 0 0,clip]{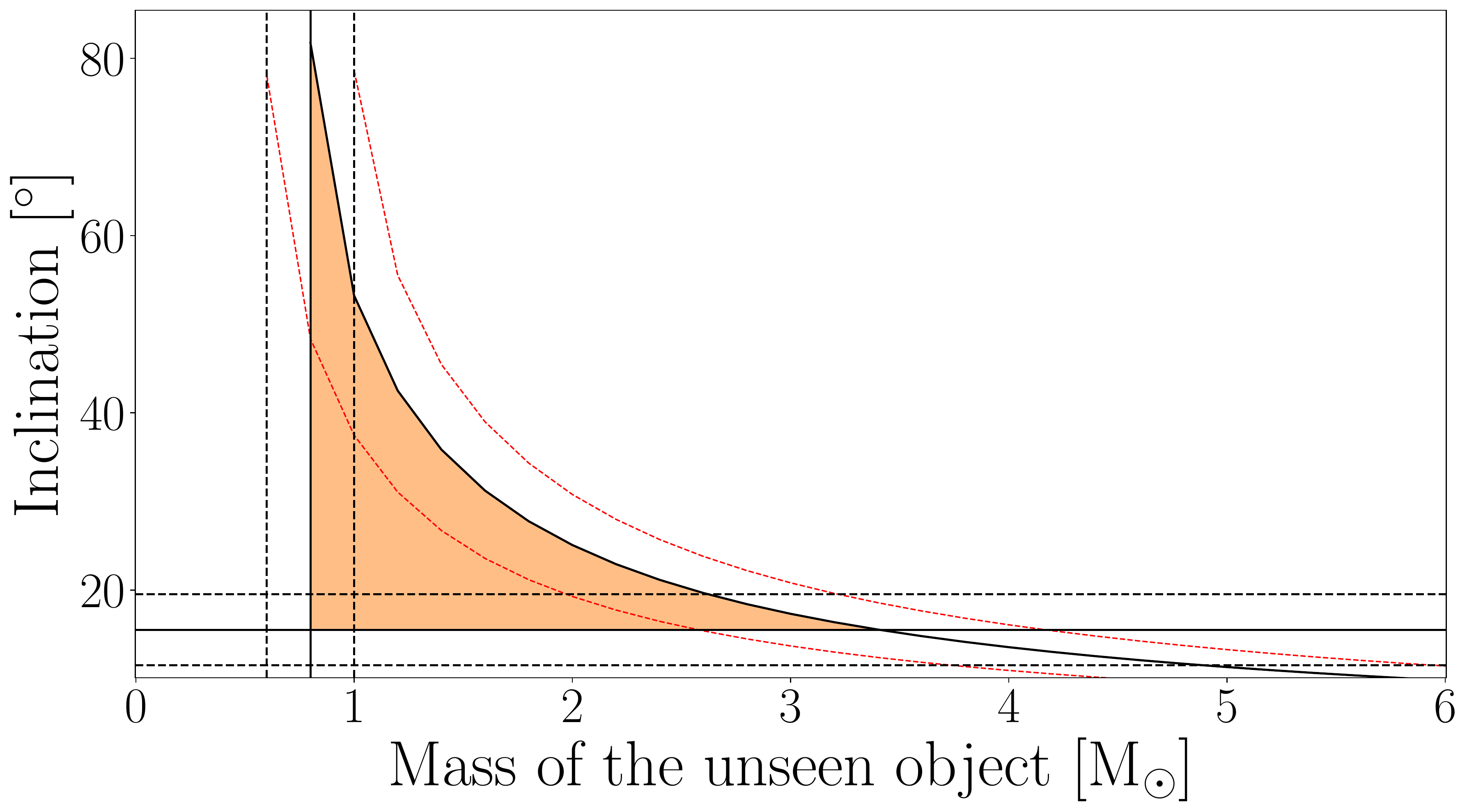}
    \caption{HD~14633}
    \end{subfigure} 
    \hfill
    \begin{subfigure}{0.33\linewidth}
    \includegraphics[width = \textwidth, trim=0 0 0 0,clip]{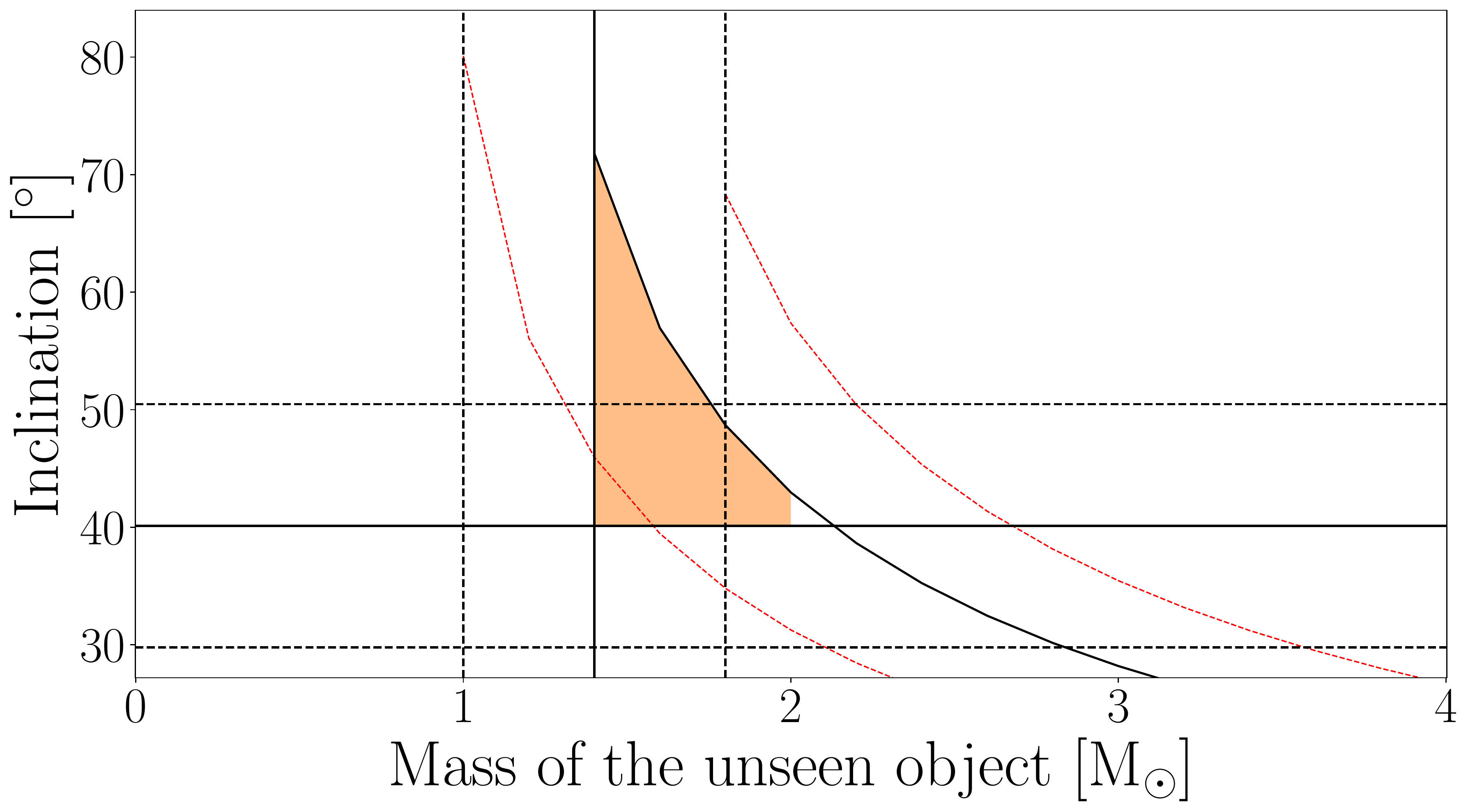}
    \caption{HD~15137}
    \end{subfigure}
    \hfill
    \begin{subfigure}{0.33\linewidth}
    \includegraphics[width = \textwidth, trim=0 0 0 0,clip]{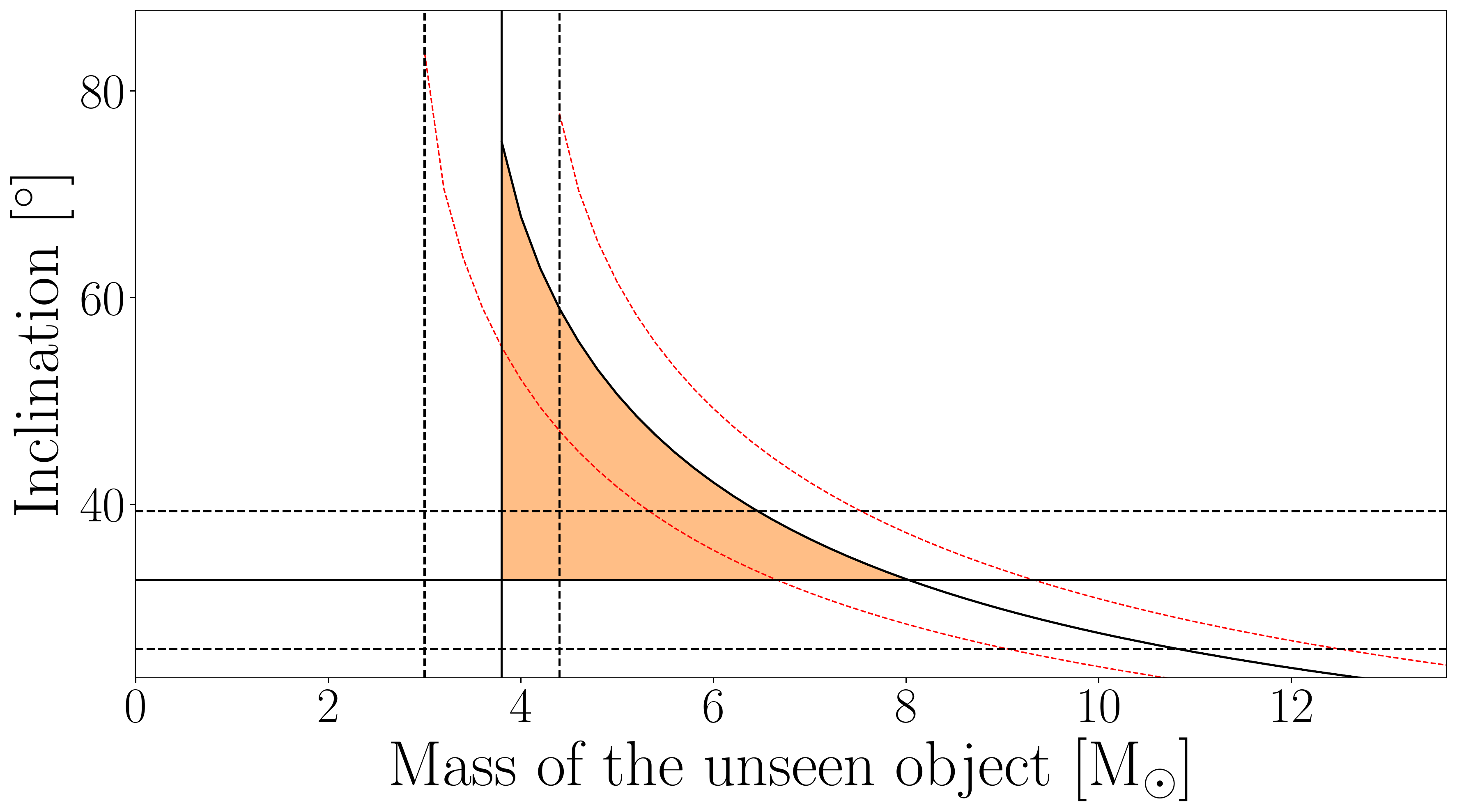}
    \caption{HD~37737}
    \end{subfigure}
    \hfill
    \begin{subfigure}{0.33\linewidth}
    \includegraphics[width = \textwidth, trim=0 0 0 0,clip]{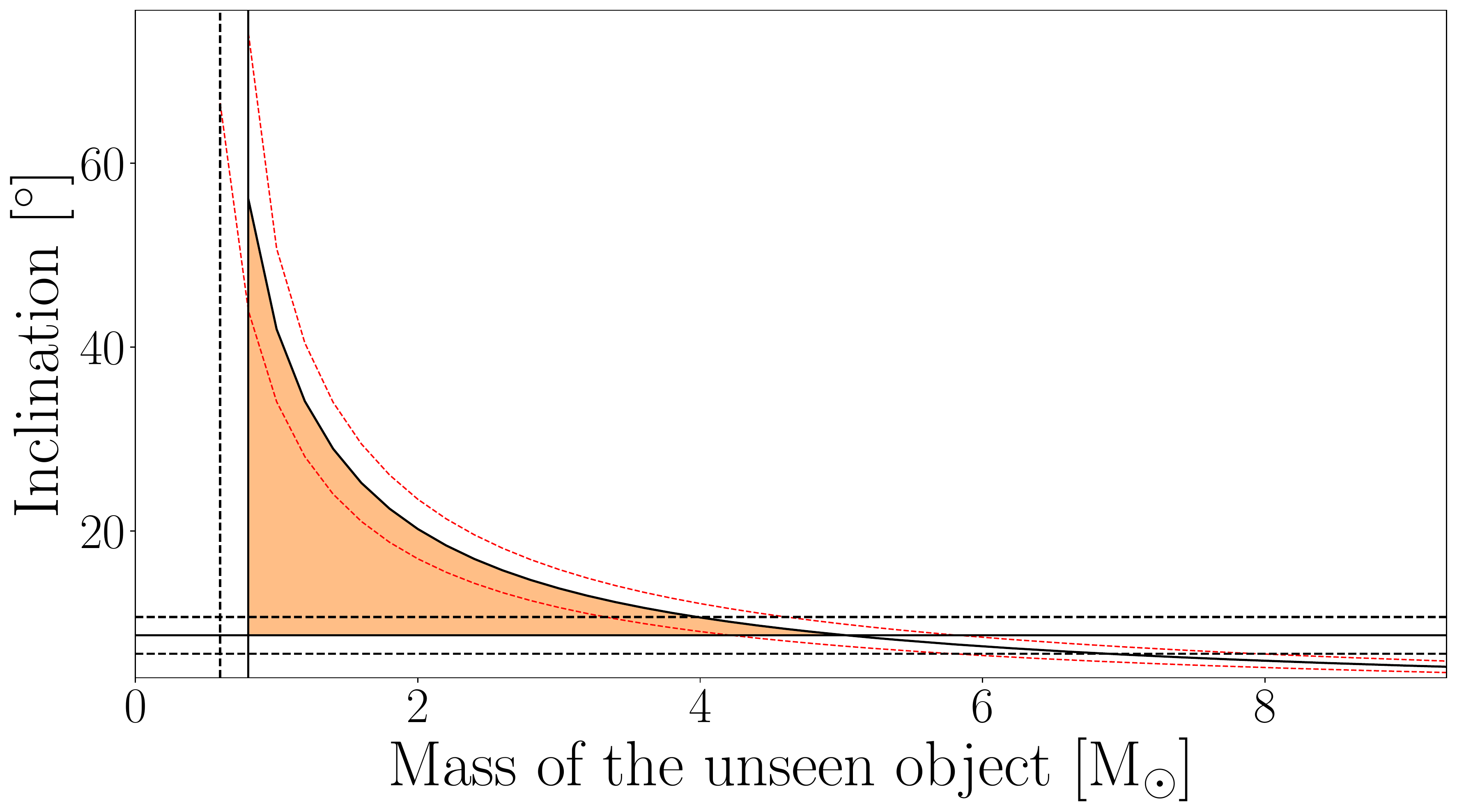}
    \caption{HD~46573}
    \end{subfigure} 
    \hfill
    \begin{subfigure}{0.33\linewidth}
    \includegraphics[width = \textwidth, trim=0 0 0 0,clip]{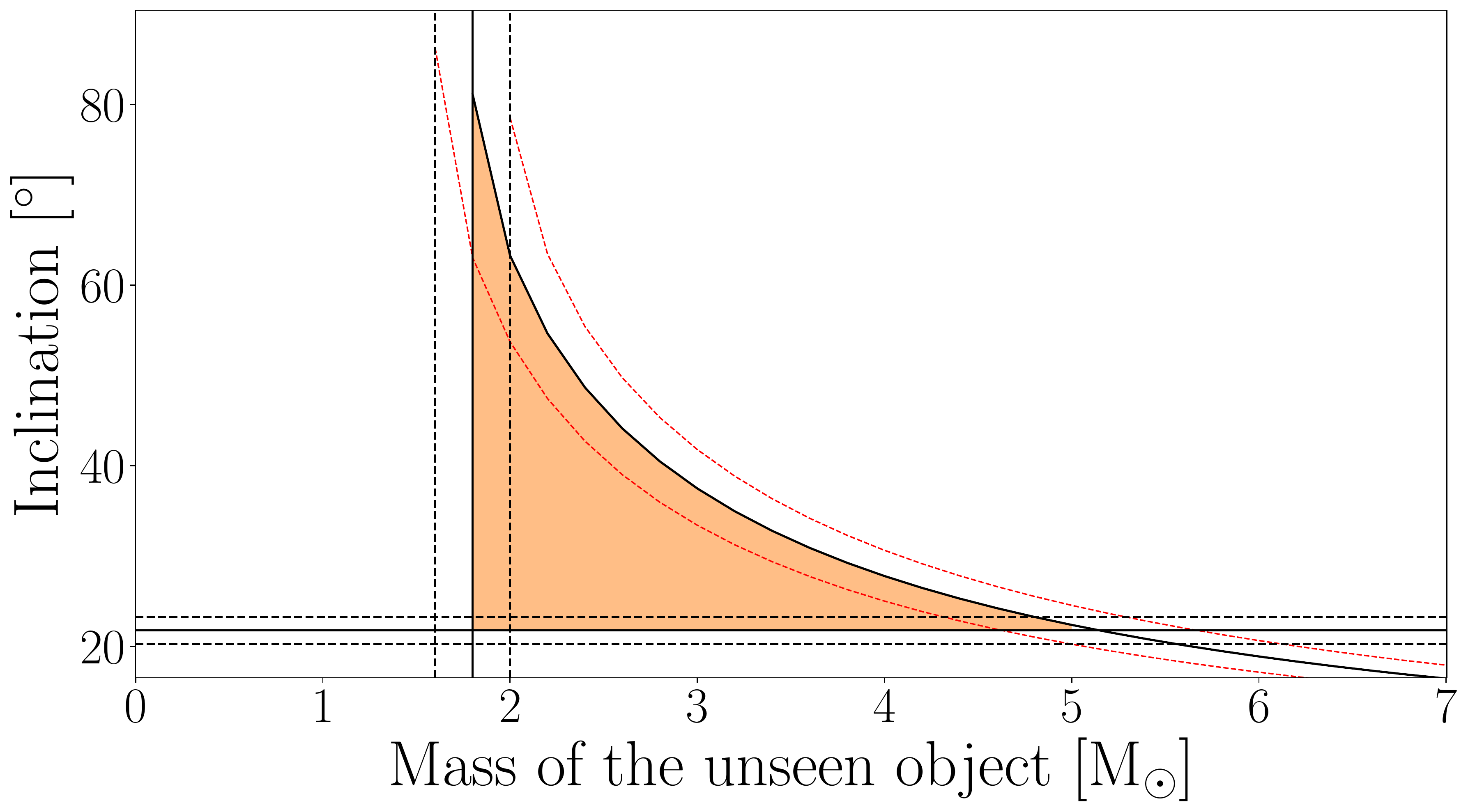}
    \caption{HD~74194}
    \end{subfigure} 
    \hfill
    \begin{subfigure}{0.33\linewidth}
    \includegraphics[width = \textwidth, trim=0 0 0 0,clip]{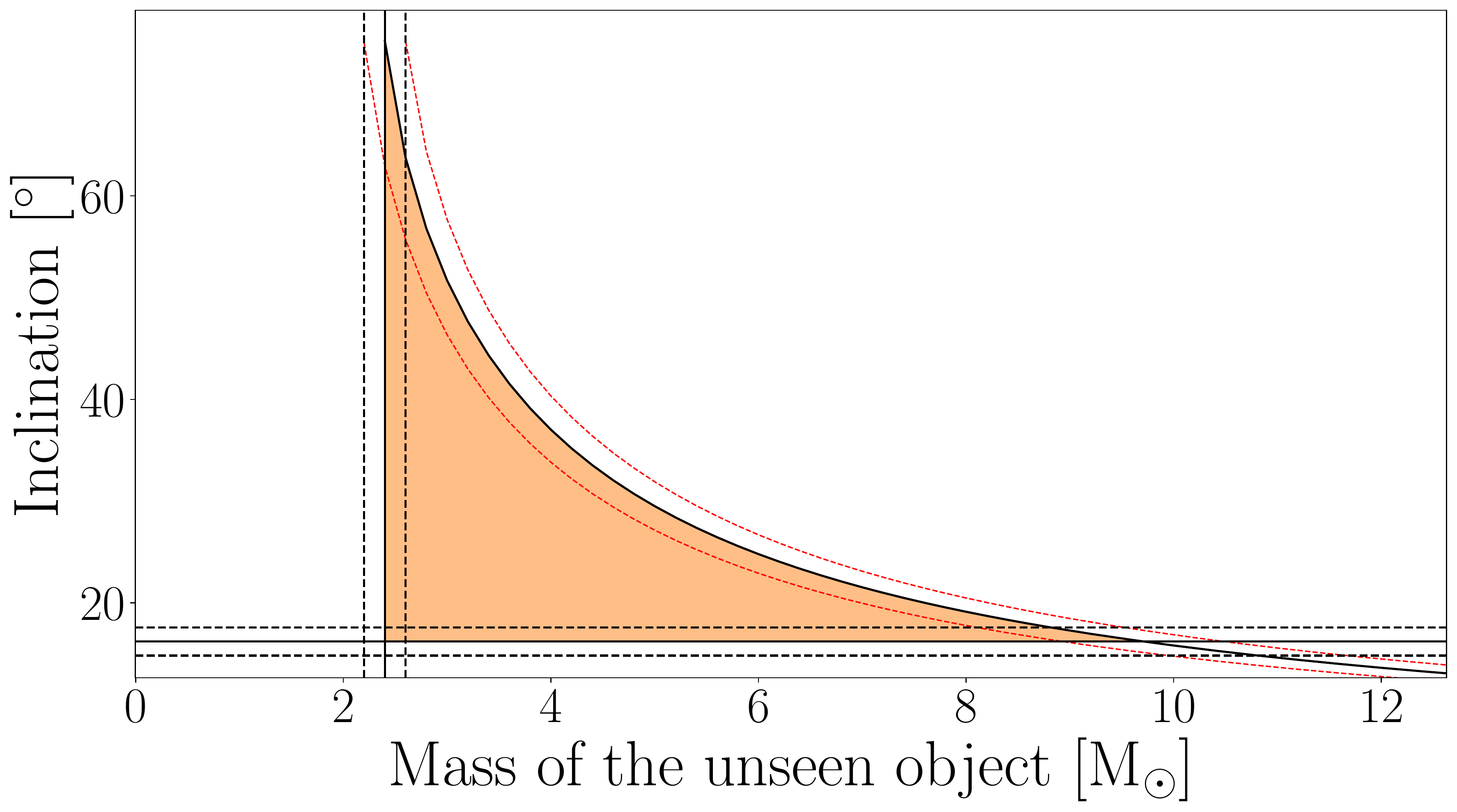} 
    \caption{HD~75211}
    \end{subfigure}
    \hfill
    \begin{subfigure}{0.33\linewidth}
    \includegraphics[width = \textwidth, trim=0 0 0 0,clip]{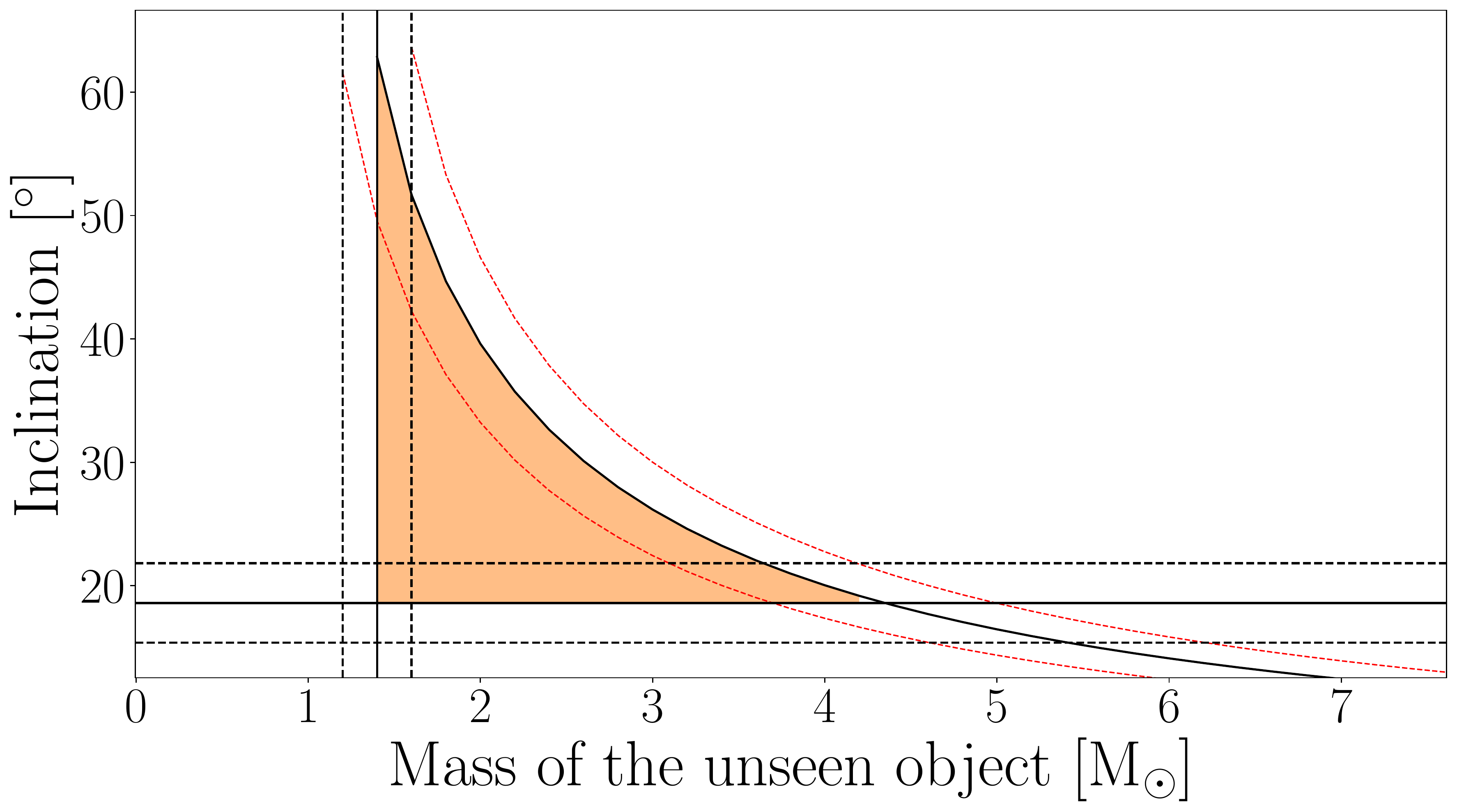}
    \caption{HD~94024}
    \end{subfigure} 
    \hfill
    \begin{subfigure}{0.33\linewidth}
    \includegraphics[width = \textwidth, trim=0 0 0 0,clip]{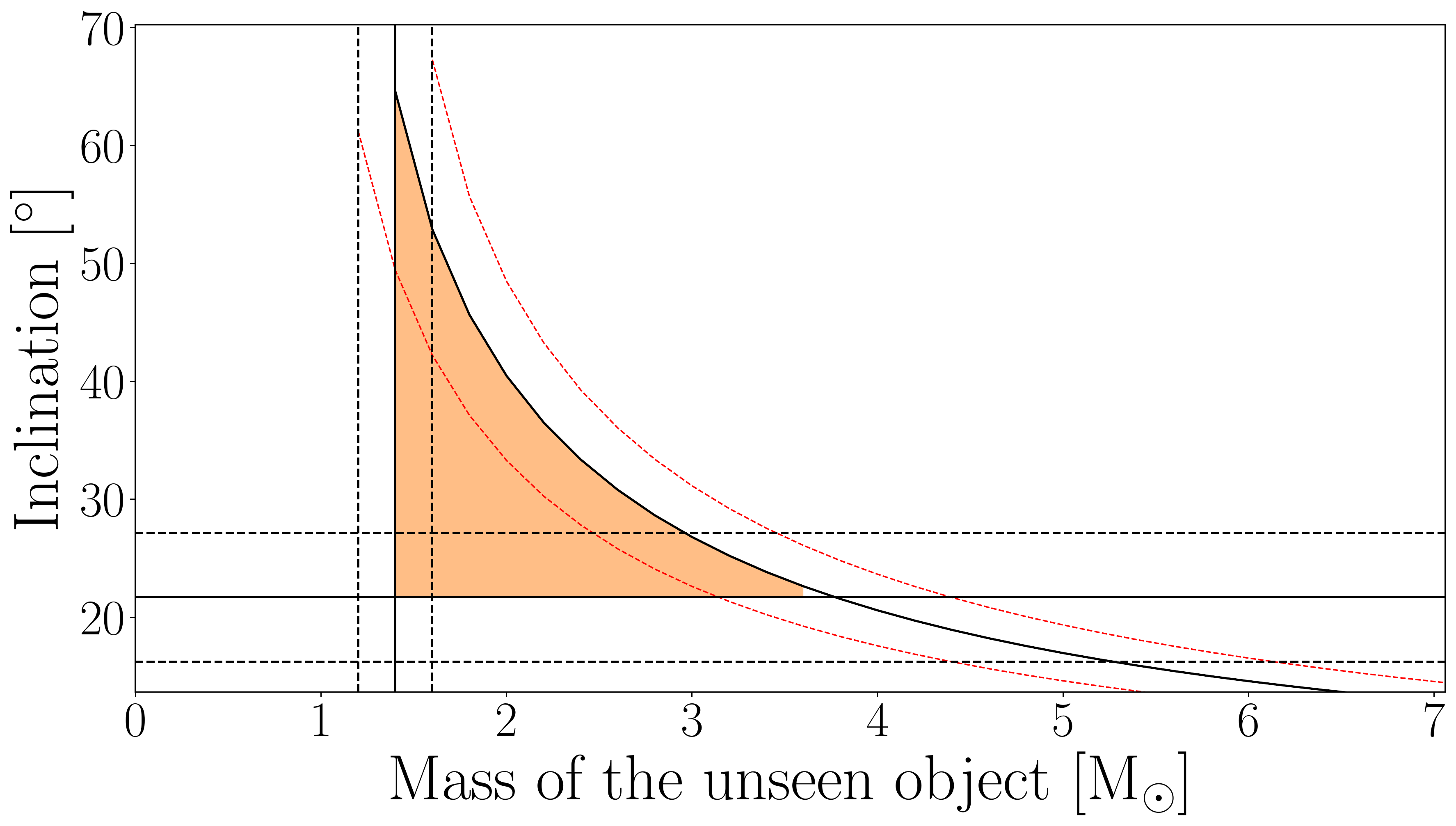}
    \caption{HD~105627}
    \end{subfigure}
    \hfill
    \begin{subfigure}{0.33\linewidth}
    \includegraphics[width = \textwidth, trim=0 0 0 0,clip]{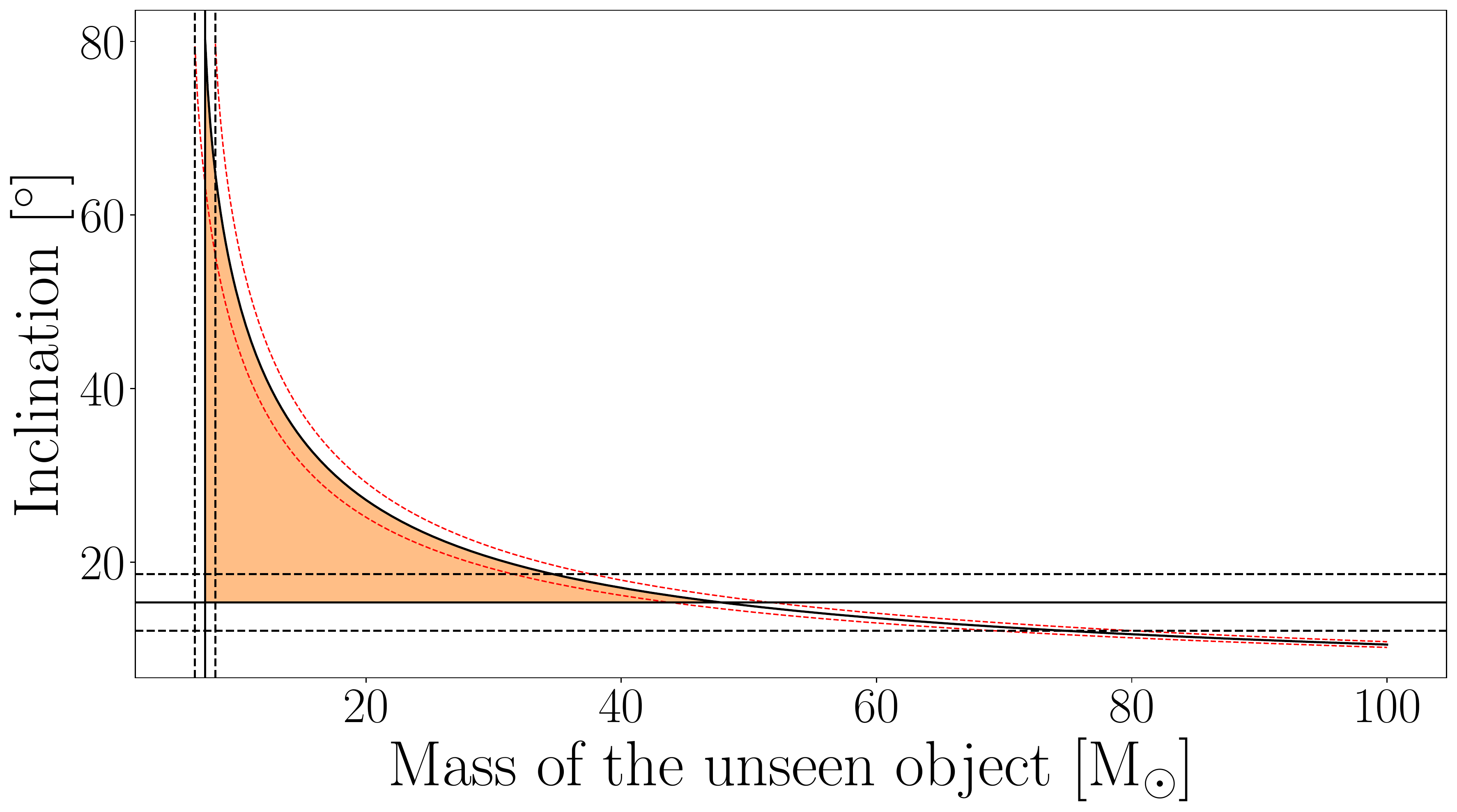}
    \caption{HD~130298}
    \end{subfigure}
    \hfill
    \begin{subfigure}{0.33\linewidth}
    \includegraphics[width = \textwidth, trim=0 0 0 0,clip]{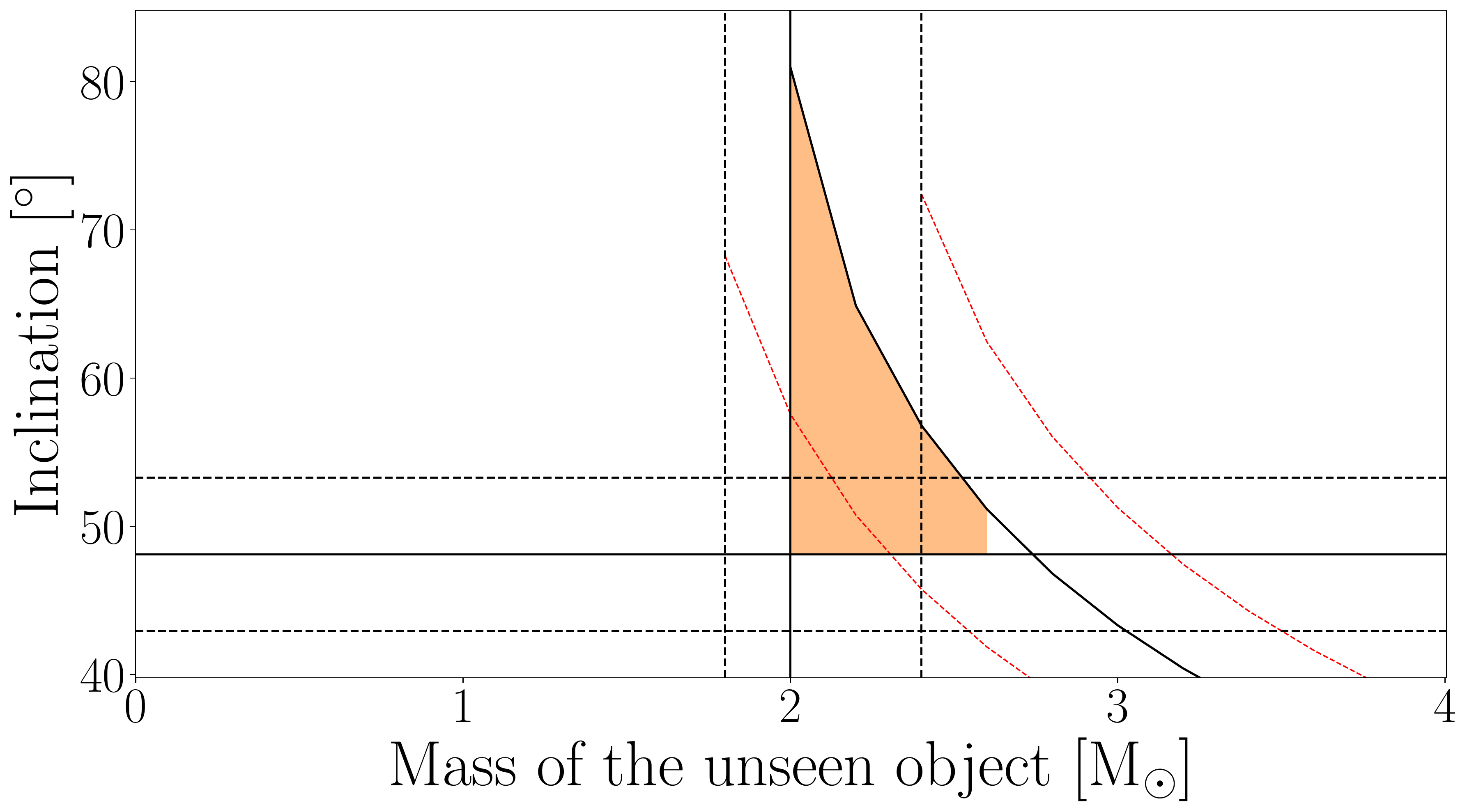}   \caption{HD~165174}
    \end{subfigure}
    \hfill
    \begin{subfigure}{0.33\linewidth}
    \includegraphics[width = \textwidth, trim=0 0 0 0,clip]{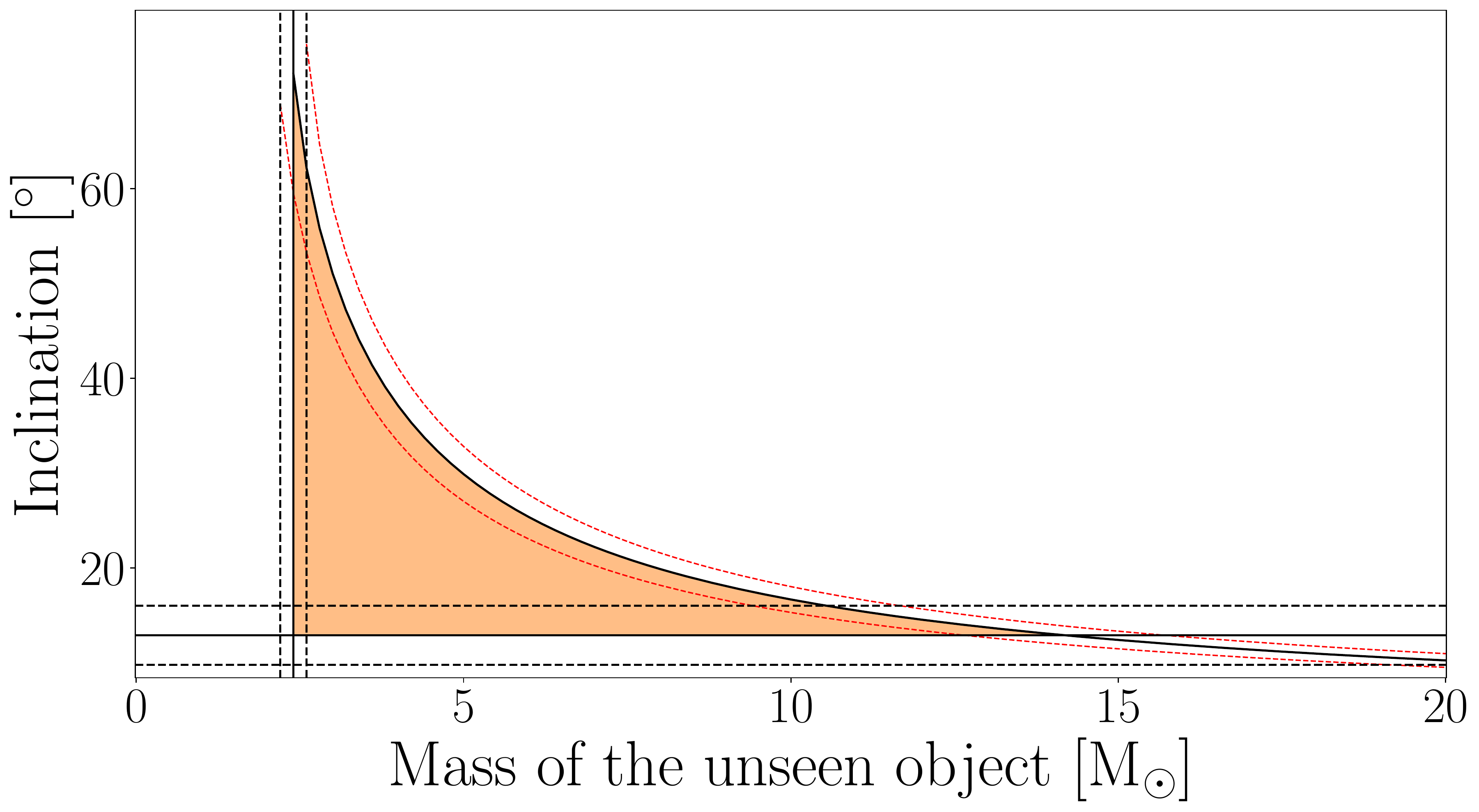}
    \caption{HD~229234}
    \end{subfigure} 
    \hfill
    \begin{subfigure}{0.33\linewidth}
    \includegraphics[width = \textwidth, trim=0 0 0 0,clip]{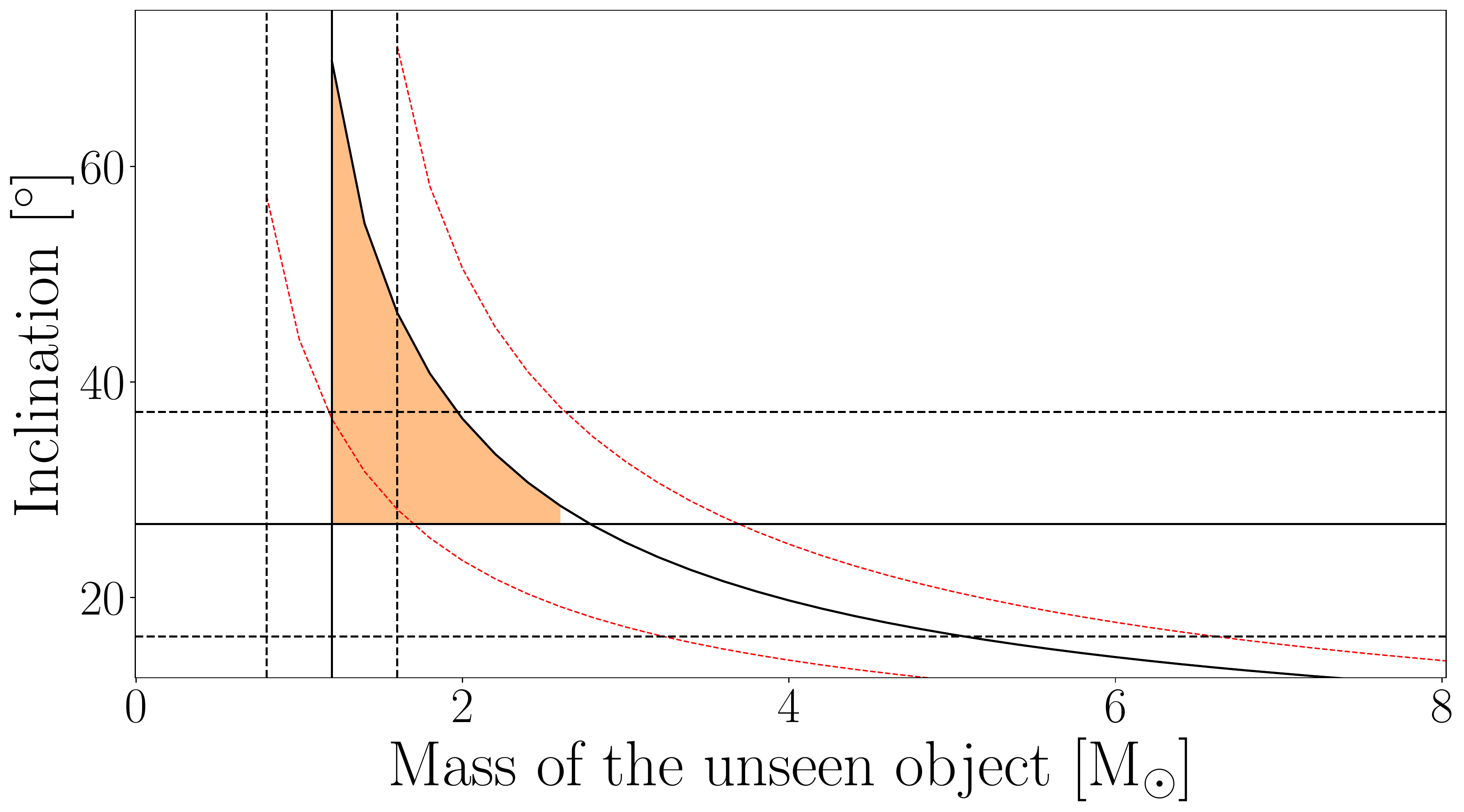}
    \caption{HD~308813}
    \end{subfigure}
    \hfill 
    \begin{subfigure}{0.33\linewidth}
    \includegraphics[width = \textwidth, trim=0 0 0 0,clip]{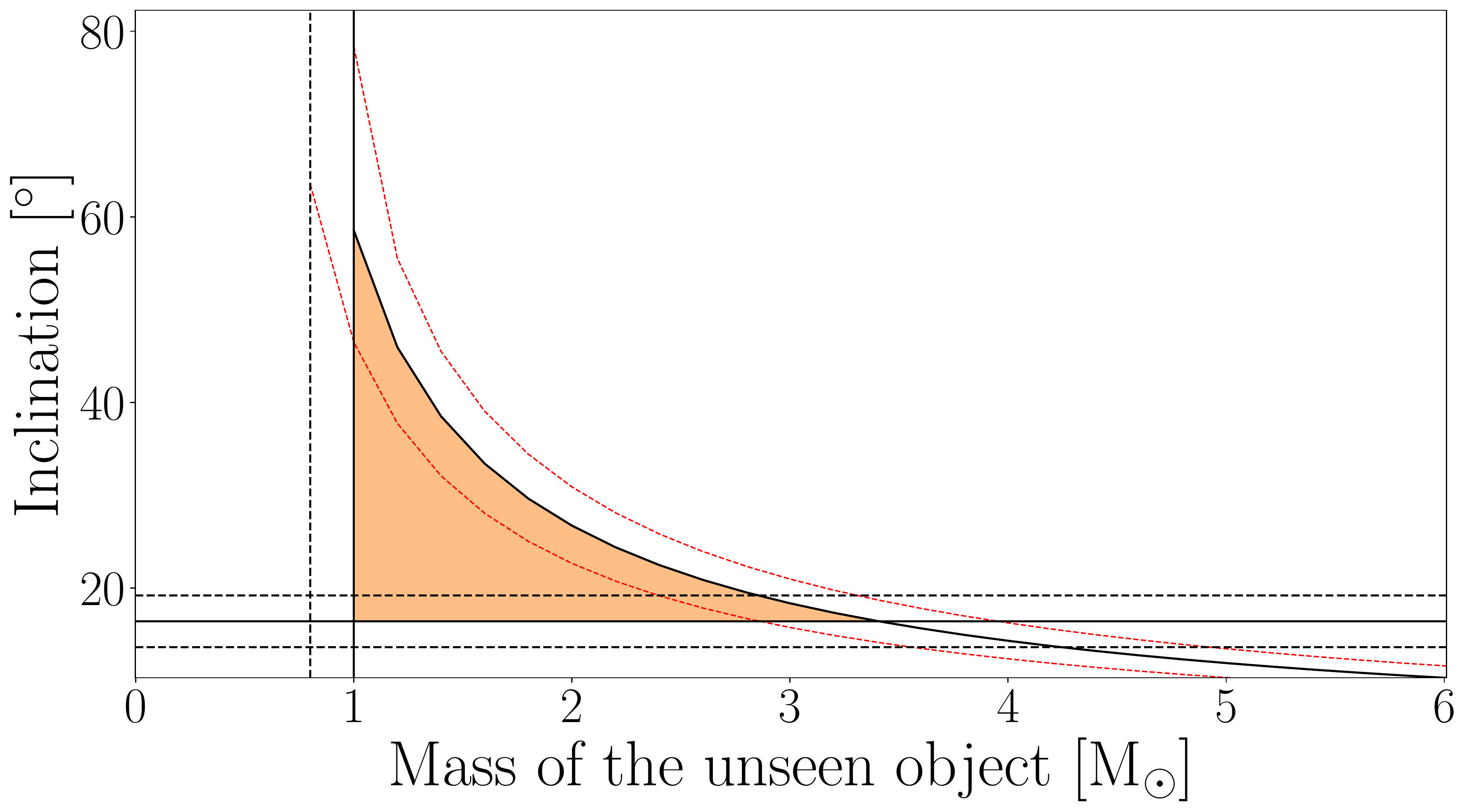}
    \caption{LS~5039}
    \end{subfigure} 
    \caption{Same as for Fig.~\ref{fig:binarymassfunction} but computed with the spectroscopic masses.}
    \label{fig:binarymassfunction_spec}
\end{figure*}

To constrain the nature of the unseen companions in the 15 SB1 systems where no spectral signatures were detected with the spectral disentangling, we computed the minimum masses by using the binary mass function:
\begin{equation}
    f \equiv
    \frac{M_u^3 \sin^3 i}{(M_u+M_P)^2} = \frac{P_{\mathrm{orb}}\,(1-e^2)^{3/2}\,K^3}{2\pi G}
\end{equation}
where $M_u$ is the mass of the unseen object, $M_P$ the mass of the primary (visible) star, $i$ the inclination of the system, $P_{\mathrm{orb}}$ the orbital period, $e$ the eccentricity, $K$ the primary RV semi-amplitude, and $G$ the gravitational constant. \\ \\
Since $0 \leq \sin i \leq 1$, it follows that
\begin{equation}
    \frac{M_u^3 \sin^3 i}{(M_u+M_P)^2} = \frac{P_{\mathrm{orb}}\,(1-e^2)^{3/2}\,K^3}{2\pi G} \leq
    \frac{M_u^3}{(M_u+M_P)^2}.
\end{equation}
By solving this inequality, we obtained the minimum mass estimates for the unseen companions, but that supposes to have a well-established mass estimate for the visible star. There are two different mass estimates that can be calculated for a single star: 1) the spectroscopic and 2) the evolutionary masses. The former was computed from the surface gravity and the radius of the star, obtained through atmosphere modelling and from the star’s absolute luminosity (Sect.\,\ref{subsec:atmosphere}). The latter is obtained through a comparison of the star physical properties to evolutionary tracks. The agreement between both mass estimates is a long-standing problem in stellar astrophysics \citep[e.g.][among others]{herrero92,burkholder97,weidner10, mahy20b, tkachenko20}. We therefore computed the minimum masses of the unseen objects by using both mass estimates. We obtained a relation between the inclinations of the systems and the masses of the unseen objects. Figure~\ref{fig:binarymassfunction} shows these relations using the evolutionary masses. The computations were also done using the spectroscopic masses and are shown in Fig.~\ref{fig:binarymassfunction_spec}.

To have an independent way of constraining the lower limit on the inclinations of the systems, we computed the critical rotational velocities of the visible stars:
\begin{equation}
    v_{\mathrm{crit}} = \sqrt{\frac{2\,GM_P}{3\,R_p}},
\end{equation}
where $M_P$ is the mass of the visible star, and $R_p$ its polar radius. As the inclinations of the systems in our sample are not known and atmosphere codes typically adopt spherical symmetry, we assumed that the radii measured through our analysis are equal to the polar radii of the visible stars \citep[see however][]{fabry22}. Assuming that the rotational axes are perpendicular to the orbital plane, and that the equatorial rotational velocities of the visible stars cannot be larger than their critical rotational velocities, one obtains
\begin{equation}
  \frac{v\,\sin i}{\sin i} \leq v_{\mathrm{crit}},
\end{equation}
which gives a minimum value on the inclination and thus provides a maximum mass for the unseen object. We note that the assumption of alignment of the rotational and orbital axes must not hold for binaries containing compact objects due to potential kicks. However, since Eq.~4 only impacts the upper limit on the mass, this has no impact on our conclusions.

Therefore, by using Eqs. 2 and 4, which rely exclusively on the measured orbital parameters and projected rotation rate, and on a mass and radius estimate of the visible star (see Sects.\,\ref{subsec:atmosphere} and \ref{sec:finalparameters}), one can derive a range of possible masses for each unseen object.

\subsection{Atmosphere modelling and spectroscopic masses}
\label{subsec:atmosphere}
The estimations of the stellar parameters, in particular of the spectroscopic and evolutionary masses, and of the surface abundances of the visible objects are a critical step to characterise the unseen objects and to understand their nature. 

We used the CMFGEN (CoMoving Frame GENeral, \citealt{hillier98}) atmosphere code. CMFGEN is a radiative-transfer code that relaxes the assumption of local thermodynamic equilibrium (non-LTE) and includes stellar winds and line-blanketing. This code solves the radiative-transfer equation for a spherically symmetric wind in the co-moving frame under the constraints of radiative and statistical equilibrium. The hydrostatic density structure is computed from mass conservation and the velocity structure is constructed from a pseudo-photosphere structure connected to a $\beta$ velocity law of the form $v = v_{\infty}(1 - R/r)^{\beta}$, where $v_{\infty}$ is the terminal velocity of the wind and $\beta$ a unitless parameter describing the shape of the wind velocity law. Our final models included the following chemical elements: \ion{H}{i}, \ion{He}{i-ii}, \ion{C}{ii-iv}, \ion{N}{ii-v}, \ion{O}{ii-v}, \ion{Al}{iii}, \ion{Ar}{iii-iv}, \ion{Mg}{ii}, \ion{Ne}{ii-iii}, \ion{S}{iii-iv}, \ion{Si}{ii-iv}, \ion{Fe}{ii-vi}, and \ion{Ni}{ii-v} with the solar composition \citep{grevesse10} unless otherwise stated. CMFGEN also uses the super-level approach to reduce the memory requirements. On average, we included about 1600 super levels for a total of 8000 levels. For the formal solution of the radiative-transfer equation that leads to the emergent spectrum, a microturbulent velocity varying linearly from 10\,\kms\ to $0.1 \times v_{\infty}$ was used. 

To derive the stellar parameters, we built a grid of synthetic solar-metallicity CMFGEN spectra  by varying $\teff{}$ in steps of $\Delta \teff{} = 1000$\,K and $\logg{}$ in steps of $\Delta \logg{} = 0.1$\,[cgs]. Our grid covers $25000 < \teff < 47000$\,K and $3.0 < \logg < 4.4$\,[cgs]. For this grid, the luminosities were assigned according to \citet{brott11} evolutionary tracks from the combination ($\teff{}, \logg{}$) by assuming an initial rotational velocity of $150$\,\kms{}. For the mass-loss rates, we used the prescriptions of \citet{vink00,vink01} with solar metallicity. The terminal wind velocities were estimated to be equal to 2.6 times the effective escape velocity from the photosphere ($v_{\rm{esc}}$, \citealt{lamers95}). The exponent $\beta$ of the velocity law was set to 1.0 and the clumping filling factor, describing the density contrast between the clumps and the equivalent smooth wind, was adopted as $f_{cl} = 0.1$. 

We generated a `master-spectrum' for each visible star by shifting the observed spectra by the primary RVs, to have them in a same reference frame, and by stacking all of these spectra. The S/N of the master-spectrum is higher than for the individual epochs (i.e. $({\rm S/N})_{\rm master} = ({\rm S/N})_{\rm obs} \cdot \sqrt{N_{\rm obs}}$, where $N_{\rm obs}$ is the number of observed spectra in our dataset).

The projected rotational velocity ($\vsini$) and the macroturbulent velocity ($\vmac$) were derived, as explained by \citet{simondiaz14}, on dedicated spectral lines, mainly the \ion{He}{i}~4713, \ion{O}{iii}~5592 or \ion{He}{i}~5876 lines. We convolved the synthetic spectra first by a rotational profile, mimicking $\vsini$, then, by a radial/transverse profile mimicking $\vmac$, and by a Gaussian mimicking the instrumental broadening. 

$\teff{}$ and $\logg{}$ were derived simultaneously from the grid of synthetic spectra. The quality of the fit is quantified by means of a $\chi^2$ analysis on the H and He lines (mainly the surface gravity is computed from the wings of the Balmer lines and the effective temperature is based on the \ion{He}{i}-\ion{He}{ii} ratio). The $\chi^2$ is computed for each model of the grid and linearly interpolated between the grid points in steps of $\Delta \teff{} = 100$\,K and $\Delta \logg{} = 0.01$\,[cgs]. The error bars in $\teff{}$ and $\logg{}$ are correlated. The uncertainties at $1$, $2$, and $3\sigma$ on $\teff{}$ and $\logg{}$ were estimated from $\Delta \chi^2 = 2.30$, $6.18$, and $11.83$ (two degrees of freedom), respectively \citep[see][for more details]{press07}.

\begin{figure*}[htbp]\centering
    \includegraphics[trim=0 0 0 10,clip,width=9cm]{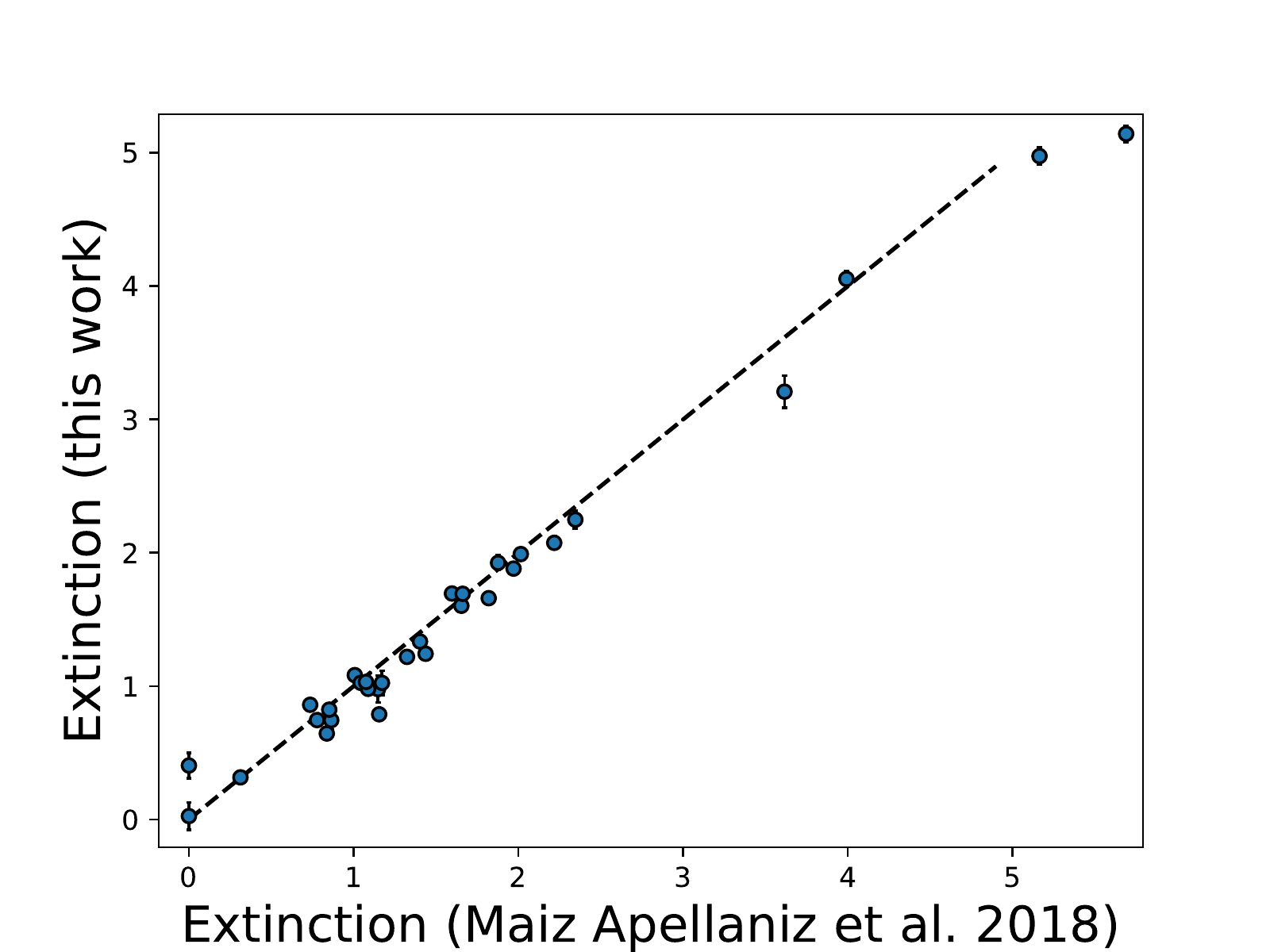}
    \includegraphics[trim=0 0 0 10,clip,width=9cm]{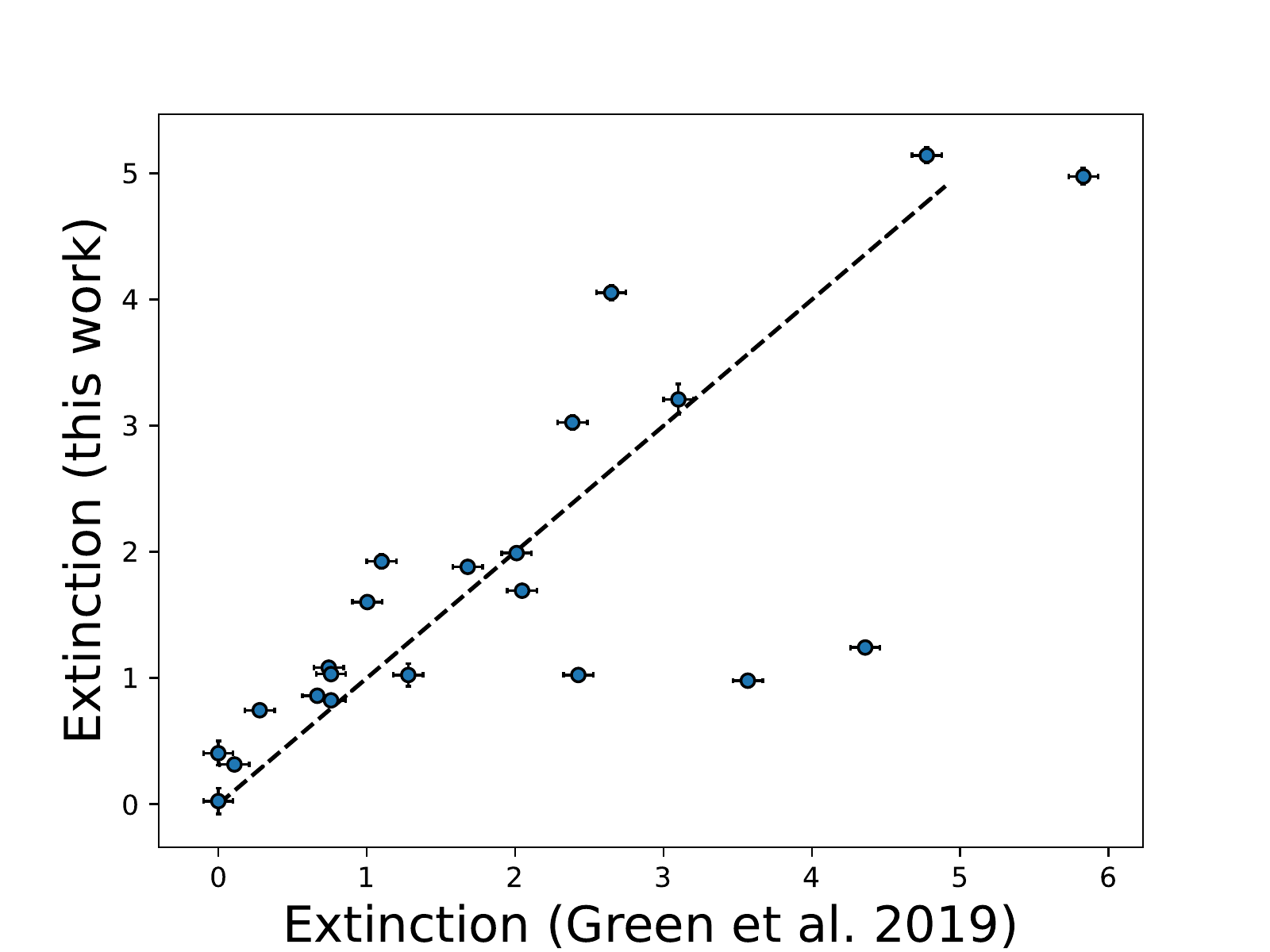}
    \caption{\label{fig:extinction} Comparison between the extinctions derived through our analysis with extinctions provided by \citet[][left panel]{maiz18} and from the 3D dust map of \citet[][right panel]{green19} assuming the Gaia eDR3 distances of the stars.}
\end{figure*}

The stellar luminosity was computed from the $V$ magnitude, extinction, bolometric correction (BC), and distance ($d$) to the stars using: 
\begin{equation}
    \log(L/L_{\odot}) = -0.4 \, (V - A_V - (5*\log(d)-5) + BC -4.75).
\end{equation} 
The extinctions were derived by fitting the spectral energy distributions (SEDs) of the systems, adopting the $\teff$ and $\logg$ obtained through the spectroscopic $\chi^2$ analysis. To build the SEDs, we used UV spectra observed with the International Ultraviolet Explorer satellite (when available), the U band magnitude given by \citet{reed03}, the BVJHK bands provided from the Naval Observatory Merged Astrometric Dataset catalogue \citep{zacharias04}, and finally the $G_{B_P}$, $G$, and $G_{R_P}$ magnitudes from the Gaia early Data Release 3 \citep[eDR3,][]{gaia16,gaia21}. The SED fitting is shown in Fig.~\ref{fig:SED} for each individual system. We considered that the two objects in each system have the same extinction. We applied the extinction law from \citet{fitzpatrick07}. We compared our extinction values with those derived by \citet[][Fig.~\ref{fig:extinction}, left panel]{maiz18} and those provided by the 3D dust map of \citet[][when available]{green19}. 

The bolometric corrections were computed using the relations based on the effective temperatures of the stars given by \citet{martins06}. We also adopted the photo-geometric distances provided by \citet{bailerjones21} using Gaia eDR3 parallaxes, unless it provides unphysical fundamental properties for the individual objects. 

For the SB1s, the luminosities inferred are attributed to the visible objects that largely dominate the $V$ band. For the newly classified SB2s, we computed the bolometric magnitudes (and thus the luminosities) of individual objects by computing the absolute magnitudes of the systems, correcting them for the brightness ratio, and we applied the bolometric correction computed from the effective temperatures of the individual components. Finally, we computed the radii of the individual objects from their effective temperature and their luminosity.

In order to discuss the evolutionary stages of the SB1s, we derive the surface abundances only for these systems, using the method described by \citet{martins15}. The choice of the diagnostic lines depends on the quality of the spectrum and on the spectral type of the star. We used a list of spectral lines from which we made the selection of the diagnostics used in the $\chi_{\rm abund}^2$ analysis:
\begin{itemize}
    \item carbon: \ion{C}{iii}~4068-70, \ion{C}{iii}~4153, \ion{C}{iii}~4156, \ion{C}{iii}~4163, \ion{C}{iii}~4187, \ion{C}{ii}~4267, \ion{C}{iii}~4325, \ion{C}{iii}~4666, \ion{C}{iii}~5246, \ion{C}{iii}~5353, \ion{C}{iii}~5272, \ion{C}{iii}~5826.
    \item nitrogen: \ion{N}{ii}~3995, \ion{N}{ii}~4004, \ion{N}{ii}~4035, \ion{N}{ii}~4041, \ion{N}{iii}~4044, \ion{N}{iii}~4196, \ion{N}{iii}~4511, \ion{N}{iii}~4515, \ion{N}{iii}~4518, \ion{N}{iii}~4524, \ion{N}{ii}~4607, \ion{N}{iv}~5200, \ion{N}{iv}~5204,
    \item oxygen: \ion{O}{ii}~4700, \ion{O}{ii}~4707, \ion{O}{iii}~5592.
\end{itemize}

The best-fit model was obtained by minimising the calculated $\chi_{\rm abund}^2$ by varying different parameters in the parameter space. To this end, we generated a non-uniform grid composed of several dozen models for each star. Once all the fundamental parameters (i.e. $\teff$, $\logg$, $\vsini$, and $\vmac$) are constrained, we ran models with different surface abundances (for He, C, N, and O). We quantitatively compared these lines to the synthetic spectra by means of a $\chi_{\rm abund}^2$ analysis from which we derived the surface abundance and their uncertainties (see \citealt{martins15} and \citealt{mahy20a} for more details).

\subsection{Physical parameters and evolutionary masses}
\label{sec:finalparameters}
\begin{figure*}[htbp]\centering
    \includegraphics[trim=10 0 10 0,clip,width=18cm]{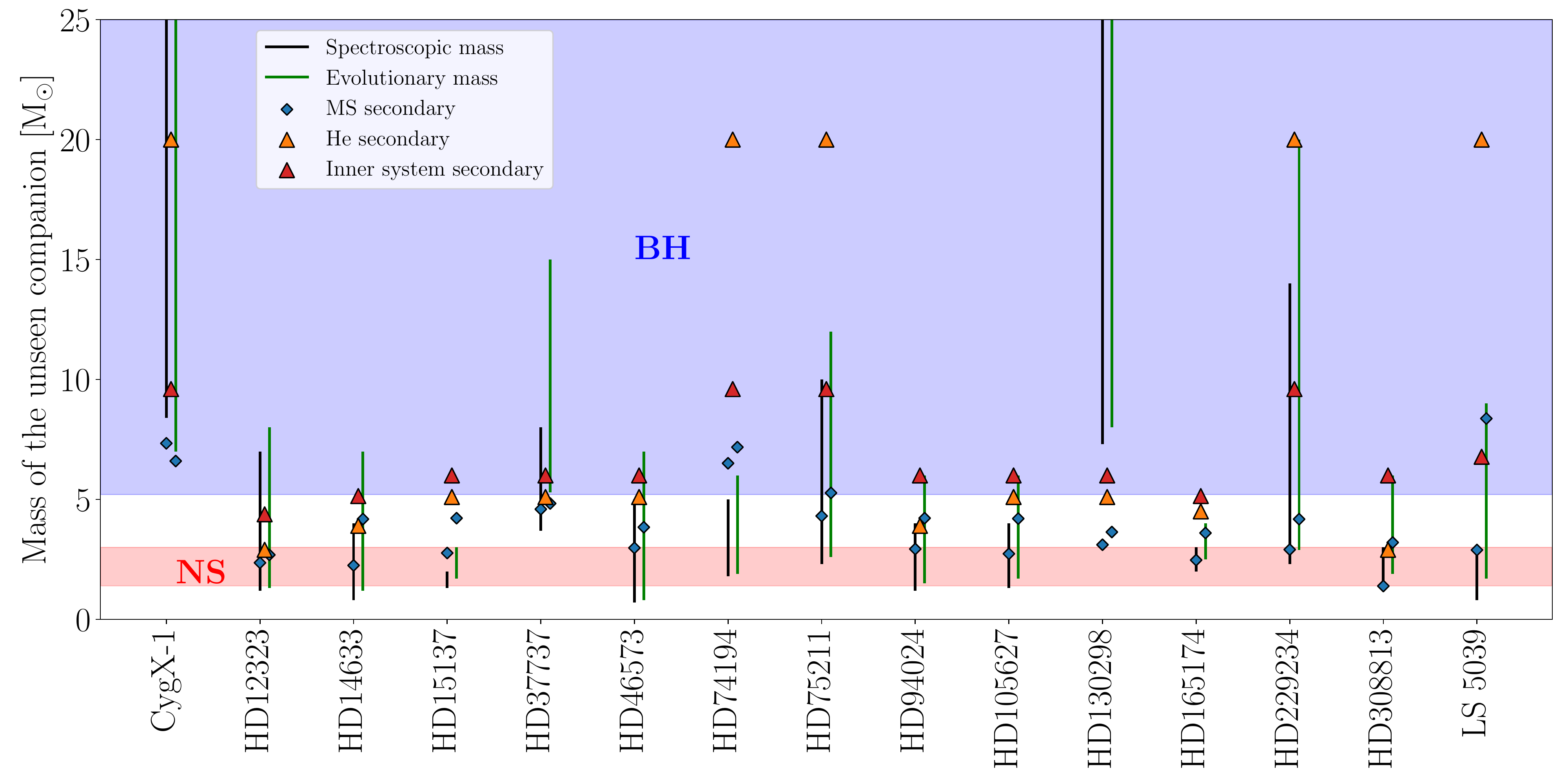}
    \caption{\label{Fig:massComp} Estimated mass ranges for the unseen companions in the SB1 systems, computed from the spectroscopic (black lines) and evolutionary (green lines) masses of the visible stars. The red area indicates the predicted mass range for Galactic NSs, and the blue area represents the predicted mass range for Galactic stellar BHs \citep{belczynski10,fryer14}, and where non-degenerate secondaries could have been retrieved. The diamonds and triangles represent the minimum mass
    that we are able to extract using spectral disentangling according to our simulations (Sect.~\ref{subsec:disentangling})}
\end{figure*}

Once we obtained the physical parameters using CMFGEN, we utilised $\log(L/L_{\odot})$, $\teff{}$, $\logg{}$, and $\vsini$ as inputs for the BONNSAI (BONN Stellar Astrophysics Interface, \citealt{schneider14,schneider17}) code to compute the evolutionary properties of the stars. BONNSAI is a Bayesian analysis tool that allows us to compare the properties of the stars with the BONN single-star evolutionary models \citep{brott11}. In this way, BONNSAI provides us with the predictions about the evolutionary masses and ages that match with our derived parameters. The stellar and predicted parameters are listed in Table\,\ref{tab:parameters_SB2} and Table\,\ref{tab:parameters} with their $1\sigma$ errors, for the SB2s and SB1s, respectively. From the estimated and predicted sets of parameters, we computed the mass ranges for the unseen companions as a function of the inclinations of the systems (Figs.~\ref{fig:binarymassfunction} and ~\ref{fig:binarymassfunction_spec}). These mass ranges are displayed in Fig.~\ref{Fig:massComp} with predicted mass ranges for NSs (in red) and BHs (in blue). We also indicate, in Fig.~\ref{Fig:massComp}, the mass limits of the unseen objects that would have been extracted with spectral disentangling, according to our simulations. 

Nine stars in our sample have a secondary companion with a mass estimate between 1 and $6\,\Msun$: HD~14633, HD~15137, HD~46573, HD~74194, HD~94024, HD~105627, HD165174, HD~308813, and LS~5039. They all have masses and brightnesses lower than what we can detect with the spectral disentangling, according to our simulations. 

For two systems (HD~12323 and HD~2292234), the mass ranges for their respective companions are from $3-12~\Msun$ and from $3-15~\Msun$, respectively. Our simulations show that secondaries with such masses could have been detected using spectral disentangling. However, we also observe, for both systems, a difference in the RVs of the visible stars of about $15-25$~\kms, respectively,  between two epochs that correspond to the same orbital phase. There is therefore a possibility that these differences might come from a variation of their systemic velocities. That would suggest that these systems might be triples, but it is too early to confirm it. 

Two objects have a companion with a mass higher than $\sim 7~M_{\odot}$: Cyg~X-1, which is known to host a $21~M_{\odot}$ stellar-mass BH \citep{Miller-Jones21}, and HD~130298. The spectral disentangling does not allow us to reveal the signatures of a secondary star for both objects. With $7~M_{\odot}$ or higher, the companions should be detectable in the composite spectra, which suggests that HD~130298 is a promising candidate to host a stellar-mass BH. We stress that no X-ray detection was reported in the Second ROSAT all-sky survey (2RXS) source catalogue \citep{boller16} for HD~130298.  

\begin{figure}[htbp]\centering
    \includegraphics[trim=0 0 0 10,clip,width=9cm]{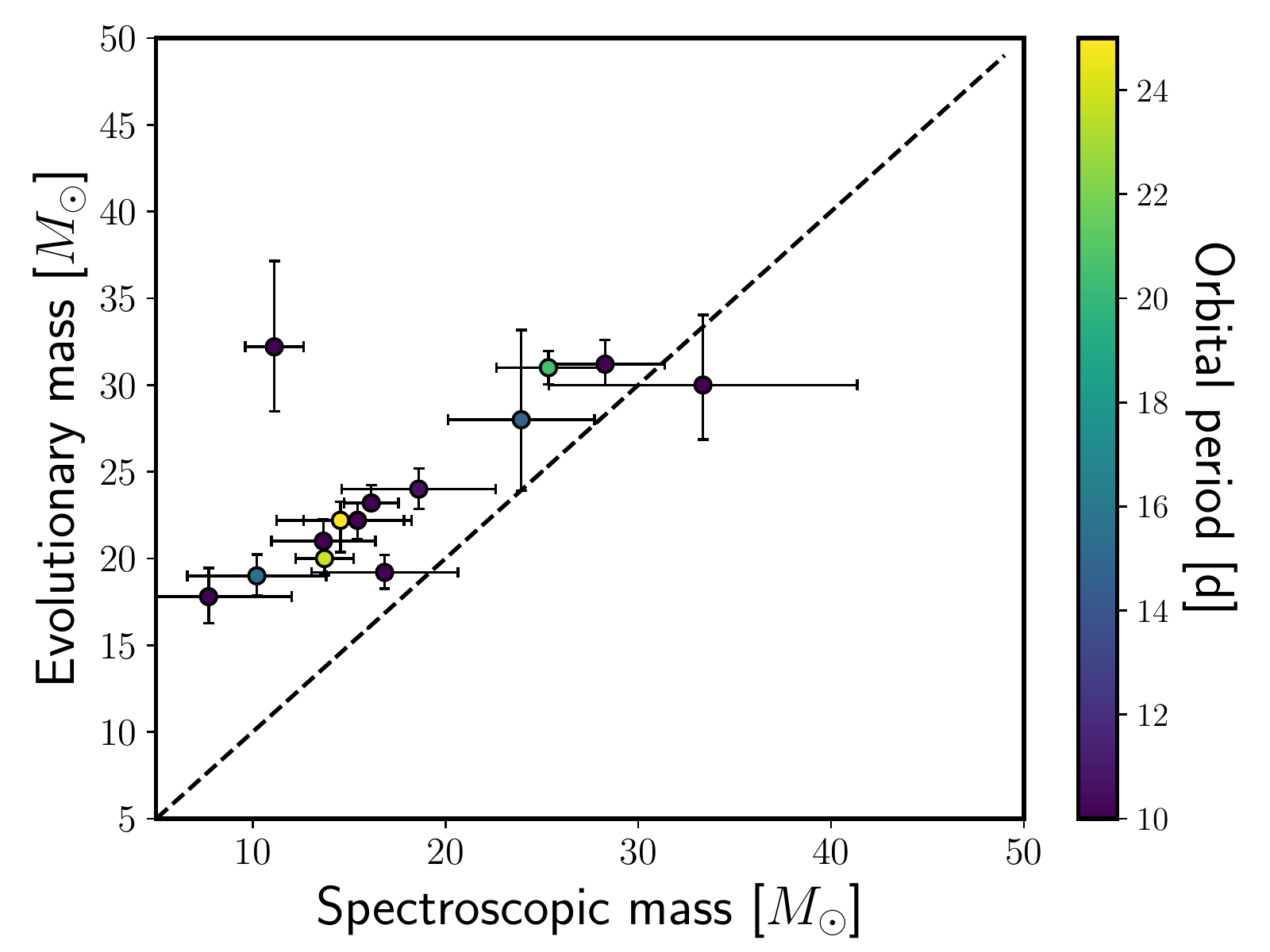}
    \caption{\label{fig:massdiscrepancy} Comparison, for the SB1 primaries, between their spectroscopic masses (derived from their stellar parameters) and their evolutionary masses (derived from their positions in the Hertzsprung-Russell diagram). The colour bar indicates the orbital periods of the systems.}
\end{figure}

Finally, we stress that the mass discrepancy is clearly present in our results (Table~\ref{tab:parameters}, Fig.~\ref{fig:massdiscrepancy}), with the spectroscopic masses being significantly smaller than the evolutionary masses in 12 out of our 15 SB1s. Interestingly, neither Cyg~X-1 and HD~74194 (which harbour a BH and a NS companion, respectively) nor HD~130298 (our main OB+BH candidate; see Sect.~\ref{sub:discussion}) are impacted by this discrepancy. This might also imply that part of the mass discrepancy could result from undetected low-mass companions that impact spectroscopic $\log g$ determination by dilution and/or by their contribution to the wings of Balmer lines, which are used as $\log g$ diagnostics. Elucidating the mass discrepancy is beyond the scope of the present work. We discuss each system individually in the Appendix (Sect.~\,\ref{sec:individual}).

\subsection{TESS photometry}
\label{subsec:tess}
Detecting putative companions or compact objects around massive stars benefits to not only focus on spectroscopy but also probe time-series photometry. Light curves can indeed be used to corroborate the presence of a non-degenerate companion in a binary system (e.g. in the case of eclipsing binaries). Searching for stellar-mass BHs, for example, in Transiting Exoplanet Survey Satellite (TESS; \citealt{ricker09,Ricker2015}) light curves, was already envisaged by \citet{masuda19}. These authors pointed out three different signals that can be detected if a quiet BH is present: 1) ellipsoidal variations \citep{gomel21b}, 2) Doppler beaming and 3) self-lensing. The two first signals produce a variability that decreases in amplitude with increasing orbital periods while the self-lensing causes pulse-like brightening only during the eclipse. High-cadence photometry is also useful for detecting modulations produced by the rotation of the star. In this case, it can provide useful information for deriving the inclinations of the stars \citep{burssens20}. 

We retrieved TESS light curves for 13 objects among the SB1 systems and 8 among the newly classified SB2 systems. The other objects have not been observed yet (HD~29763, HD~163892, HD~164438, HD~164536, HD~165174, HD~167263, HD~167264, and LS~5039), suffer from contamination of other stars in their neighbourhood (HD~93028), or were within the TESS sectors but fall just on the edge of the detector so that no light curve can be extracted (HD~152405 and HD~152723). The 2-min cadence light curves were retrieved from the Mikulski Archive for Space Telescopes archive as light curves. The light curves are those in the pre-conditioned form (PDCSAP, Pre-search Data Conditioning Simple Aperture Photometry). The 30-min cadence light curves were extracted from the full- frame images (FFIs). Aperture photometry was performed on image cutouts of $50 \times 50$ pixels using the \textsc{PYTHON} package \textsc{LIGHTKURVE} \citep{lightkurve18}. The source mask was defined from pixels above a given threshold (generally from 3 to 10 depending on the target). The background mask was defined by pixels with fluxes below the median flux, thereby avoiding nearby field sources. 

\begin{figure*}
    \centering
    \begin{subfigure}{0.33\linewidth}
    \includegraphics[width = \textwidth, trim=0 15 0 0,clip]{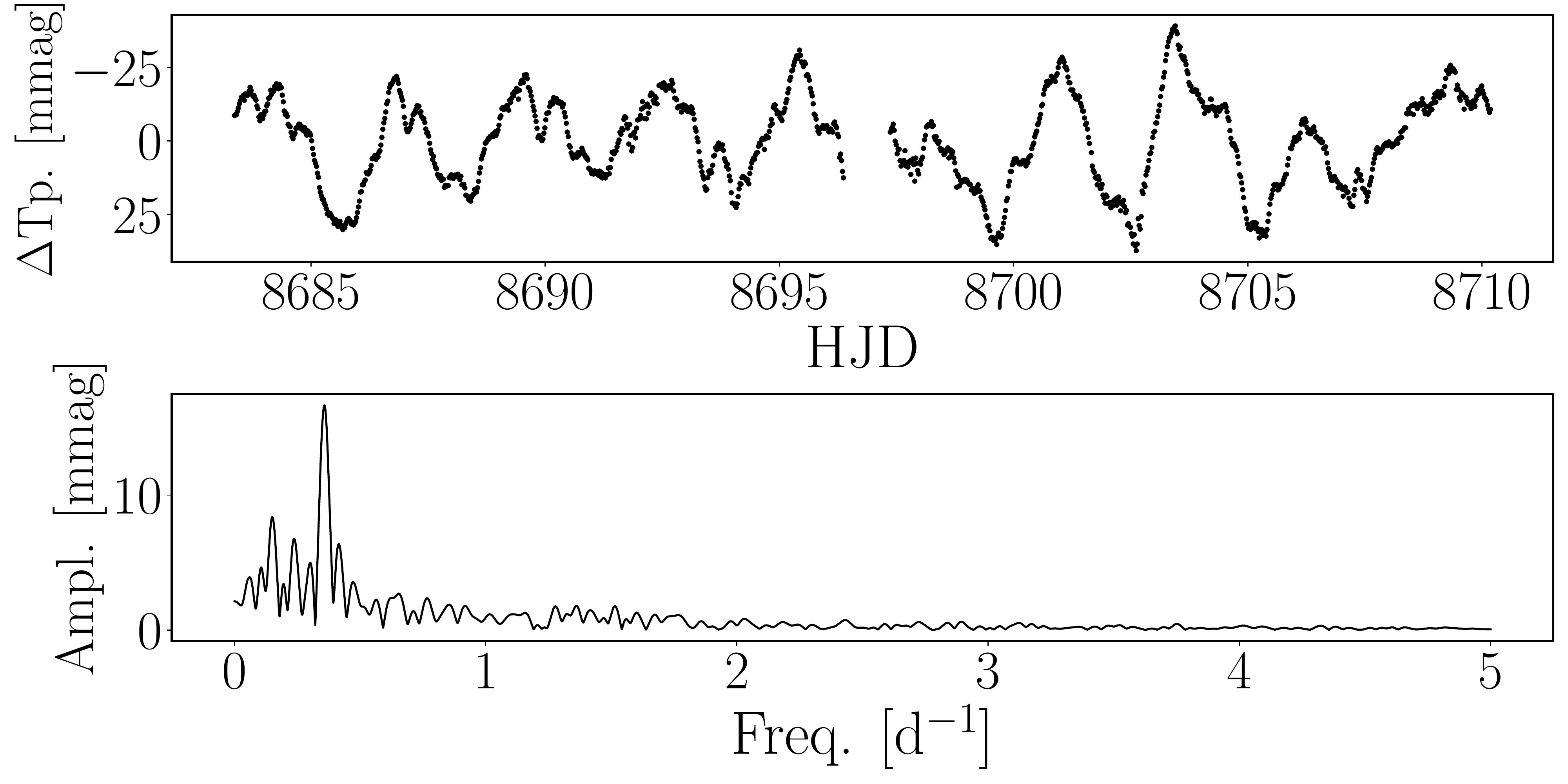}
    \caption{Cyg~X-1}
    \end{subfigure}
    \begin{subfigure}{0.33\linewidth}
    \includegraphics[width = \textwidth, trim=0 15 0 0,clip]{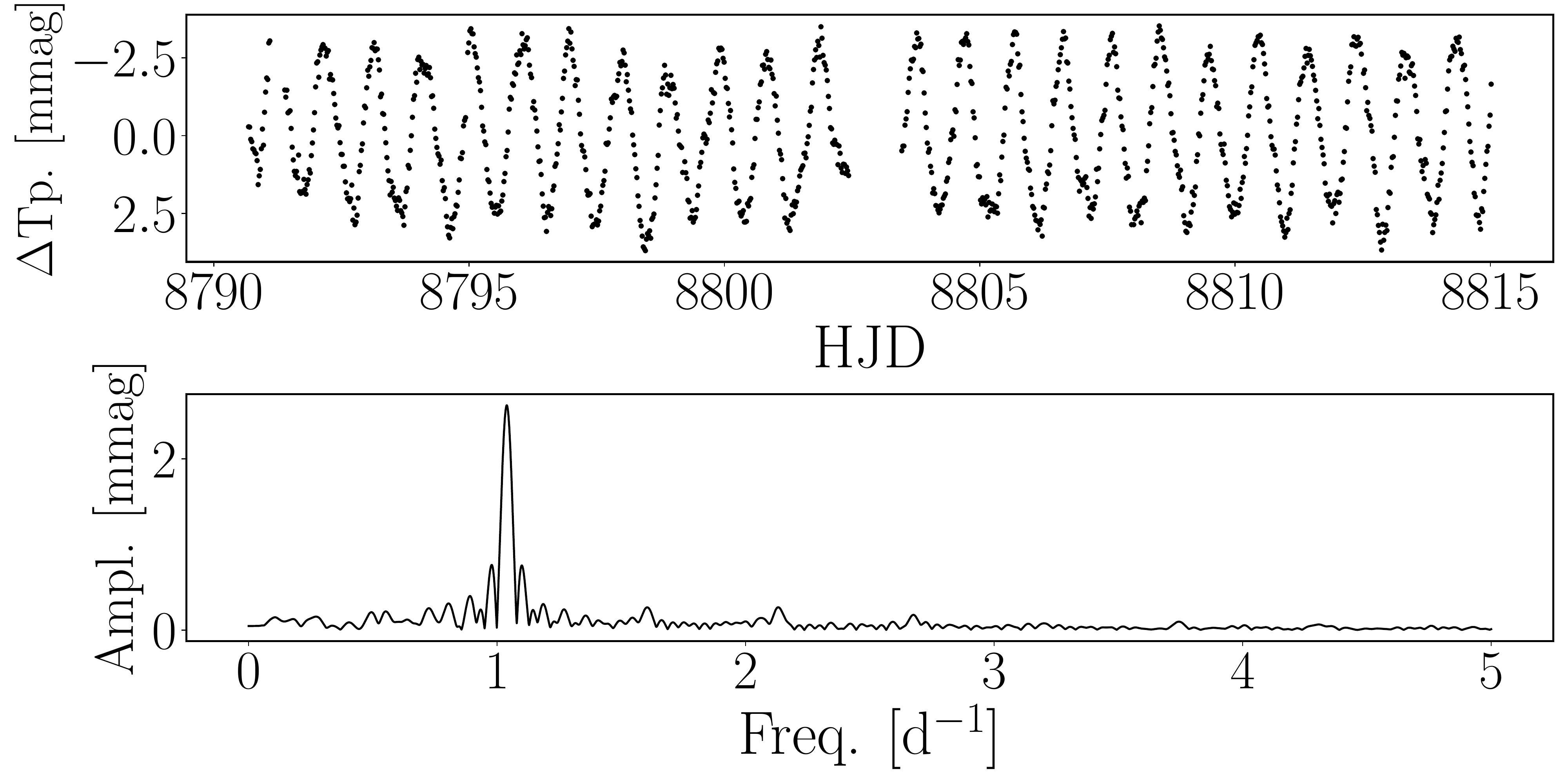}
    \caption{HD~12323}
    \end{subfigure}
    \begin{subfigure}{0.33\linewidth}
    \includegraphics[width = \textwidth, trim=0 15 0 0,clip]{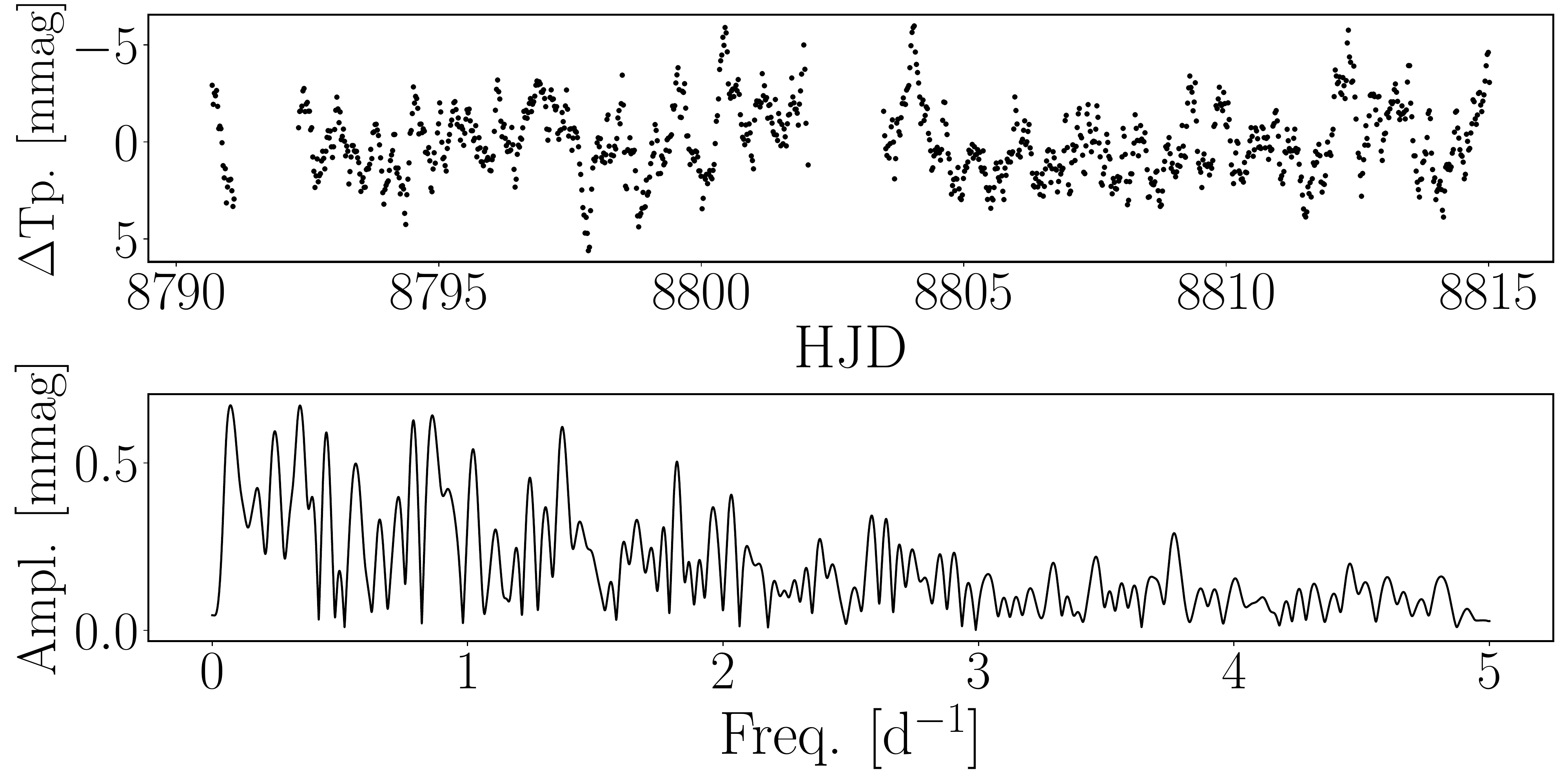}
    \caption{HD~14633}
    \end{subfigure} 
    \begin{subfigure}{0.33\linewidth}
    \includegraphics[width = \textwidth, trim=0 15 0 0,clip]{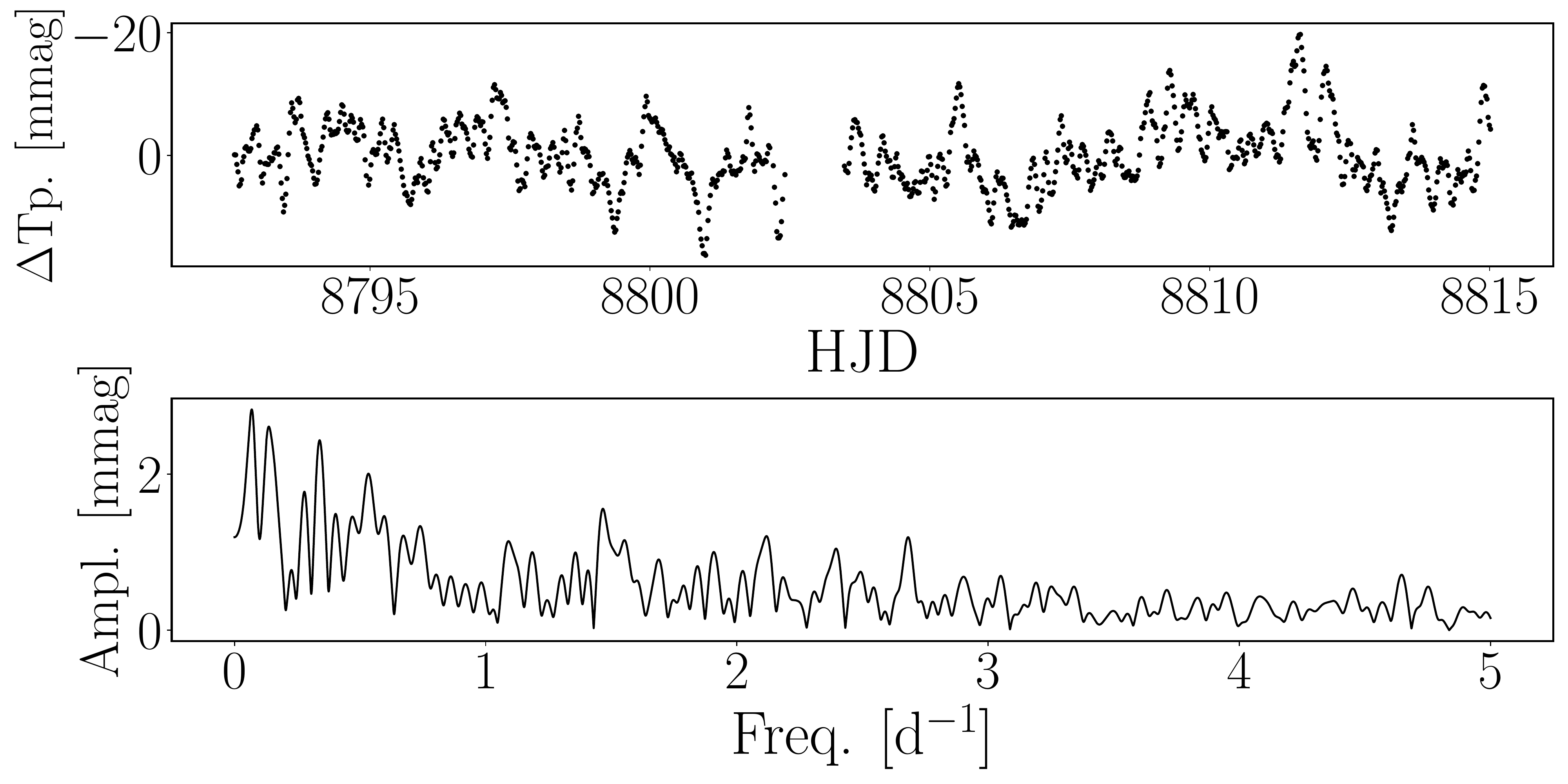}
    \caption{HD~15137}
    \end{subfigure}
    \begin{subfigure}{0.33\linewidth}
    \includegraphics[width = \textwidth, trim=0 15 0 0,clip]{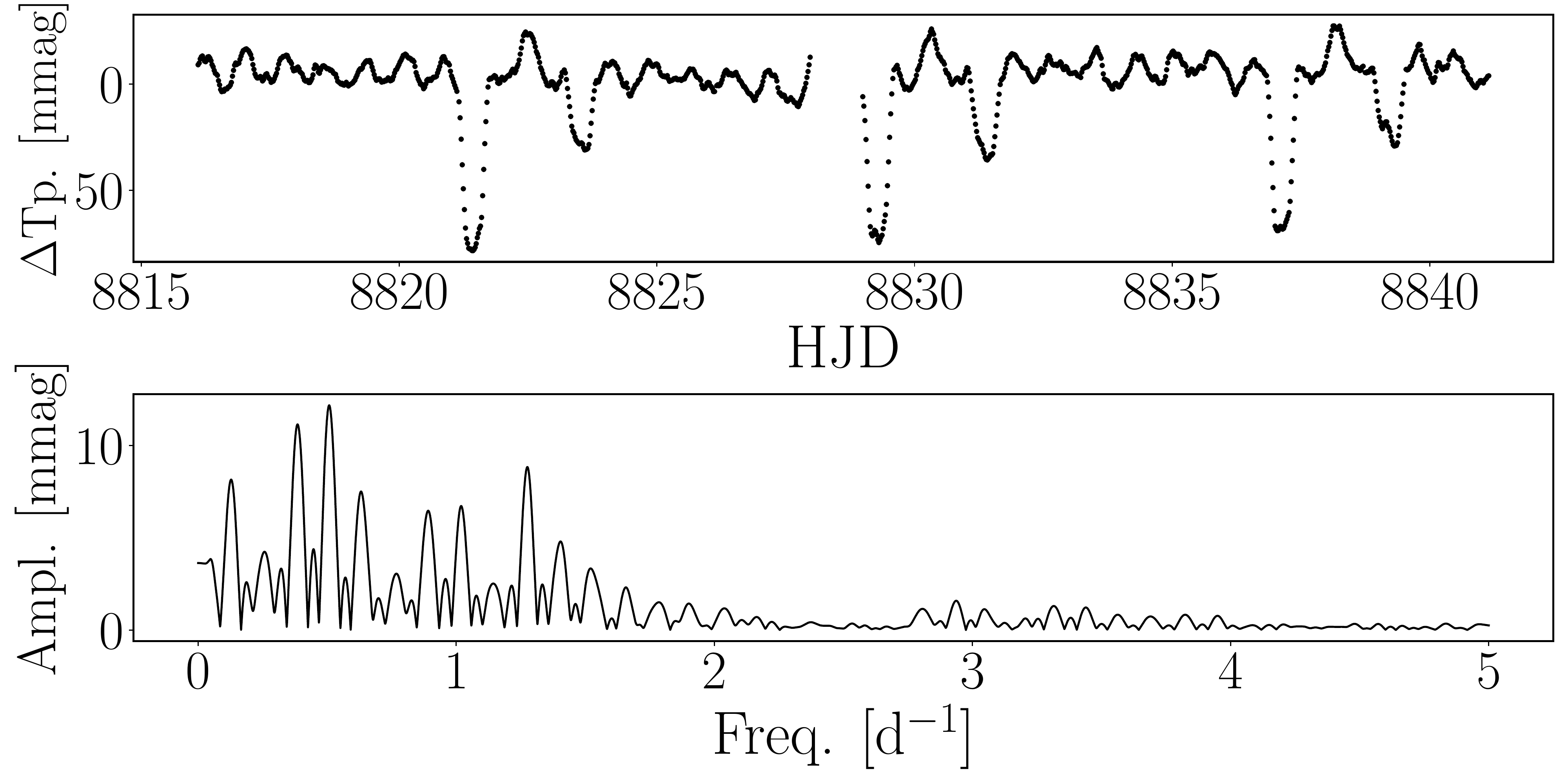}
    \caption{HD~37737}
    \end{subfigure}
    \begin{subfigure}{0.33\linewidth}
    \includegraphics[width = \textwidth, trim=0 15 0 0,clip]{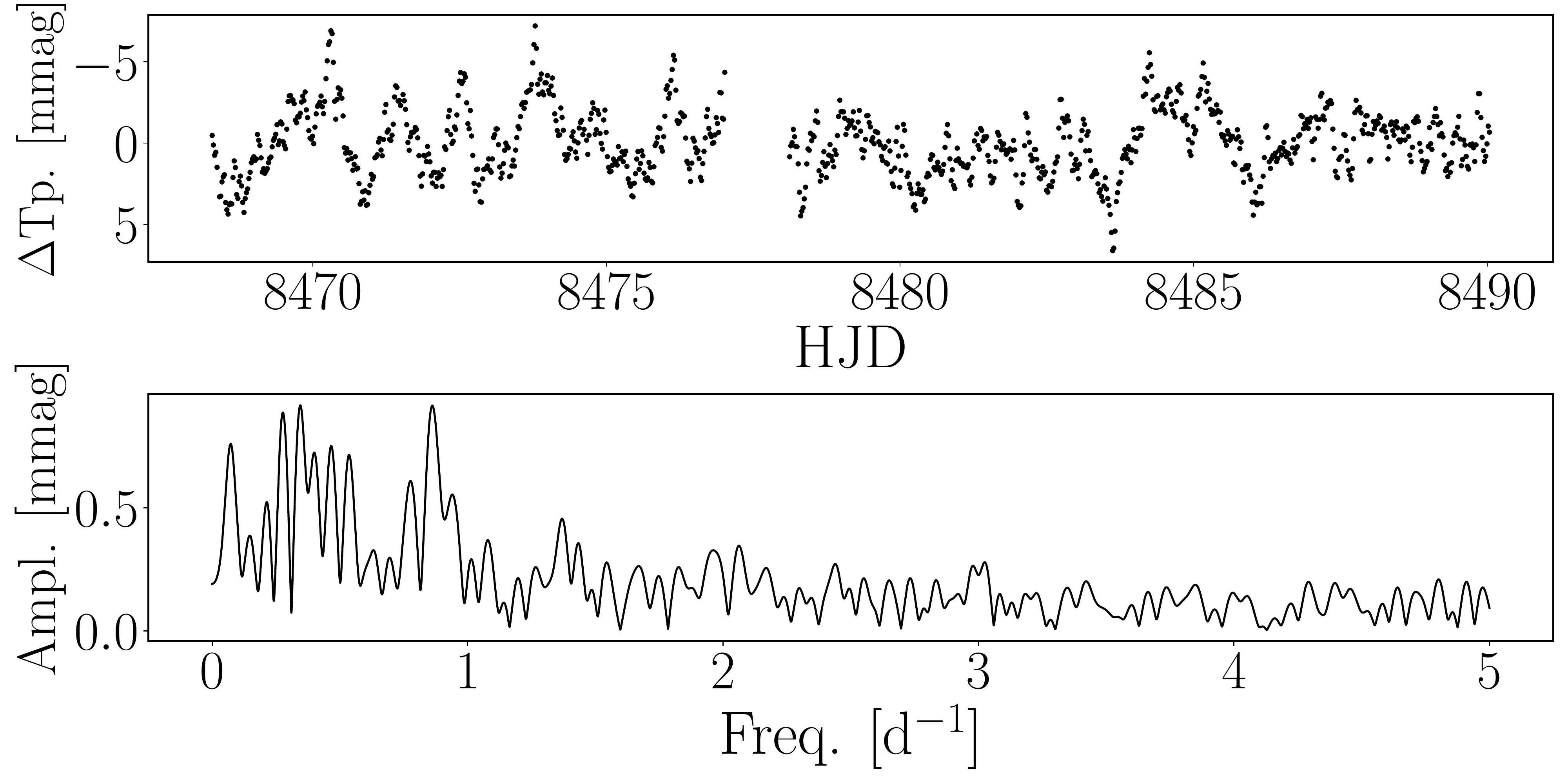}
    \caption{HD~46573}
    \end{subfigure} 
    \begin{subfigure}{0.33\linewidth}
    \includegraphics[width = \textwidth, trim=0 15 0 0,clip]{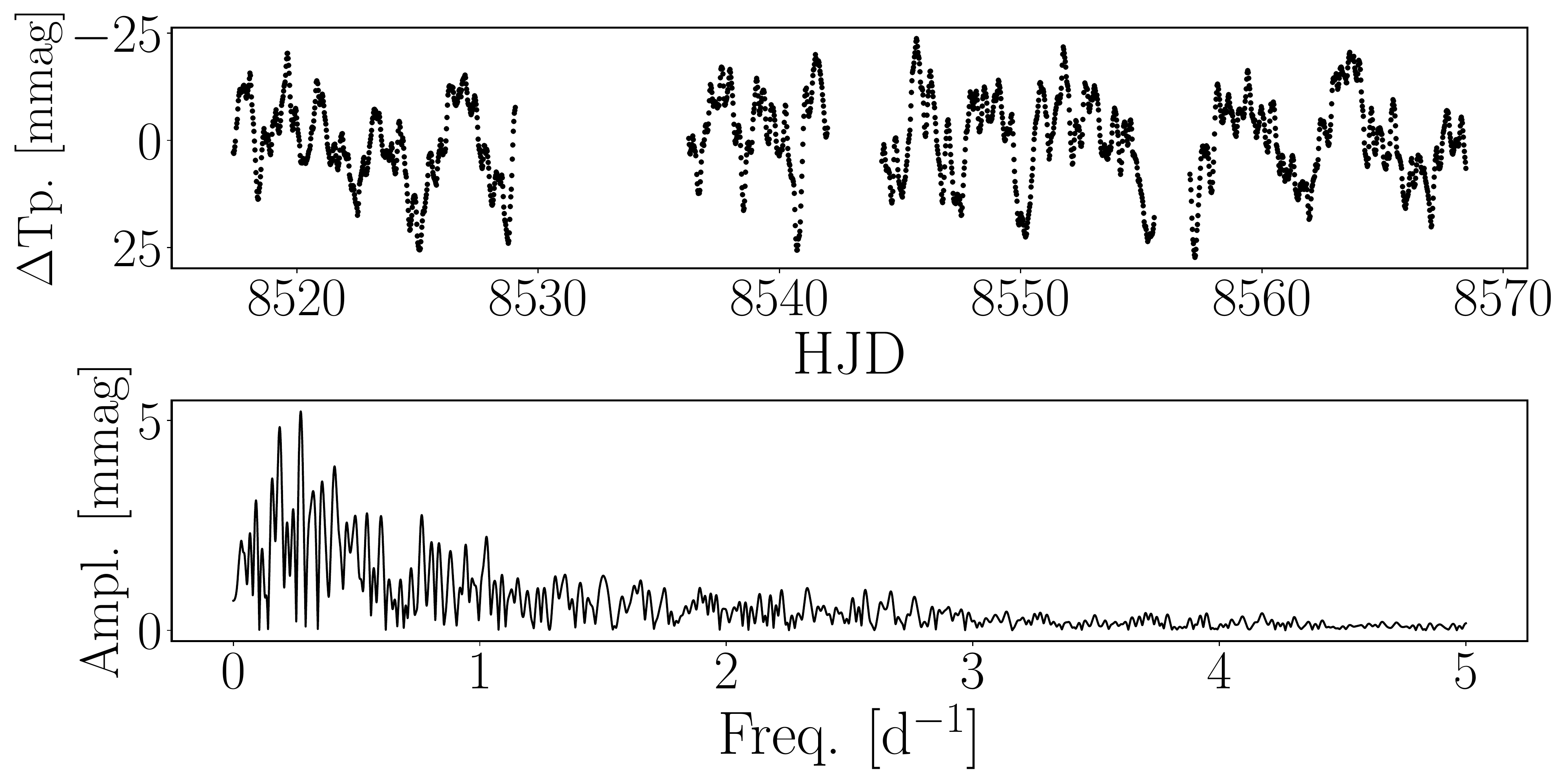}
    \caption{HD~74194}
    \end{subfigure}   
    \begin{subfigure}{0.33\linewidth}
    \includegraphics[width = \textwidth, trim=0 15 0 0,clip]{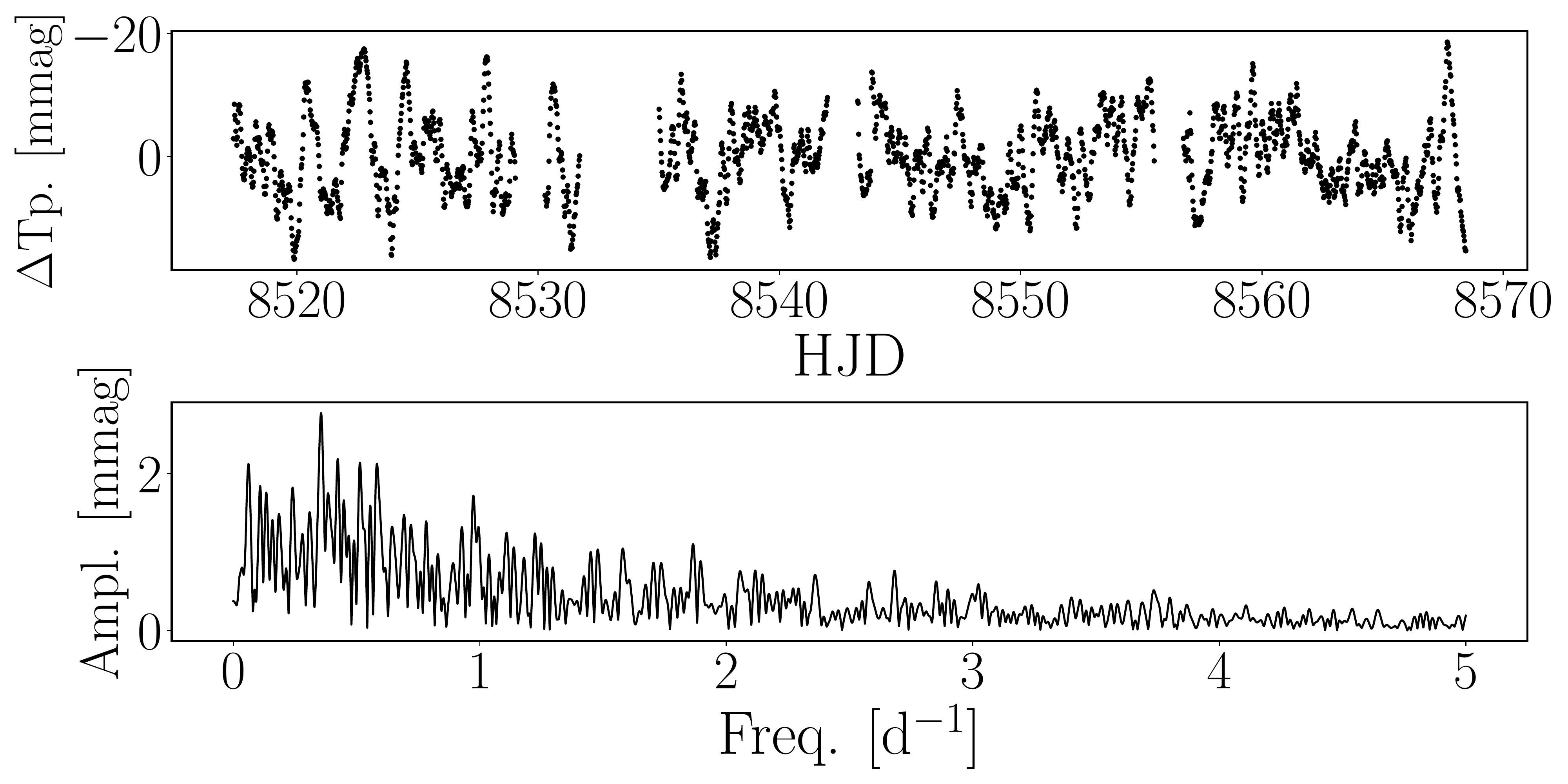}        
    \caption{HD~75211}
    \end{subfigure}
    \begin{subfigure}{0.33\linewidth}
    \includegraphics[width = \textwidth, trim=0 15 0 0,clip]{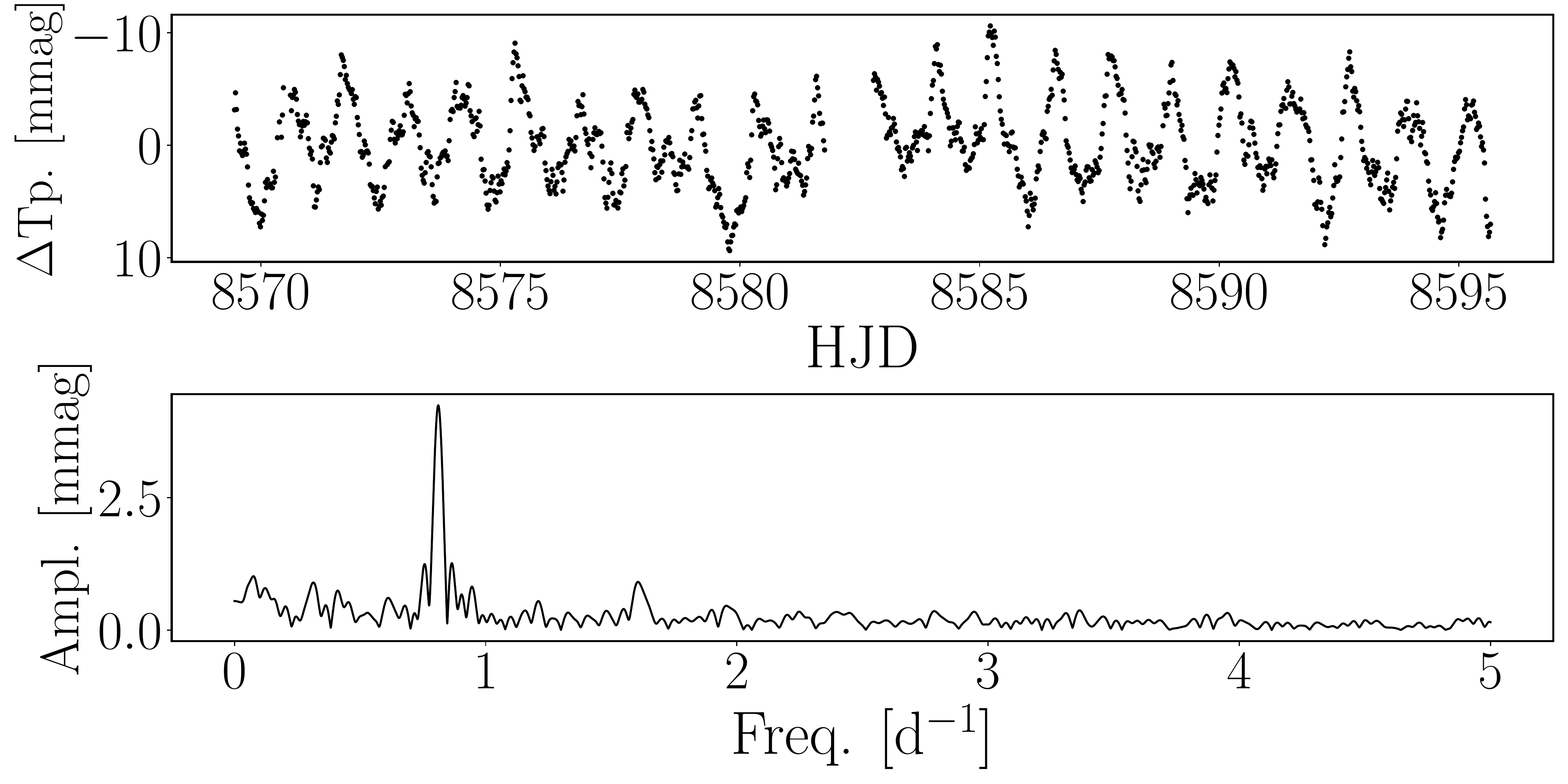}
    \caption{HD~94024}
    \end{subfigure} 
    \begin{subfigure}{0.33\linewidth}
    \includegraphics[width = \textwidth, trim=0 15 0 0,clip]{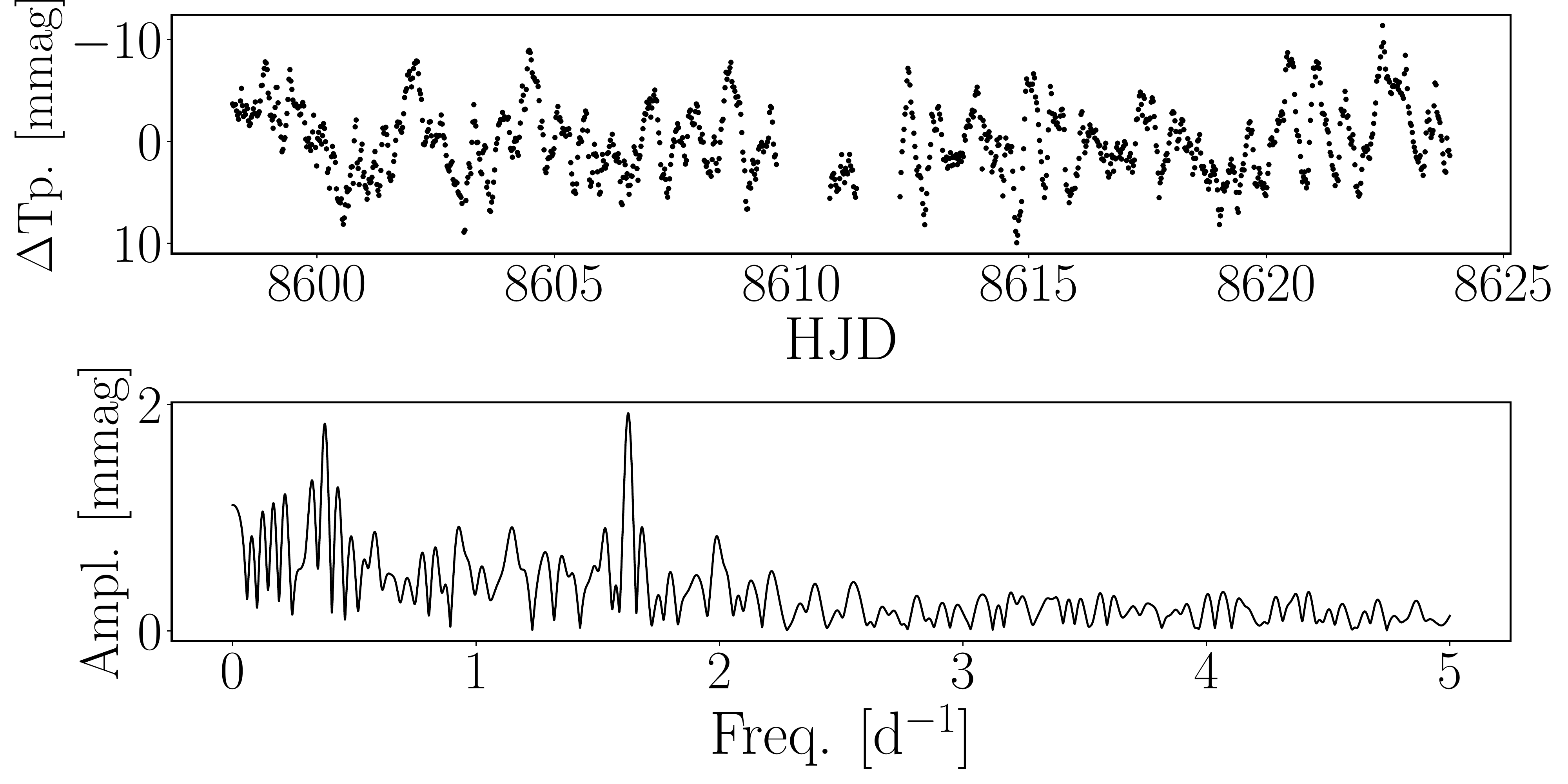}
    \caption{HD~105627}
    \end{subfigure}
    \begin{subfigure}{0.33\linewidth}
    \includegraphics[width = \textwidth, trim=0 15 0 0,clip]{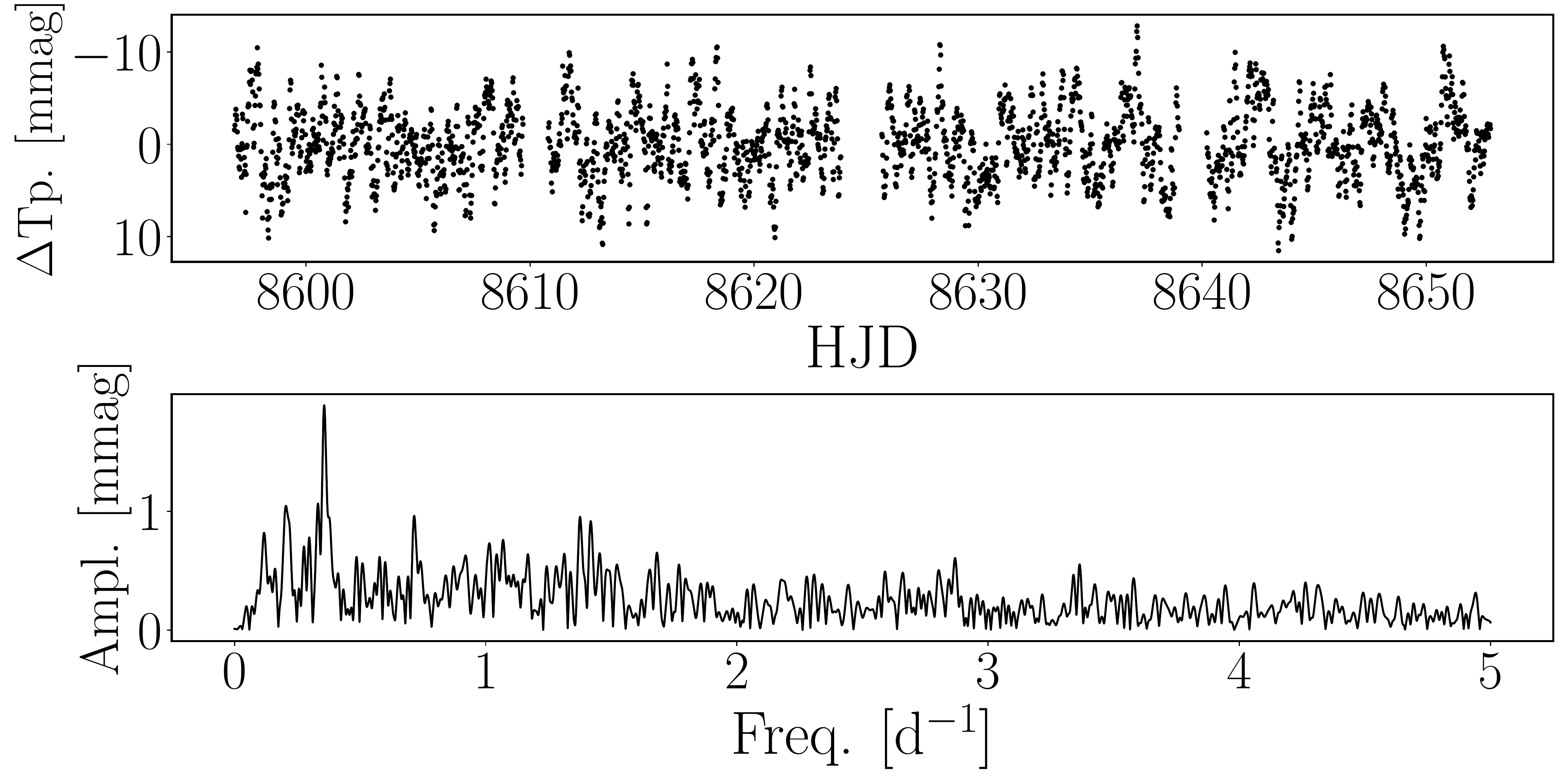}
    \caption{HD~130298}
    \end{subfigure}
    \begin{subfigure}{0.33\linewidth}
    \includegraphics[width = \textwidth, trim=0 15 0 0,clip]{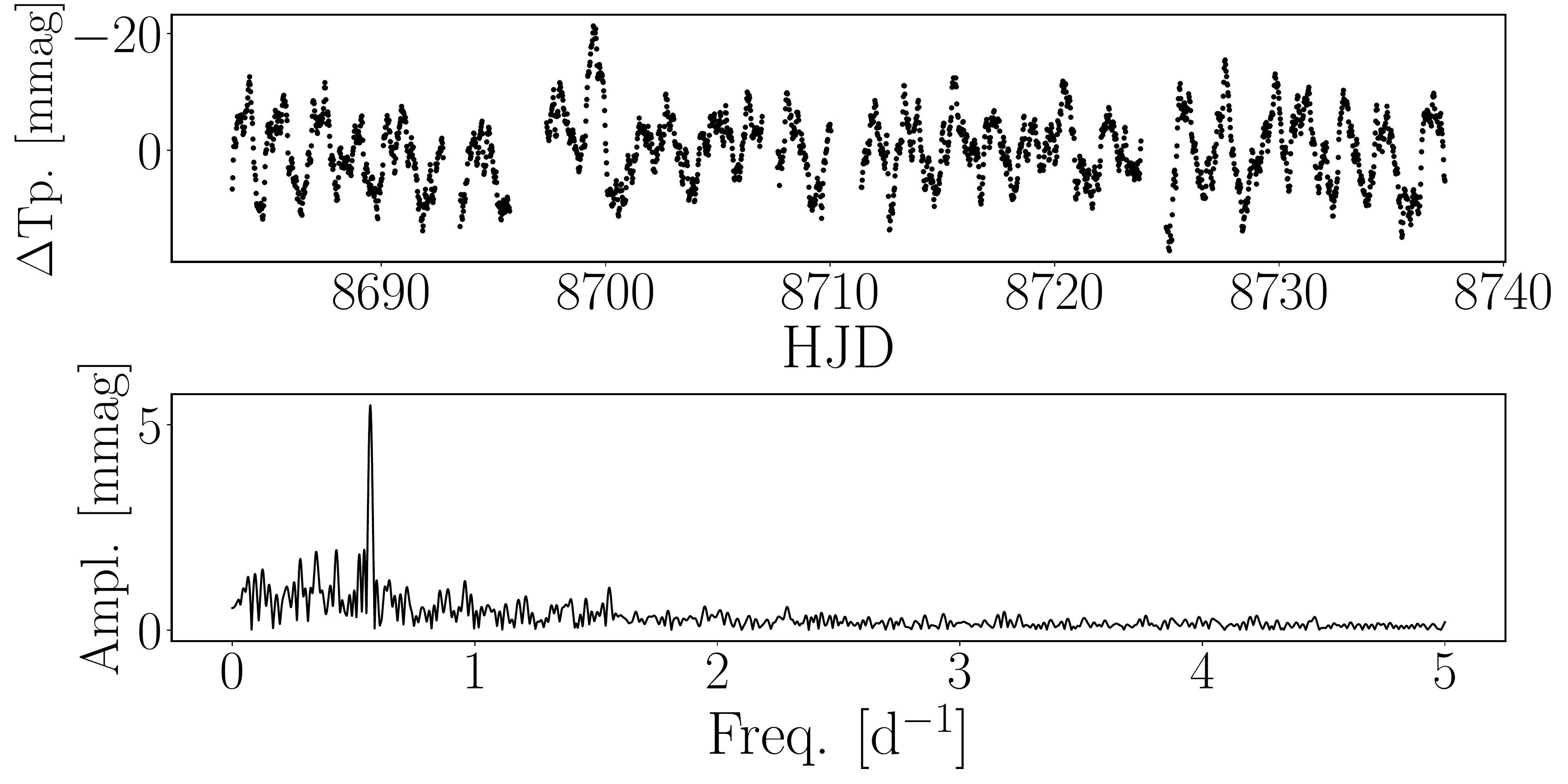}
    \caption{HD~229234}
    \end{subfigure} 
    \begin{flushleft}
    \begin{subfigure}{0.33\linewidth}
    \includegraphics[width = \textwidth, trim=0 15 0 0,clip]{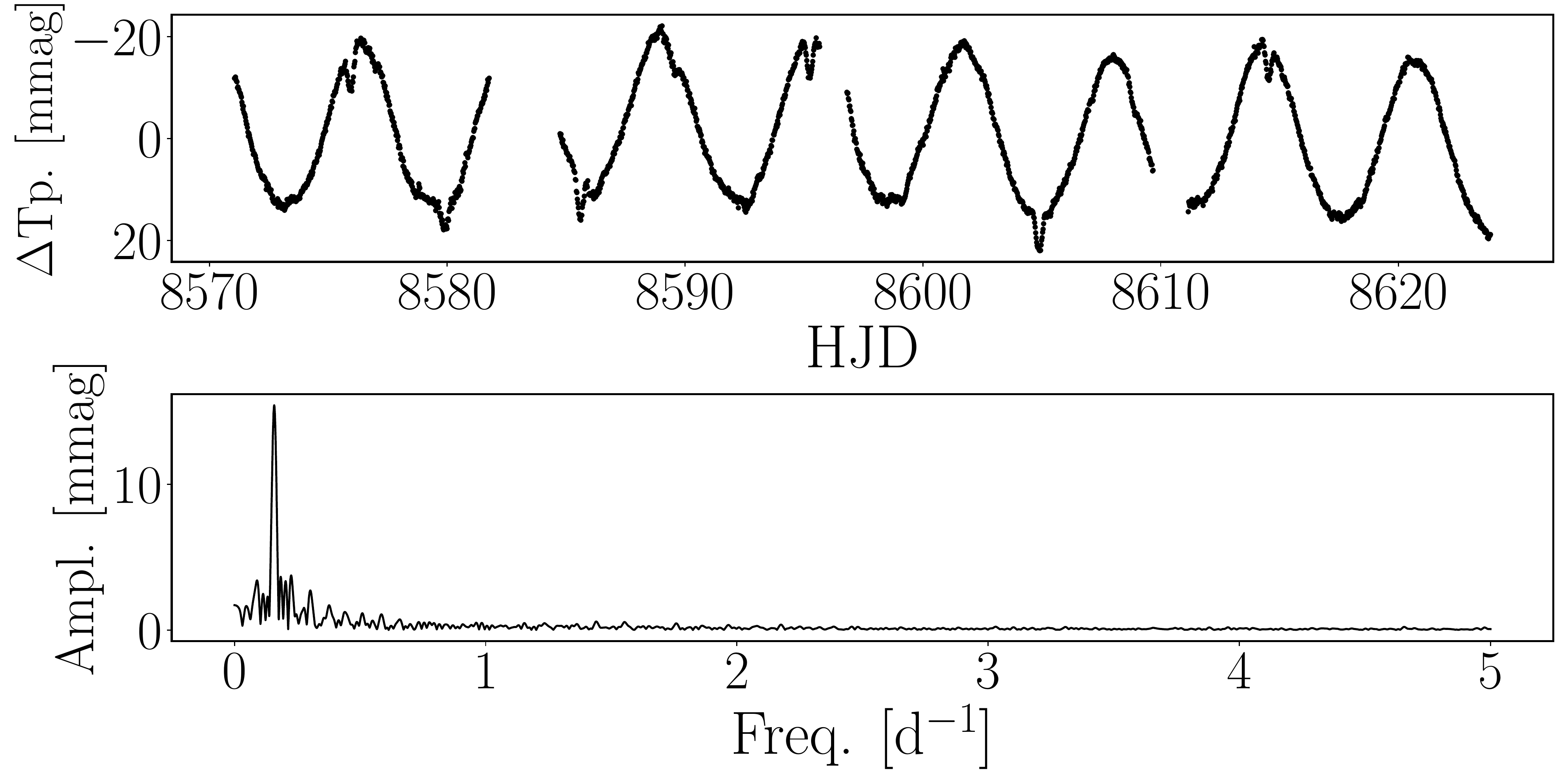}
    \caption{HD~308813}
    \end{subfigure}
    \end{flushleft}        
    \caption{TESS light curves (top panels of the sub-figures) and their corresponding HMM periodograms (bottom panels of the sub-figures) for the SB1 systems. No significant variability is seen beyond 5 d$^{-1}$. The heliocentric Julian date (HJD) corresponds to HJD$- 2~450~000$.}
    \label{fig:lightcurve}
\end{figure*}

\begin{figure*}
    \centering
    \begin{subfigure}{0.33\linewidth}
    \includegraphics[width = \textwidth, trim=0 120 0 0,clip]{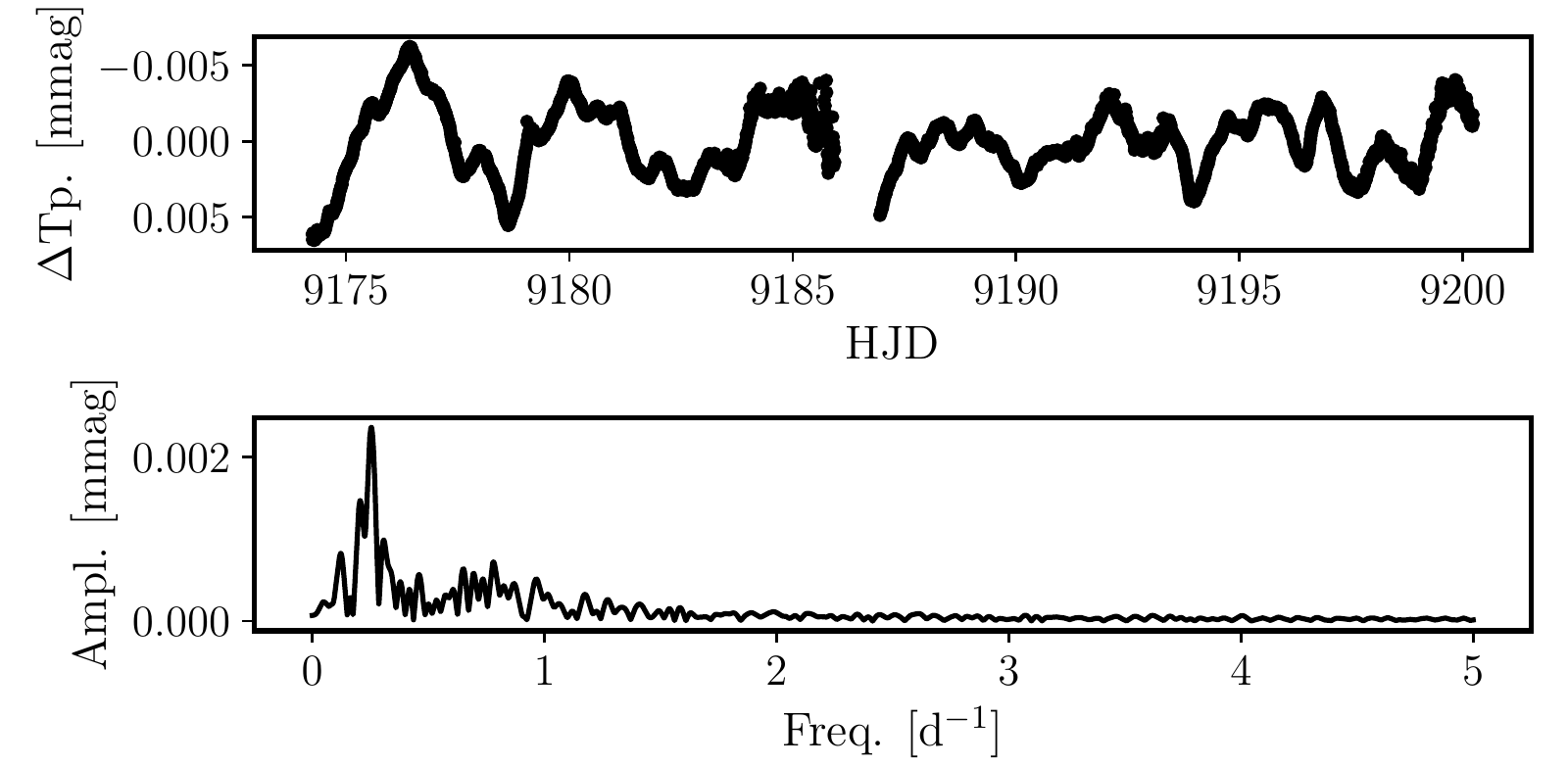}
    \caption{HD~30836}
    \end{subfigure}
    \begin{subfigure}{0.33\linewidth}
    \includegraphics[width = \textwidth, trim=0 120 0 0,clip]{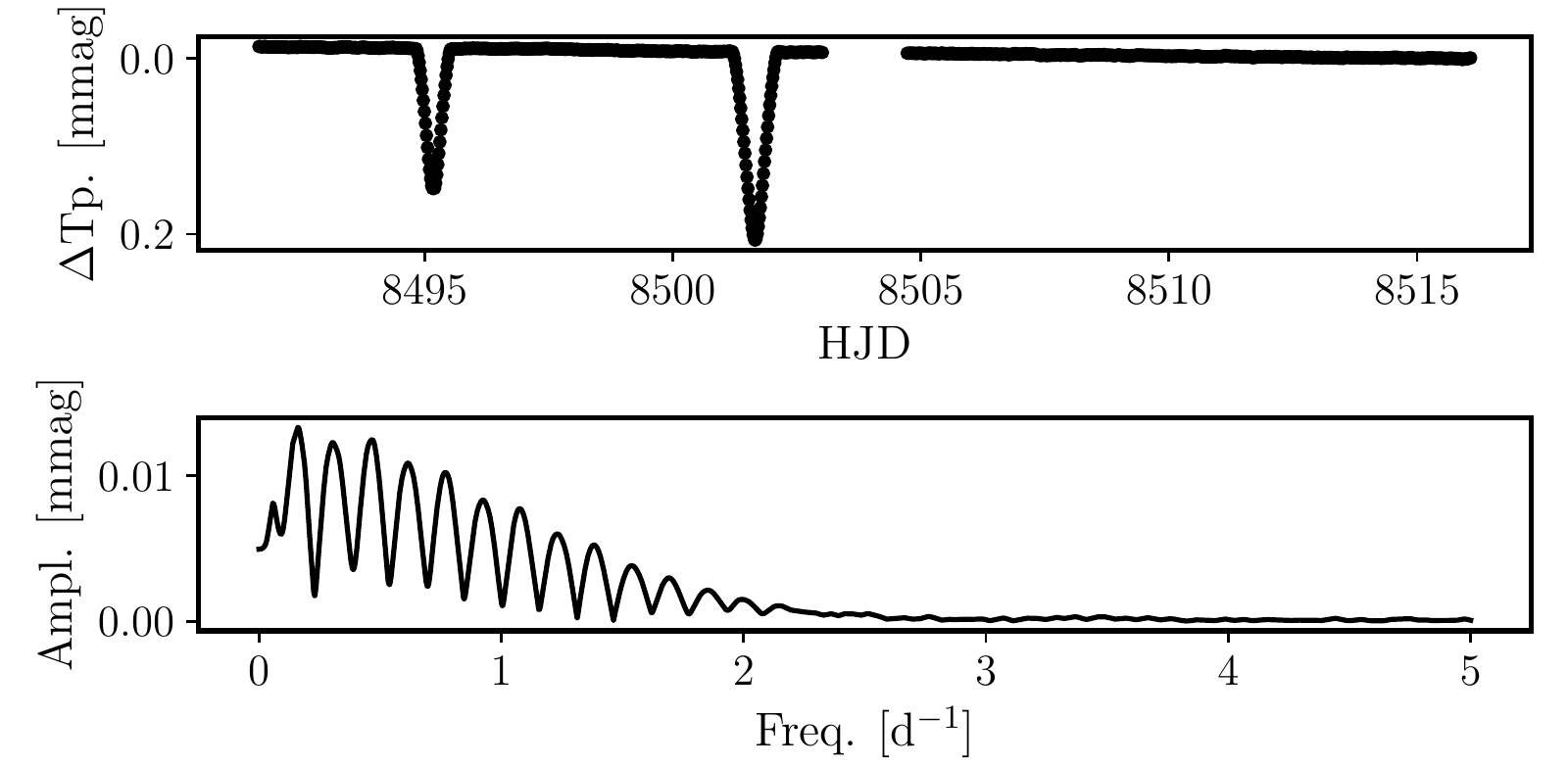}
    \caption{HD~52533}
    \end{subfigure} 
    \begin{subfigure}{0.33\linewidth}
    \includegraphics[width = \textwidth, trim=0 120 0 0,clip]{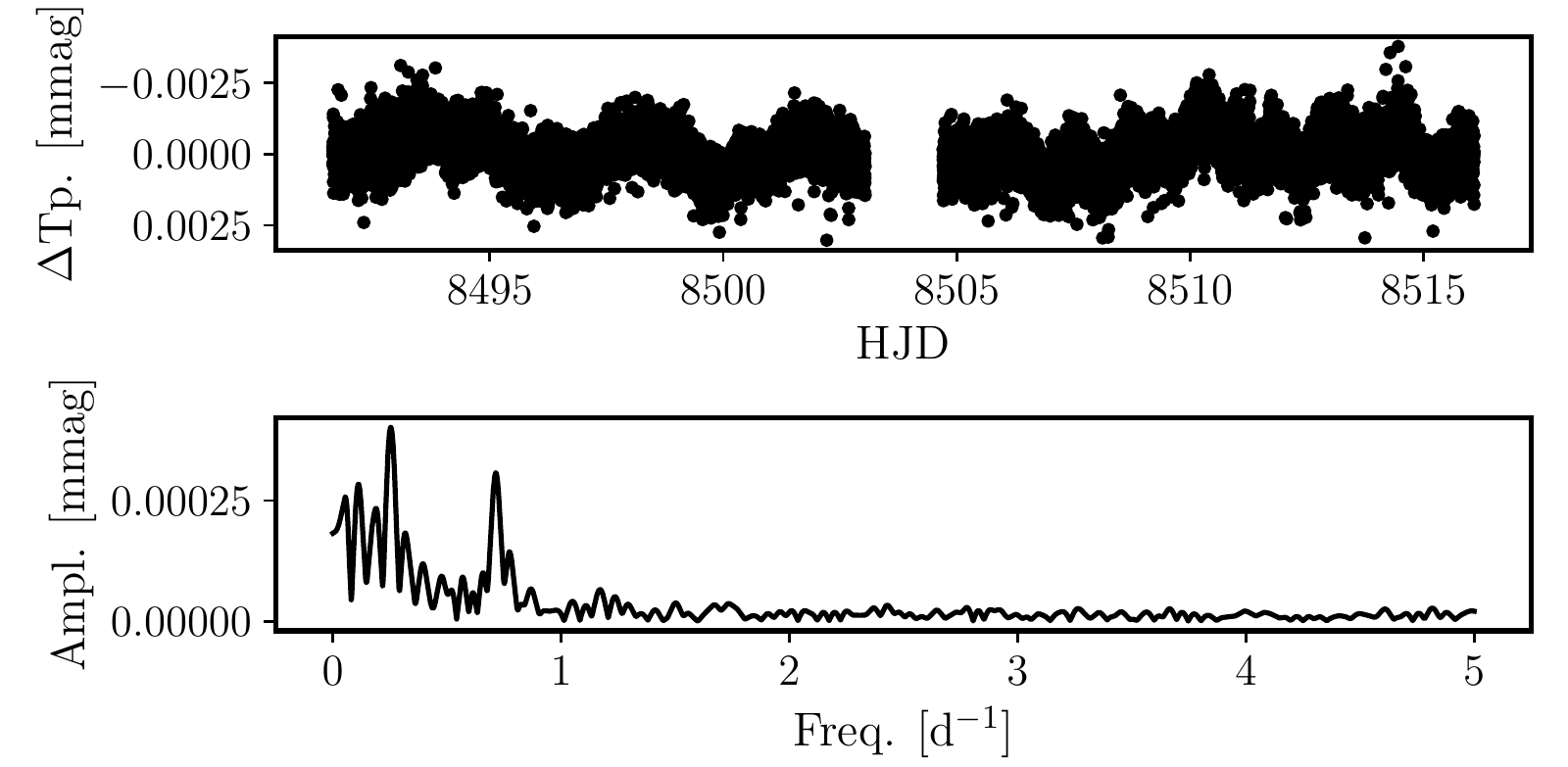}
    \caption{HD~57236}
    \end{subfigure}  
    \begin{subfigure}{0.33\linewidth}
    \includegraphics[width = \textwidth, trim=0 120 0 0,clip]{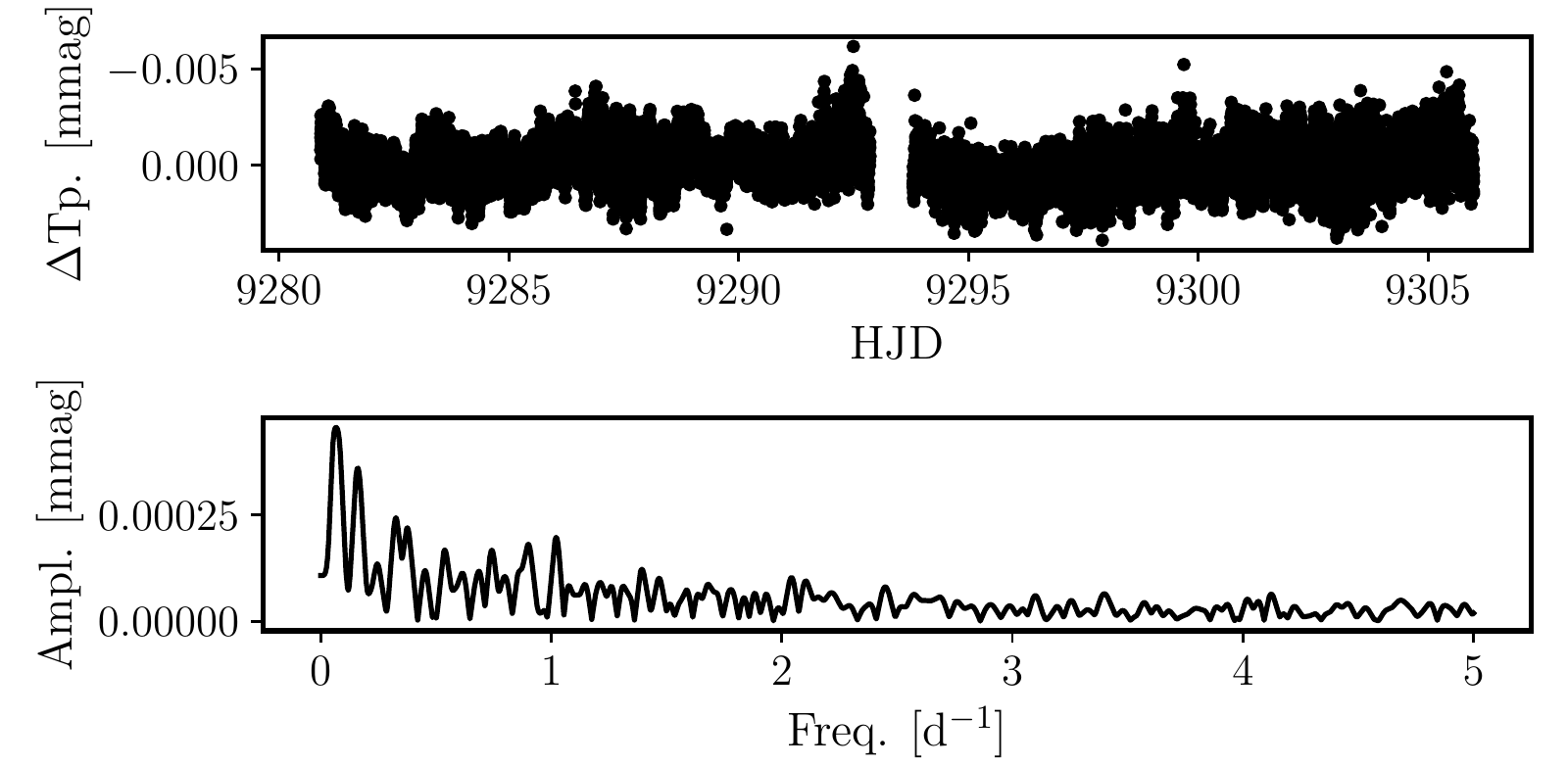}
    \caption{HD~91824}
    \end{subfigure}
        \begin{subfigure}{0.33\linewidth}
    \includegraphics[width = \textwidth, trim=0 120 0 0,clip]{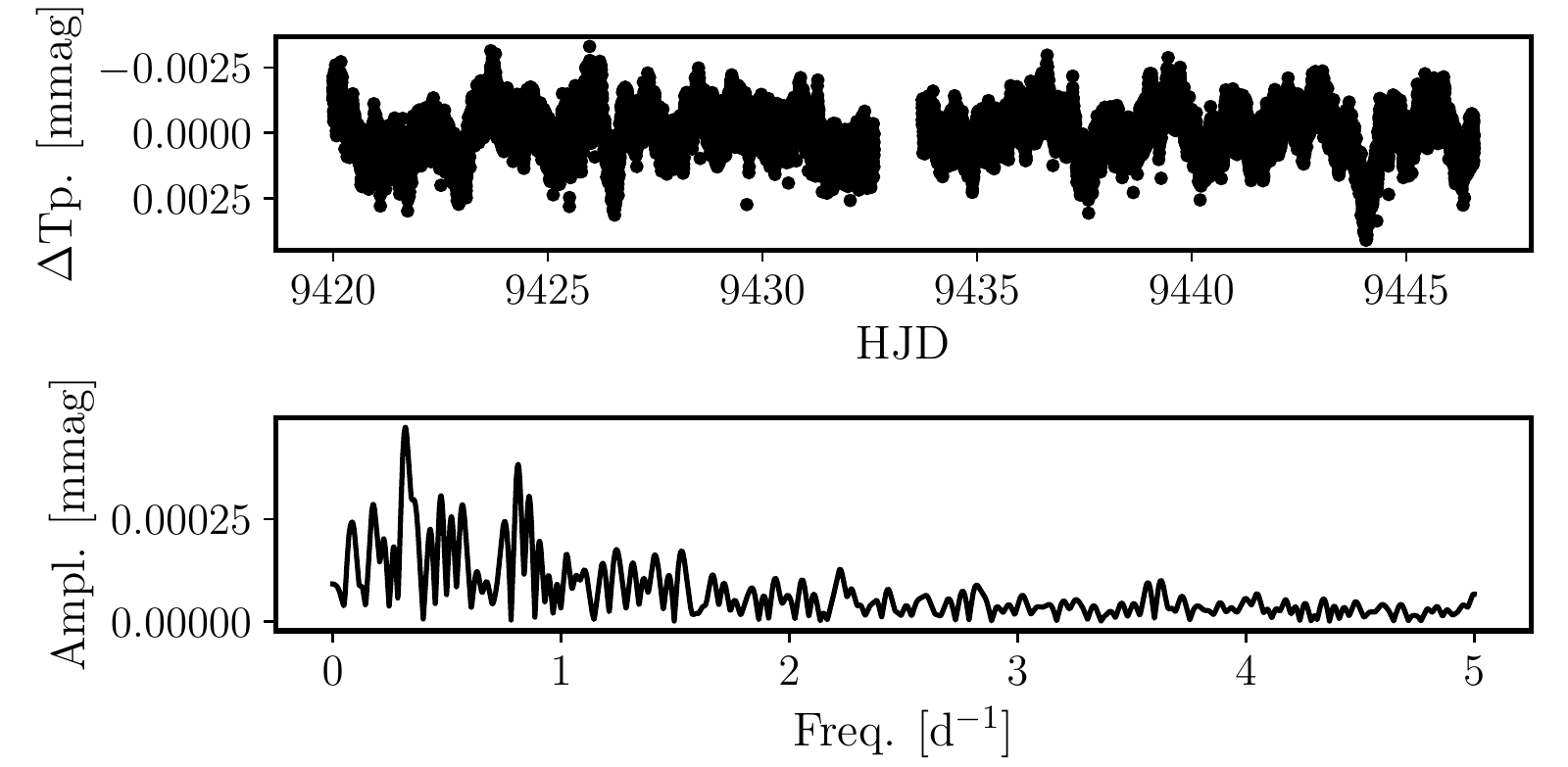}
    \caption{HD~192001}
    \end{subfigure}
    \begin{subfigure}{0.33\linewidth}
    \includegraphics[width = \textwidth, trim=0 120 0 0,clip]{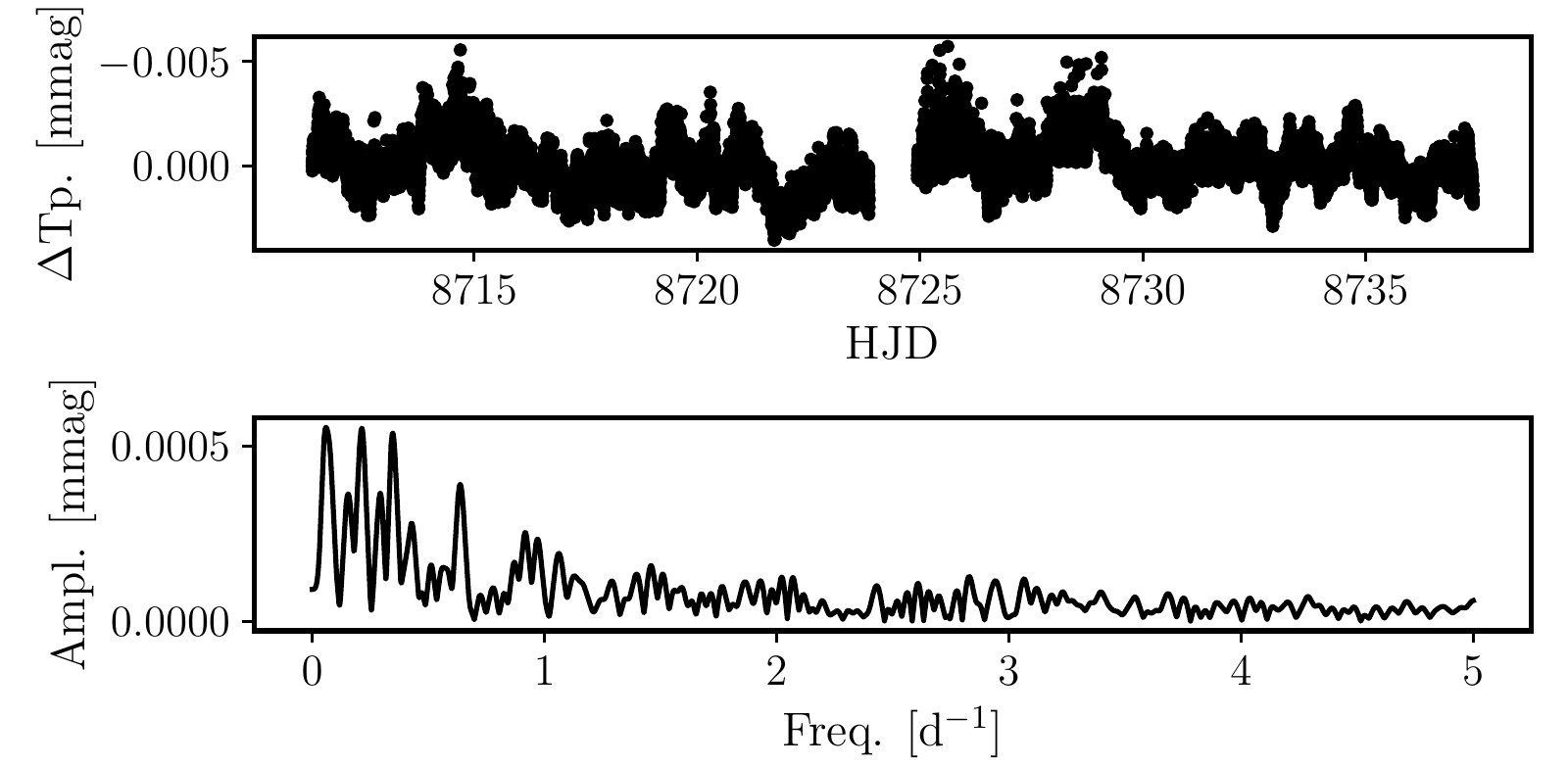}
    \caption{HD~199579}
    \end{subfigure}
     \begin{flushleft}
    \begin{subfigure}{0.33\linewidth}
    \includegraphics[width = \textwidth, trim=0 120 0 0,clip]{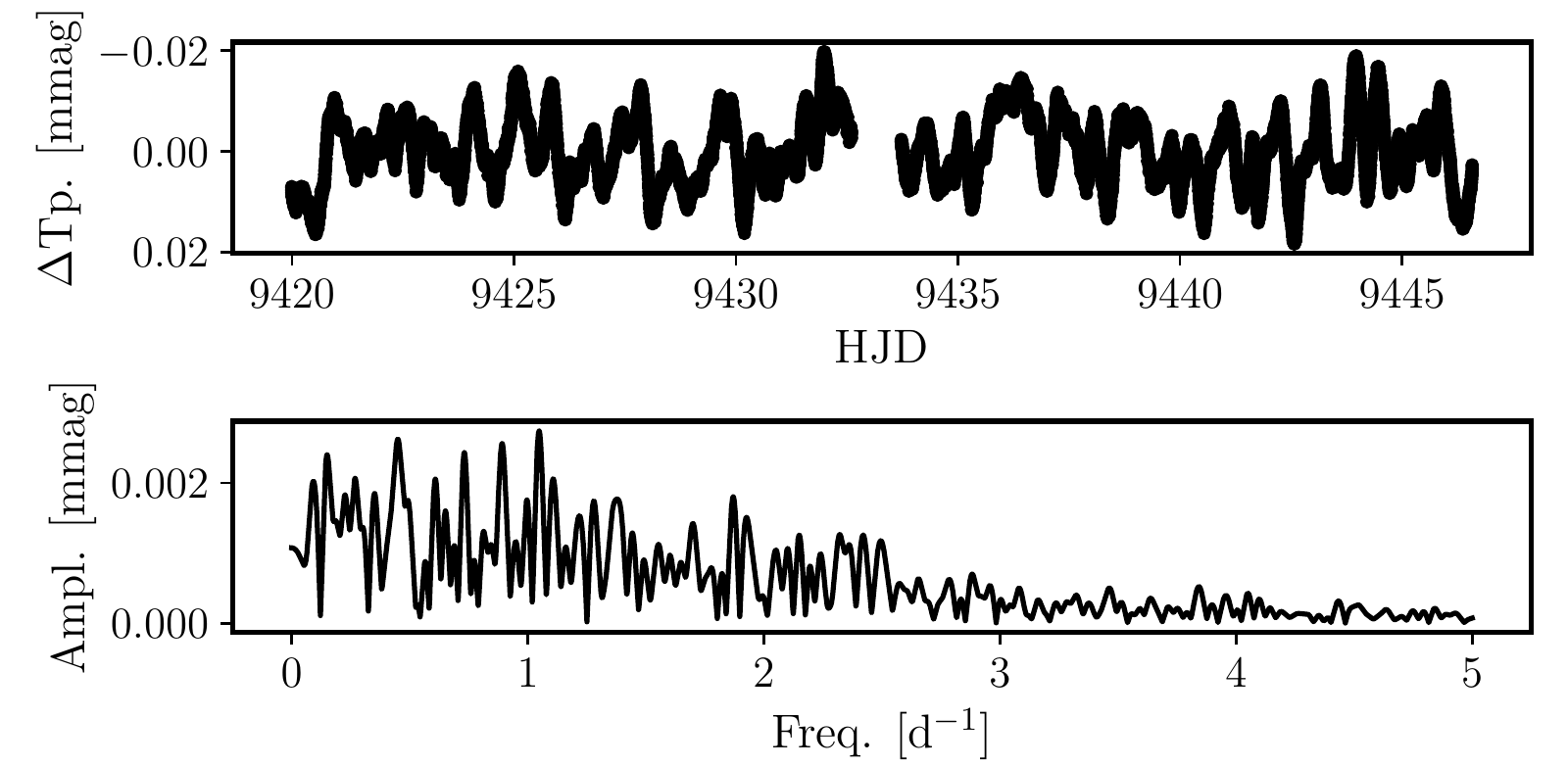}
    \caption{Schulte~11}
    \end{subfigure} 
    \begin{subfigure}{0.33\linewidth}
    \includegraphics[width = \textwidth, trim=0 120 0 0,clip]{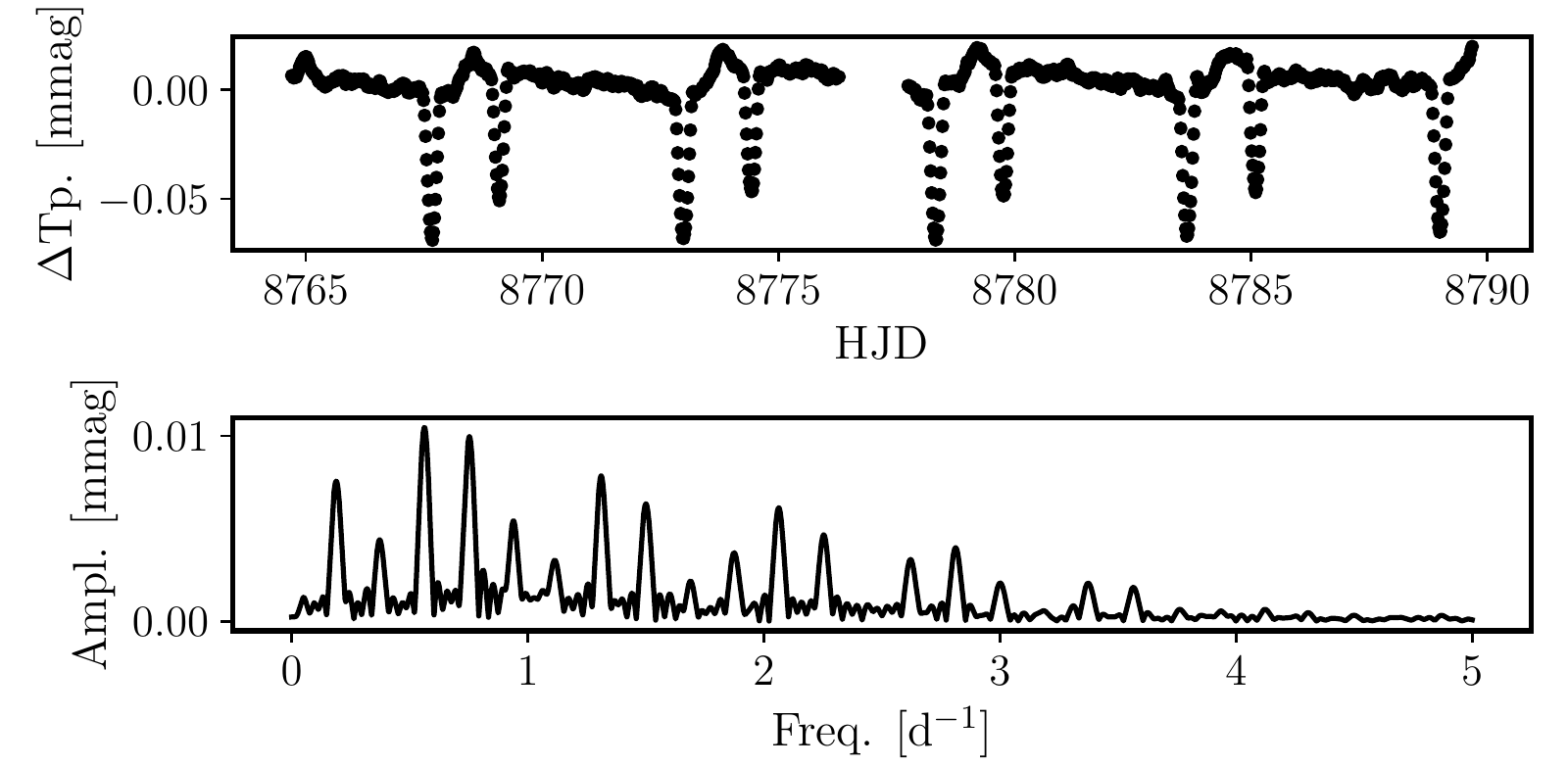}        \caption{V747~Cep}
    \end{subfigure}
    \end{flushleft}
    \caption{TESS light curves (top panels of the sub-figures) and their corresponding HMM periodograms (bottom panels of the sub-figures) for the newly detected SB2 systems. No significant variability is seen beyond 5 d$^{-1}$. HJD given in these figures corresponds to HJD$- 2~450~000$. There are no TESS light curves for HD~164438, HD~164536, HD~29763, HD~93028, HD~152405, HD~152723, HD~167263, or HD~167264.}
    \label{fig:lightcurveSB2}
\end{figure*}

All the light curves were detrended by using low-order polynomials and we looked for periodic signals using the HMM technique (see Sect~\ref{subsec:orbitalsolution}). While for deriving the orbital periods of the binary systems, we focused on the highest peak in the periodogram, for the analysis of the light curves, we used the iterative criterion given by \citet{mahy11} to define the significance of the different peaks in the periodograms. The light curves of the SB1 systems, with their respective periodograms, are shown in Fig.~\ref{fig:lightcurve} while those for the SB2 systems are shown in Fig.~\ref{fig:lightcurveSB2}.

\begin{figure}[htbp]\centering
    \includegraphics[width=9cm]{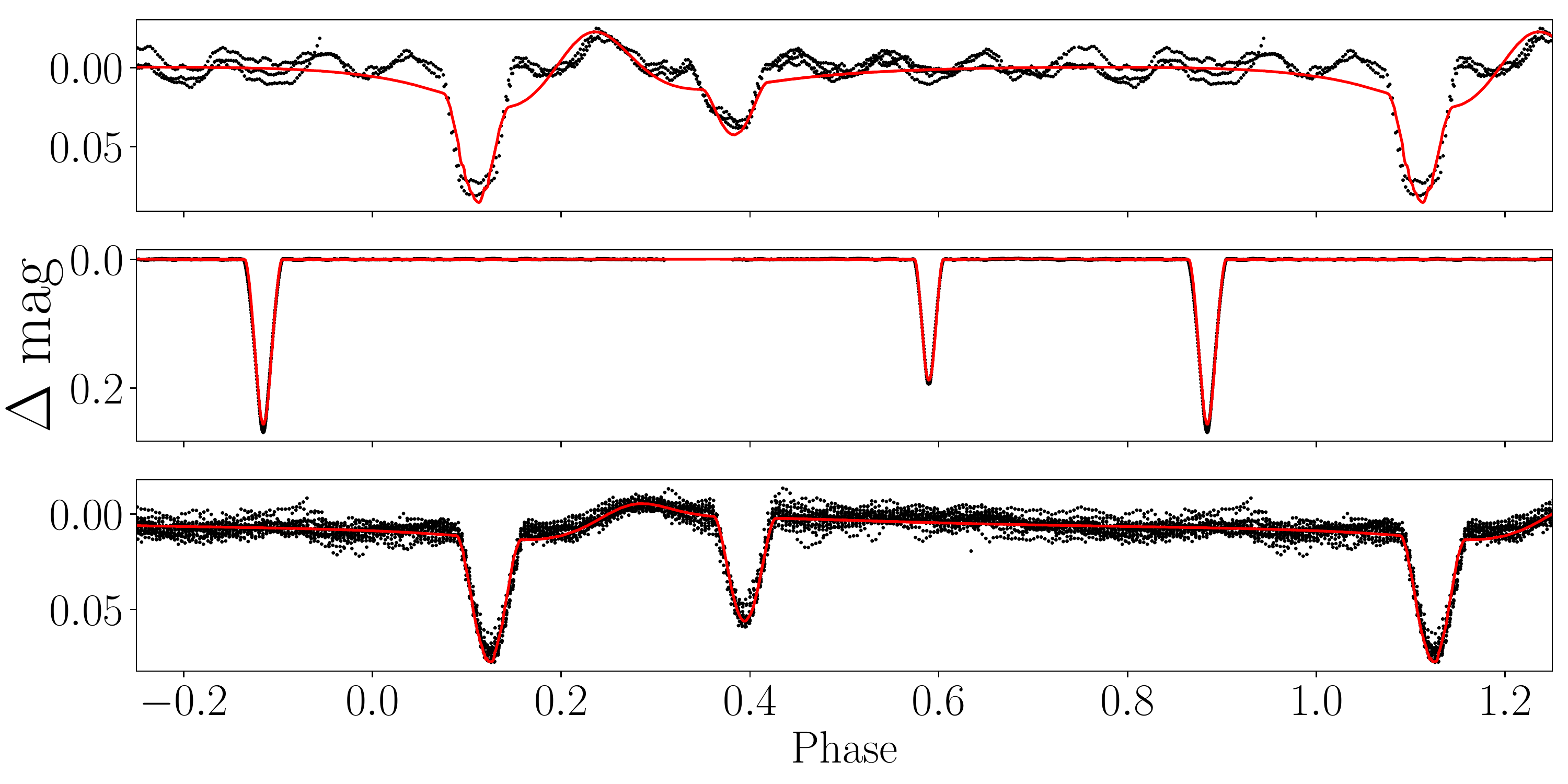}
    \caption{\label{Fig:phoebe} Best PHOEBE models (in red) compared to the TESS light curves (in black) of HD~37737 (top), HD~52533 (middle), and V747~Cep (bottom). }
\end{figure}

Once the list of significant frequencies has been generated for each object, we looked for signals that can be related to the orbital motion. HD~37737, HD~52533, and V747~Cep show light curves that display clear eclipses and can be (re-)classified as SB1E. For HD~37737, we detected 16 harmonics generated from its orbital frequency in the periodogram (the highest peak being at one-fourth the orbital period). The light curve also shows a pulse-like excess between the two eclipses that can be due to heartbeat variability \citep{trigueros21}. Given the presence of eclipses, it is ruled out that the secondary in HD~37737 is a compact object. The same conclusion can be drawn for HD~52533 and V747~Cep. We use PHOEBE (PHysics Of Eclipsing BinariEs, v2.3, \citealt{prsa05,conroy20}) to model the three light curves and derive the fundamental parameters of the individual components. We adopt bolometric albedos and gravity darkening coefficients equal to 1.0, and the square root law for the limb darkening \citep{mahy17,mahy20b}. The parameters are given in Table~\ref{tab:parameters_SB2} and the comparisons between the best PHOEBE models and the TESS light curves are displayed in Fig.~\ref{Fig:phoebe}. By comparing the masses with the detection predictions displayed in Fig.~\ref{Fig:massComp}, the masses of the secondary in HD~37737 is at the limit of detection. A secondary with a mass of $3.4~\Msun$ would even not be detected with our method. These rough estimations, however, need to be spectroscopically confirmed with additional higher-quality spectra.

HD~94024, HD~229234, HD~12323, and Cyg~X-1 show signals in their light curves that correspond to half their orbital period, suggesting ellipsoidal variations. Ellipsoidal variations might occur in the closest OB$+$BH binaries due to the deformation of the visible OB star (as it is stated above and in \citealt{masuda19}) but that does not guarantee that the companion is a degenerate object.

No clear frequencies were found in the periodograms of HD~14633, HD~15137 and HD~46573. For the other objects, we systematically considered the significant frequencies. Other mechanisms can be responsible for the signals in these light curves such as stochastic low-frequency variability (SLF; \citealt{bowman19a,bowman19b, bowman20}) or rotational modulations \citep{burssens20}. Assuming rotation as a possible cause for the detected frequencies, we can roughly deduce possible mass estimates for the secondaries in those systems. For that purpose, we used the projected rotational velocities and estimated radii (both from atmosphere modelling and from evolutionary models) of the visible star, and we used the significant frequencies detected from the light curves. This also assumes that the rotational axes of both stars are perpendicular to the orbital planes. These inclinations are then used to speculate on the possible mass ranges of the secondaries in those systems. A discussion object by object is given in Sect.~\ref{sec:individual}.

\section{Discussion}
~\label{sec:discussion}
\subsection{Nature of the unseen companions in SB1s}
\label{sub:discussion}

Our analysis has shown that we could retrieve the properties of stellar companions down to a mass ratio of 0.13--0.15 and a brightness ratio of $\sim 0.01-0.02$ but we are limited with the quality and the number of composite spectra in our dataset. The systems that we have selected for the present study are also limited in terms of orbital period. For longer-period systems, dedicated monitoring over several years need to take place, and, in that sense, Gaia will also help to unveil those systems \citep{janssens22}. 

From a large grid of detailed binary evolution models computed at Large Magellanic Cloud (LMC) metallicity with initial primary masses between 10 and $40\,\Msun$, \citet{langer20} predicted that about 3\% of the LMC late-O and early-B stars in binaries are expected to possess a stellar-mass BH companion. Even though these results were produced at LMC metallicity, there is no reason to believe that the fraction is significantly different at Galactic metallicity. According to these predictions, a high fraction of OB$+$BH systems are expected with orbital periods close to 6 days and RV semi-amplitudes around 100\,\kms\ if the BH progenitor filled its Roche lobe and interacted with its companion during the MS (Case A evolution), or orbital periods of the order of 1~yr and RV semi-amplitudes around 35\,\kms\ if they went through Case B mass transfer. 

\begin{figure}[htbp]\centering
    \includegraphics[width=9cm]{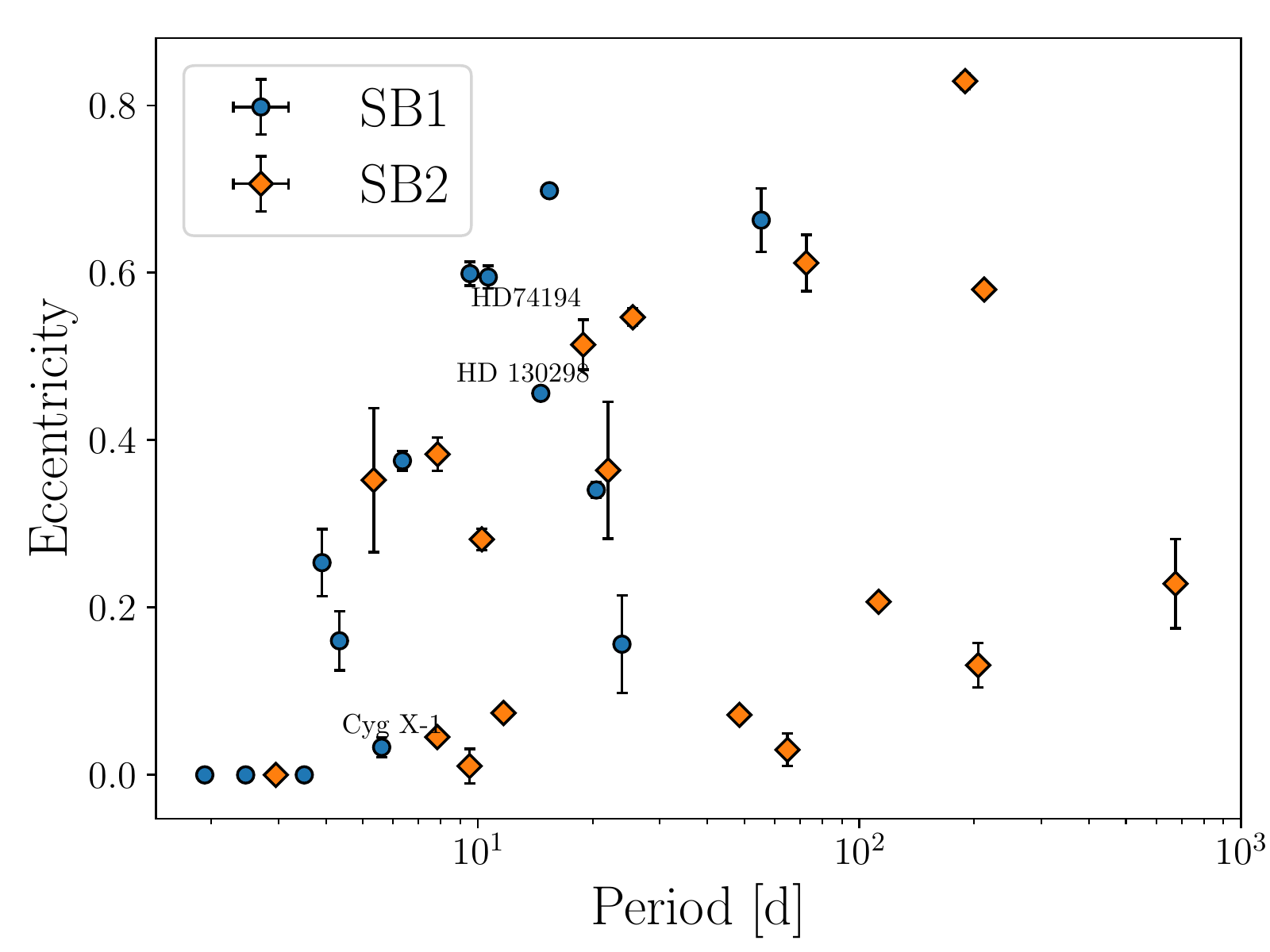}
    \includegraphics[width=9cm]{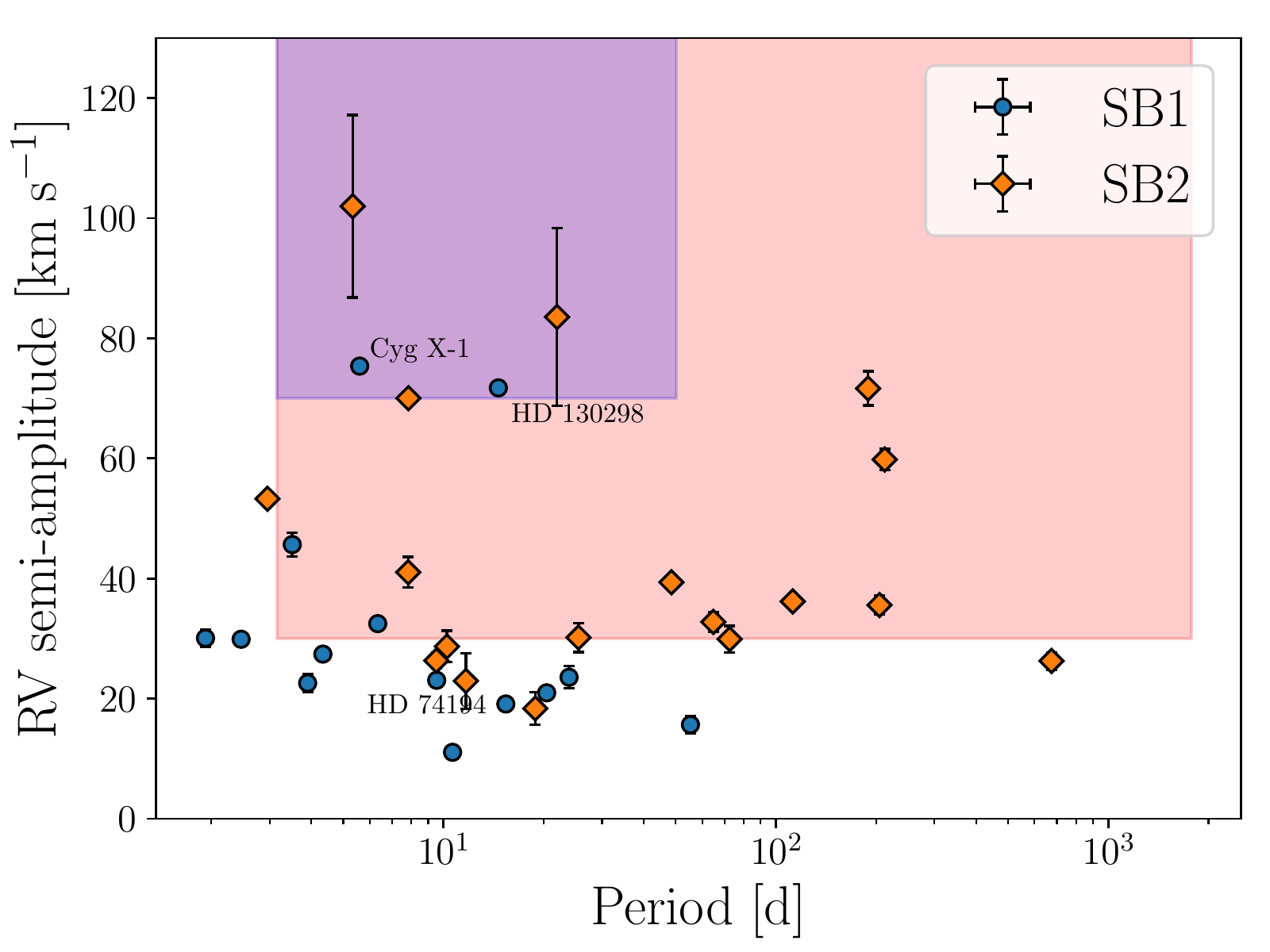}
    \caption{\label{Fig:P-e-K} {\it Top:} Period versus eccentricity diagram for systems in our sample. Blue circles represent the SB1 systems, orange diamonds the SB2s. {\it Bottom:} Period versus primary RV semi-amplitude diagram for systems in our sample. The colour code is the same as in the top panel. We also plotted the parameter space that corresponds to the predictions of \citet{langer20} for case A (blue) and case B (red) mass transfers.}
\end{figure}

In the top panel of Fig.\,\ref{Fig:P-e-K}, we show the period-eccentricity diagram for all the systems in our sample. The SB1s have in general shorter orbital periods than the newly classified SB2s. This is not expected from a homogeneous sample of binaries (Shenar et al., 2022, A\&A, in prep.). Here, we selected already-reported SB1s and exclude the already-known SB2s, biasing our sample. There are several SB1 systems having orbital periods shorter than 10 days and eccentricities higher than 0.2. The bottom panel of Fig.\,\ref{Fig:P-e-K} shows the RV semi-amplitudes of the visible stars as a function of the orbital periods of the system. We also plotted the parameter space corresponding to the predictions of \citet{langer20} for case A (blue) and case B (red) mass transfers. Comparing our SB1 population with these predictions gives us 2 possible OB$+$BH systems if the stellar-mass BH is formed after a case B and 2 if it is formed from a case A mass transfer, respectively. Those systems are: Cyg~X-1, and HD~130298 for case A (appearing in the blue box of the bottom panel of Fig.~\ref{Fig:P-e-K}), and HD~308813, and HD~229234 for case B (in the red box of the bottom panel of Fig.~\ref{Fig:P-e-K}). 

\subsubsection{OB$+$BH candidates}

Using the binary mass function to derive the mass estimates of the companions, we found three objects (Cyg~X-1, HD~130298, and HD~37737) for which the companion should be classified as B5 or earlier ($M > 5~\,\Msun$), and therefore should be detected in the composite spectra. These systems are clearly candidates to host a (X-ray quiet) stellar-mass BH, except HD~37737 for which the light curve shows eclipses. As expected, no evidence of a massive non-degenerate companion was found for Cyg\,X-1. This object is indeed known for hosting a stellar-mass BH with a mass of approximately $14\,\Msun$ \citep{orosz11} up to $21\,\Msun$ \citep{Miller-Jones21}. 

Another interesting candidate is HD~130298, which, in contrast to Cyg~X-1 exhibits a high eccentricity of $e=0.47$. We find a minimum mass of $7.7 \pm 1.5\,\Msun$ for the companion. However, we did not detect any signatures in the composite and disentangled spectra. No X-ray detections were reported from the Second ROSAT all-sky survey (2RXS) source catalog \citep{boller16}. The fact that we do not detect the signature of a companion in HD~130298 suggests that it could be either an X-ray quiet stellar-mass BH or a stripped helium star. At this stage, the possibility of having a stripped star more massive than $\sim 7~\Msun{}$ cannot be fully excluded. However, it seems very unlikely as \citet{gotberg18} showed that such systems (MS O-type star and massive stripped helium star) would be detectable even from the optical bands as the stripped star would outshine the companion especially in the \ion{He}{ii} lines but we do not detect such features in the composite spectra of HD~130298. Furthermore, stripped stars more massive than $7.5 \pm 1.5\,\Msun$ are expected to appear as Wolf-Rayet stars, as estimated in \citet{shenar20}. There is no doubt that we would detect such a star in the case of HD~130298. This strongly points to the presence of a quiet stellar-mass BH as companion of HD~130298 and emphasises the importance of acquiring new observations in different wavelength domains to firmly confirm this important detection. 

Finally, HD~75211 and HD~229234 are also candidates but the likelihood is lower. For HD~75211, its companion has an expected mass between 3 and $12\,\Msun$. We can however rule out the presence of a companion more massive than $5-6\,\Msun$ from our simulations, but not lower.  It seems therefore very unlikely that its companion is a stellar-mass BH. We can also rule out that this object form a hierarchical triple system where the O star is the outer object. Our data are, however, not good enough to reject the possibility that the companion is a stripped star. For HD~229234, the mass of the secondary is higher than $2.6 \pm 0.3~\,\Msun$ and could reach $\sim 20~\,\Msun$. The secondary, if still on the MS, is therefore at the limit of detection (see Fig.~\ref{fig:DetectionLimit}). A stripped helium star would not be detected from our data, nor would an inner close system if its mass were not higher than $10~\,\Msun$. This latter case is, however, unlikely since ellipsoidal variations are detected in the light curve. These ellipsoidal variations strongly indicate that one or both objects are distorted by the tidal influence of the orbiting companions. We also detected systematic differences in the systemic velocity of this system through the different epochs, suggesting that it could be a triple system, similar to HD~96670 \citep{Gomez21}.

\subsubsection{OB$+$NS candidates}

Nine SB1s have companions that have mass estimates between $1~\,\Msun$ and $5~\,\Msun$. If these companions are degenerate, this range is similar to the expected mass estimate of NSs, but cannot exclude low-mass stellar BHs \citep{belczynski10,fryer14,zevin20}. We can also not exclude non-degenerate low-mass companions from the data that we have acquired. In addition to HD~130298, five SB1s are also reported as runaway stars, those are stars that have a space velocity of $30$~\kms\ or higher: HD~12323, HD~14633, HD~15137, HD~46573, and HD~94024. Both dynamical interactions within a cluster and a supernova in a close binary can produce runaway SB1 systems. Therefore, the nature of the companions cannot be inferred from the runaway status of these objects. 

\subsubsection{Physical properties}

\begin{figure}[htbp]\centering
    \includegraphics[width=9cm]{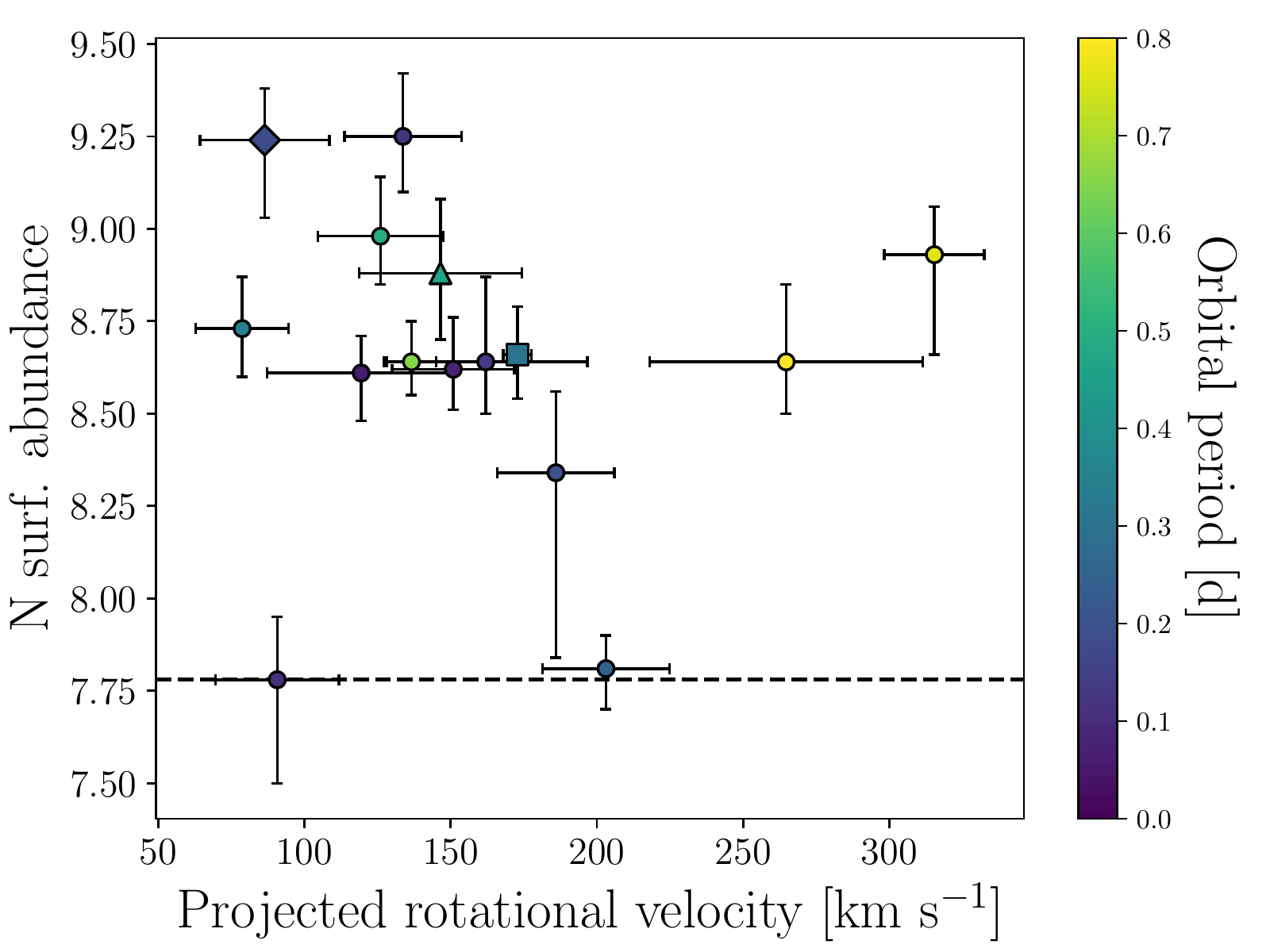}
    \caption{\label{Fig:hunter} Projected rotational velocity versus nitrogen surface abundances of the SB1s in our sample. The colour bar indicates the orbital periods of the systems. The baseline value for the nitrogen enrichment is equal to 7.78 and is marked by a dashed line. Cyg~X-1, HD~130298, and HD~74194 are marked by a diamond, a triangle, and a square, respectively. }
\end{figure}

OB stars in post-interaction OB+BH binaries are expected to be rapid rotators and enriched in helium and nitrogen. Depending on whether the mass transfer occurred through case A or case B, the expected enrichment will be different \citep{langer20}. In the case of case B mass transfer, the OB stars remain mostly unenriched because only small amounts of mass (about 10\% of their initial mass) are accreted, while, for case A mass transfer, much more mass, coming from deeper layers of the mass donor, directly falls onto the surface of the mass gainer and is accreted. In Fig.~\ref{Fig:hunter}, we show the distribution of nitrogen surface abundance as a function of the projected rotational velocity (i.e. the Hunter diagram) for all the SB1s in our sample. Four SB1s show a nitrogen enrichment that could be produced by rotational mixing \citep{maeder00}. Two objects have a projected rotational velocity lower than 200~\kms\ with no significant enrichment. The remaining ten objects have a lower $\vsini$ but show a high nitrogen enrichment at their surface. The fact that we do not know the inclinations of the systems might be a bias to explain the causes of these enrichments but a possibility would be that these enrichments are due to binary interactions \citep{demink13}. However, if these ten objects are mass gainers from conservative mass transfer, it is expected that they show rapid rotation, and one would therefore expect that they are seen under a low inclination. 

\begin{figure}[htbp]\centering
    \includegraphics[width=9cm]{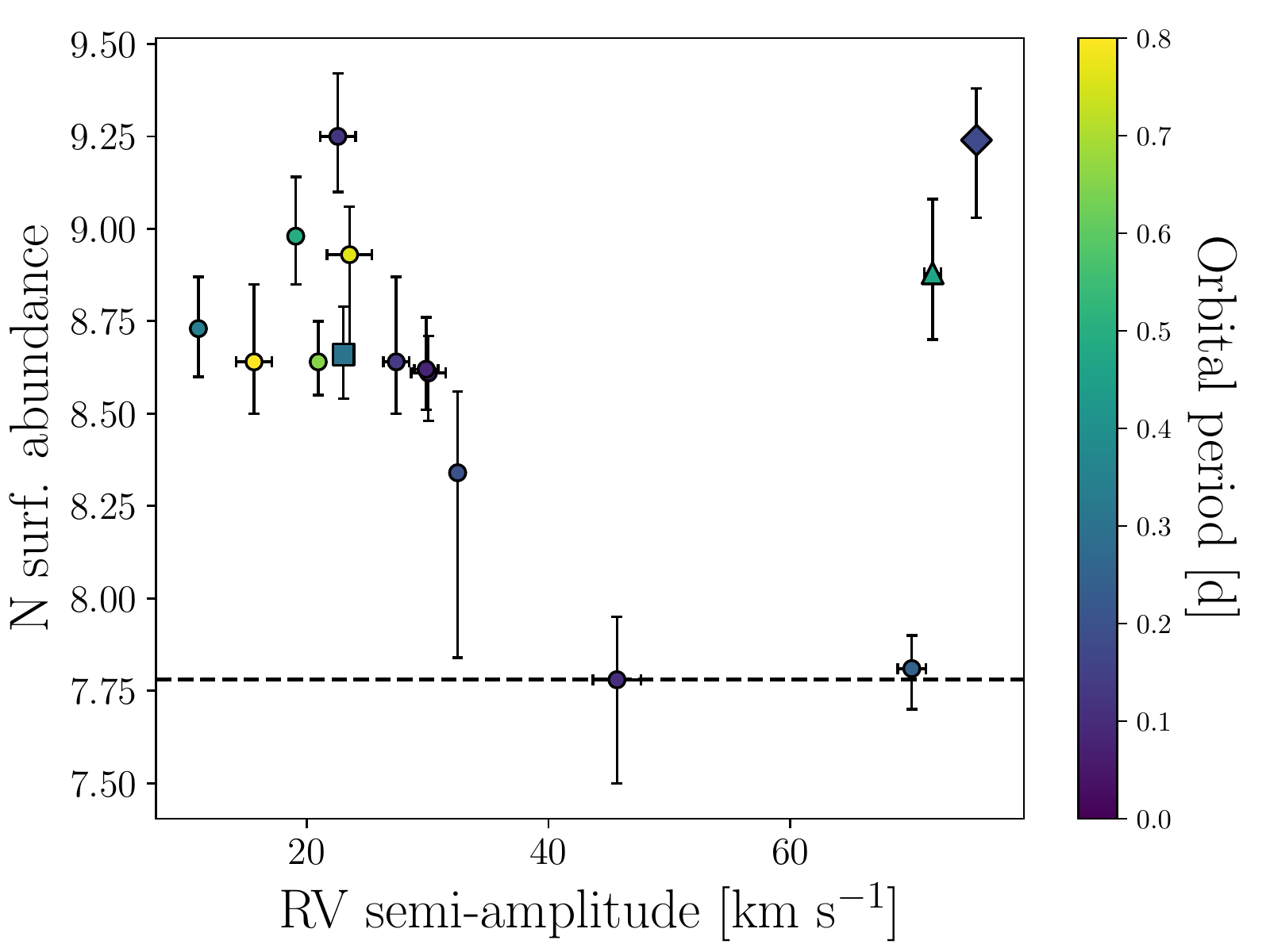}
    \caption{\label{Fig:distrib_N} Diagram showing the RV semi-amplitude of the systems and the nitrogen enrichment of the visible star. The colour bar indicates their orbital period. Cyg~X-1, HD~130298, and HD~74194 are marked by a diamond, a triangle, and a square, respectively. The baseline value is equal to 7.78.  }
\end{figure}

In Fig.~\ref{Fig:distrib_N} we show the distribution of nitrogen enrichment as a function of the RV semi-amplitude of the visible stars in our SB1 populations. Most of the systems that show nitrogen enrichment at their surface have $K_1<30$~\kms\ and two systems have $K_1>70$~\kms (among them Cyg~X-1). The other systems for which no significant enrichment has been measured have $30 < K_1 < 70$~\kms.

From Figs.~\ref{Fig:hunter} and~\ref{Fig:distrib_N}, there is no significant difference in terms of nitrogen enrichment between the SB1s that are supposed to evolve through case A ($P_{\rm orb} \lesssim 10$~d) or case B ($P_{\rm orb} \gtrsim 10$~d) mass transfer. The similarities regarding the nitrogen surface abundances between these stars and Cyg~X-1 or HD~74194 (which are marked by a diamond and a square in Fig.~\ref{Fig:hunter}, and are known to host a BH or a NS, respectively) are striking. 

\subsection{X-ray emission}

X-ray detections were reported for six objects: Cyg~X-1, HD~74194, LS~5039, HD~14633, HD~15137, and HD~12323. No X-ray detections have been reported for the other stars in the literature. Whether or not they are X-ray emitters thus requires further dedicated monitoring. 

The interaction of the primary's wind with the unseen companions in our SB1 systems may give rise to X-ray emission. This is most evidently so in the case where the companion is a stellar-mass BH or NS, where the deep potential heats any accreted matter to X-ray emitting temperatures. However, X-ray emission may also arise in the presence of a non-degenerate companion due to the thermalisation of the fast O star wind. The physical processes involved in either accreted or braked stellar winds are complex, and it is beyond our means to compute them in detail. Instead, we derive some order-of-magnitude estimates using suitable but simplified analytic approximations.  

\subsubsection{Wind accreting black holes}
The case of a wind accreting BH may potentially produce the highest X-ray
luminosities and is thus given most room in our consideration. However, even in this case, high levels of X-ray emission are only expected if the
in-falling wind material can form an accretion disk. For the case of a  BH companion, we expect the BH to accrete matter from the stellar wind of the O star via Bondi-Hoyle accretion \citep{Bondi1944}. When the accreted matter has sufficient angular momentum, it can form an accretion disk around the BH. Such a disk is expected to radiate energy mostly in X-rays \citep{Frank2002}. To estimate whether an accretion disk can form, and the corresponding X-ray luminosity, we follow the work of \citet{Sen2021}. We take the maximum possible unseen companion mass as the mass of the BH, which increases the likelihood of accretion disk formation. For our general case, we assume a non-rotating BH \citep{Qin2018}. We apply a standard $\beta$-law for the wind velocity  and calculate the wind mass-loss rate following the prescription of  \citep{vink00}, using the luminosity, effective temperature and spectroscopic mass of the O stars derived in this work (Table\,\ref{tab:parameters}). 

Only for HD\,229234 do we obtain  $j_{\rm acc}/j_{\rm ISCO} > 1$ (Table\,\ref{X-ray}), where $j_{\rm acc}$ is the specific angular momentum of the accreted matter and $j_{\rm ISCO}$ is the specific angular momentum of a particle in the innermost stable circular orbit of the BH. This implies that with a 14\,$M_{\odot}$ BH companion, the accretion flow is expected to form an accretion disk, giving rise to an X-ray luminosity of about 600\,$L_{\odot}$. As this X-ray luminosity is large enough to be detectable by current non-focussing all-sky X-ray monitoring telescopes \citep{Priedhorsky1996}, a 14\,$M_{\odot}$ 
BH companion can be safely excluded. However, we cannot exclude the existence of a 3\,$M_{\odot}$ BH companion (corresponding to the minimum unseen companion mass of this system) as our analysis predicts that an accretion disk does not form around the BH if its mass was 3\,$M_{\odot}$. 

For all other systems, an accretion disk is not expected to form within 
our standard assumptions. However, in two systems, Cyg\,X-1 and HD\,94024, the angular momentum of the accreted wind matter is so high that a disk may be expected for the case of a spinning BH \citep{Kerr1963,McClintock2006,Visser2007}. In fact, Cyg X-1 is known to contain a maximally spinning BH of 21.2 $M_{\odot}$ \citep{Miller-Jones21}. Assuming that to be the case, the method of \citet{Sen2021} does predict an accretion disk radiating about 700\,$L_{\odot}$, a factor of a few smaller than the observed average of $\sim$2600\,$L_{\odot}$ \citep{orosz11}. Due to the absence of bright X-rays from HD\,94024, a 4\,$M_{\odot}$ Kerr BH companion can be excluded. 

When assuming unimpeded strictly radial in-fall onto the BH, the level of the thermal bremsstrahlung escaping from the accreted adiabatically heated plasma is many orders of magnitude below the X-ray emission of an accretion disk, considering the same accretion rate \citep{Shapiro1986}. While turbulence, magnetic fields, and non-radial accretion may all enhance the X-ray emission \citep{Sharma2007}, the expected X-ray flux is still well below that of an equivalent accretion disk. In any case, it is difficult to constrain the three mentioned processes. Therefore, for any of the SB1s considered here, the absence of BHs cannot be ruled out based on the absence of detected X-ray emission. This holds even for HD\,130298 assuming a 48\,M$_{\odot}$ BH companion. 

\subsubsection{Wind collision from a MS companion}
If the unseen companion in our SB1 systems is a MS star, we may  expect some level of X-ray emission due to the braking of the primary's wind. If the companion is massive enough to emit a significant wind by itself, it may collide with the primary's wind. Here we consider HD130298 assuming an equal-mass O star companion. While we would have likely detected such a companion through our spectral analysis, it may serve here as an example giving an order of magnitude estimate for the most favourable situation. 

We assume that the winds of the two O stars interact to create an optically  thin, fully ionised shock front from which X-rays are emitted via thermal bremsstrahlung. We calculate the density and temperature of the shocked material using the Rankine-Hugonoit jump condition for an ideal gas with adiabatic exponent $\gamma = 5/3$, and a Mach number of the un-shocked wind $\gg 1$ \citep{Regev2016}. Then, the integrated X-ray emissivity (i.e. energy per unit volume per unit time) of the shocked material is calculated as in \citet{Courvoisier2013}. For an orbital separation $a$, we assume the volume $V$ of the shock front as $V=\left( \frac{a}{2} \right)^3$. This gives an X-ray luminosity of the order of 10 L$_{\odot}$ for our example (Table\,\ref{X-ray}). This number is similar to the observed X-ray luminosity of colliding wind binaries resembling our example \citep{Gagne2012}, and broadly agrees with results from multi-dimensional hydrodynamic calculations \citep{Pittard2018}. 

For mass ratios well below one, the wind of the unseen companion is too weak to prevent the direct impact of the primary's wind on its surface \citep{sana04}. As an example, we use HD\,74194 as its 28.2\,$M_{\odot}$ O\,star emits a strong wind. We adopt a mass for the companion of 5\,$M_{\odot}$, which corresponds roughly to the maximum possible companion mass. We calculate the X-ray luminosity here by assuming that the wind kinetic energy of the O star enclosed in the solid angle subtended the MS companion gets completely converted to X-rays, which is surely an upper limit. This results in an X-ray luminosity of the order of 0.05\,$L_{\odot}$. This is just about twice the X-ray luminosity expected from HD\,74194 if it were a single star. Phase-locked variations of the X-ray flux is, however, expected given that X-ray is expected to be emitted only from the surface of the secondary star facing the primary \citep{sana04}. This will decrease the apparent average X-ray flux, rendering the process even more difficult to detect. 

\subsection{Characterisation of the detected higher-mass companions }

\begin{figure}[htbp]\centering
    \includegraphics[width=9cm]{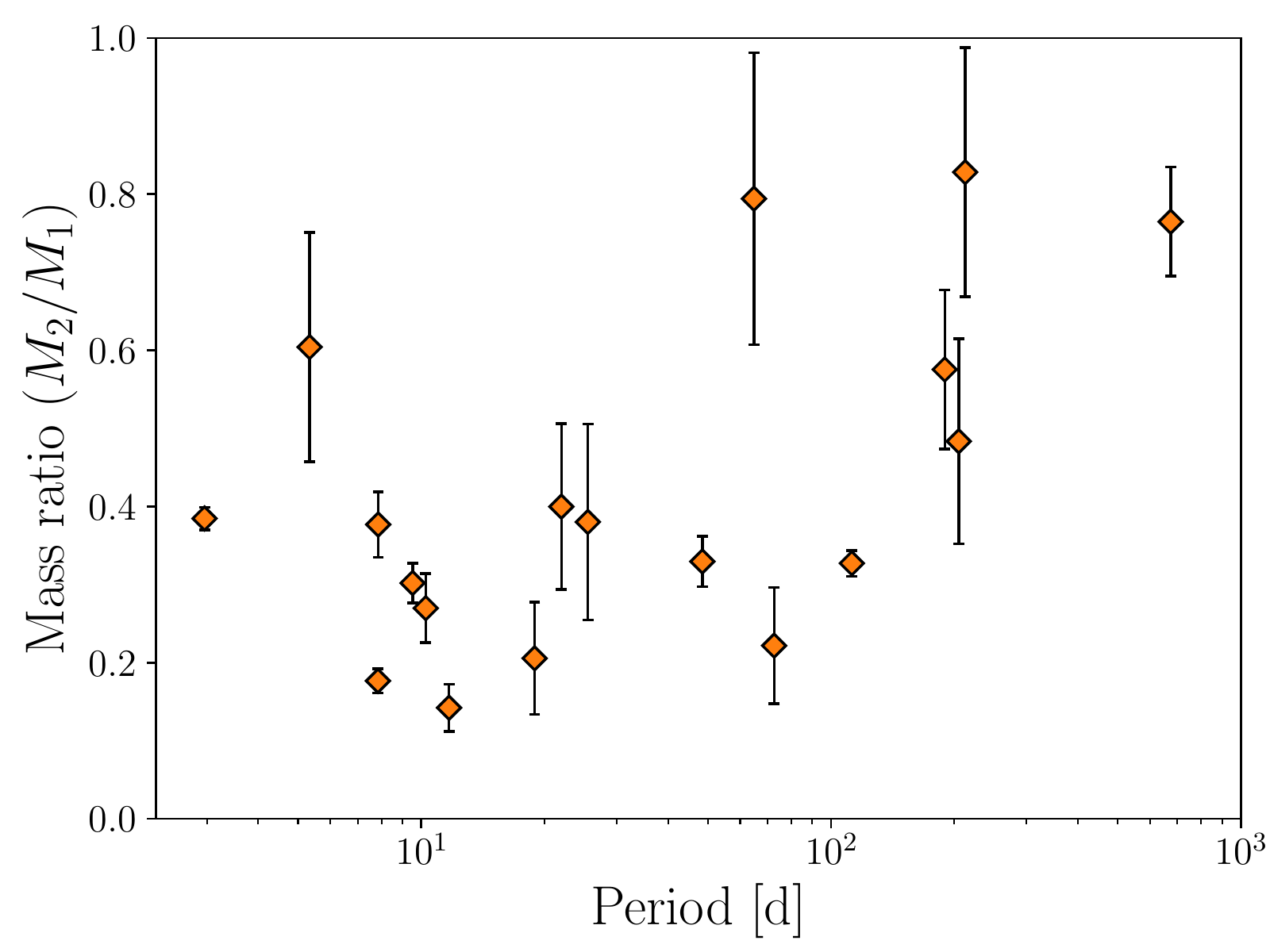}
    \includegraphics[width=9cm]{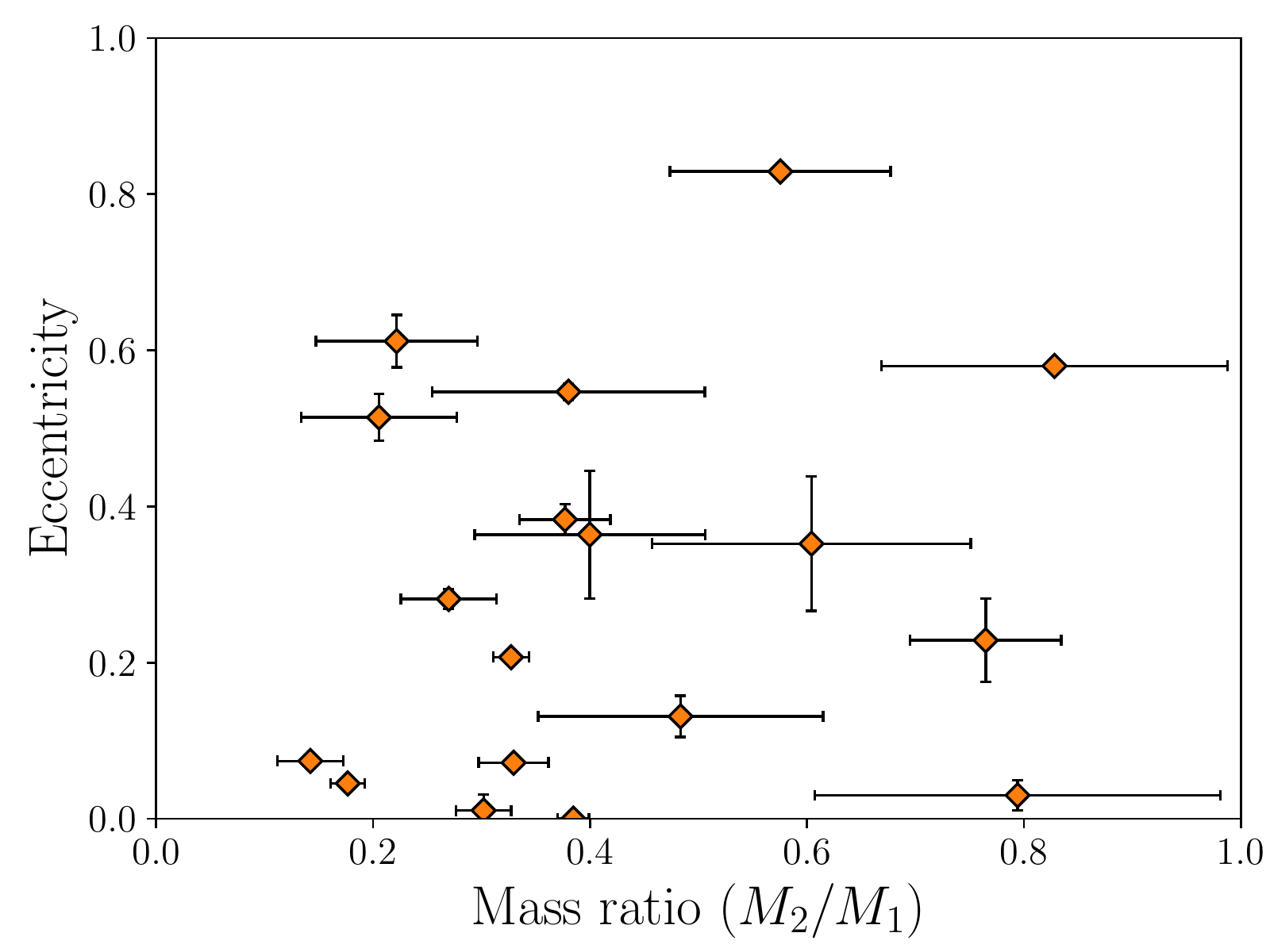}
    \caption{\label{Fig:P-q} {\it Top:} Orbital period-mass ratio diagram of the SB2 systems. {\it Bottom:} Same as for the top panel but comparing the eccentricity to the mass ratio.}
\end{figure}

Spectral disentangling revealed the nature of the secondary companions for 17 systems in our sample, allowing us to characterise the physical properties of the companions down to mass and brightness ratios of 0.15 and 0.02, respectively. Among these systems, most of them have orbital periods longer than 10 days, eccentricities up to 0.8, and mass ratios down to $\sim 0.15$. Figures~\ref{Fig:P-e-K} and \ref{Fig:P-q} show the positions of the SB2s in our sample in the period ($P_{\rm orb}$) - mass-ratio ($q$) - eccentricity ($e$) parameter space. We have a dearth of systems with short periods and high mass ratios and with long periods and low mass ratios. The short-period high mass-ratio binaries are indeed easier to characterise as SB2s and were therefore not selected for our analysis. For the systems with a long orbital period ($P_{\rm orb} > 20$ days) and a low mass ratio ($q<0.3$), they are more difficult to detect because they require long-term monitoring and high S/N data. 

\begin{figure}[htbp]\centering
    \includegraphics[width=9cm]{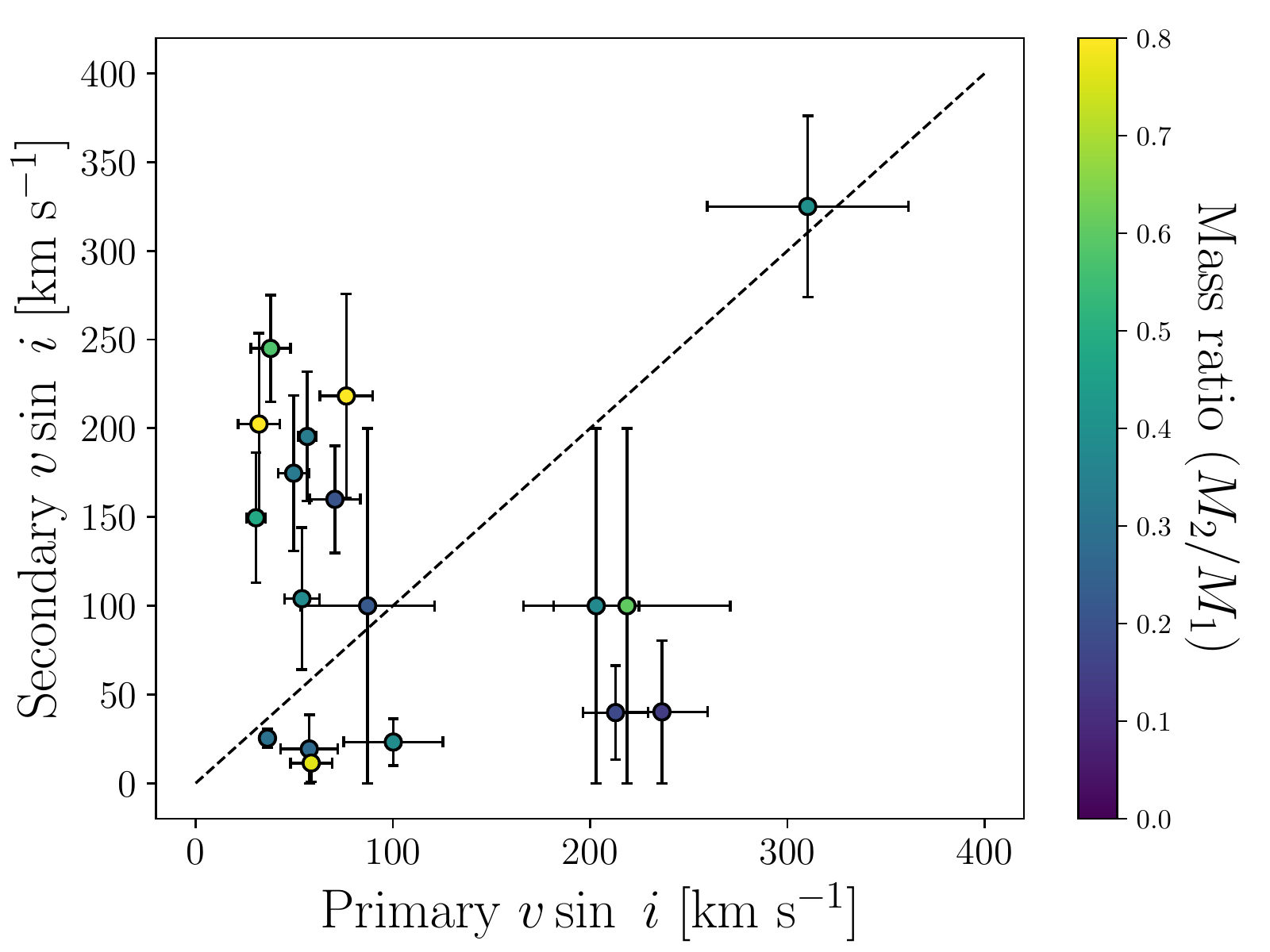}
    \caption{\label{Fig:ratio_vsini} Comparison between the projected rotational velocities of the primaries and the secondaries in our SB2 sample.}
\end{figure}

In Fig.~\ref{Fig:ratio_vsini}, we display the projected rotational velocities measured for the primaries and the secondaries, together with the mass ratios of the different SB2s. Most the primaries are slow rotators while their secondaries rotate on average faster. The high rotation of the secondaries is one of the reasons to explain that some systems were classified as SB1s, even though their secondaries are massive stars. That shows the difficulty to extract the spectral features of the secondary without using state-of-the-art techniques such as spectral disentangling. The dilution of secondary spectra due to high rotation was already pointed out to explain the non-detection of secondaries in systems like LB-1, or HR~6819 \citep[see e.g.][for more details]{Shenar2020LB1,AbdulMasih2020LB1,bodensteiner20}. Of interest are the strongly asynchronous spins, which might point towards past mass-transfer events for these systems; they therefore would deserve further investigation. 

\begin{figure}[htbp]\centering
    \includegraphics[width=9cm]{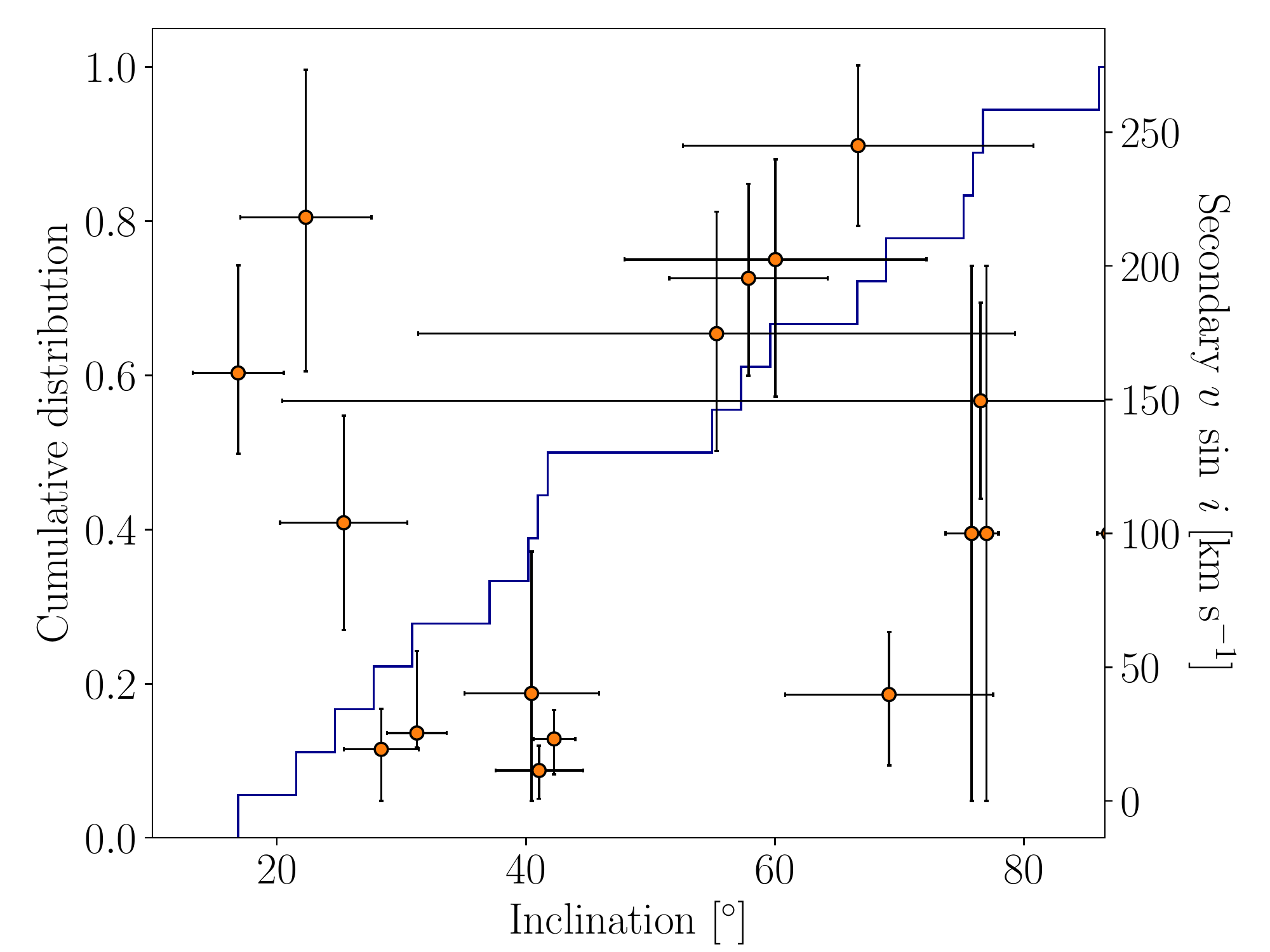}
    \caption{\label{fig:inclination} Comparison between the cumulative distribution of the inclinations estimated for the SB2s in our sample and the projected rotational velocities of the secondaries.}
\end{figure}

All the SB2 systems are discussed individually in Sect.~\ref{sec:individual}. We applied the CMFGEN atmosphere code to derive the individual parameters, such as their spectroscopic and evolutionary masses. By comparing the minimum masses, the spectroscopic and the evolutionary masses of the primary stars (we excluded the secondaries given notably the uncertainties on the $K2$), we can derive a rough estimation of the inclinations of the systems (Table~\ref{tab:parameters_SB2}). We do not derive the surface abundances of these components because discussing the evolution of these systems is beyond the scope of this paper. Figure~\ref{fig:inclination} displays the cumulative distribution of the inclination of the SB2 systems and the projected rotational velocities of the secondaries. Half of our SB2s have an inclination higher than $50^{\circ}$. Except for some outliers (and some secondary for which we were not able to compute the $\vsini$ and for which we took the standard value of 100~\kms), there seems, however, to be no correlation between the inclinations of the systems and the projected rotational velocities of the secondaries. 

\section{Conclusion}
\label{sec:conclusion}
For this analysis, we combined time series of high-resolution high signal-to-noise spectra and high-cadence photometry to characterise the nature of unseen companions in massive Galactic SB1 systems. For that purpose, we performed spectral disentangling to extract the spectral features of faint companions. For half of our sample, we revealed, for the first time, the stellar classification of their companions, down to a mass ratio of about 0.15. Some systems have high mass ratios, but their SB2 nature was hard to constrain because of the high projected rotational velocity of the secondary companions.

For the other half of our sample, we could not extract any spectral features of a putative faint companion. We combined atmosphere modelling to derive the fundamental parameters of the visible stars, the binary mass function, and the critical rotation to provide mass ranges for the secondary stars. In addition to Cyg~X-1, which is known to host a stellar-mass BH, we found two other candidates in our sample. One is HD~229234, which shares the same characteristics as HD~96670 \citep[][an SB1 system with a possible tertiary star, and a mass range for the visible star similar to that of a stellar-mass BH]{Gomez21}, and HD~130298, where the expected mass of the secondary component (higher than $7\,\Msun$) and the fact that we did not detect the spectral features of the secondary make it a suitable candidate to host a quiet stellar-mass BH. 

Finally, we found nine systems where the mass estimates for the secondaries are in the same range as the predicted masses for NSs. However, optical data alone are not sufficient to confirm their compact nature. Additional multi-wavelength observations are crucial for understanding all the evolutionary phases in between binary systems with massive stars on the MS and in binary BH systems. 

\begin{acknowledgements}
We are grateful to the anonymous referee for his/her important advice that improved the quality of the paper. L.M. thanks the European Space Agency (ESA) and the Belgian Federal Science Policy Office (BELSPO) for their support in the framework of the PRODEX Programme. H.S.\ acknowledges support from the FWO\_Odysseus program under project G0F8H6N. The research leading to these results has received funding from the European Research Council (ERC) under the European Union's Horizon 2020 research and innovation programme (grant agreement numbers 772225: MULTIPLES). T.S. acknowledges support from the European Union’s Horizon 2020 under the Marie Skłodowska-Curie grant agreement No 101024605. M.A.M. and J.B. are supported by an ESO fellowship. D.M.B., T.V.R. and S.J. gratefully acknowledge senior and junior postdoctoral fellowships, and PhD fellowship from the Research Foundation Flanders (FWO), with grant agreement numbers: 1286521N, 12ZB620N and 11E1721N, respectively. We also thank J. Hillier for making CMFGEN publicly available.
The Mercator telescope is operated by the Flemish Community on the island of La Palma at the Spain Observatory del Roche de los Muchachos of the Instituto de Astrofisica de Canarias. Mercator and Hermes are supported by the Funds for Scientific Research of Flanders (FWO), the Research Council of KU Leuven, the Fonds National de la Recherche Scientifique (FNRS), the Royal Observatory of Belgium, the Observatoire de Gen{\`e}ve, and the Th{\"u}ringer Landessternwarte Tautenburg. We thank all the observers who collected the data with Hermes.
The TESS data presented in this paper were obtained from the Mikulski Archive for Space Telescopes (MAST) at the Space Telescope Science Institute (STScI), which is operated by the Association of Universities for Research in Astronomy, Inc., under NASA contract NAS5-26555. Support to MAST for these data is provided by the NASA Office of Space Science via grant NAG5-7584 and by other grants and contracts. Funding for the TESS mission is provided by the NASA Explorer Program. 
This research has made use of the SIMBAD database, operated at CDS, Strasbourg, France and of NASA's Astrophysics Data System Bibliographic Services. We are grateful to the staff of the ESO Paranal Observatory for their technical support. This paper is based in part on spectroscopic observations made with the Southern African Large Telescope (SALT) under programme 2021-1-SCI-014 (PI: Manick). We are grateful to our SALT colleagues for maintaining the telescope facilities and conducting the observations. This work has also made use of data from the European Space Agency (ESA) mission {\it Gaia} (\url{https://www.cosmos.esa.int/gaia}), processed by the {\it Gaia} Data Processing and Analysis Consortium (DPAC, \url{https://www.cosmos.esa.int/web/gaia/dpac/consortium}). Funding for the DPAC has been provided by national institutions, in particular the institutions participating in the {\it Gaia} Multilateral Agreement. This research was achieved using the POLLUX database(\url{http://pollux.oreme.org}) operated at LUPM  (Universit{\'e} Montpellier - CNRS, France) with the support of the PNPS and INSU. This research also made use of Lightkurve, a Python package for Kepler and TESS data analysis \citep{lightkurve18}. The data used in this article will be shared on reasonable request to the corresponding authors.
\end{acknowledgements}

\bibliographystyle{aa} 
\bibliography{BH} 

\begin{appendix}

\section{Additional tables}
\FloatBarrier 
\begin{sidewaystable*}[h!]
\caption{\label{tab:parameters_SB2} Stellar parameters of the individual components in the new SB2 systems. Error bars correspond to $1\sigma$.}
\centering
\begin{tabular}{lrrrrrrrrrrr}
\hline\hline
Star & Comp. &   $\log(L/L_{\odot})$ &   $\teff$  & $R$  &  $\logg_c$ &  $\vsini$  &   $\vmac$   &  M$_{\rm{spec}}$  &  M$_{\rm{evol}}$ & Flux perc. & Incl. \\
        &                             &    & [kK] & [$R_{\odot}$] & [cgs] & [\kms]  & [\kms]  & [$\Msun$] & [$\Msun$] & [\%] & [$^{\circ}$]\\
\hline
HD29763 & Primary & $2.75_{-0.06}^{+0.06}$ & $16.6_{- 0.5}^{+ 0.5}$ & $ 2.9_{- 0.2}^{+ 0.2}$ & $4.26_{-0.05}^{+0.05}$ & $100_{- 25}^{+ 27}$ & $  7_{-  7}^{+ 20}$ & $ 5.5_{- 0.9}^{+ 0.9}$ & $ 4.8_{- 0.1}^{+ 0.2}$ & $93 \pm 5$ & $42.3 \pm 1.7$\\[1pt]
HD29763 & Secondary & $1.02_{-0.06}^{+0.06}$ & $10.0_{- 2.0}^{+ 2.0}$ & $ 1.1_{- nan}^{+ nan}$ & $4.50_{-0.50}^{+0.50}$ & $ 23_{- 13}^{+ 10}$ & $  1_{-  9}^{+  7}$ & $ 1.4_{- 0.9}^{+ 0.9}$ & $ - $ & $7 \pm 5$& \\[1pt]
HD30836 & Primary & $4.05_{-0.10}^{+0.10}$ & $20.3_{- 2.1}^{+ 2.1}$ & $ 8.6_{- 0.8}^{+ 0.8}$ & $3.40_{-0.25}^{+0.25}$ & $ 36_{-  2}^{+  2}$ & $ 30_{-  2}^{+  2}$ & $ 6.8_{- 2.1}^{+ 2.1}$ & $ 9.4_{- 0.7}^{+ 0.7}$ & $95 \pm 5$ & $31.2  \pm 2.4$ \\[1pt]
HD30836 & Secondary & $2.24_{-0.06}^{+0.06}$ & $13.0_{- 2.0}^{+ 2.0}$ & $ 2.6_{- 0.2}^{+ 0.2}$ & $3.90_{-0.50}^{+0.50}$ & $ 25_{-  5}^{+ 30}$ & $  1_{-  1}^{+  1}$ & $ 2.0_{- 2.0}^{+ 2.1}$ & $ 3.3_{- 0.2}^{+ 0.2}$ & $5 \pm 5$& \\[1pt]
HD37737 & Primary & $4.74_{-0.08}^{+0.08}$ & $29.2_{- 0.5}^{+ 0.5}$ & $ 9.2_{- 1.4}^{+ 1.4}$ & $3.57_{-0.05}^{+0.05}$ & $203_{- 21}^{+ 14}$ & $ 65_{- 65}^{+ 69}$ & $11.4_{- 1.3}^{+ 1.3}$ & $19.0_{- 1.1}^{+ 0.7}$ & $95 \pm 5$& $ 75.8 \pm 2.1 $ \\[1pt]
HD37737 & Secondary & $3.02_{-0.12}^{+0.12}$ & $19.6_{- 2.5}^{+ 2.5}$ & $ 2.8_{- 1.4}^{+ 1.4}$ & $4.07_{-0.30}^{+0.30}$ & $  0_{-  0}^{+  0}$ & $  0_{-  0}^{+  0}$ & $ 3.4_{- 1.0}^{+ 1.0}$ & $ 6.2_{- 0.5}^{+ 0.4}$ & $5 \pm 5$& \\[1pt]
HD52533 & Primary & $5.07_{-0.10}^{+0.10}$ & $35.0_{- 0.5}^{+ 0.5}$ & $ 9.3_{- 0.6}^{+ 0.6}$ & $4.03_{-0.10}^{+0.10}$ & $286_{- 25}^{+ 25}$ & $  0_{- 10}^{+ 10}$ & $33.6_{- 1.2}^{+ 1.2}$ & $23.8_{- 1.2}^{+ 1.4}$ & $85 \pm 3$ & $86.8  \pm 0.9$\\[1pt]
HD52533 & Secondary & $4.40_{-0.16}^{+0.16}$ & $31.0_{- 1.4}^{+ 1.4}$ & $ 5.5_{- 1.1}^{+ 1.1}$ & $4.08_{-0.10}^{+0.10}$ & $325_{- 51}^{+ 56}$ & $101_{-101}^{+103}$ & $13.3_{- 2.1}^{+ 2.1}$ & $16.4_{- 1.4}^{+ 1.3}$ & $15 \pm 3$ & \\[1pt]
HD57236 & Primary & $4.90_{-0.05}^{+0.05}$ & $36.4_{- 0.8}^{+ 0.9}$ & $ 7.1_{- 0.5}^{+ 0.5}$ & $4.09_{-0.10}^{+0.10}$ & $ 25_{-  6}^{+  7}$ & $ 50_{-  7}^{+  4}$ & $22.6_{-10.2}^{+10.2}$ & $23.6_{- 0.9}^{+ 1.0}$ & $64 \pm 6$& $60.0  \pm 12.1$ \\[1pt]
HD57236 & Secondary & $4.61_{-0.10}^{+0.10}$ & $32.5_{- 1.5}^{+ 1.5}$ & $ 6.3_{- 1.0}^{+ 1.0}$ & $4.12_{-0.15}^{+0.15}$ & $202_{- 51}^{+ 37}$ & $12_{-10}^{+ 38}$ & $19.4_{- 7.4}^{+ 7.4}$ & $18.2_{- 1.2}^{+ 1.4}$ & $36 \pm 6$& \\[1pt]
HD91824 & Primary & $4.81_{-0.06}^{+0.06}$ & $38.2_{- 1.0}^{+ 1.0}$ & $ 5.8_{- 0.3}^{+ 0.3}$ & $4.00_{-0.10}^{+0.10}$ & $ 42_{-  7}^{+  7}$ & $ 42_{-  7}^{+  6}$ & $12.3_{- 6.7}^{+ 6.7}$ & $22.2_{- 1.1}^{+ 1.0}$ & $84 \pm 4$& $55.3  \pm 24.0$\\[1pt]
HD91824 & Secondary & $3.84_{-0.06}^{+0.06}$ & $28.0_{- 1.0}^{+ 1.0}$ & $ 3.5_{- 0.3}^{+ 0.3}$ & $4.33_{-0.20}^{+0.20}$ & $174_{- 43}^{+ 45}$ & $100_{- 90}^{+ 90}$ & $ 9.6_{- 6.7}^{+ 6.7}$ & $12.4_{- 0.6}^{+ 0.5}$ & $16 \pm 4$& \\[1pt]
HD93028 & Primary & $4.84_{-0.08}^{+0.08}$ & $35.0_{- 0.5}^{+ 0.5}$ & $ 7.1_{- 0.6}^{+ 0.6}$ & $4.00_{-0.10}^{+0.10}$ & $ 33_{-  3}^{+  3}$ & $ 45_{-  4}^{+  3}$ & $18.6_{- 6.3}^{+ 6.3}$ & $21.8_{- 0.8}^{+ 1.0}$ & $85 \pm 2$& $76.5  \pm 56.1$ \\[1pt]
HD93028 & Secondary & $3.58_{-0.08}^{+0.08}$ & $23.0_{- 1.2}^{+ 1.2}$ & $ 3.9_{- 0.6}^{+ 0.6}$ & $4.27_{-0.15}^{+0.15}$ & $149_{- 36}^{+ 36}$ & $ 10_{- 10}^{+ 10}$ & $10.3_{- 6.3}^{+ 6.3}$ & $ 8.6_{- 0.6}^{+ 0.4}$ & $15 \pm 2$& \\[1pt]
HD152405 & Primary & $5.08_{-0.05}^{+0.05}$ & $30.7_{- 0.5}^{+ 0.5}$ & $12.3_{- 0.7}^{+ 0.7}$ & $3.45_{-0.05}^{+0.05}$ & $ 53_{-  8}^{+  8}$ & $ 70_{- 10}^{+  9}$ & $15.4_{- 2.2}^{+ 2.2}$ & $23.4_{- 0.9}^{+ 1.2}$ & $88 \pm 4$& $25.4  \pm 5.1$\\[1pt]
HD152405 & Secondary & $3.82_{-0.05}^{+0.05}$ & $22.0_{- 3.0}^{+ 3.0}$ & $ 5.6_{- 0.7}^{+ 0.7}$ & $3.87_{-0.30}^{+0.30}$ & $104_{- 40}^{+ 40}$ & $ 12_{- 10}^{+ 10}$ & $ 8.4_{- 2.2}^{+ 2.2}$ & $ 8.8_{- 0.5}^{+ 0.9}$ & $12 \pm 4$& \\[1pt]
HD152723 & Primary & $5.72_{-0.17}^{+0.17}$ & $37.7_{- 1.0}^{+ 1.4}$ & $17.1_{- 3.4}^{+ 3.4}$ & $3.84_{-0.11}^{+0.16}$ & $ 70_{- 12}^{+ 11}$ & $ 98_{- 14}^{+ 13}$ & $73.9_{-28.2}^{+28.2}$ & $38.6_{- 5.1}^{+ 6.3}$ & $85 \pm 5$ &$ 16.9 \pm 3.7$\\[1pt]
HD152723 & Secondary & $4.57_{-0.17}^{+0.17}$ & $27.0_{- 2.0}^{+ 2.0}$ & $ 8.8_{- 3.4}^{+ 3.4}$ & $4.02_{-0.25}^{+0.25}$ & $160_{- 30}^{+ 40}$ & $ 30_{- 20}^{+ 20}$ & $29.6_{-28.2}^{+28.2}$ & $14.4_{- 1.4}^{+ 1.7}$ & $15 \pm 5$& \\[1pt]
HD163892 & Primary & $4.77_{-0.04}^{+0.04}$ & $31.8_{- 1.2}^{+ 1.2}$ & $ 8.0_{- 0.3}^{+ 0.3}$ & $3.83_{-0.10}^{+0.10}$ & $212_{- 16}^{+ 11}$ & $ 47_{- 40}^{+ 39}$ & $16.0_{- 2.5}^{+ 2.5}$ & $19.0_{- 0.7}^{+ 0.9}$ & $95 \pm 4$&$ 69.2 \pm 8.3$ \\[1pt]
HD163892 & Secondary & $2.67_{-0.04}^{+0.04}$ & $16.0_{- 1.0}^{+ 1.0}$ & $ 2.8_{- 0.3}^{+ 0.3}$ & $4.00_{-0.10}^{+0.10}$ & $ 39_{- 26}^{+ 23}$ & $  1_{-  1}^{+  4}$ & $ 2.9_{- 2.5}^{+ 2.5}$ & $ 4.4_{- 0.2}^{+ 0.2}$ & $5 \pm 4$& \\[1pt]
HD164438 & Primary & $4.94_{-0.03}^{+0.03}$ & $30.6_{- 0.8}^{+ 0.8}$ & $10.5_{- 0.4}^{+ 0.4}$ & $3.56_{-0.05}^{+0.05}$ & $ 57_{- 14}^{+ 13}$ & $102_{- 14}^{+ 12}$ & $14.4_{- 4.0}^{+ 4.0}$ & $20.2_{- 0.6}^{+ 0.6}$ & $94 \pm 5$& $ 28.4 \pm 3.0 $\\[1pt]
HD164438 & Secondary & $3.11_{-0.03}^{+0.03}$ & $18.0_{- 1.0}^{+ 1.0}$ & $ 3.7_{- 0.4}^{+ 0.4}$ & $3.80_{-0.25}^{+0.25}$ & $ 19_{- 19}^{+ 15}$ & $  1_{-  1}^{+  1}$ & $ 3.1_{- 3.0}^{+ 4.0}$ & $ 5.6_{- 0.1}^{+ 0.4}$ & $6 \pm 5$& \\[1pt]
HD164536 & Primary & $5.09_{-0.15}^{+0.15}$ & $34.4_{- 1.1}^{+ 1.1}$ & $ 9.8_{- 1.6}^{+ 1.6}$ & $3.84_{-0.16}^{+0.16}$ & $236_{- 23}^{+ 15}$ & $ 33_{- 53}^{+ 64}$ & $24.6_{-14.7}^{+14.7}$ & $23.8_{- 1.9}^{+ 2.5}$ & $97 \pm 2$& $ 40.5 \pm 5.4 $\\[1pt]
HD164536 & Secondary & $2.73_{-0.15}^{+0.15}$ & $17.0_{- 1.0}^{+ 1.0}$ & $ 2.7_{- 1.6}^{+ 1.6}$ & $4.10_{-0.50}^{+0.50}$ & $ 40_{- 40}^{+ 53}$ & $ 60_{- 60}^{+ 78}$ & $ 3.3_{-14.7}^{+14.7}$ & $ 4.8_{- 0.4}^{+ 0.3}$ & $ 3 \pm 2$& \\[1pt]
HD167263 & Primary & $5.25_{-0.26}^{+0.26}$ & $33.0_{- 0.5}^{+ 0.5}$ & $12.9_{- 9.4}^{+ 9.4}$ & $3.90_{-0.10}^{+0.10}$ & $ 81_{- 13}^{+  9}$ & $ 60_{- 23}^{+ 22}$ & $48.8_{-65.9}^{+65.9}$ & $24.8_{- 1.4}^{+ 1.8}$ & $75 \pm 10$& $ 22.3 \pm 5.3 $ \\[1pt]
HD167263 & Secondary & $4.77_{-0.26}^{+0.26}$ & $30.0_{- 1.5}^{+ 1.5}$ & $ 9.0_{- 9.4}^{+ 9.4}$ & $3.94_{-0.10}^{+0.10}$ & $218_{- 57}^{+ 55}$ & $ 50_{- 50}^{+ 50}$ & $25.6_{-25.5}^{+65.9}$ & $18.2_{- 1.2}^{+ 1.3}$ & $25 \pm 10$&  \\[1pt]
HD167264 & Primary & $5.40_{-0.19}^{+0.19}$ & $29.5_{- 0.8}^{+ 0.8}$ & $19.2_{- 4.1}^{+ 4.1}$ & $3.33_{-0.05}^{+0.05}$ & $ 58_{- 10}^{+  9}$ & $ 78_{- 12}^{+ 10}$ & $28.4_{- 9.3}^{+ 9.3}$ & $29.2_{- 3.0}^{+ 3.7}$ & $90 \pm 5$& $ 41.1 \pm 3.5 $\\[1pt]
HD167264 & Secondary & $4.30_{-0.19}^{+0.19}$ & $26.0_{- 2.0}^{+ 2.0}$ & $ 6.9_{- 4.1}^{+ 4.1}$ & $4.10_{-0.20}^{+0.20}$ & $ 11_{- 10}^{+  9}$ & $ 45_{- 12}^{+ 10}$ & $22.0_{- 9.3}^{+ 9.3}$ & $11.8_{- 1.1}^{+ 1.5}$ & $10 \pm 5$& \\[1pt]
HD192001 & Primary & $4.85_{-0.05}^{+0.05}$ & $33.3_{- 0.8}^{+ 0.8}$ & $ 8.0_{- 0.4}^{+ 0.4}$ & $3.97_{-0.10}^{+0.10}$ & $ 38_{- 10}^{+ 10}$ & $ 69_{- 20}^{+ 20}$ & $22.0_{- 6.6}^{+ 6.6}$ & $20.6_{- 0.7}^{+ 0.9}$ & $85 \pm 7$&  $ 66.7 \pm 14.1 $ \\[1pt]
HD192001 & Secondary & $4.07_{-0.05}^{+0.05}$ & $28.6_{- 1.6}^{+ 1.7}$ & $ 4.4_{- 0.4}^{+ 0.4}$ & $4.10_{-0.13}^{+0.14}$ & $245_{- 30}^{+ 30}$ & $  0_{- 10}^{+ 10}$ & $ 9.0_{- 6.6}^{+ 6.6}$ & $12.4_{- 0.6}^{+ 0.7}$ & $15 \pm 7$& \\[1pt]
HD199579 & Primary & $5.18_{-0.03}^{+0.03}$ & $39.0_{- 0.5}^{+ 0.5}$ & $ 8.5_{- 0.3}^{+ 0.3}$ & $3.90_{-0.10}^{+0.10}$ & $ 56_{-  4}^{+  4}$ & $ 82_{-  4}^{+  4}$ & $21.2_{- 3.8}^{+ 3.8}$ & $30.2_{- 0.9}^{+ 0.9}$ & $92 \pm 6$& $ 57.9 \pm 6.4 $\\[1pt]
HD199579 & Secondary & $3.72_{-0.03}^{+0.03}$ & $28.0_{- 1.0}^{+ 1.0}$ & $ 3.1_{- 0.3}^{+ 0.3}$ & $4.25_{-0.10}^{+0.10}$ & $195_{- 36}^{+ 35}$ & $  0_{- 53}^{+ 55}$ & $ 6.1_{- 3.8}^{+ 3.8}$ & $10.8_{- 0.6}^{+ 0.7}$ & $8 \pm 6$& \\[1pt]
Schulte 11 & Primary & $5.82_{-0.04}^{+0.04}$ & $40.8_{- 0.7}^{+ 0.7}$ & $16.4_{- 0.7}^{+ 0.7}$ & $3.85_{-0.05}^{+0.05}$ & $ 87_{- 33}^{+ 22}$ & $ 36_{- 28}^{+ 25}$ & $69.7_{- 7.6}^{+ 7.6}$ & $56.2_{- 3.1}^{+ 2.7}$ & $98 \pm 5$& $ 31.1 \pm 8.9 $ \\[1pt]
Schulte 11 & Secondary & $-$ & $-$ & $-$ & $ -$ & $ -$ & $-$ & $-$ & $-$ & $2 \pm 2$& \\[1pt]
V747 Cep & Primary & $5.28_{-0.03}^{+0.03}$ & $40.0_{- 1.0}^{+ 1.0}$ & $ 9.1_{- 0.3}^{+ 0.3}$ & $4.05_{-0.10}^{+0.10}$ & $158_{-158}^{+ 86}$ & $  0_{-101}^{+103}$ & $33.5_{- 1.2}^{+ 1.2}$ & $33.6_{- 1.4}^{+ 1.3}$ & $94 \pm 4$& $ 77.0 \pm 1.0 $\\[1pt]
V747 Cep & Secondary & $3.56_{-0.10}^{+0.10}$ & $21.6_{- 2.5}^{+ 2.5}$ & $ 4.0_{- 0.6}^{+ 0.6}$ & $4.08_{-0.30}^{+0.30}$ & $  0_{-  0}^{+  0}$ & $  0_{-  0}^{+  0}$ & $ 7.9_{- 1.8}^{+ 1.8}$ & $ 8.0_{- 0.6}^{+ 0.6}$ & $6\pm 4$ & \\
\hline
\end{tabular}
\end{sidewaystable*}

\FloatBarrier 
\begin{sidewaystable*}[h!]
\caption{\label{tab:parameters} Stellar parameters and surface abundance of the visible stars in the SB1 systems. Error bars correspond to $1\sigma$.}
\centering
\begin{tabular}{llrrrrrrrrrrrrr}
\hline\hline
Star & Spec. Type &  $\log(L/L_{\odot})$ &   $\teff$  & $R$  &  $\logg_c$ &  $\vsini$  &   $\vmac$   &  $M_{\rm{spec}}$  &  $M_{\rm{evol}}$ & $M_2$ min. & He/H & $\epsilon_{\rm{C}}$ & $\epsilon_{\rm{N}}$ &  $\epsilon_{\rm{O}}$\\
        &                         &          & [kK] & [$R_{\odot}$] & [cgs] & [\kms]  & [\kms]  & [$\Msun$] & [$\Msun$]  & [$\Msun$] & & & &\\
\hline
CygX-1 & O9.7Iabpvar & $5.48_{-0.06}^{+0.06}$ & $29.8_{- 0.2}^{+ 0.2}$ & $20.7_{-1.2}^{+1.2}$ & $3.33_{-0.05}^{+0.05}$ & $ 86_{- 22}^{+ 15}$ & $ 82_{- 29}^{+ 28}$ & $33.4_{- 8.0}^{+ 8.0}$ & $30.0_{- 3.1}^{+ 4.0}$ & $6.8 \pm 1.7$ & $0.15_{-0.01}^{+0.01} $ & $8.14_{-0.14}^{+0.11}$ & $9.24_{-0.21}^{+0.14}$ & $8.59_{-0.03}^{+0.02}$ \\ [1pt] 

HD~12323 & ON9.2V & $4.70_{-0.07}^{+0.07}$ & $33.2_{- 1.0}^{+ 1.0}$ & $6.8_{-0.6}^{+0.6}$ & $3.99_{-0.12}^{+0.12}$ & $119_{-32}^{+ 34}$ & $ 57_{- 24}^{+ 28}$ & $17.1_{- 3.8}^{+ 3.8}$ & $19.2_{- 0.9}^{+ 1.0}$  & $1.3 \pm 0.3$ & $0.15_{-0.02}^{+0.02}$ & $7.74_{-0.26}^{+0.16}$ & $8.61_{-0.13}^{+0.10}$ & $8.40_{-0.10}^{+0.08}$ \\ [1pt] 

HD~14633 & ON8.5V & $4.60_{-0.10}^{+0.10}$ & $33.9_{- 1.1}^{+ 1.2}$ & $5.8_{-0.7}^{+0.7}$ & $3.93_{-0.20}^{+0.22}$ & $126_{-21}^{+ 12}$ & $ 48_{- 48}^{+ 52}$ & $10.6_{- 3.6}^{+ 3.6}$ & $19.0_{- 1.1}^{+ 1.2}$  & $1.0 \pm 0.3$ & $0.16_{-0.02}^{+0.02}$	 & $7.77_{-0.29}^{+0.19}$ & $8.98_{-0.13}^{+0.16}$ & $8.40_{-0.14}^{+0.11}$ \\ [1pt] 

HD~15137 & O9.5II-IIIn & $4.97_{-0.08}^{+0.08}$ & $30.5_{- 0.8}^{+ 0.8}$ & $10.9_{-1.0}^{+1.0}$ & $3.53_{-0.04}^{+0.04}$ & $264_{-46}^{+ 46}$ & $103_{-103}^{+101}$ & $14.9_{- 3.3}^{+ 3.3}$ & $22.2_{- 1.8}^{+ 1.1}$  & $1.5 \pm 0.4$ & $0.15_{-0.03}^{+0.03}$ & $7.78_{-0.30}^{+0.17}$ & $8.64_{-0.14}^{+0.21}$ & $8.58_{-0.06}^{+0.05}$ \\ [1pt] 

HD~37737 & O9.5II-III(n) & $4.81_{-0.12}^{+0.12}$ & $29.2_{- 0.5}^{+ 0.5}$ & $10.1_{-1.4}^{+1.4}$ & $3.49_{-0.05}^{+0.05}$ & $203_{-22}^{+ 15}$ & $ 65_{- 65}^{+ 69}$ & $11.3_{- 2.9}^{+ 2.9}$ & $21.0_{- 1.6}^{+ 1.2}$  & $4.5 \pm 0.8$ & $0.10_{-0.02}^{+0.02}$&$ 8.30_{-0.12}^{+0.10}$&$ 7.81_{-0.11}^{+0.09}$&$8.66_{-0.03}^{+0.03}$\\ [1pt] 

HD~46573 & O7V((f))z & $5.01_{-0.04}^{+0.04}$ & $35.3_{- 1.4}^{+ 1.4}$ & $8.6_{-0.3}^{+0.3}$ & $3.85_{-0.15}^{+0.16}$ & $ 78_{- 16}^{+ 12}$ & $ 74_{- 20}^{+ 20}$ & $18.9_{- 4.0}^{+ 4.0}$ & $24.0_{- 1.1}^{+ 1.2}$ & $0.7 \pm 0.1$ & $0.14_{-0.02}^{+0.02}$ & $8.00_{-0.07}^{+0.06}$ & $8.73_{-0.13}^{+0.14}$ & $8.58_{-0.05}^{+0.04}$ \\ [1pt] 

HD~74194 & O8.5Ib-II(f)p & $5.41_{-0.04}^{+0.04}$ & $32.1_{- 0.5}^{+ 0.5}$ & $16.5_{-0.8}^{+0.8}$ & $3.45_{-0.05}^{+0.05}$ & $172_{-4}^{+4}$ & $ 49_{- 17}^{+ 17}$ & $28.2_{- 3.1}^{+ 3.1}$ & $31.2_{-1.2}^{+1.4}$  & $1.8 \pm 0.2$ & $0.12_{-0.02}^{+0.02}$ & $8.01_{-0.13}^{+0.16}$& $8.66_{-0.12}^{+0.13}$&$8.63_{-0.03}^{+0.03}$ \\ [1pt] 

HD~75211 & O8.5II((f)) & $5.36_{-0.03}^{+0.03}$ & $34.5_{- 0.5}^{+ 0.5}$ & $13.4_{-0.4}^{+0.4}$ & $3.59_{-0.07}^{+0.08}$ & $136_{-  8}^{+  6}$ & $ 66_{- 19}^{+ 20}$ & $25.3_{- 2.7}^{+ 2.7}$ & $31.0_{- 1.0}^{+ 1.0}$  & $2.5 \pm 0.2$ & $0.14_{-0.02}^{+0.02}$ & $8.21_{-0.09}^{+0.09}$ & $8.64_{-0.09}^{+0.11}$ & $8.23_{-0.08}^{+0.11}$ \\ [1pt] 

HD~94024 & O8IV & $4.95_{-0.05}^{+0.05}$ & $33.7_{- 1.2}^{+ 1.2}$ & $8.7_{-0.4}^{+0.4}$ & $3.75_{-0.12}^{+0.12}$ & $150_{-21}^{+ 13}$ & $ 62_{- 45}^{+ 51}$ & $15.6_{- 2.8}^{+ 2.8}$ & $22.2_{- 1.1}^{+ 1.0}$  & $1.4 \pm 0.2$ & $0.12_{-0.02}^{+0.02}$ &	$8.28_{-0.05}^{+0.05}$ & $8.62_{-0.11}^{+0.14} $& $8.63_{-0.04}^{+0.04}$ \\ [1pt] 
HD~105627 & O9III & $4.91_{-0.07}^{+0.07}$ & $32.5_{- 1.0}^{+ 1.0}$ & $9.0_{-0.7}^{+0.7}$ & $3.67_{-0.10}^{+0.10}$ & $162_{-35}^{+ 19}$ & $ 37_{- 37}^{+ 78}$ & $13.8_{- 2.7}^{+ 2.7}$ & $21.0_{- 1.0}^{+ 1.3}$  & $1.4 \pm 0.3$ & $0.11_{-0.01}^{+0.01}$ & $8.11_{-0.11}^{+0.10}$ & $8.64_{-0.14}^{+0.23}$	& $8.59_{-0.04}^{+0.04}$\\ [1pt] 

HD~130298 & O6.5III(n)(f) & $5.22_{-0.04}^{+0.04}$ & $37.2_{- 1.4}^{+ 1.0}$ & $10.0_{-0.5}^{+0.5}$ & $3.82_{-0.10}^{+0.12}$ & $146_{-28}^{+ 14}$ & $ 69_{- 54}^{+ 59}$ & $24.2_{- 3.8}^{+ 3.8}$ & $28.0_{- 4.1}^{+ 5.2}$  & $7.7 \pm 1.5$ & $0.12_{-0.02}^{+0.02}$ & $7.52_{-0.26}^{+0.31}$ & $8.88_{-0.18}^{+0.20}$ & $8.18_{-0.18}^{+0.12}$ \\ [1pt] 


HD~165174 & O9.7IIn & $4.87_{-0.05}^{+0.05}$ & $30.6_{- 0.7}^{+ 0.7}$ & $9.7_{-0.5}^{+0.5}$ & $3.60_{-0.05}^{+0.05}$ & $315_{-17}^{+ 19}$ & $ 24_{- 24}^{+100}$ & $13.7_{- 1.5}^{+ 1.5}$ & $20.0_{- 1.0}^{+ 1.1}$  & $2.2 \pm 0.4$ & $0.17_{-0.02}^{+0.02}$ & $8.23_{-0.13}^{+0.09}$& $8.93_{-0.27}^{+0.13}$ & $8.66_{-0.05}^{+0.05}$ \\ [1pt] 

HD~229234 & O9III & $5.12_{-0.03}^{+0.03}$ & $31.2_{- 0.8}^{+ 0.8}$ & $12.4_{-0.4}^{+0.4}$ & $3.46_{-0.04}^{+0.04}$ & $ 90_{- 21}^{+ 13}$ & $ 76_{- 28}^{+ 29}$ & $16.1_{- 1.4}^{+ 1.4}$ & $23.2_{- 0.4}^{+ 1.0}$  & $2.6 \pm 0.3$ & $0.09_{-0.01}^{+0.01}$ & $8.36_{-0.07}^{+0.06}$ & $7.78_{-0.28}^{+0.17}$ & $8.66_{-0.05}^{+0.05}$ \\ [1pt] 

HD~308813 & O9.7IV(n) & $4.77_{-0.20}^{+0.20}$ & $30.3_{- 1.0}^{+ 1.0}$ & $10.7_{-2.3}^{+2.3}$ & $3.81_{-0.10}^{+0.10}$ & $186_{-20}^{+ 20}$ & $ 30_{- 10}^{+ 10}$ & $10.7_{- 4.3}^{+ 4.3}$ & $17.8_{- 1.5}^{+ 1.7}$  & $1.6 \pm 0.5$ & $0.10_{-0.02}^{+0.02}$ & $8.11_{-0.06}^{+0.09}$ & $8.34_{-0.50}^{+0.22}$ & $8.66_{-0.03}^{+0.03}$ \\ [1pt] 

LS~5039 & ON6V((f))z & $4.90_{-0.04}^{+0.04}$ & $38.7_{- 1.0}^{+ 1.0}$ & $6.3_{-0.3}^{+0.3}$ & $3.89_{-0.08}^{+0.08}$ & $133_{-20}^{+ 12}$ & $ 81_{- 33}^{+ 37}$ & $11.1_{- 1.5}^{+ 1.5}$ & $32.2_{- 3.7}^{+ 5.0}$ & $1.3 \pm 0.3$ & $0.13_{-0.02}^{+0.04}$ & $6.75_{-0.71}^{+0.84}$ & $9.25_{-0.15}^{+0.17}$ & $8.48_{-0.05}^{+0.04}$\\ [1pt] 




\hline
\end{tabular}
\end{sidewaystable*}
\FloatBarrier

\begin{table*}[h!]
\centering
\caption{Expected stellar wind mass accretion rate ($\dot{M}_{\rm acc}$), specific angular momentum accretion ($j_{\rm acc}$) and X-ray luminosity (L$_{\rm x}$), for our SB1 binaries, using the specified mass of the unseen companion. The top part of the table assumes a non-spinning BH companion, except for the two X-ray luminosities in italic (see text). The other two parts assume a MS companion.
}
\label{X-ray}
\begin{tabular}{l r r r r r r r}
\hline
\hline
Name & Period & R$_{\rm O\,star}$ & M$_{\rm O\,star}$ & M$_{\rm unseen}$ & $\dot{M}_{\rm acc}$ & $j_{\rm acc}/j_{\rm ISCO}$ & log L$_{\rm x}$ \\ 
     & (days) & ($R_{\odot}$) & ($M_{\odot}$) & ($M_{\odot}$) & ($M_{\odot}$/yr) & & (erg/s) \\
\hline
BH accretion & & & & & & & \\
Cyg X-1  &   5.59 &   18.61 &   40.60 &   21.2$^{a}$ &  9.43e-11 &   0.45       & {\it 36.43} \\
HD12323  &   1.92 &    6.80 &   17.10 &    7.0       &  6.34e-11 &   0.34       &      -- \\
HD14633  &  15.41 &    5.80 &   10.60 &    4.0       &  9.29e-13 &   0.01       &      -- \\
HD15137  &  55.34 &   10.91 &   14.90 &    2.0       &  6.28e-14 &   0.00       &      -- \\
HD37737  &   7.85 &   10.00 &   11.30 &    8.0       &  3.55e-11 &   0.24       &      -- \\
HD46573  &  10.65 &    8.60 &   18.90 &    5.0       &  1.50e-12 &   0.02       &      -- \\
HD74194  &   9.54 &   16.51 &   28.20 &    5.0       &  4.19e-12 &   0.07       &      -- \\
HD75211  &  20.44 &   13.41 &   25.30 &   10.0       &  2.44e-12 &   0.03       &      -- \\
HD94024  &   2.46 &    8.70 &   15.60 &    4.0       &  4.78e-11 &   0.42       & {\it 36.13} \\
HD105627 &   4.34 &    9.00 &   13.80 &    4.0       &  1.73e-11 &   0.18       &      -- \\
HD130298 &  14.63 &   10.00 &   24.20 &   48.0       &  2.81e-11 &   0.09       &      -- \\
HD165174 &  23.87 &    9.70 &   13.70 &    3.0       &  4.82e-13 &   0.01       &      -- \\
HD229234 &   3.51 &   12.41 &   16.10 &   14.0       &  4.57e-10 &   1.70       &   36.34 \\
HD308813 &   6.35 &   10.71 &   10.70 &    3.0       &  1.66e-11 &   0.22       &      -- \\
LS5039   &   3.91 &    6.29 &   11.10 &    3.0       &  7.34e-12 &   0.08       &     --  \\
\hline
Wind-wind collision & & & & & & & \\
HD130298 &  14.63   &   10.00   &   24.20   &   24.0$^{b}$  &     --      &    --    &   33.82 \\
\hline
Direct impact & & & & & & & \\
HD74194  &   9.54   &   16.51   &   28.20   &    5.0  &     --      &    --    &   32.31 \\
\hline
\hline
\end{tabular}
\newline
\newline
\tablefoot{ (a) Measured BH mass is adopted \citep{Miller-Jones21}. 
(b) Maximum MS companion mass.  
}
\end{table*}
\FloatBarrier 

\section{Journal of observations}
\begin{sidewaystable*}[h!]
\caption{Journal of observations, with the instruments used to collect the data, the number of spectra and the mean S/N.}              
\label{table:database}      
\centering                                      
\begin{tabular}{c c c c | r r r r r r r r r r r}          
\hline\hline                        
Star & Instruments & \# Spec. & <S/N> & $U$ & $B$ & $V$ & $G_{B_P}$  & $G$  & $G_{R_P}$  & $J$  & $H$  & $K$  & $A_V$  & Distance \\    
\hline                                   
Cyg~X-1 & HERMES & 52 & 75 & $9.38$ & $9.60$ & $8.91$ & $9.07$ & $8.54$ & $7.81$ & $6.87$ & $6.65$ & $6.50$ & $3.21 \pm 0.04$ & $2164.89_{-74.64}^{+74.60}$\\ [1pt]       
HD~12323 & HERMES/ELODIE & 24  & 78 & $7.95$ & $8.68$ & $8.79$ & $8.85$ & $8.90$ & $8.93$ & $9.06$ & $9.16$ & $9.17$ & $0.86 \pm 0.03$ & $2375.74_{-192.92}^{+164.55}$\\ [1pt] 
HD~14633 & HERMES/SOPHIE/ESPaDOnS & 10  & 188 & $6.14$ & $7.23$ & $7.44$ & $7.28$ & $7.41$ & $7.62$ & $7.92$ & $8.05$ & $8.12$ & $0.32 \pm 0.04$ & $1420.66_{-155.88}^{+135.84}$\\ [1pt]
HD~15137 & HERMES/ELODIE & 22      & 128 & $7.00$ & $7.88$ & $7.88$ & $7.82$ & $7.83$ & $7.76$ & $7.80$ & $7.83$ & $7.84$ & $1.08 \pm 0.02$ & $2142.79_{-184.84}^{+168.42}$\\ [1pt] 
HD~29763 & HERMES & 21      & 153 & $ - $ & $4.14$ & $4.25$ & $4.18$ & $4.32$ & $4.34$ & $4.79$ & $4.57$ & $4.62$ & $0.03 \pm 0.10$ & $122.02_{-5.81}^{+5.05}$\\ [1pt] 
HD~30836 & HERMES & 27      & 74 &  $2.70$ & $3.50$ & $3.65$ & $3.63$ & $3.67$ & $3.76$ & $4.11$ & $4.15$ & $4.13$ & $0.41 \pm 0.10$ & $260.53_{-22.97}^{+14.37}$\\ [1pt]
HD~37737 & HERMES & 22      & 95 & $8.02$ & $8.32$ & $8.07$ & $8.11$ & $7.96$ & $7.64$ & $7.32$ & $7.32$ & $7.25$ & $1.93 \pm 0.05$ & $1413.56_{-192.54}^{+158.03}$\\ [1pt] 
HD~46573 & HERMES/FEROS & 46      & 110 & $7.61$ & $8.22$ & $7.96$ & $8.00$ & $7.83$ & $7.49$ & $7.20$ & $7.17$ & $7.13$ & $1.88 \pm 0.03$ & $1322.47_{-39.77}^{+38.05}$\\ [1pt] 
HD~52533 & HERMES/FEROS & 11      & 94 & $6.66$ & $7.60$ & $7.69$ & $7.60$ & $7.70$ & $7.73$ & $7.83$ & $7.92$ & $7.93$ & $0.75 \pm 0.02$ & $1688.69_{-141.81}^{+117.13}$\\ [1pt] 
HD~57236 & HERMES/FEROS & 26 & 74 & $8.21$ & $8.90$ & $8.88$ & $8.78$ & $8.69$ & $8.46$ & $8.29$ & $8.25$ & $8.20$ & $1.56 \pm 0.02$ & $2507.80_{-115.21}^{+125.62}$\\ [1pt] 
HD~74194 & FEROS & 22 & 245 & $7.05$ & $7.72$ & $7.56$ & $7.56$ & $7.45$ & $7.19$ & $6.93$ & $6.89$ & $6.81$ & $1.66 \pm 0.05$ & $2202.62_{-77.98}^{+88.49}$\\ [1pt] 
HD~75211 & FEROS/XShooter & 13 & 312 & $7.35$ & $7.84$ & $7.55$ & $7.62$ & $7.40$ & $7.01$ & $6.58$ & $6.54$ & $6.40$ & $2.08 \pm 0.05$ & $1542.02_{-34.40}^{+35.57}$\\ [1pt]
HD~91824 & FEROS/UVES & 14 & 185 & $7.17$ & $8.08$ & $8.15$ & $8.07$ & $8.12$ & $8.13$ & $8.21$ & $8.30$ & $8.35$ & $0.74 \pm 0.01$ & $1826.03_{-92.83}^{+89.23}$\\ [1pt] 
HD~93028 & FEROS & 10 & 162 & $7.29$ & $8.29$ & $8.42$ & $8.29$ & $8.34$ & $8.38$ & $8.47$ & $8.53$ & $8.61$ & $0.65 \pm 0.03$ & $2584.75_{-208.18}^{+183.76}$\\ [1pt] 
HD~94024 & FEROS & 8 & 155 &   $8.03$ & $8.84$ & $8.78$ & $8.73$ & $8.68$ & $8.51$ & $8.45$ & $8.49$ & $8.48$ & $1.22 \pm 0.01$ & $2586.30_{-121.58}^{+99.32}$\\ [1pt] 
HD~105627 & FEROS & 7 & 166 & $7.27$ & $8.18$ & $8.18$ & $8.11$ & $8.11$ & $8.03$ & $7.98$ & $8.03$ & $8.07$ & $0.98 \pm 0.03$ & $2206.69_{-135.92}^{+105.02}$\\ [1pt]
HD~130298 & FEROS/SALT & 12 & 141 & $9.14$ & $9.64$ & $9.26$ & $9.36$ & $9.12$ & $8.69$ & $8.27$ & $8.19$ & $8.11$ & $2.25 \pm 0.04$ & $2425.43_{-75.07}^{+80.53}$\\ [1pt] 
HD~152405 & FEROS & 10 & 146 & $6.71$ & $7.29$ & $7.20$ & $7.19$ & $7.13$ & $6.97$ & $6.83$ & $6.86$ & $6.80$ & $1.34 \pm 0.02$ & $1670.64_{-86.71}^{+84.64}$\\ [1pt] 
HD~152723  & FEROS & 15 & 101 & $6.43$ & $7.35$ & $7.26$ & $7.24$ & $7.18$ & $6.97$ & $6.80$ & $6.81$ & $6.76$ & $1.70 \pm 0.04$ & $2335.59_{-264.41}^{+435.59}$\\ [1pt]
HD~163892 & HERMES/FEROS & 32 & 135 & $6.81$ & $7.58$ & $7.47$ & $7.46$ & $7.39$ & $7.21$ & $7.08$ & $7.10$ & $7.08$ & $1.24 \pm 0.02$ & $1264.29_{-38.27}^{+34.19}$\\ [1pt] 
HD~164438 & HERMES & 35  & 162 & $7.37$ & $7.74$ & $7.50$ & $7.53$ & $7.36$ & $7.02$ & $6.73$ & $6.65$ & $6.62$ & $1.99 \pm 0.05$ & $1166.10_{-31.64}^{+29.87}$\\ [1pt] 
HD~164536 & HERMES & 30  & 123 & $6.46$ & $7.07$ & $7.14$ & $7.06$ & $7.11$ & $7.08$ & $7.12$ & $7.15$ & $7.13$ & $1.03 \pm 0.04$ & $1532.77_{-246.83}^{+192.02}$\\ [1pt]
HD~165174 & HERMES/FEROS/UVES & 20      & 177 & $5.23$ & $6.08$ & $6.15$ & $6.06$ & $6.09$ & $6.09$ & $6.15$ & $6.22$ & $6.24$ & $0.82 \pm 0.05$ & $964.79_{-45.83}^{+37.58}$\\ [1pt] 
HD~167263 & HERMES/FEROS/ESPaDOnS & 21 & 145 & $5.25$ & $5.94$ & $5.96$ & $5.90$ & $5.90$ & $5.85$ & $5.87$ & $5.91$ & $5.88$ & $0.98 \pm 0.02$ & $2965.54_{-756.53}^{+810.34}$\\ [1pt] 
HD~167264 & HERMES/FEROS/UVES & 42      & 96 & $4.64$ & $5.34$ & $5.34$ & $5.28$ & $5.28$ & $5.19$ & $5.21$ & $5.21$ & $5.16$ & $1.03 \pm 0.02$ & $1227.63_{-245.46}^{+208.47}$\\ [1pt] 
HD~192001 & HERMES & 30      & 111 & $7.97$ & $8.50$ & $8.28$ & $8.34$ & $8.20$ & $7.90$ & $7.68$ & $7.68$ & $7.68$ & $1.69 \pm 0.03$ & $1679.40_{-66.89}^{+63.84}$\\ [1pt] 
HD~199579 & HERMES & 112      & 190 & $5.16$ & $5.98$ & $5.96$ & $5.91$ & $5.89$ & $5.81$ & $5.80$ & $5.83$ & $5.86$ & $1.02 \pm 0.01$ & $855.94_{-28.75}^{+29.03}$\\ [1pt] 
HD~229234  & HERMES & 57      & 86 & $9.50$ & $9.58$ & $8.98$ & $9.11$ & $8.67$ & $8.04$ & $7.34$ & $7.18$ & $7.10$ & $3.03 \pm 0.05$ & $1670.99_{-28.86}^{+35.98}$\\ [1pt] 
HD~308813 & FEROS & 10      & 162 & $8.46$ & $9.23$ & $9.22$ & $9.24$ & $9.24$ & $9.16$ & $9.15$ & $9.16$ & $9.14$ & $0.79 \pm 0.03$ & $2379.76_{-106.06}^{+93.03}$\\ [1pt] 
LS~5039  & FEROS/UVES & 10      & 65 & $12.02$ & $12.53$ & $11.47$ & $11.45$ & $10.80$ & $10.00$ & $9.02$ & $8.75$ & $8.60$ & $4.06 \pm 0.06$ & $1901.23_{-69.66}^{+54.20}$\\ [1pt] 
Schulte~11 & HERMES & 30 & 31 & $11.74$ & $11.22$ & $10.12$ & $10.25$ & $9.25$ & $8.24$ & $6.65$ & $6.23$ & $5.99$ & $5.14 \pm 0.06$ & $1692.75_{-46.59}^{+90870.75}$\\ [1pt] 
V747~Cep & HERMES & 11 & 34 & $11.60$ & $11.19$ & $10.10$ & $10.26$ & $9.33$ & $8.36$ & $6.99$ & $6.63$ & $6.42$ & $4.98 \pm 0.07$ & $988.87_{-12.34}^{+12.91}$\\ [1pt] 
\hline                                             
\end{tabular}
\end{sidewaystable*}
\FloatBarrier 

\section{Spectral disentangling of the newly detected SB2 systems}
\FloatBarrier 
\begin{figure*}[h!]
    \centering
    \begin{subfigure}{0.46\linewidth}
    \includegraphics[width = \textwidth, trim=0 0 35 35,clip]{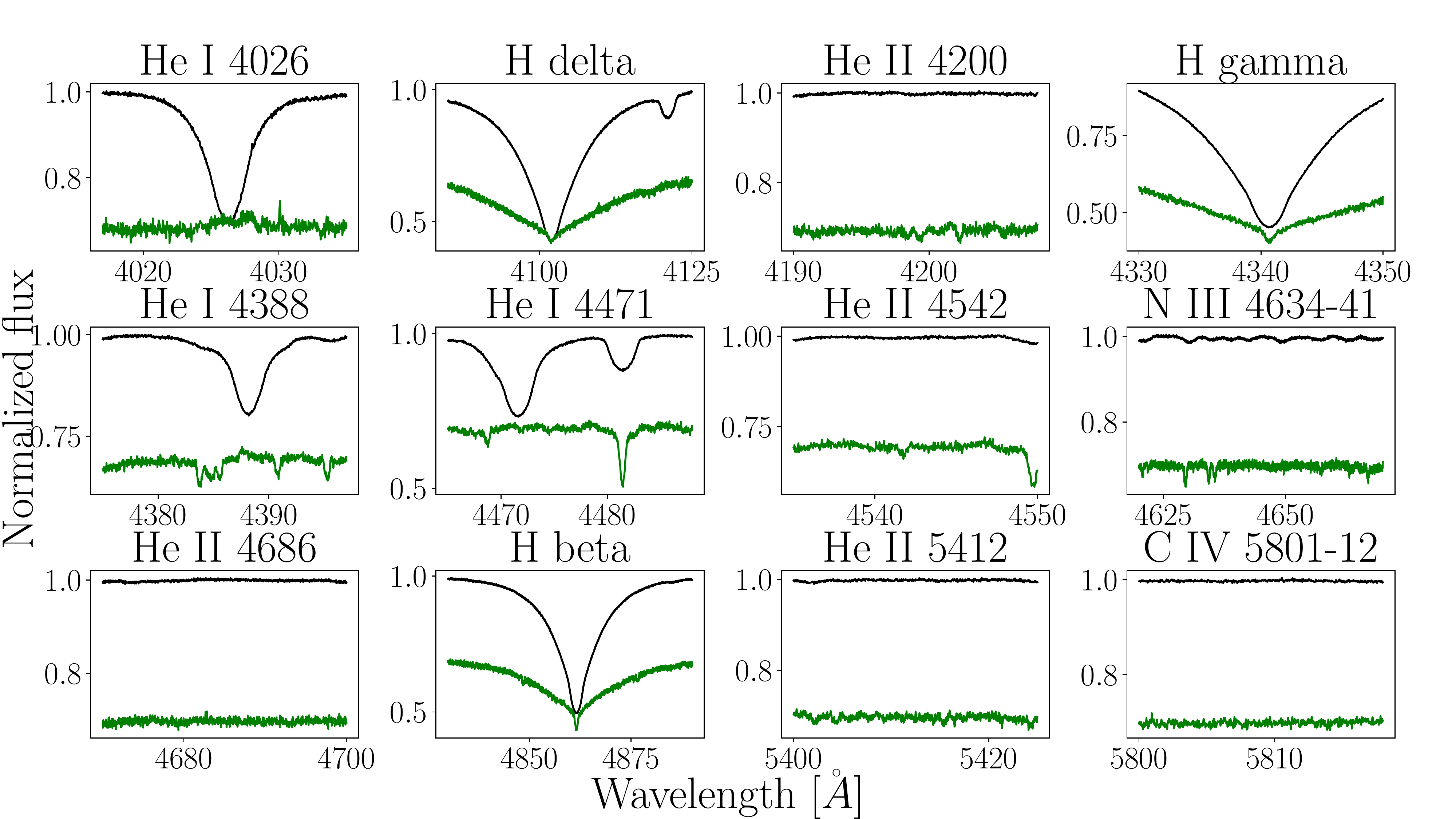}        
    \caption{HD~29763}
    \end{subfigure}
    \hfill
    \begin{subfigure}{0.46\linewidth}
    \includegraphics[width = \textwidth, trim=0 0 35 35,clip]{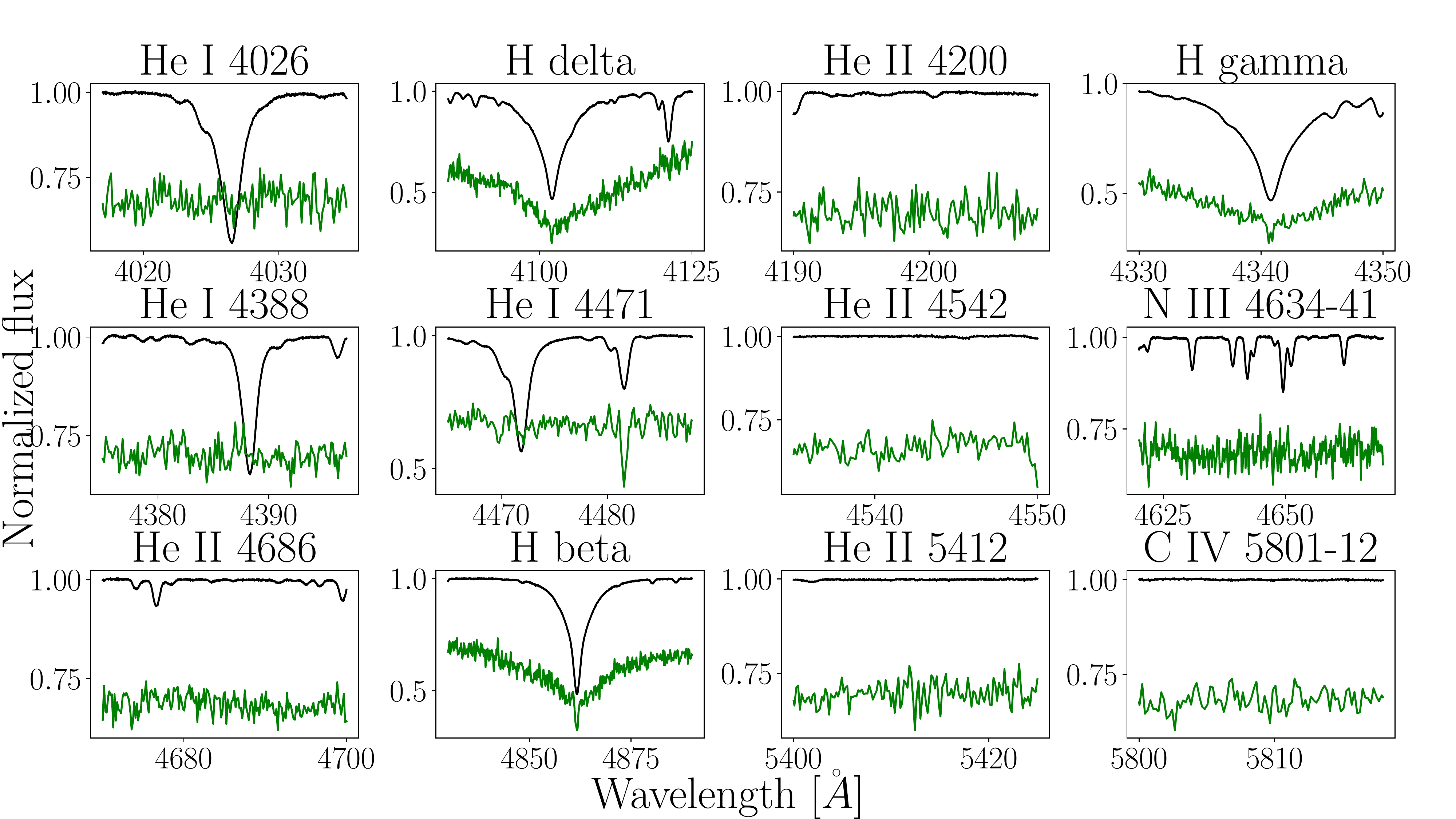}
    \caption{HD~30836}
    \end{subfigure}
    \hfill
    \begin{subfigure}{0.46\linewidth}
    \includegraphics[width = \textwidth, trim=0 0 35 35,clip]{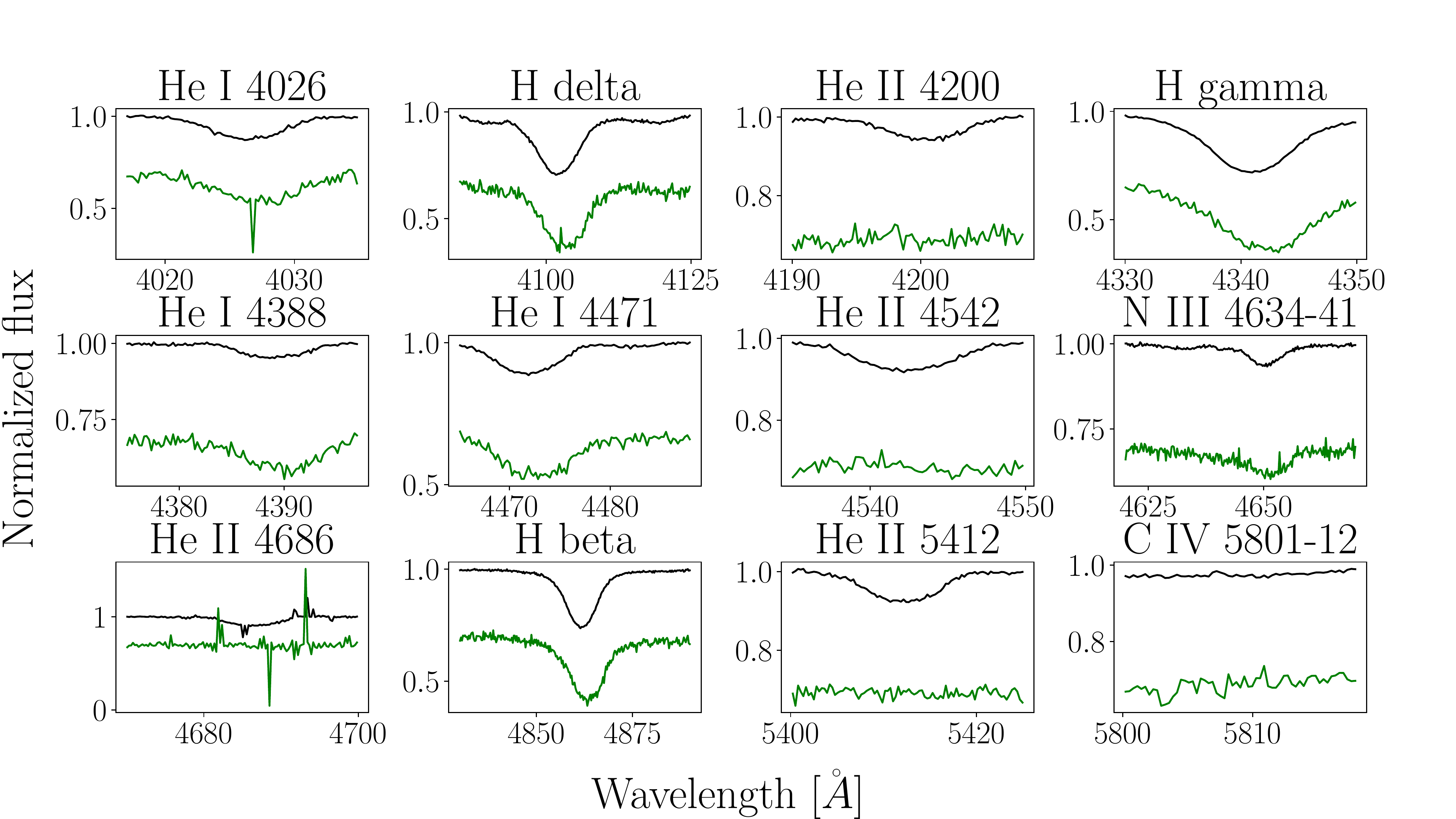}
    \caption{HD~52533}
    \end{subfigure} 
    \hfill
    \begin{subfigure}{0.46\linewidth}
    \includegraphics[width = \textwidth, trim=0 0 35 35,clip]{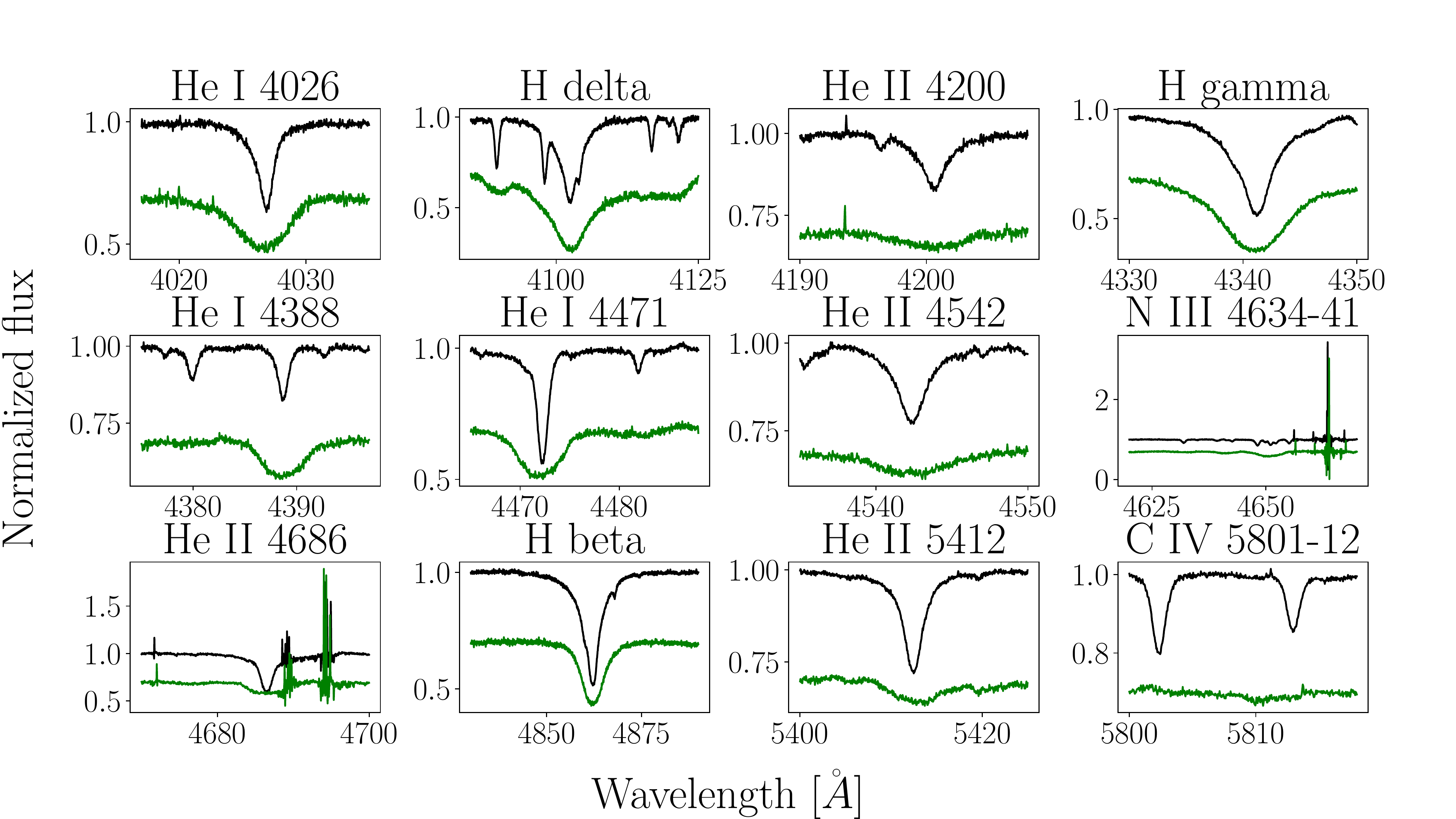}
    \caption{HD~57236}
    \end{subfigure} 
    \hfill
    \begin{subfigure}{0.46\linewidth}
    \includegraphics[width = \textwidth, trim=0 0 35 35,clip]{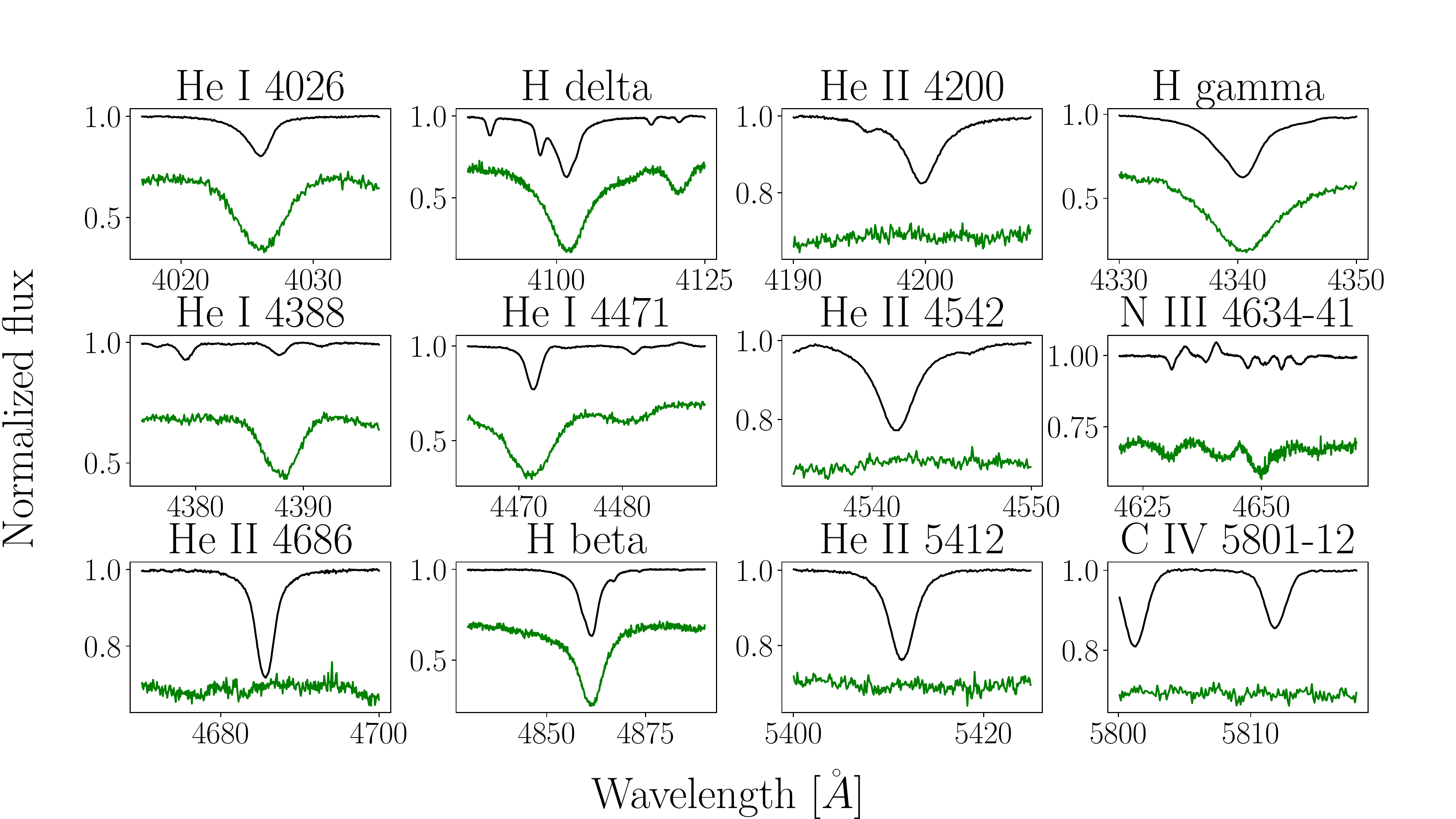}
    \caption{HD~91824}
    \end{subfigure}
    \hfill
    \begin{subfigure}{0.46\linewidth}
    \includegraphics[width = \textwidth, trim=0 0 35 35,clip]{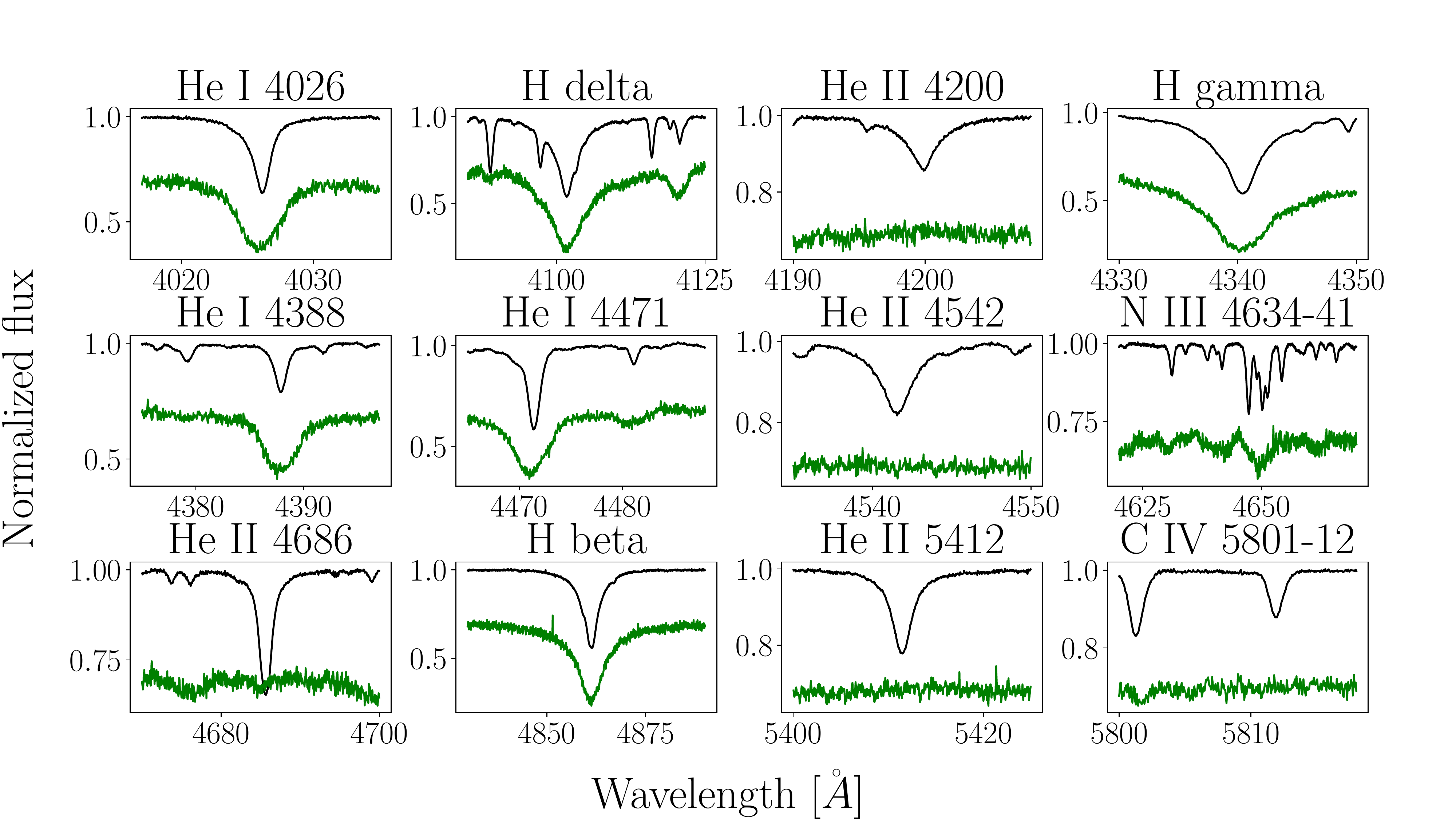}        
    \caption{HD~93028}
    \end{subfigure} 
    \hfill
    \begin{subfigure}{0.46\linewidth}
    \includegraphics[width = \textwidth, trim=0 0 35 35,clip]{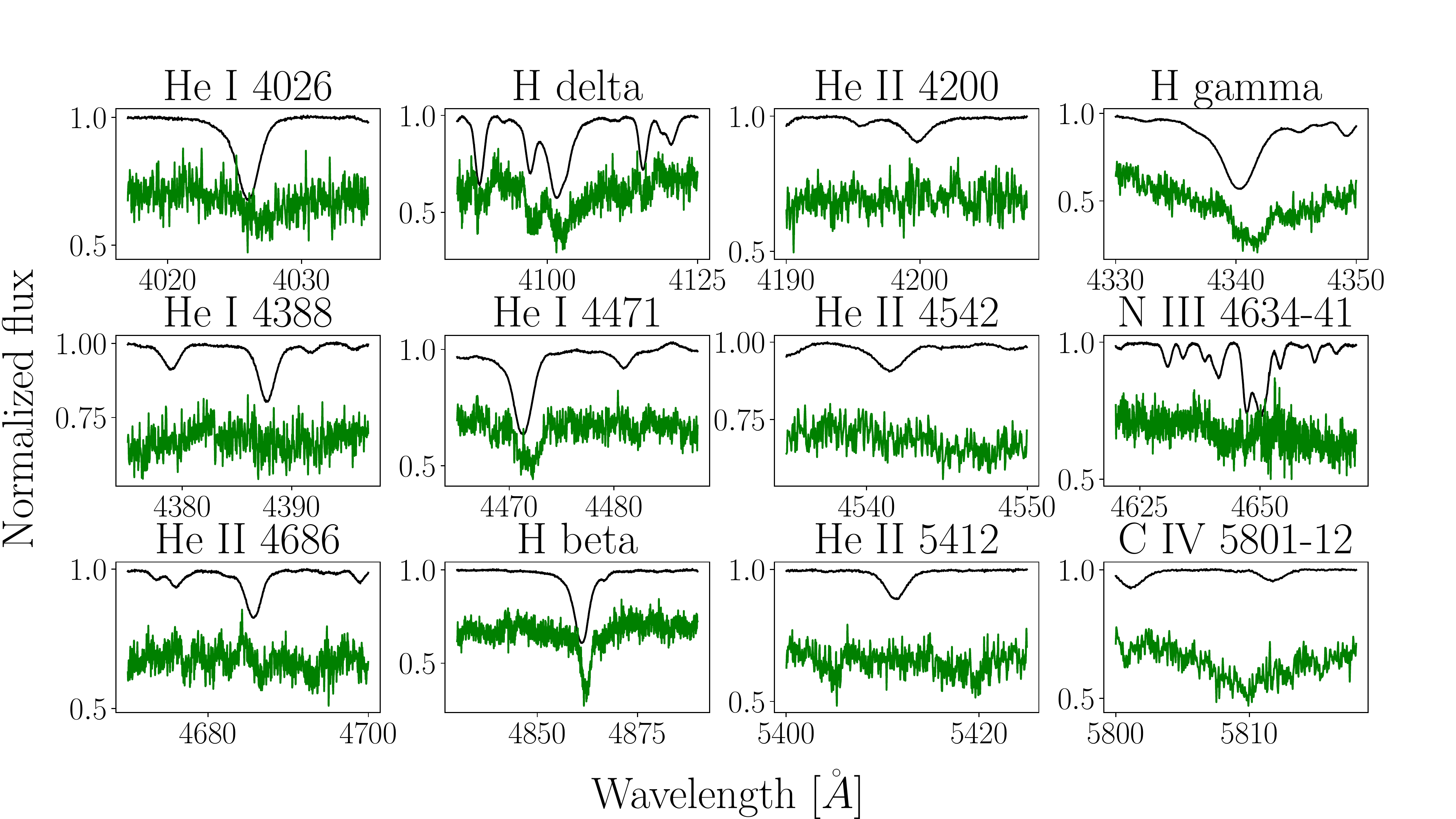}        
    \caption{HD~152405}
    \end{subfigure}
    \hfill
    \begin{subfigure}{0.46\linewidth}
    \includegraphics[width = \textwidth, trim=0 0 35 35,clip]{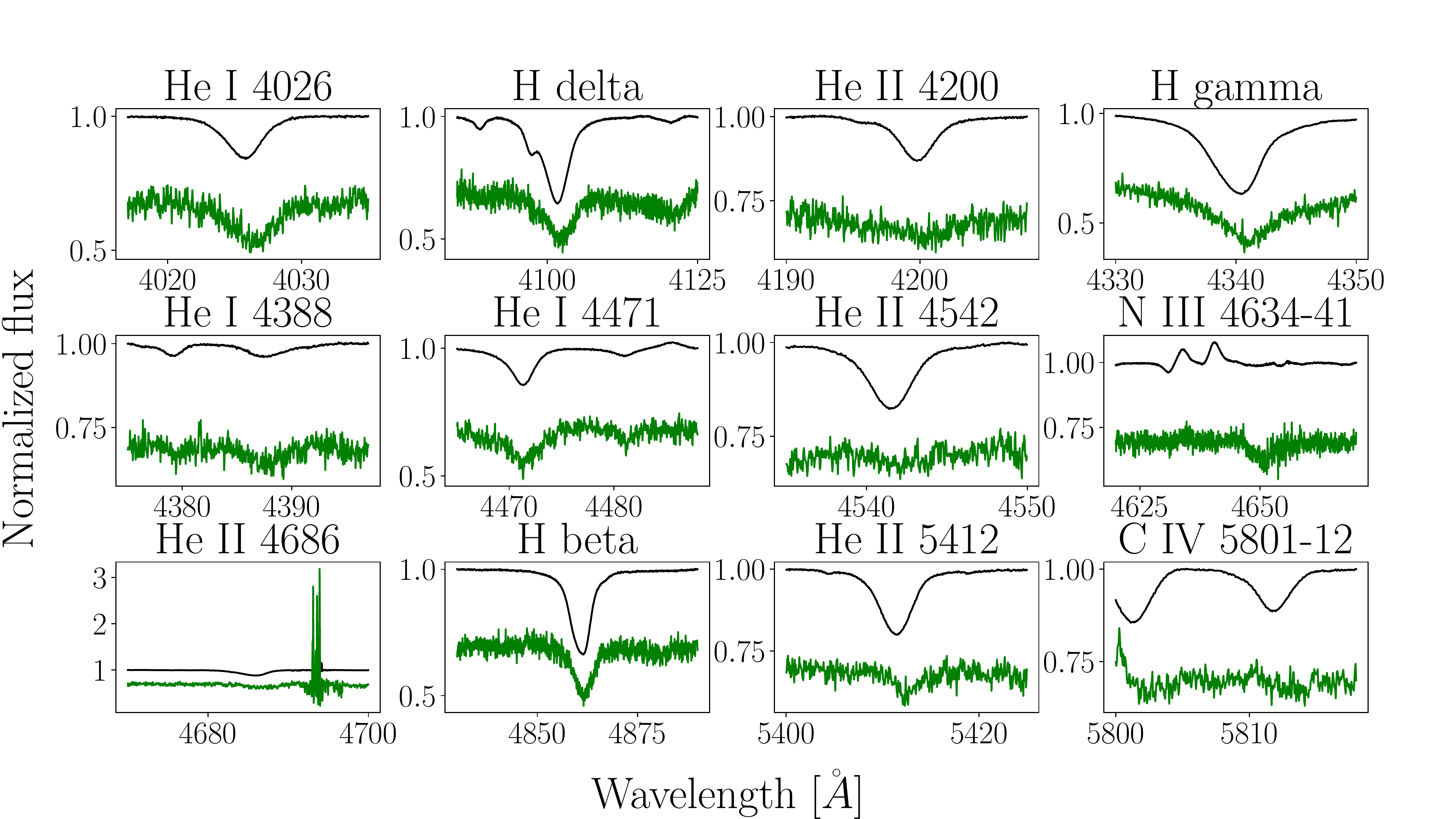}        
    \caption{HD~152723}
    \end{subfigure}
    \caption{Disentangled spectra of the newly-detected SB2 sample. Black (green) spectra are the primaries (secondaries). The secondary spectra were shifted for clarity. }
\end{figure*}
\FloatBarrier
\begin{figure*}[h!]
\ContinuedFloat 
    \begin{subfigure}{0.46\linewidth}
    \includegraphics[width = \textwidth, trim=0 0 35 35,clip]{Images/disent_HD163892.pdf}        
    \caption{HD~163892}
    \end{subfigure}
    \hfill
    \begin{subfigure}{0.46\linewidth}
    \includegraphics[width = \textwidth, trim=0 0 35 35,clip]{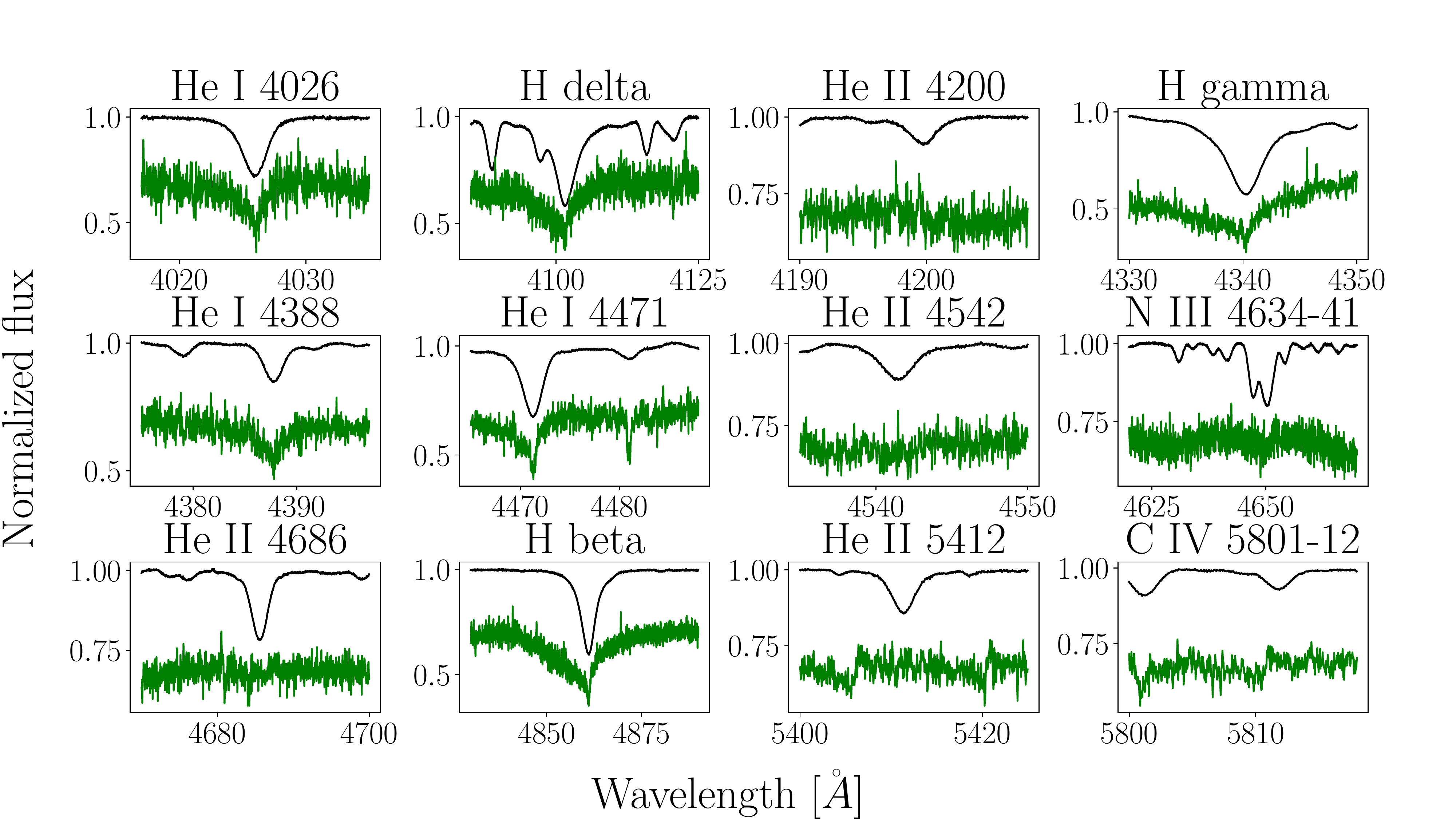}        
    \caption{HD~164438}
    \end{subfigure}
    \hfill
    \begin{subfigure}{0.46\linewidth}
    \includegraphics[width = \textwidth, trim=0 0 35 35,clip]{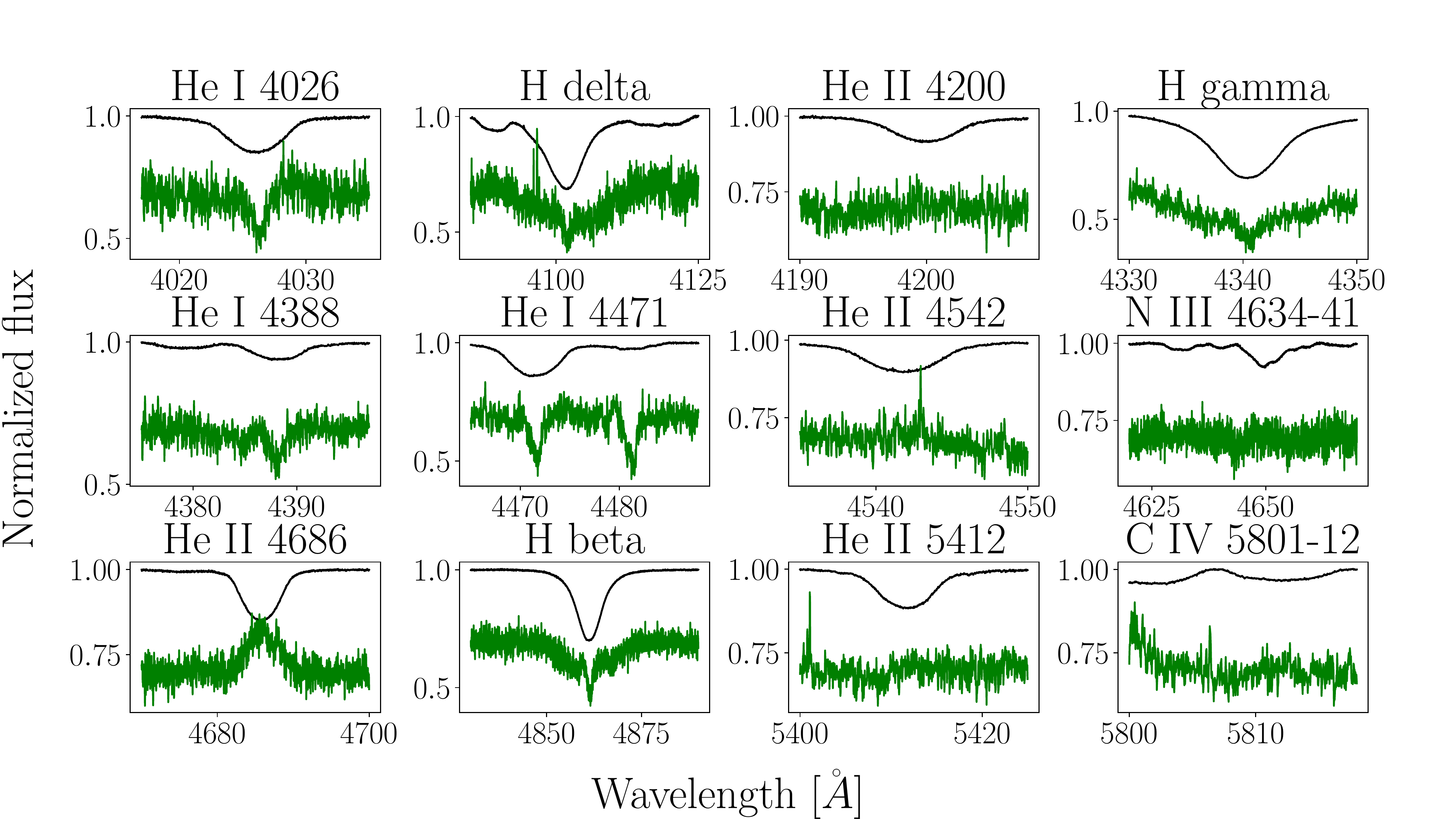}        
    \caption{HD~164536}
    \end{subfigure}
    \hfill
    \begin{subfigure}{0.46\linewidth}
    \includegraphics[width = \textwidth, trim=0 0 35 35,clip]{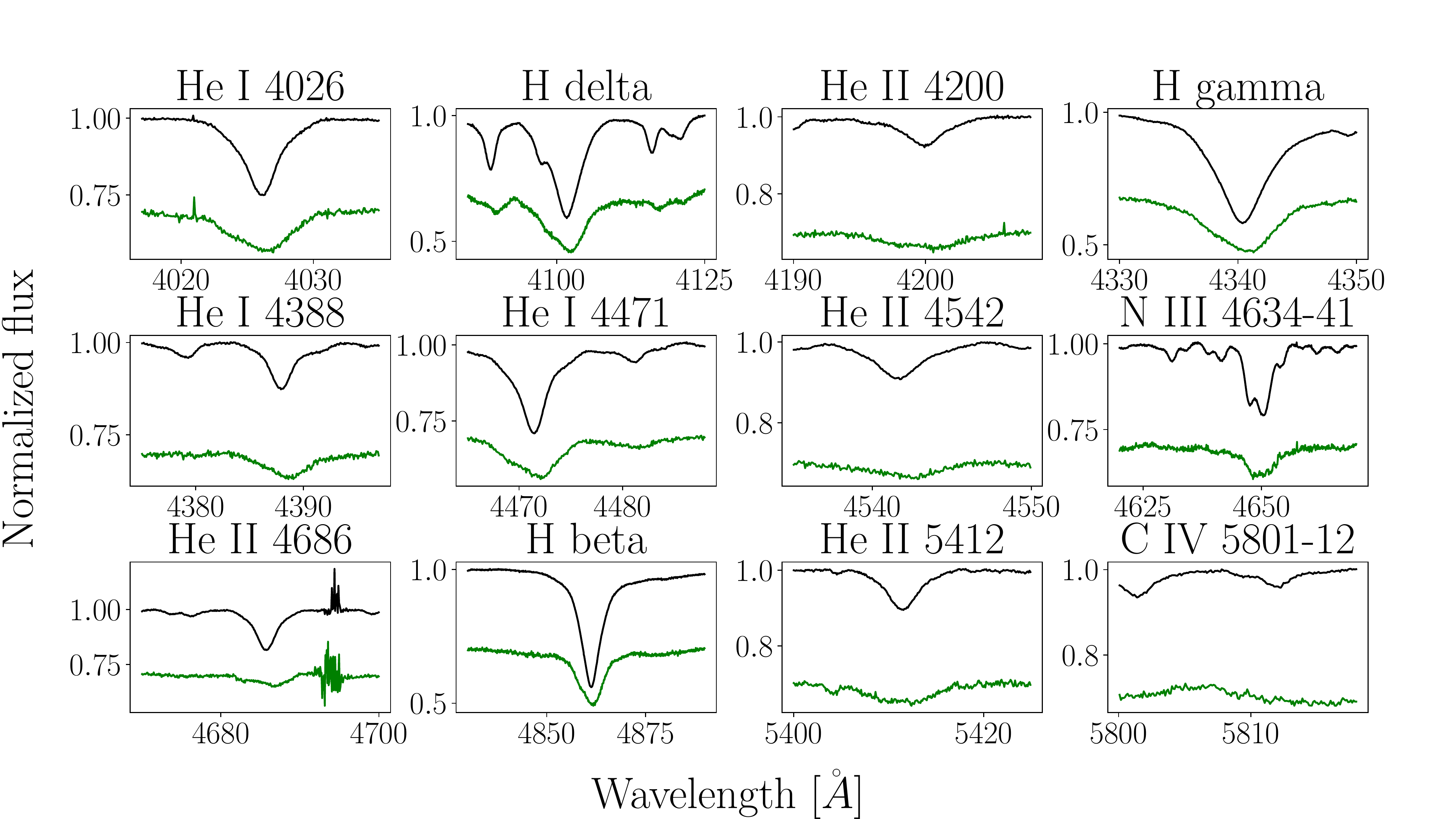}        
    \caption{HD~167263}
    \end{subfigure}  
    \hfill
    \begin{subfigure}{0.46\linewidth}
    \includegraphics[width = \textwidth, trim=0 0 35 35,clip]{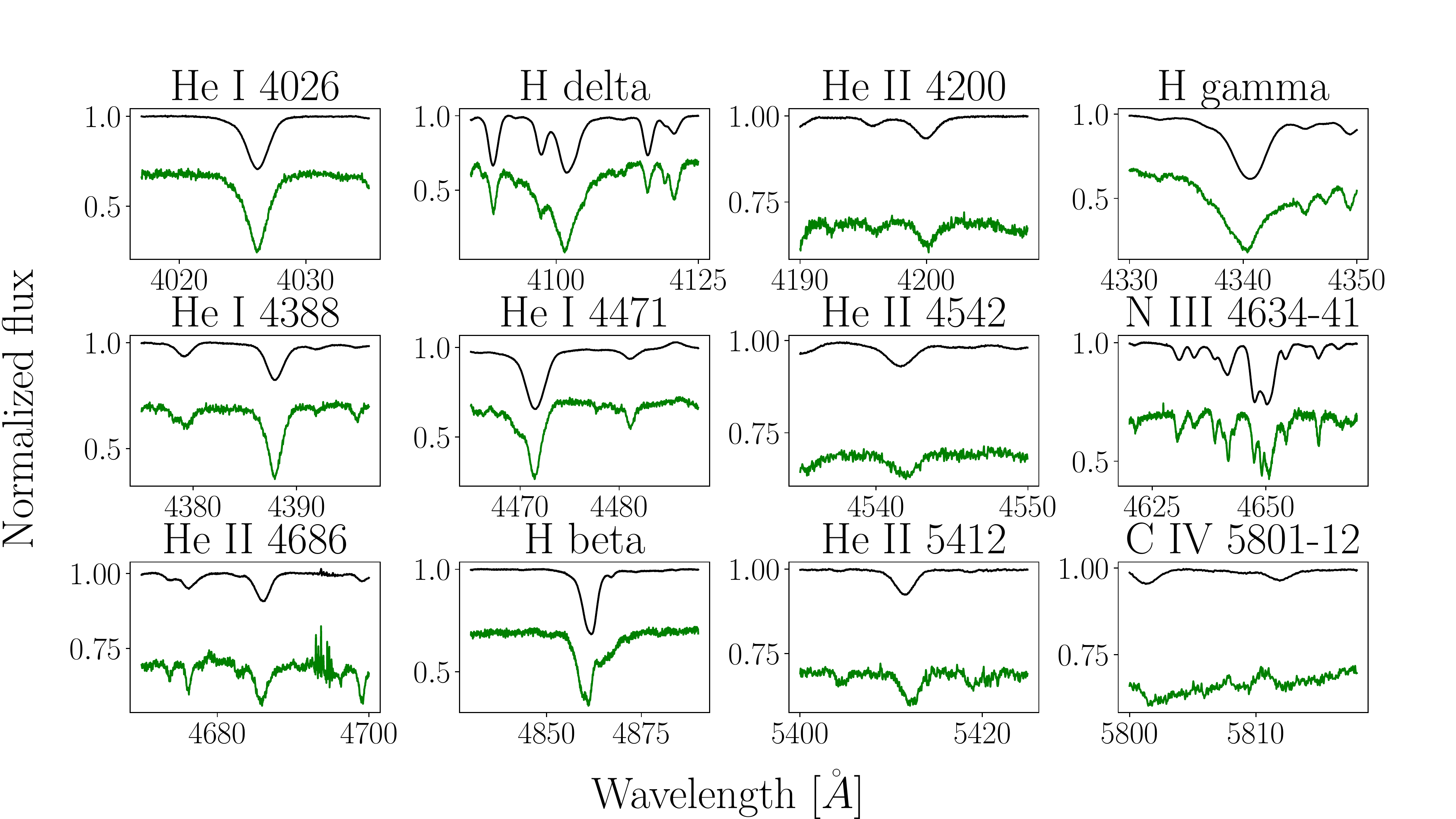}        
    \caption{HD~167264}
    \end{subfigure}
    \hfill
    \begin{subfigure}{0.46\linewidth}
    \includegraphics[width = \textwidth, trim=0 0 35 35,clip]{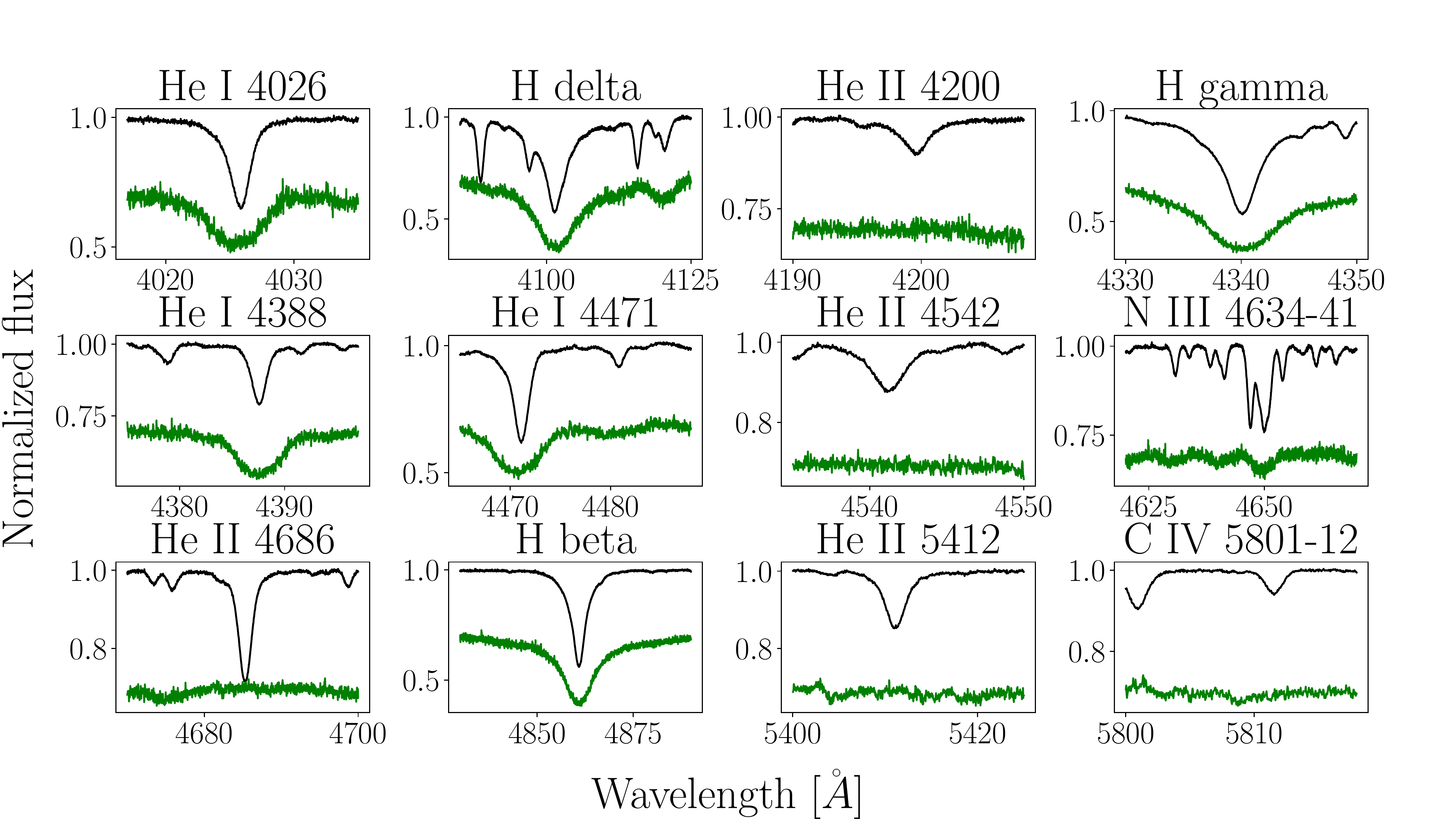}
    \caption{HD~192001}
    \end{subfigure}
    \hfill
    \begin{subfigure}{0.46\linewidth}
    \includegraphics[width = \textwidth, trim=0 0 35 35,clip]{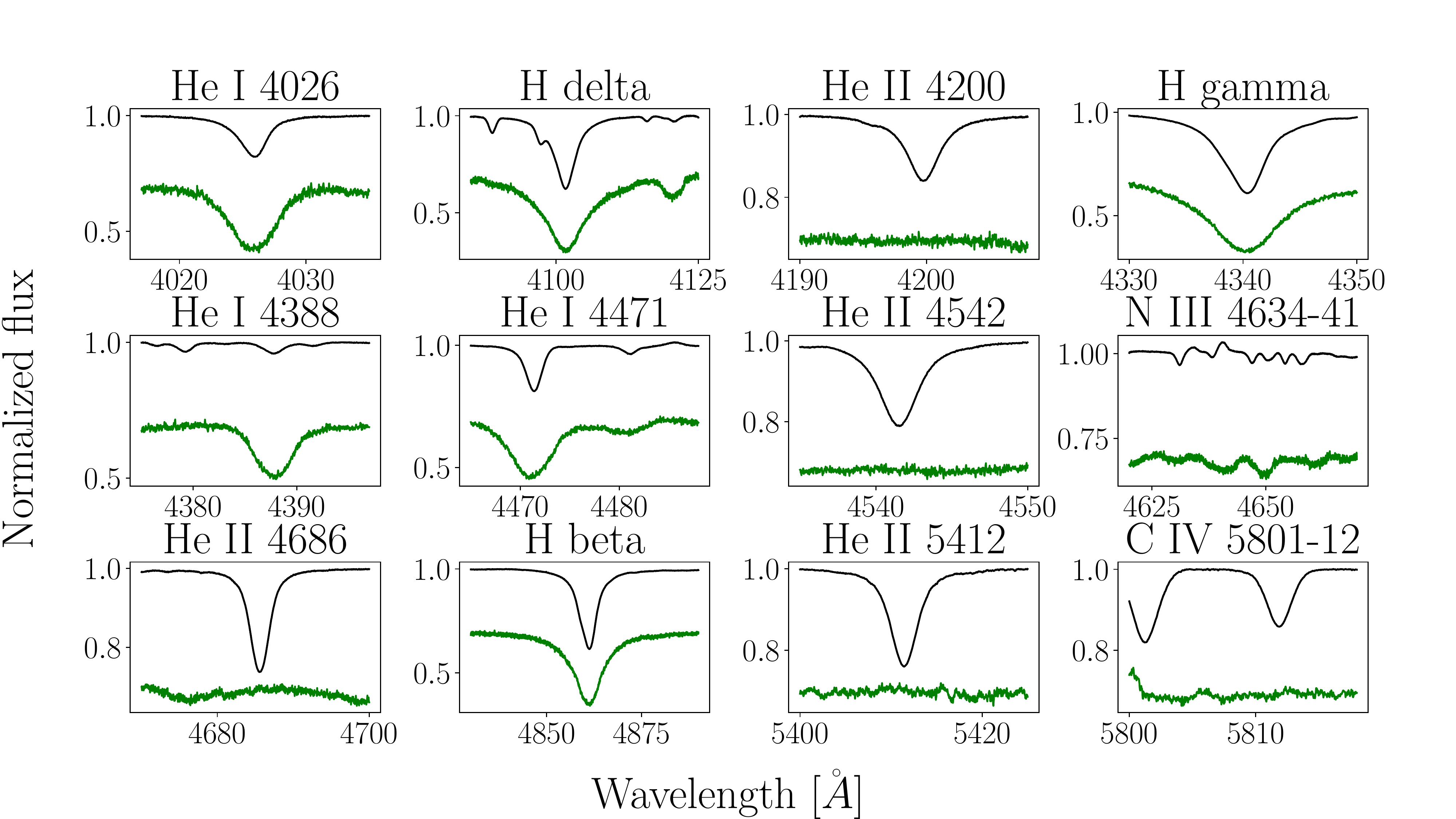}
    \caption{HD~199579}
    \end{subfigure} 
    \hfill
    \begin{subfigure}{0.46\linewidth}
    \includegraphics[width = \textwidth, trim=0 0 35 35,clip]{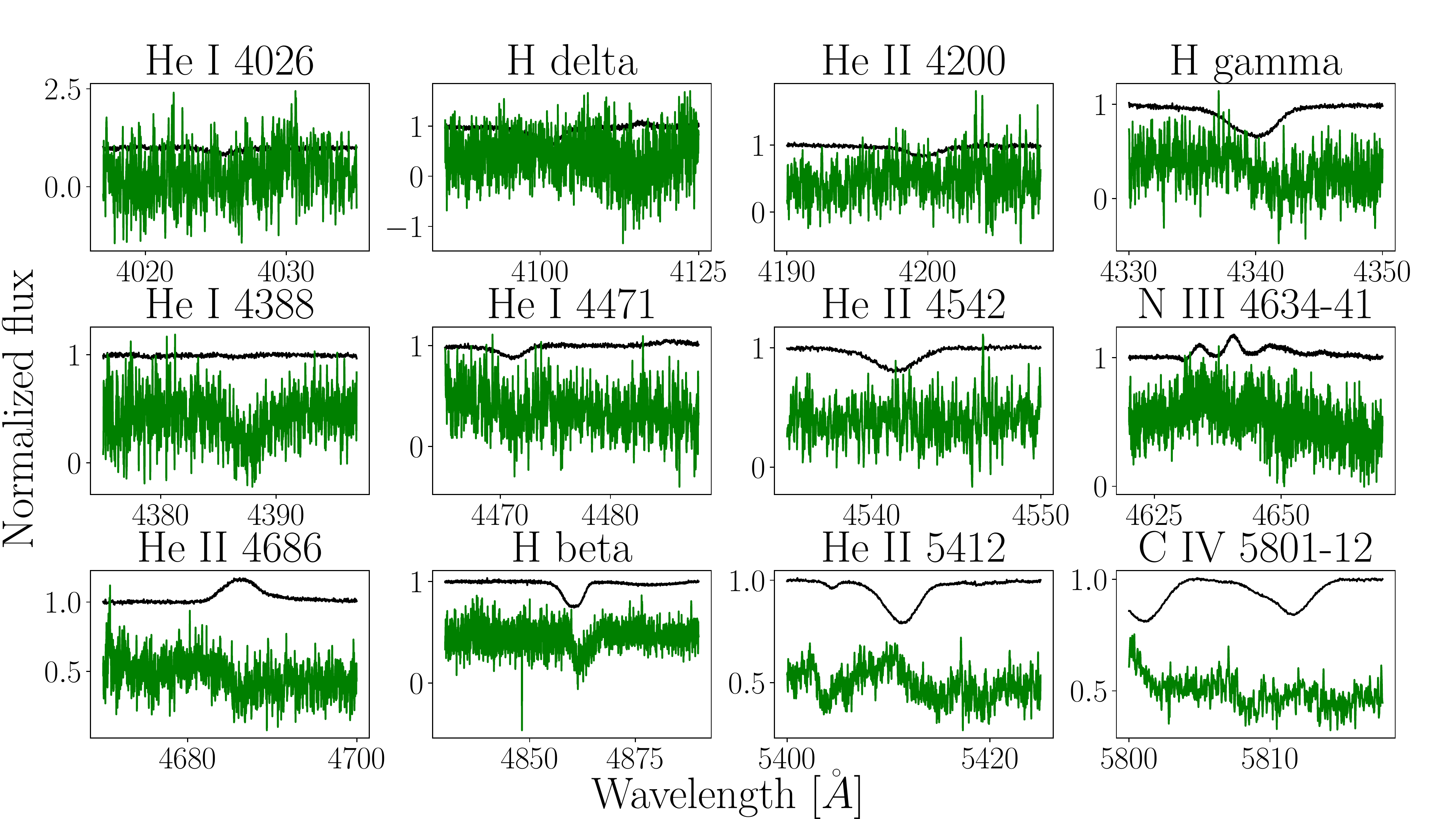}
    \caption{Schulte~11}
    \end{subfigure}   
    \caption{continued}
\end{figure*}
\FloatBarrier
\begin{figure*}[h!]
\ContinuedFloat 
    \begin{subfigure}{0.46\linewidth}
    \includegraphics[width = \textwidth, trim=0 0 35 35,clip]{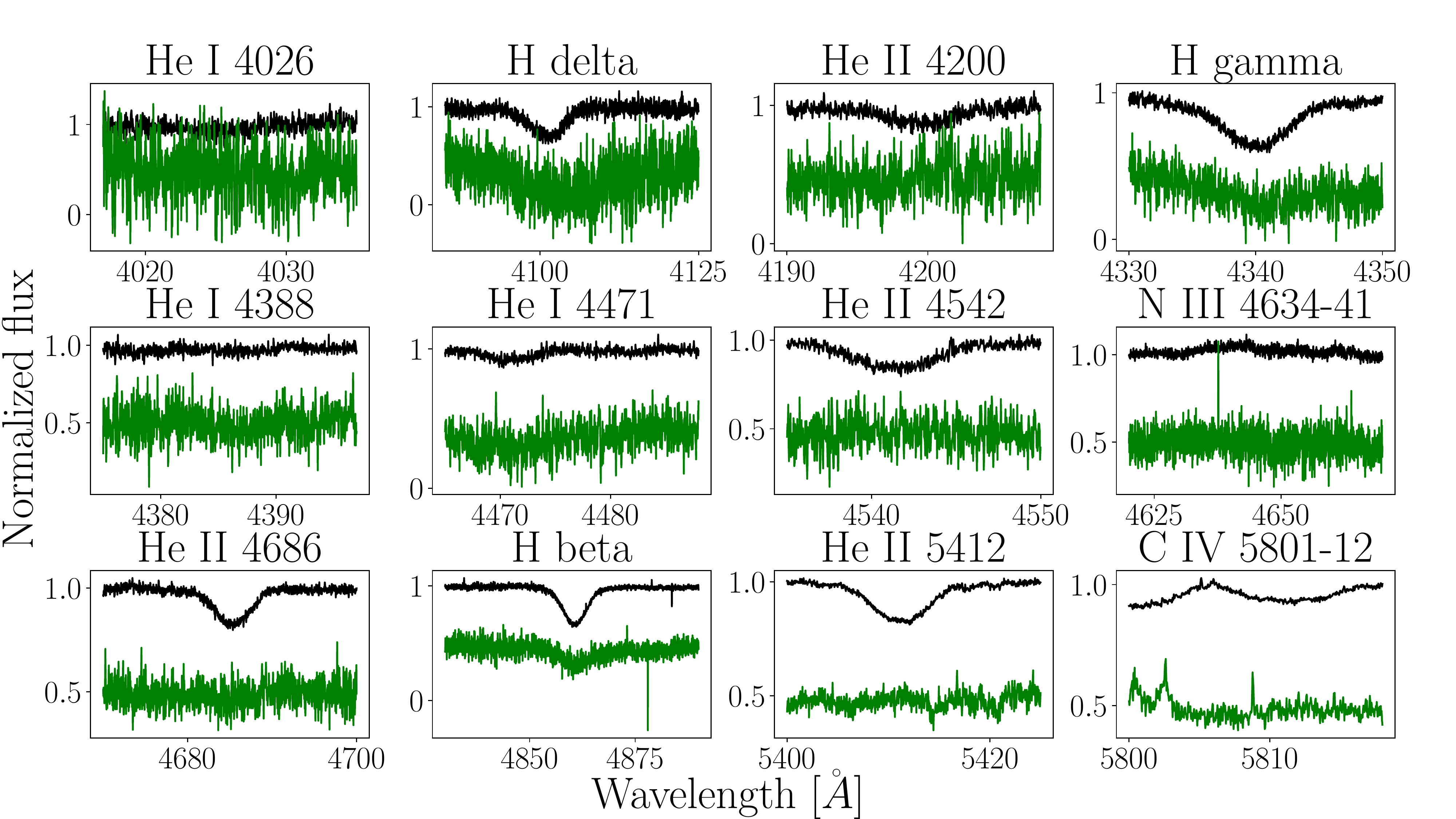}        
    \caption{V747~Cep}
    \end{subfigure}
    \caption{continued}
    \label{fig:disent_all}
\end{figure*}

\FloatBarrier

\section{Spectral Energy Distribution}
\FloatBarrier 
\begin{figure*}[h!]
    \begin{subfigure}{0.46\linewidth}
    \includegraphics[width = \textwidth, trim=0 0 0 0,clip]{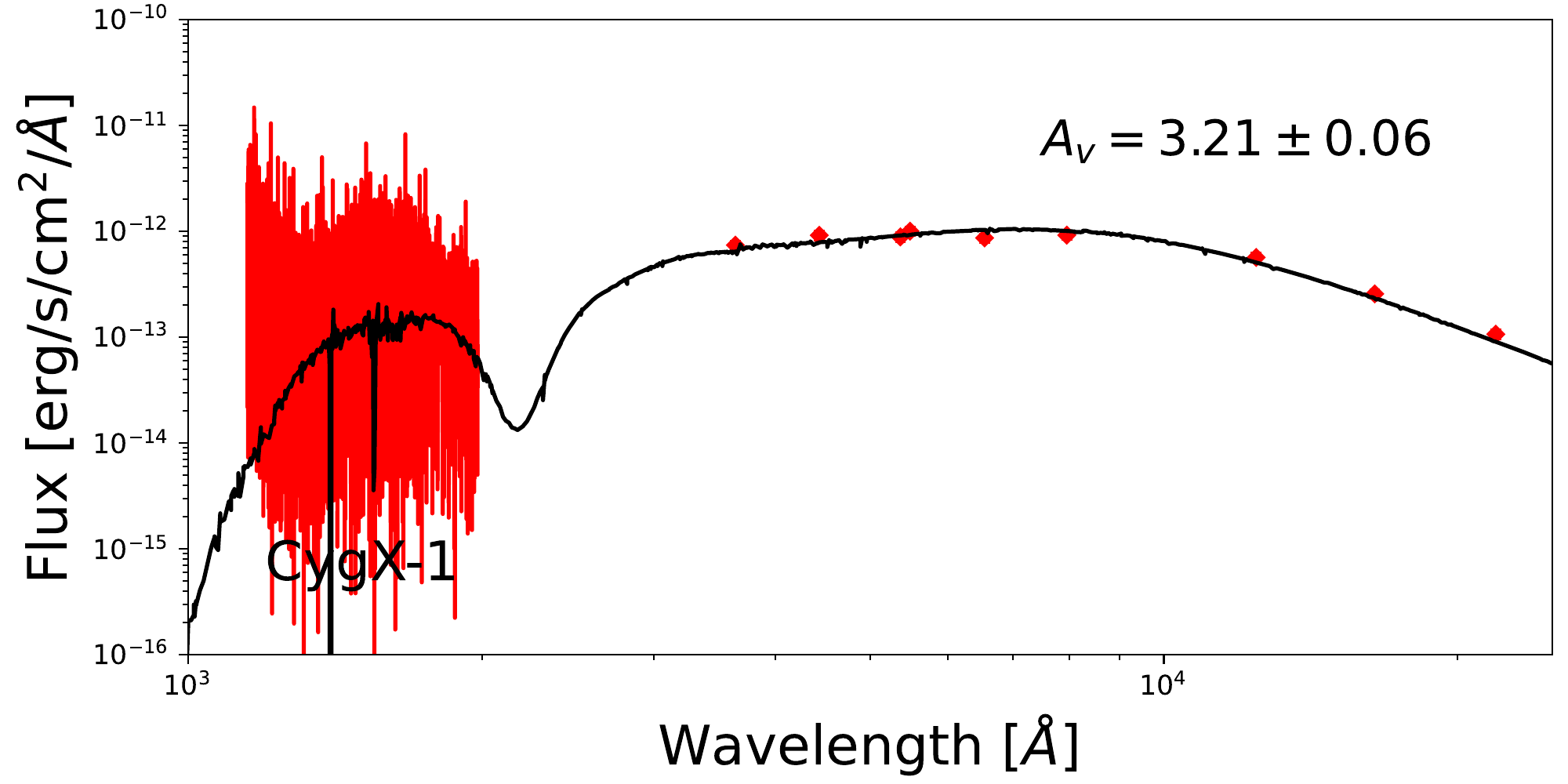} 
    \caption{Cyg~X-1}
    \end{subfigure}
    \hfill
    \begin{subfigure}{0.46\linewidth}
    \includegraphics[width = \textwidth, trim=0 0 0 0,clip]{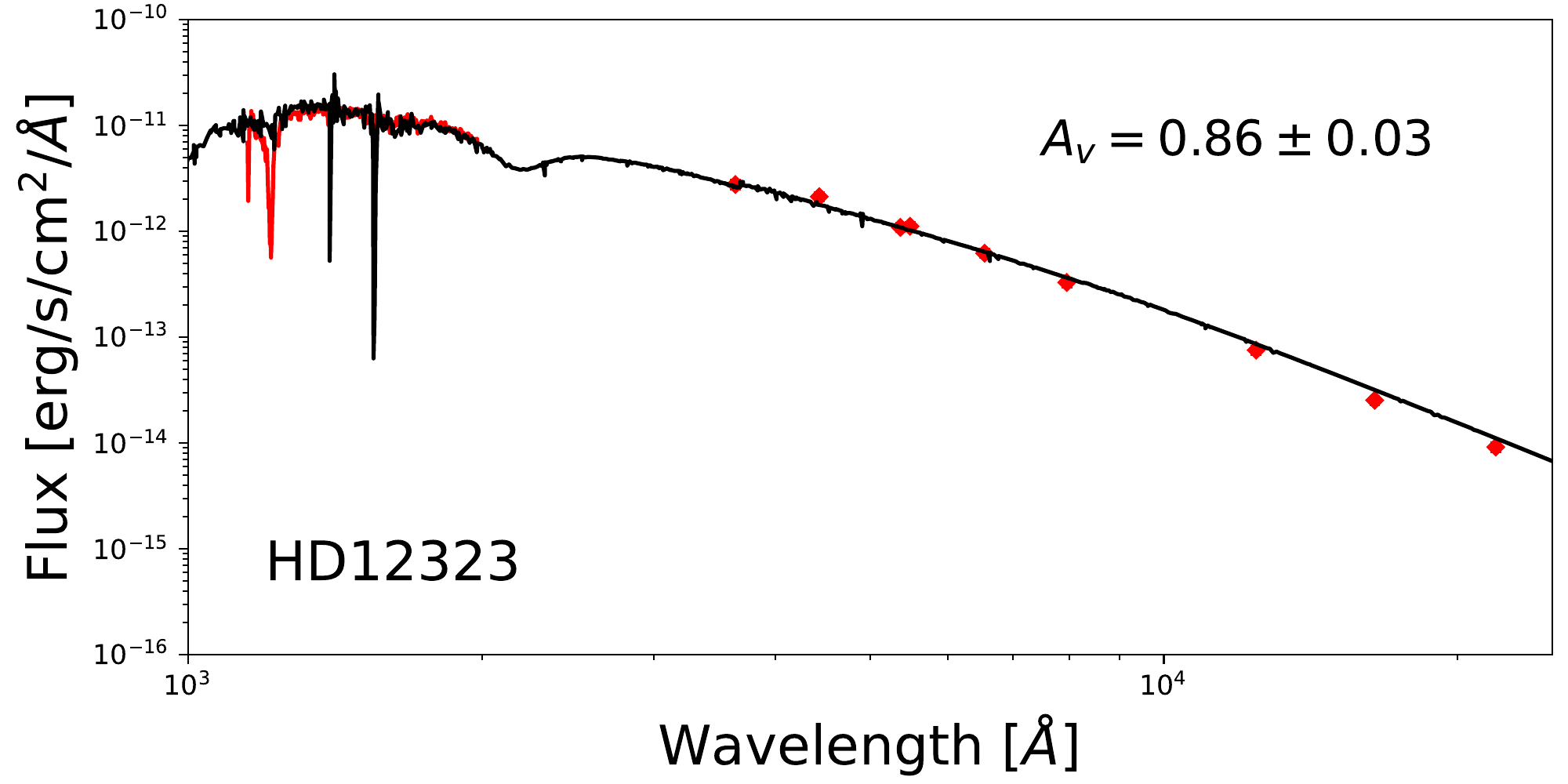}
    \caption{HD~12323}
    \end{subfigure}
    \hfill
    \begin{subfigure}{0.46\linewidth}
    \includegraphics[width = \textwidth, trim=0 0 0 0,clip]{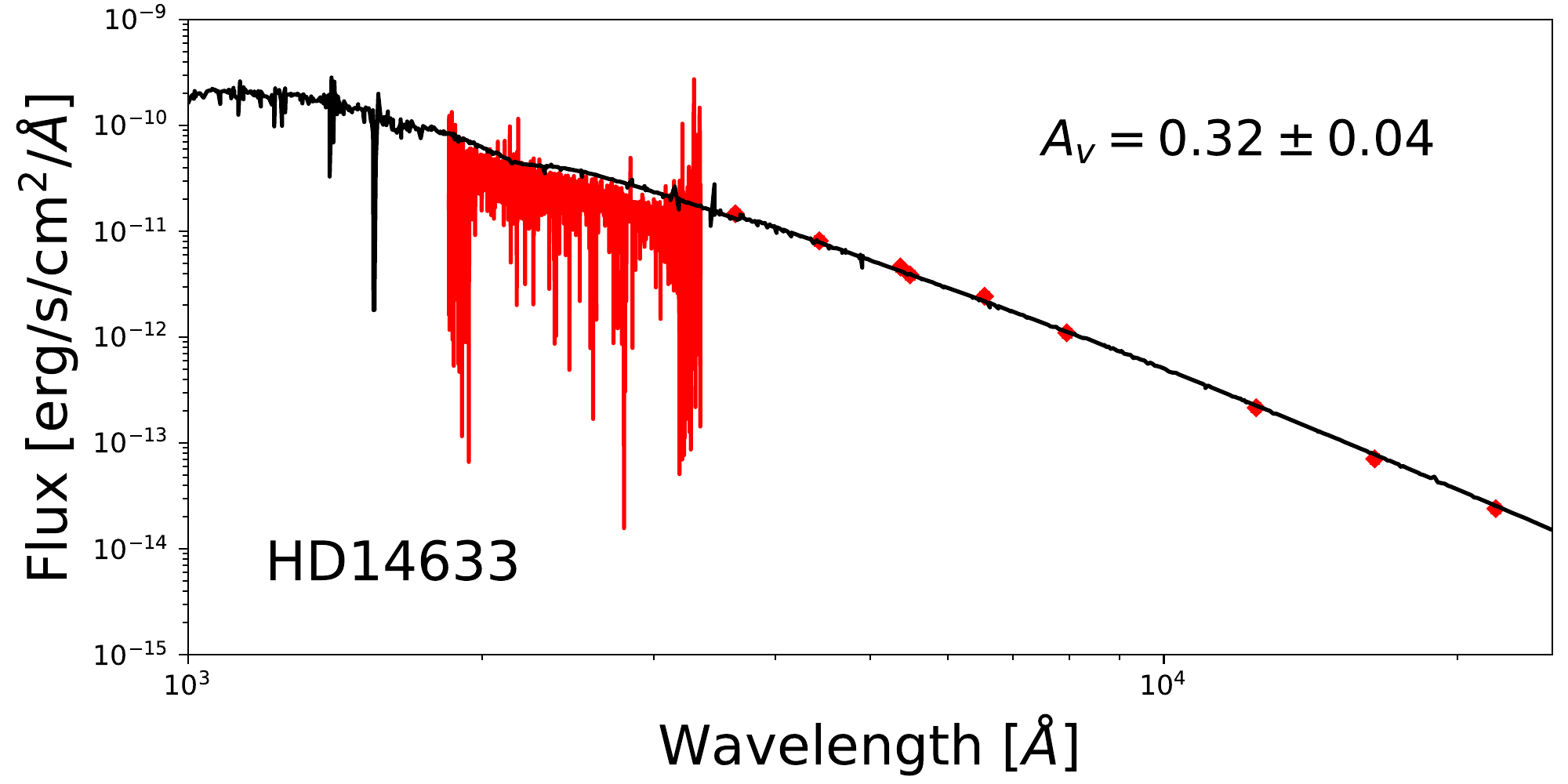} \caption{HD~14633}
    \end{subfigure}
    \hfill
    \begin{subfigure}{0.46\linewidth}
    \includegraphics[width = \textwidth, trim=0 0 0 0,clip]{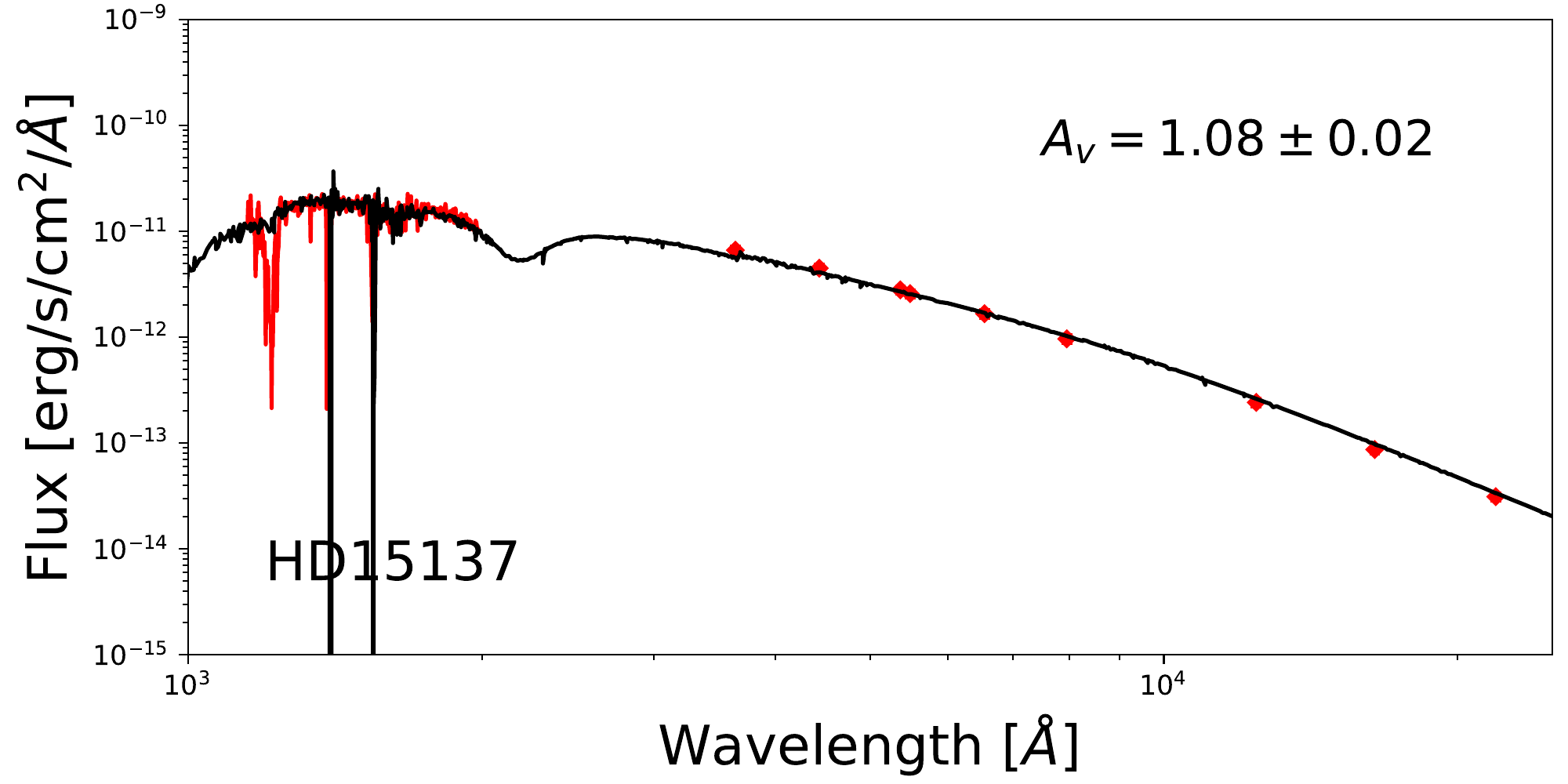} 
    \caption{HD~15137}
    \end{subfigure}
    \hfill
    \begin{subfigure}{0.46\linewidth}
    \includegraphics[width = \textwidth, trim=0 0 0 0,clip]{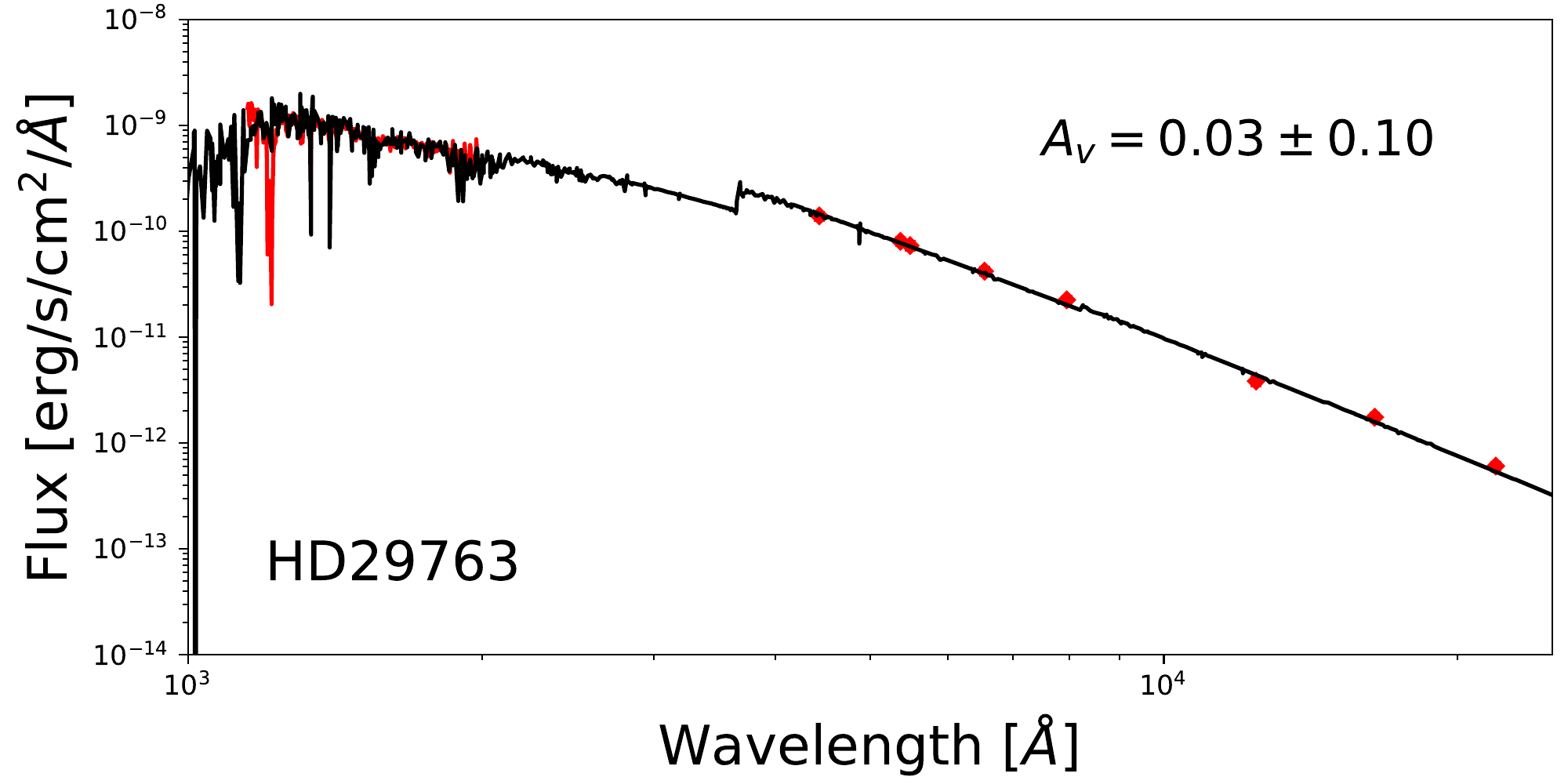}
    \caption{HD~29763}
    \end{subfigure}
    \hfill
    \begin{subfigure}{0.46\linewidth}
    \includegraphics[width = \textwidth, trim=0 0 0 0,clip]{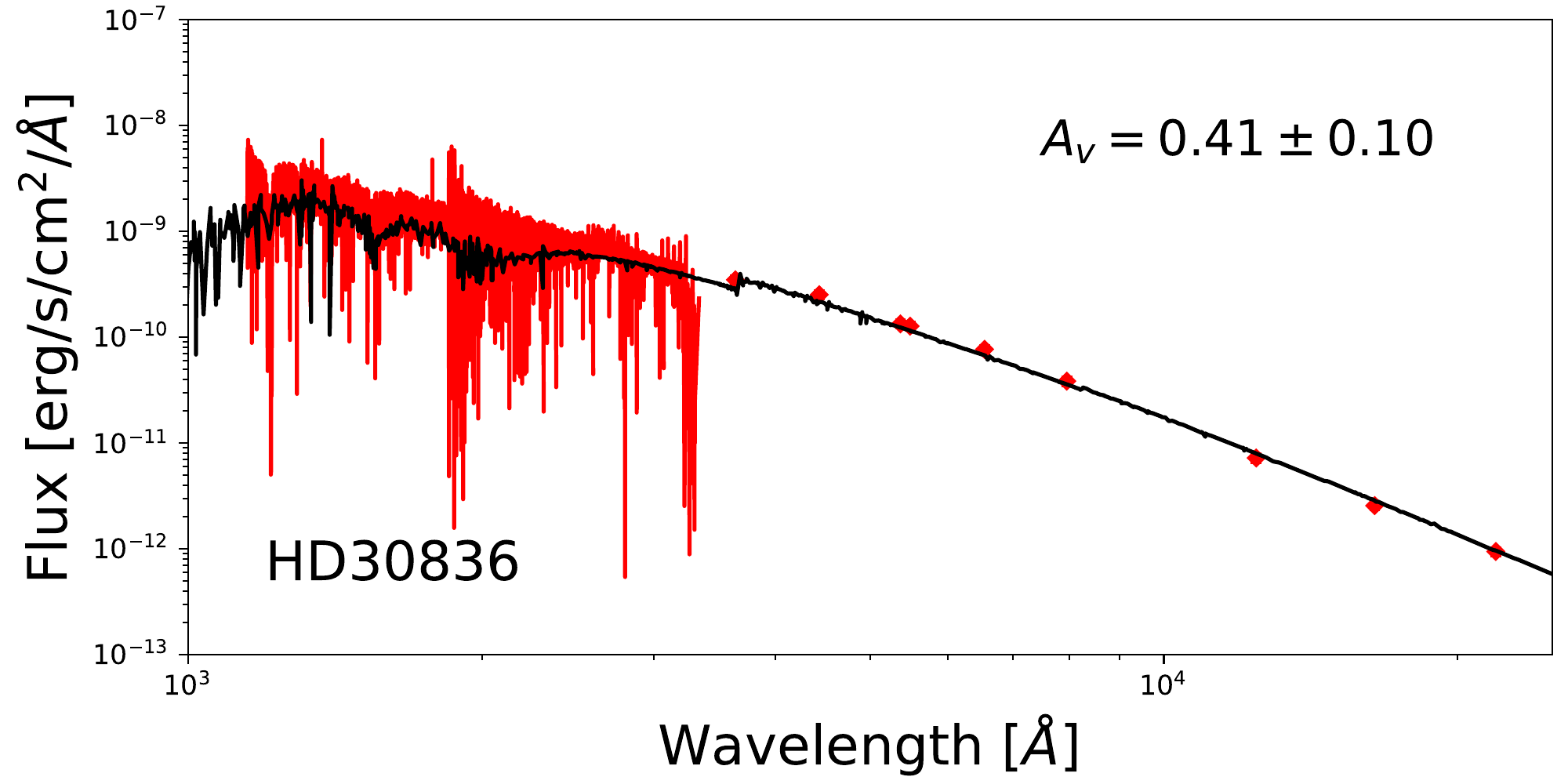} 
    \caption{HD~30836}
    \end{subfigure}
    \hfill
    \begin{subfigure}{0.46\linewidth}
    \includegraphics[width = \textwidth, trim=0 0 0 0,clip]{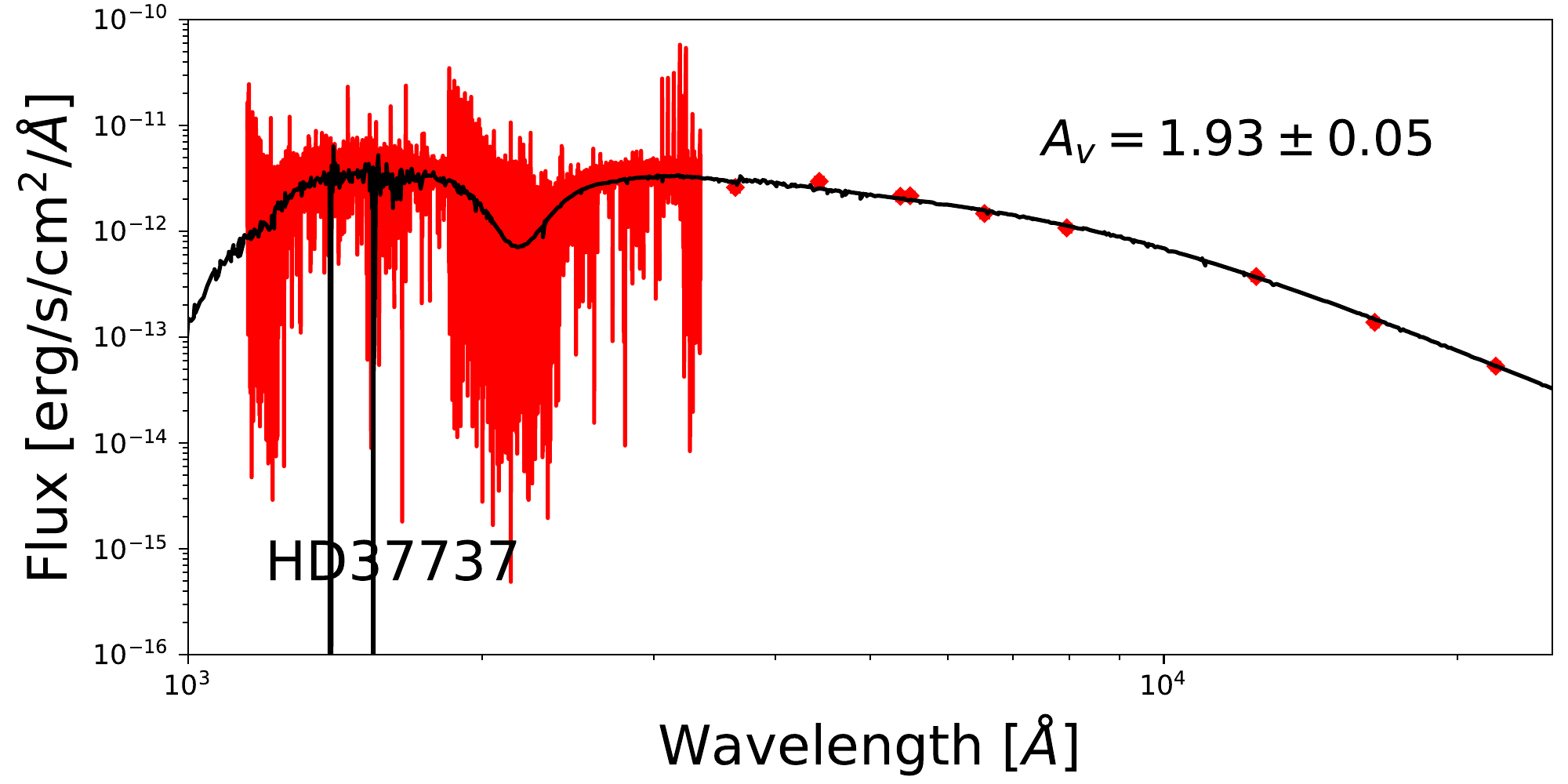} 
    \caption{HD~37737}
    \end{subfigure}
    \hfill
    \begin{subfigure}{0.46\linewidth}
    \includegraphics[width = \textwidth, trim=0 0 0 0,clip]{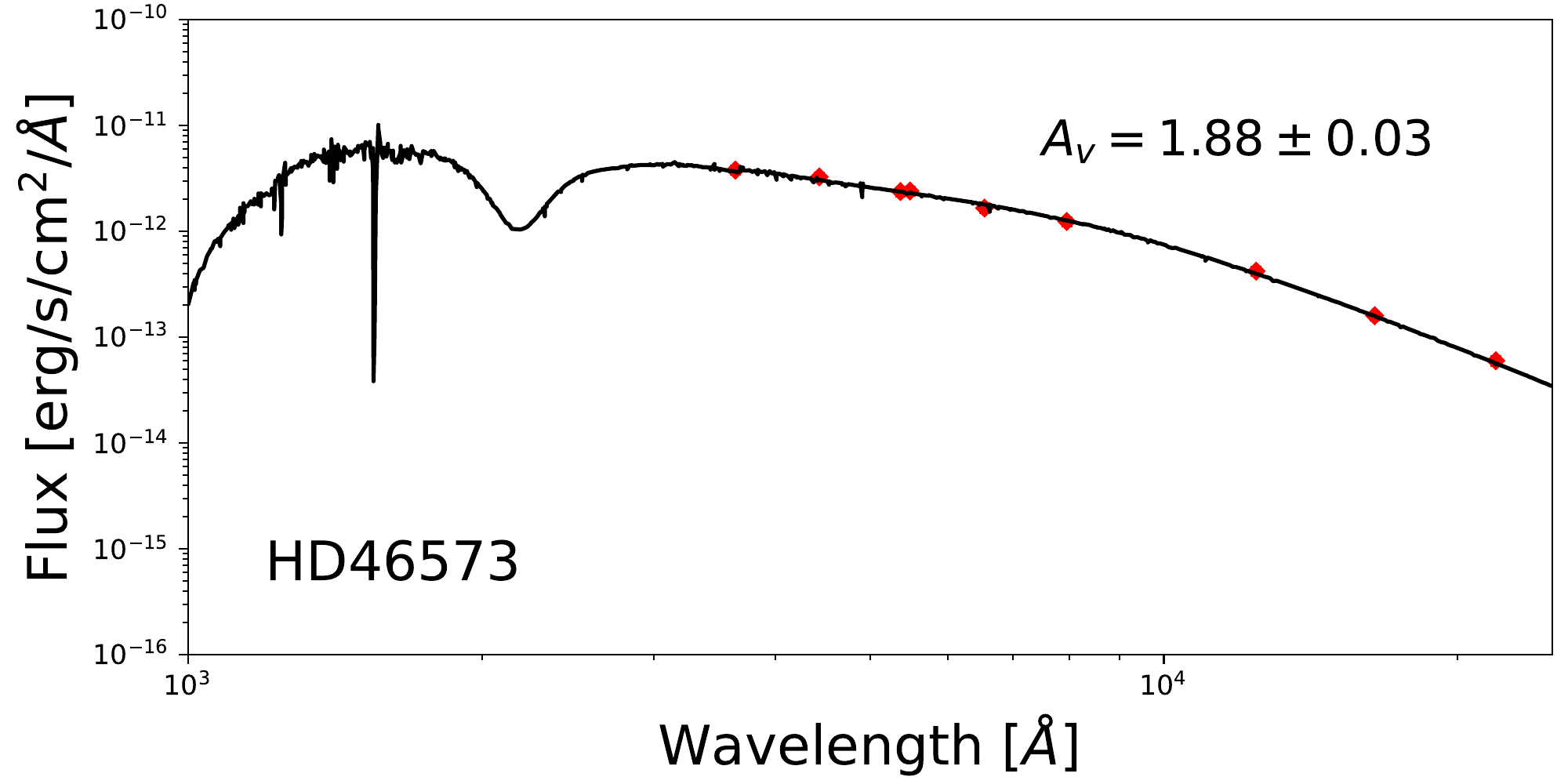}  \caption{HD~46573}
    \end{subfigure}
    \hfill
    \begin{subfigure}{0.46\linewidth}
    \includegraphics[width = \textwidth, trim=0 0 0 0,clip]{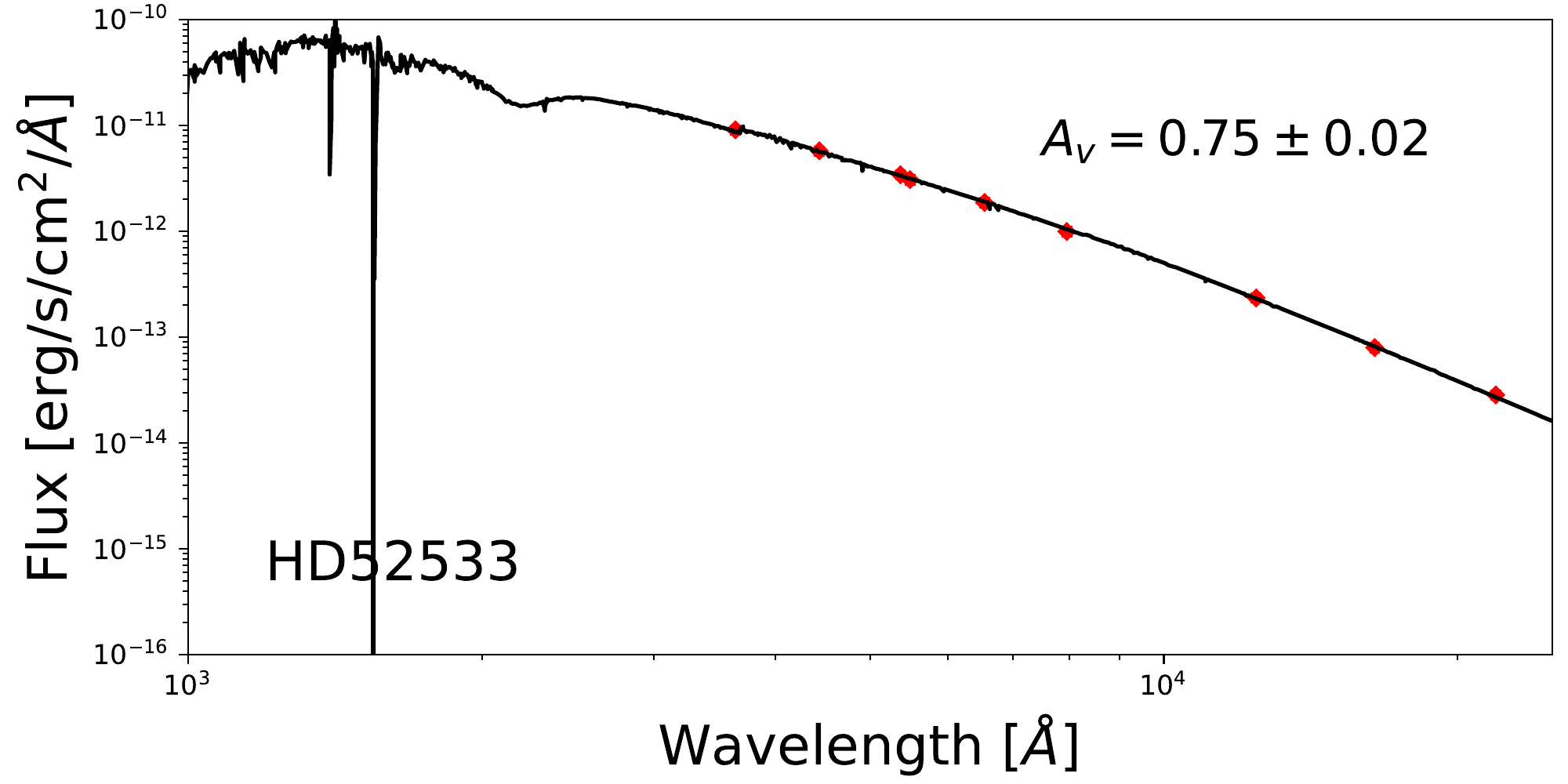}  \caption{HD~52533}
    \end{subfigure}
    \hfill
    \begin{subfigure}{0.46\linewidth}
    \includegraphics[width = \textwidth, trim=0 0 0 0,clip]{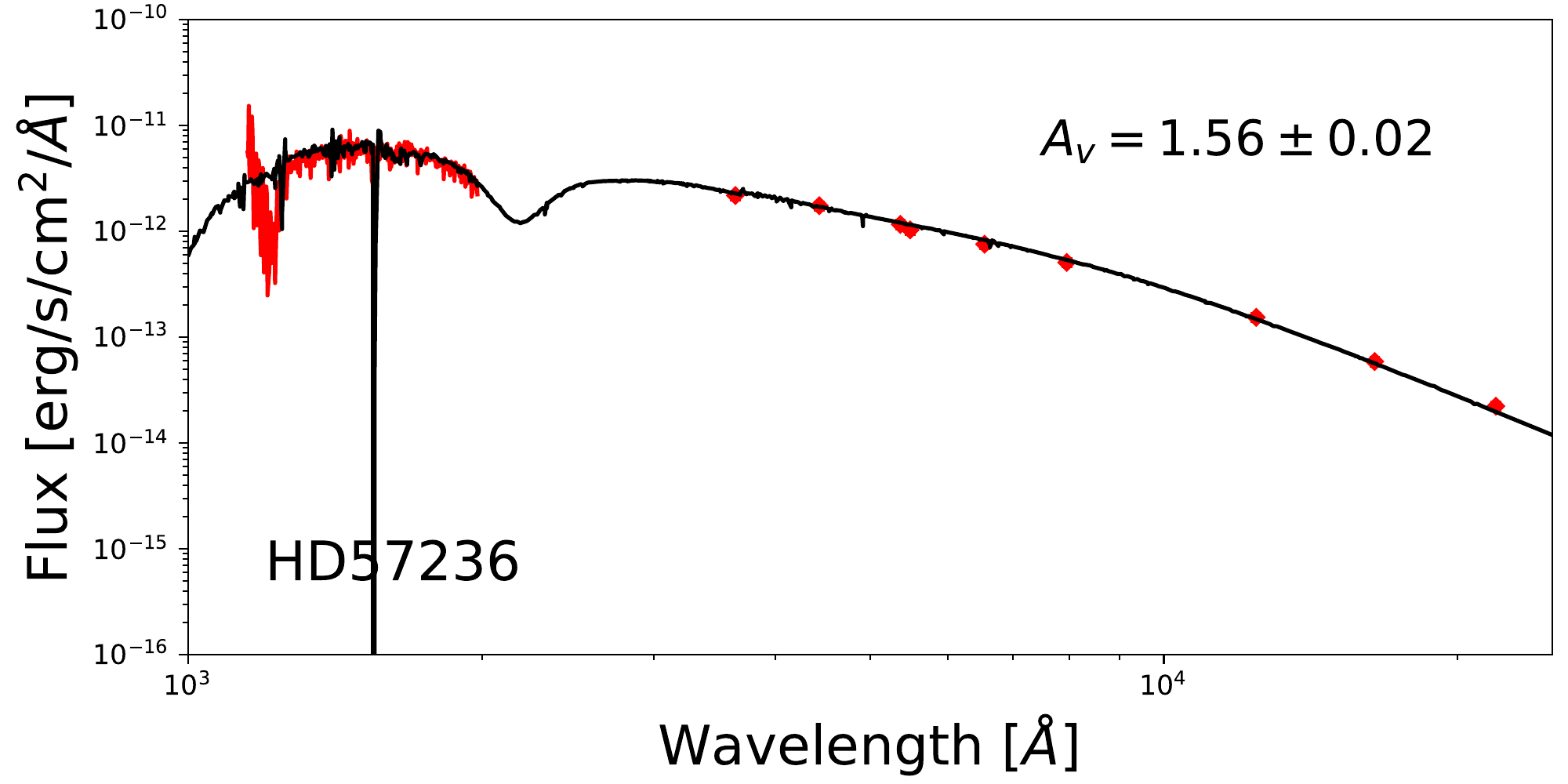}  \caption{HD~57236}
    \end{subfigure}
    \caption{Spectral Energy Distribution of all the systems in our sample }
    \label{fig:SED}
\end{figure*}
\FloatBarrier
\begin{figure*}[h!]
\ContinuedFloat 
    \centering
    \begin{subfigure}{0.46\linewidth}
    \includegraphics[width = \textwidth, trim=0 0 0 0,clip]{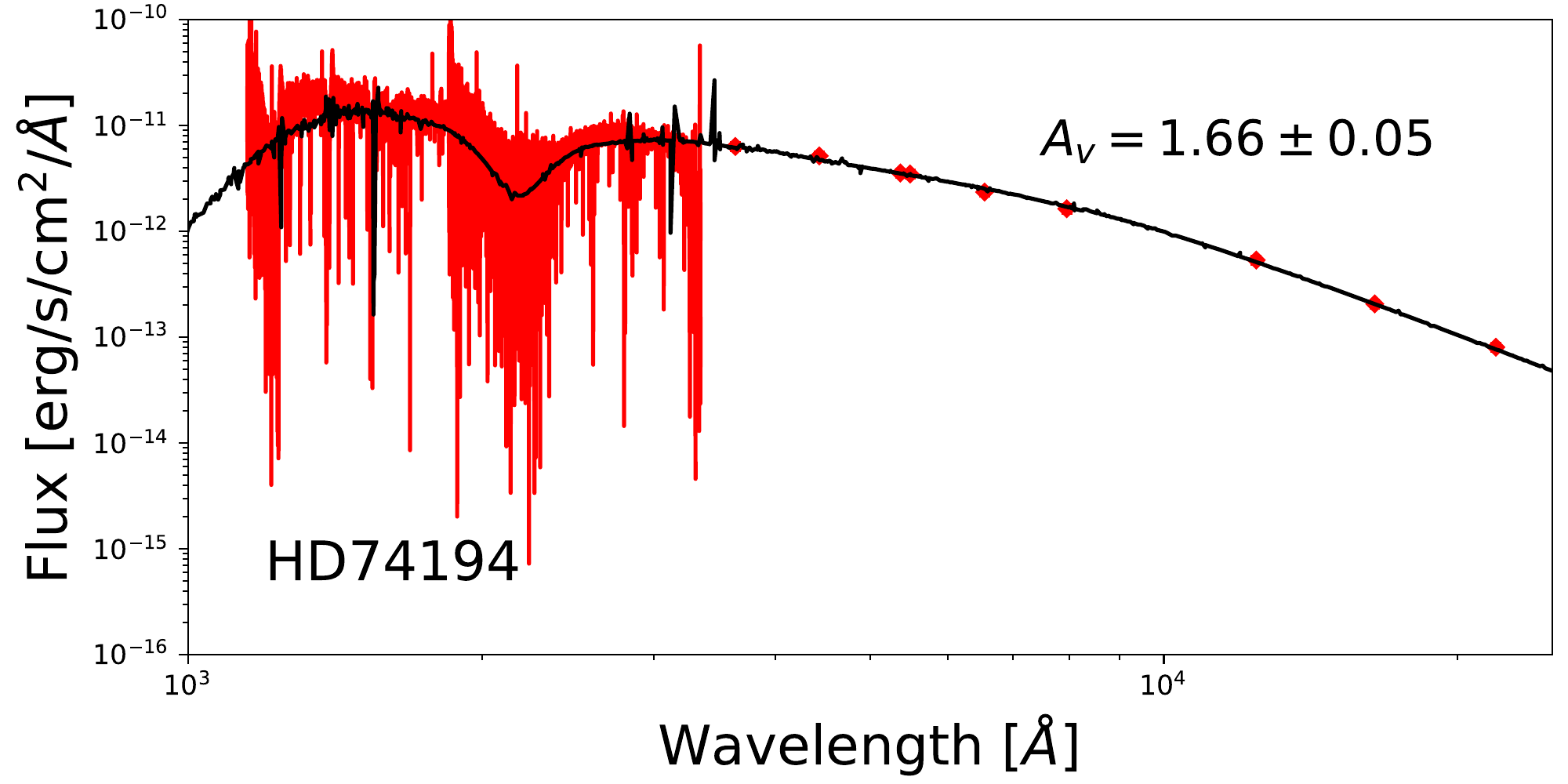}  \caption{HD~74194}
    \end{subfigure}
    \hfill
    \begin{subfigure}{0.46\linewidth}
    \includegraphics[width = \textwidth, trim=0 0 0 0,clip]{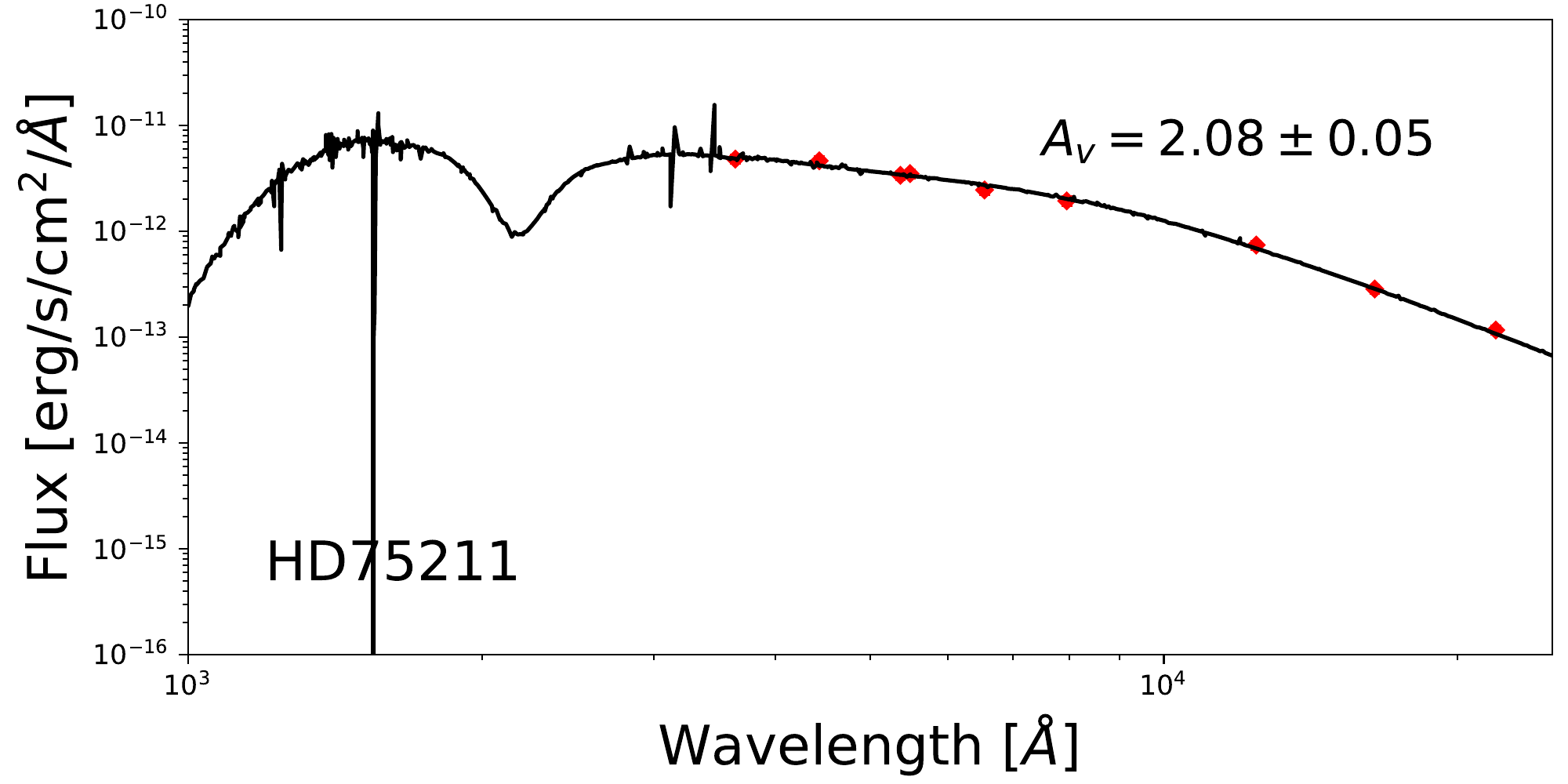}  \caption{HD~75211}
    \end{subfigure}
    \hfill
    \begin{subfigure}{0.46\linewidth}
    \includegraphics[width = \textwidth, trim=0 0 0 0,clip]{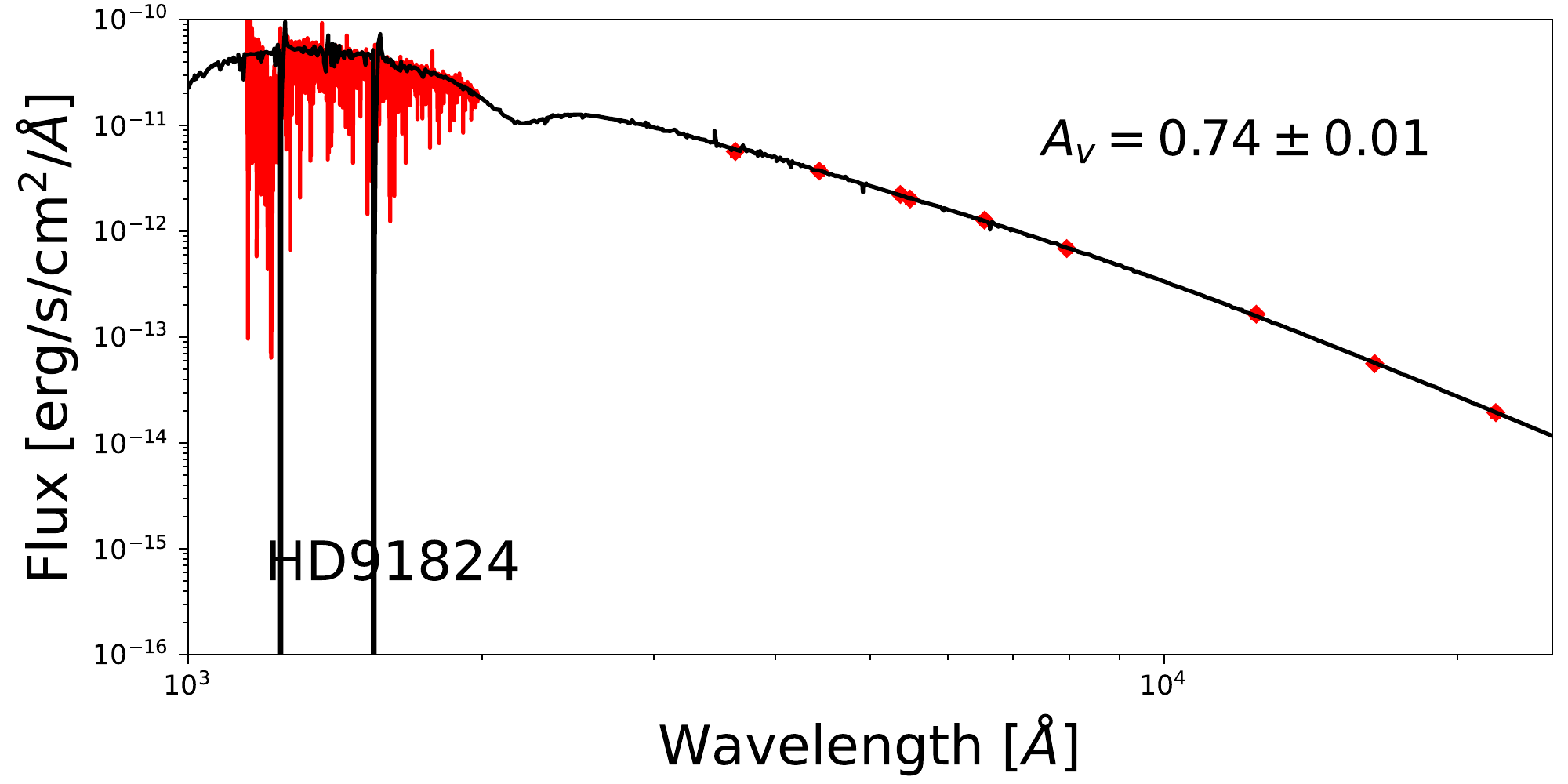} 
    \caption{HD~91824}
    \end{subfigure}
    \hfill
    \begin{subfigure}{0.46\linewidth}
    \includegraphics[width = \textwidth, trim=0 0 0 0,clip]{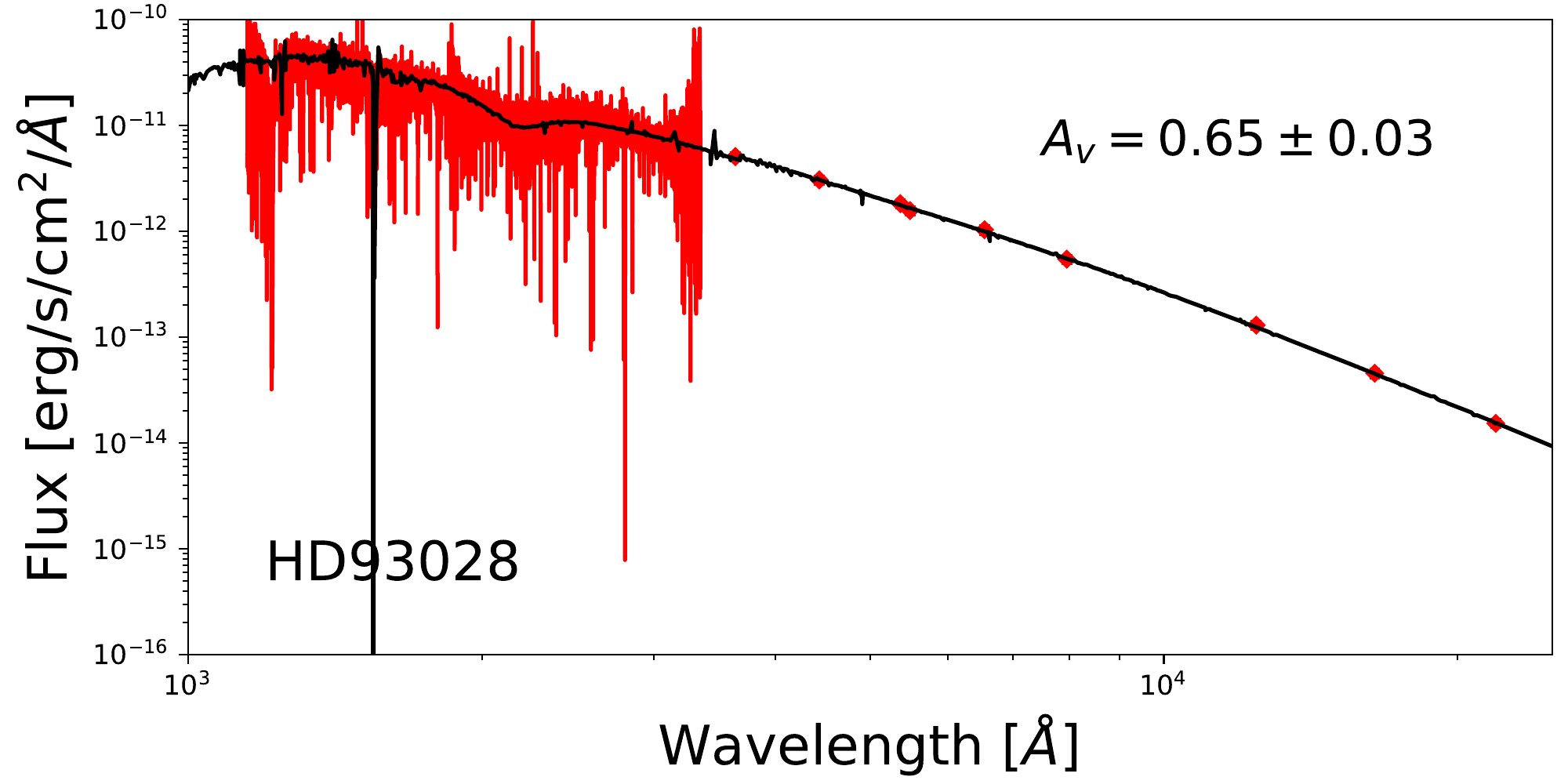} \caption{HD~93028}
    \end{subfigure}
    \hfill
    \begin{subfigure}{0.46\linewidth}
    \includegraphics[width = \textwidth, trim=0 0 0 0,clip]{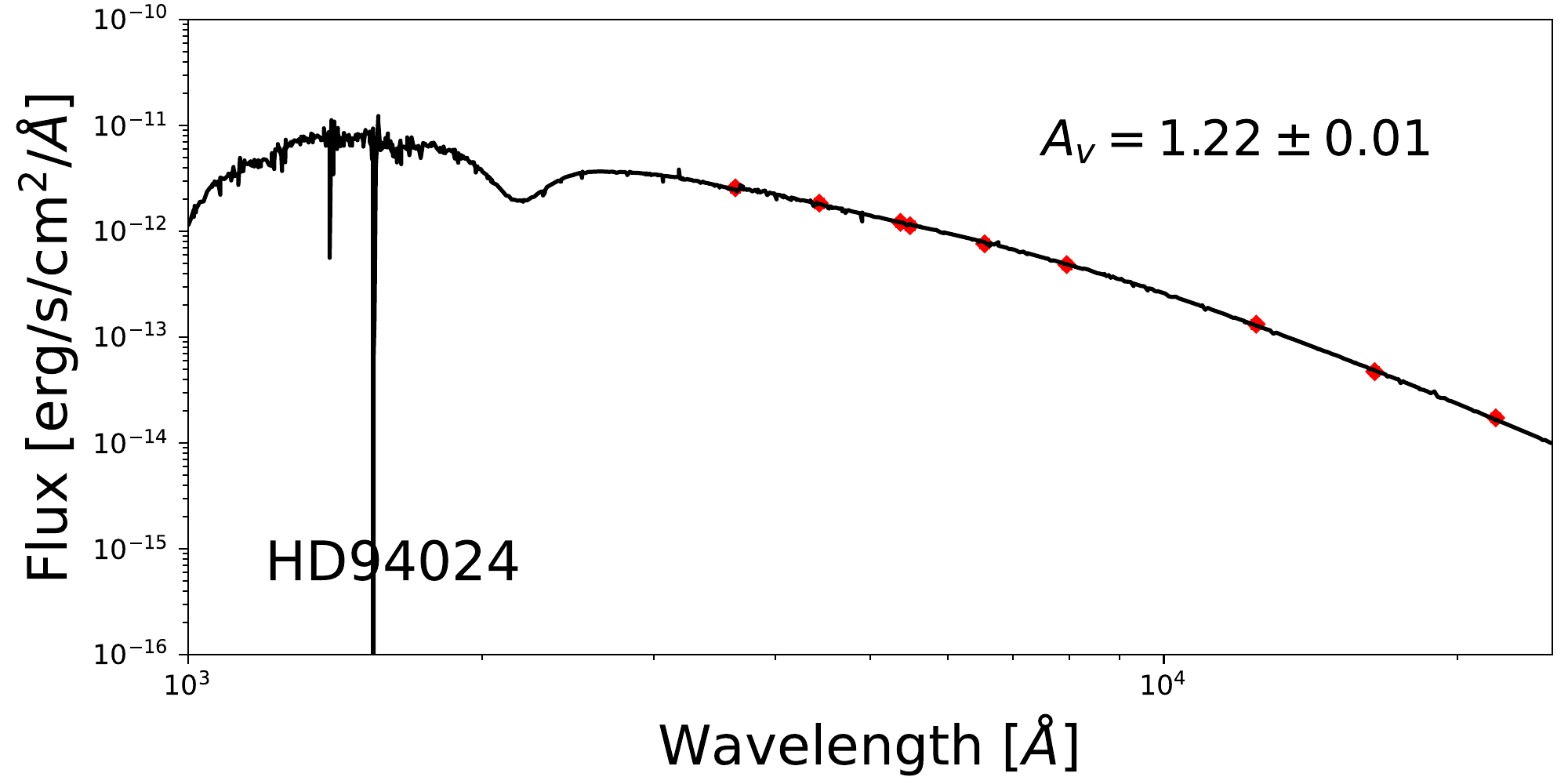}  \caption{HD~94024}
    \end{subfigure}
    \hfill
    \begin{subfigure}{0.46\linewidth}
    \includegraphics[width = \textwidth, trim=0 0 0 0,clip]{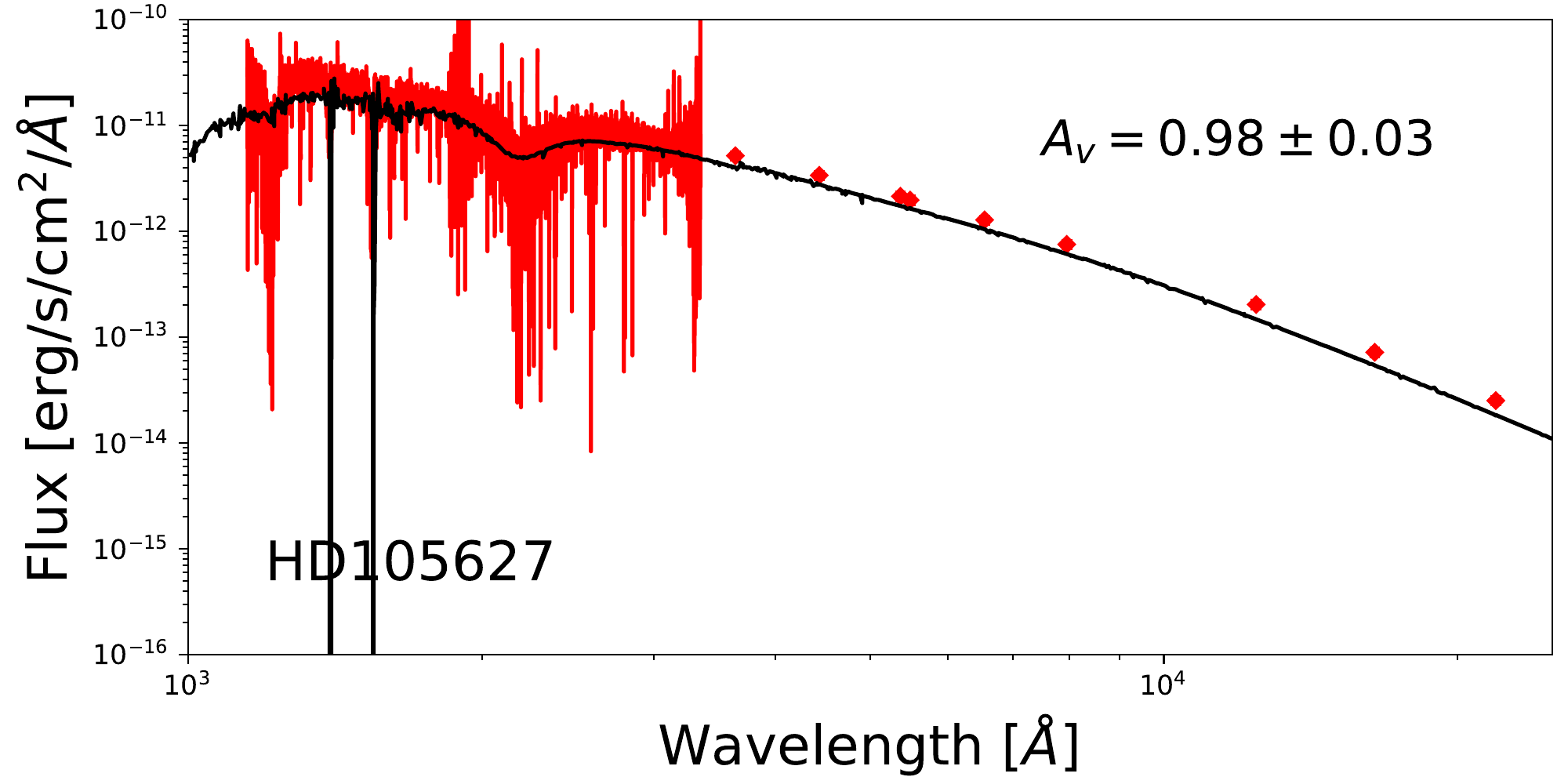} \caption{HD~105627}
    \end{subfigure}
    \hfill
    \begin{subfigure}{0.46\linewidth}
    \includegraphics[width = \textwidth, trim=0 0 0 0,clip]{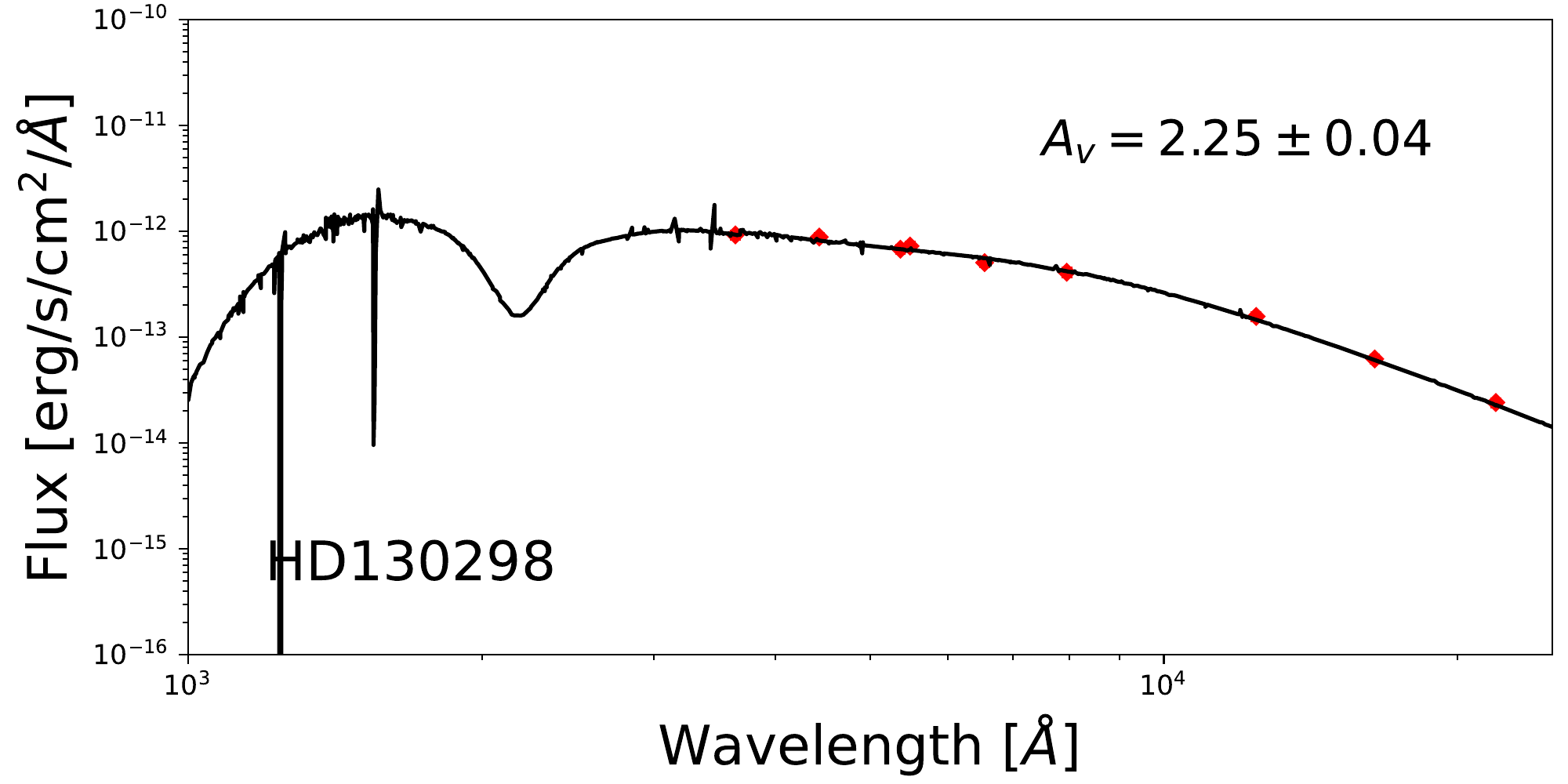}
    \caption{HD~130298}
    \end{subfigure}
    \hfill
    \begin{subfigure}{0.46\linewidth}
    \includegraphics[width = \textwidth, trim=0 0 0 0,clip]{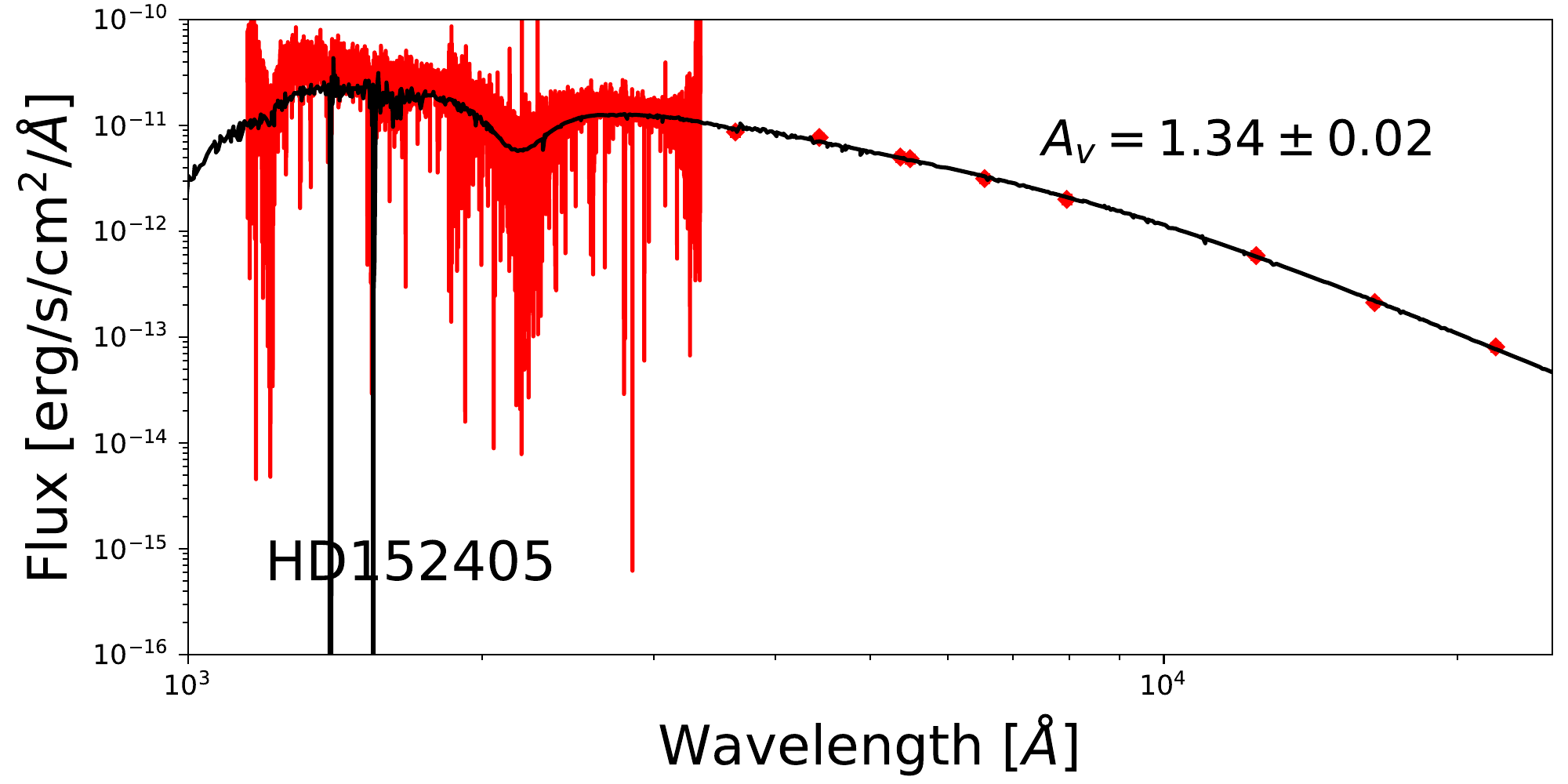}
    \caption{HD~152405}
    \end{subfigure}
    \hfill
    \begin{subfigure}{0.46\linewidth}
    \includegraphics[width = \textwidth, trim=0 0 0 0,clip]{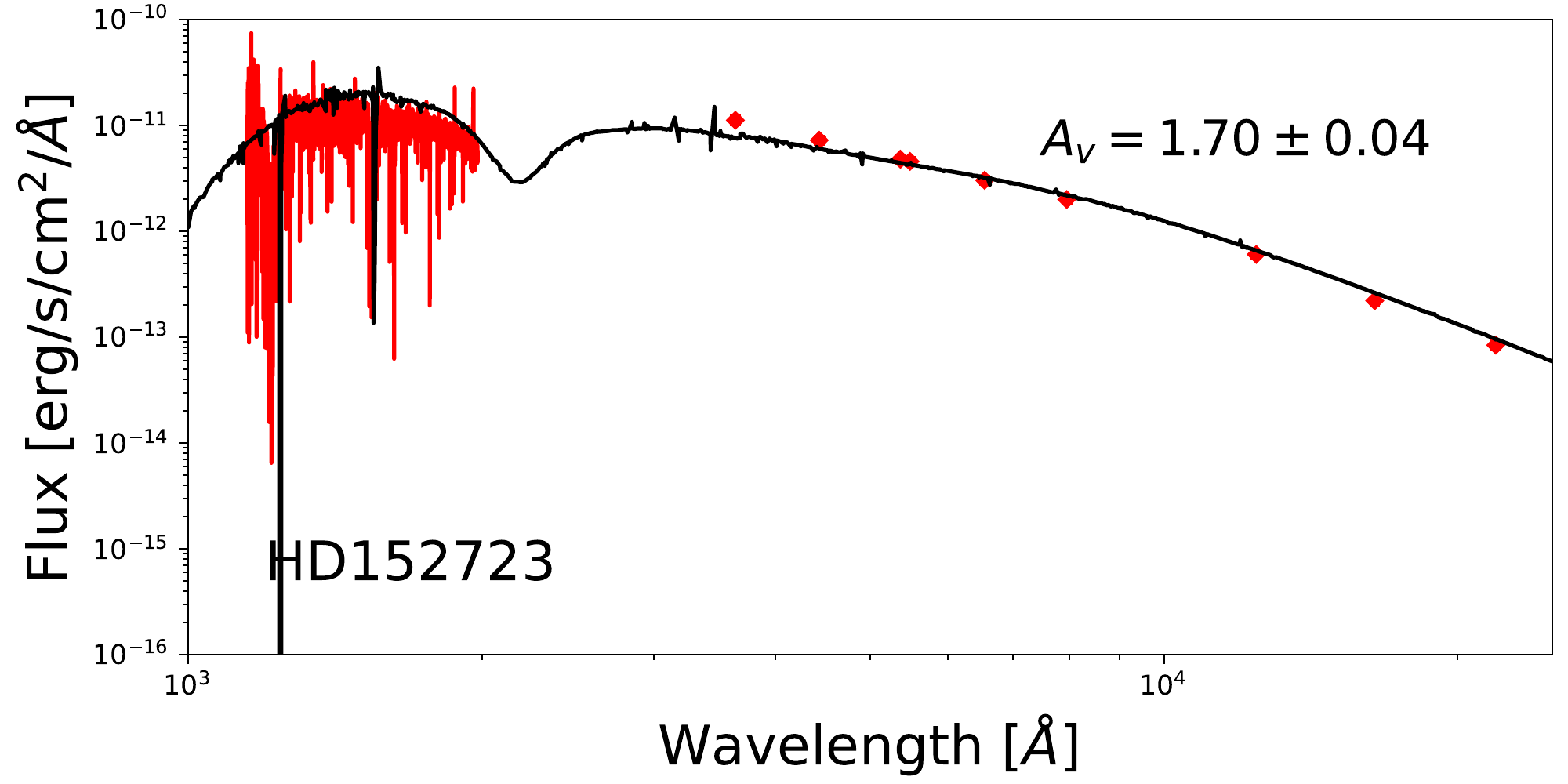}
    \caption{HD~152723}
    \end{subfigure}
    \hfill
    \begin{subfigure}{0.44\linewidth}
    \includegraphics[width = \textwidth, trim=0 0 0 0,clip]{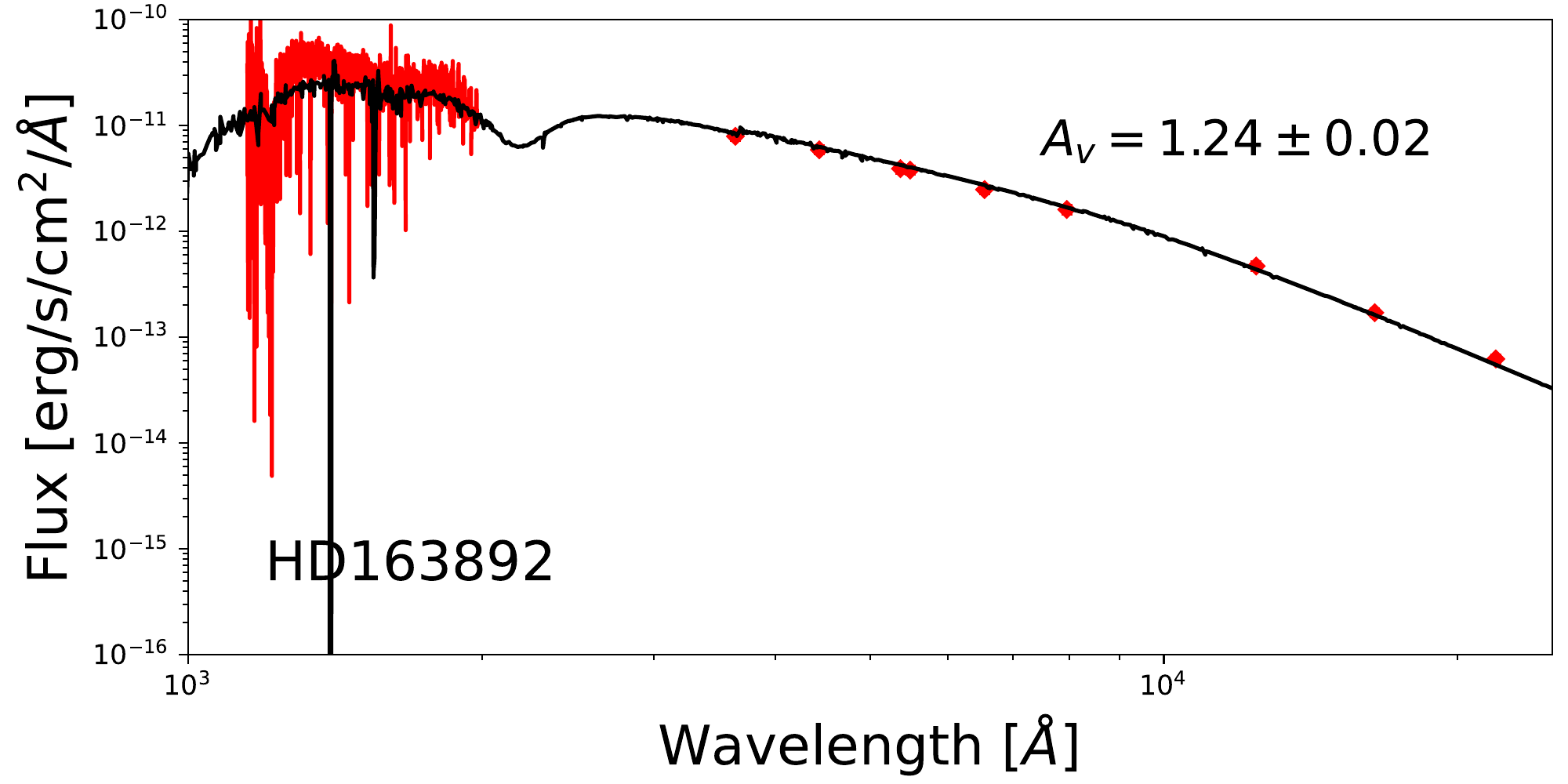}
    \caption{HD~163892}
    \end{subfigure}

    \caption{continued }
\end{figure*}

\begin{figure*}[h!]
\ContinuedFloat 
    \centering
    \begin{subfigure}{0.46\linewidth}
    \includegraphics[width = \textwidth, trim=0 0 0 0,clip]{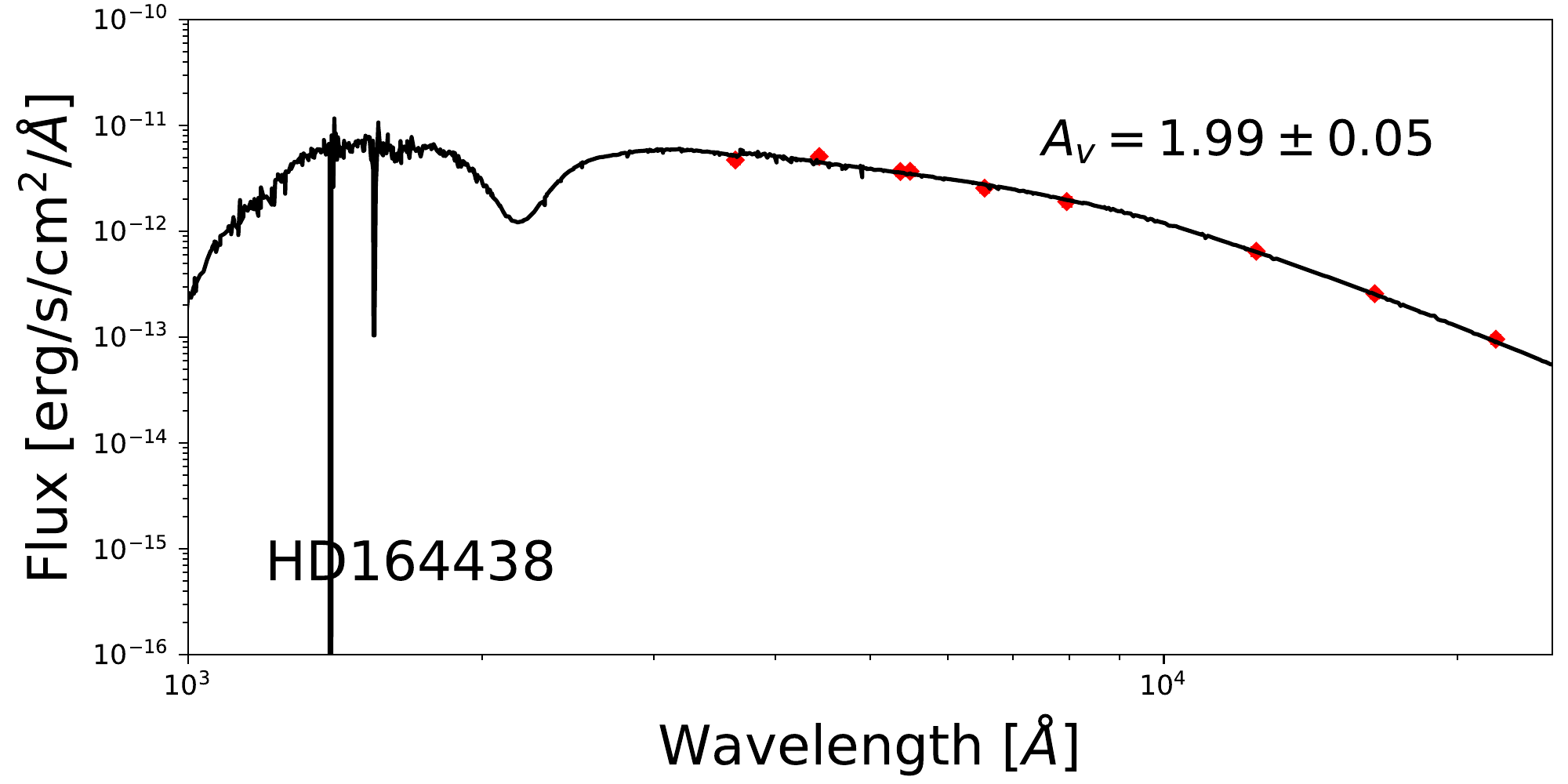} \caption{HD~164438}
    \end{subfigure}
    \hfill
    \begin{subfigure}{0.46\linewidth}
    \includegraphics[width = \textwidth, trim=0 0 0 0,clip]{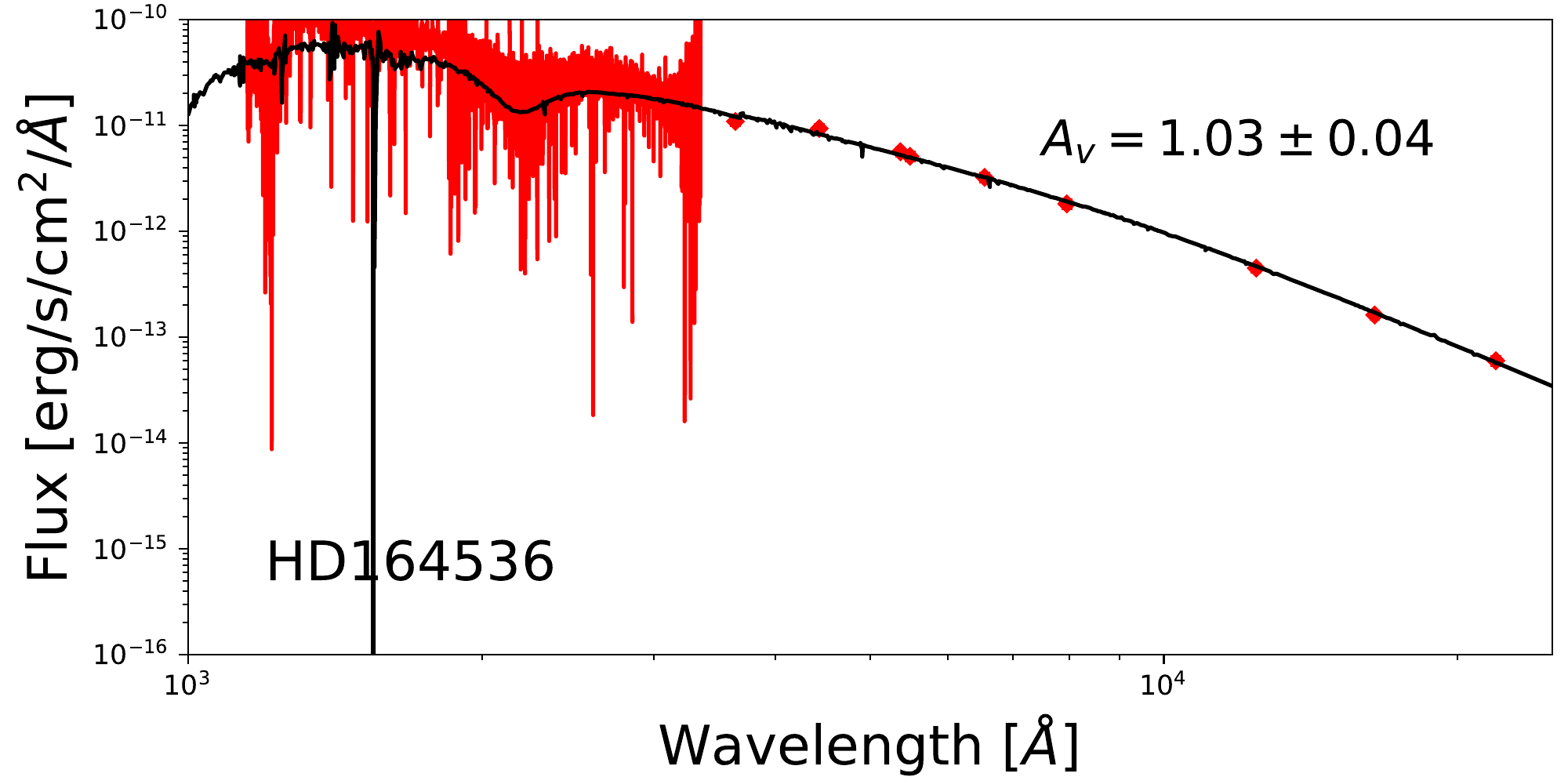}
    \caption{HD~164536}
    \end{subfigure}
    \hfill
    \begin{subfigure}{0.46\linewidth}
    \includegraphics[width = \textwidth, trim=0 0 0 0,clip]{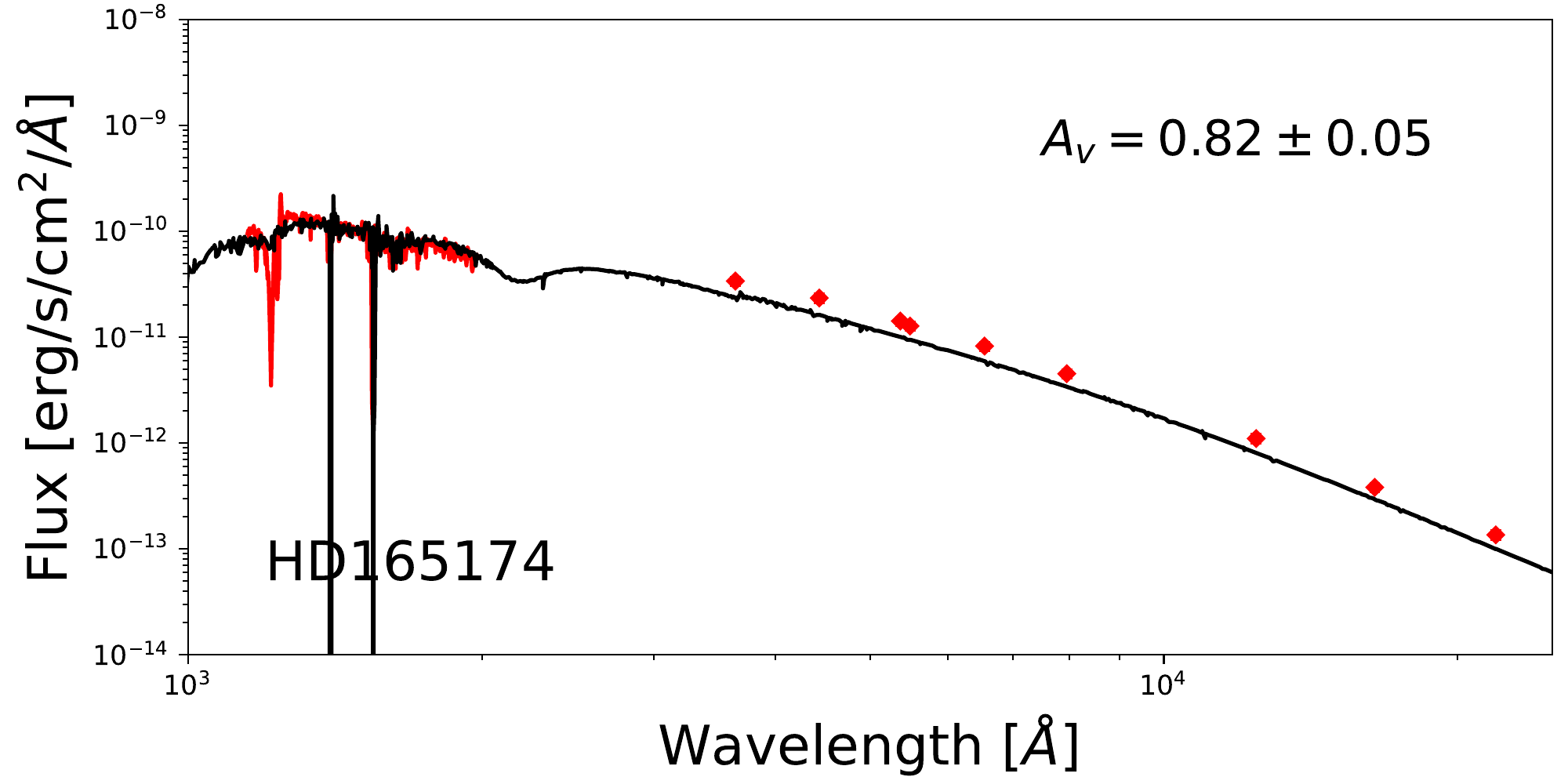} \caption{HD~165174}
    \end{subfigure}
    \hfill
    \begin{subfigure}{0.46\linewidth}
    \includegraphics[width = \textwidth, trim=0 0 0 0,clip]{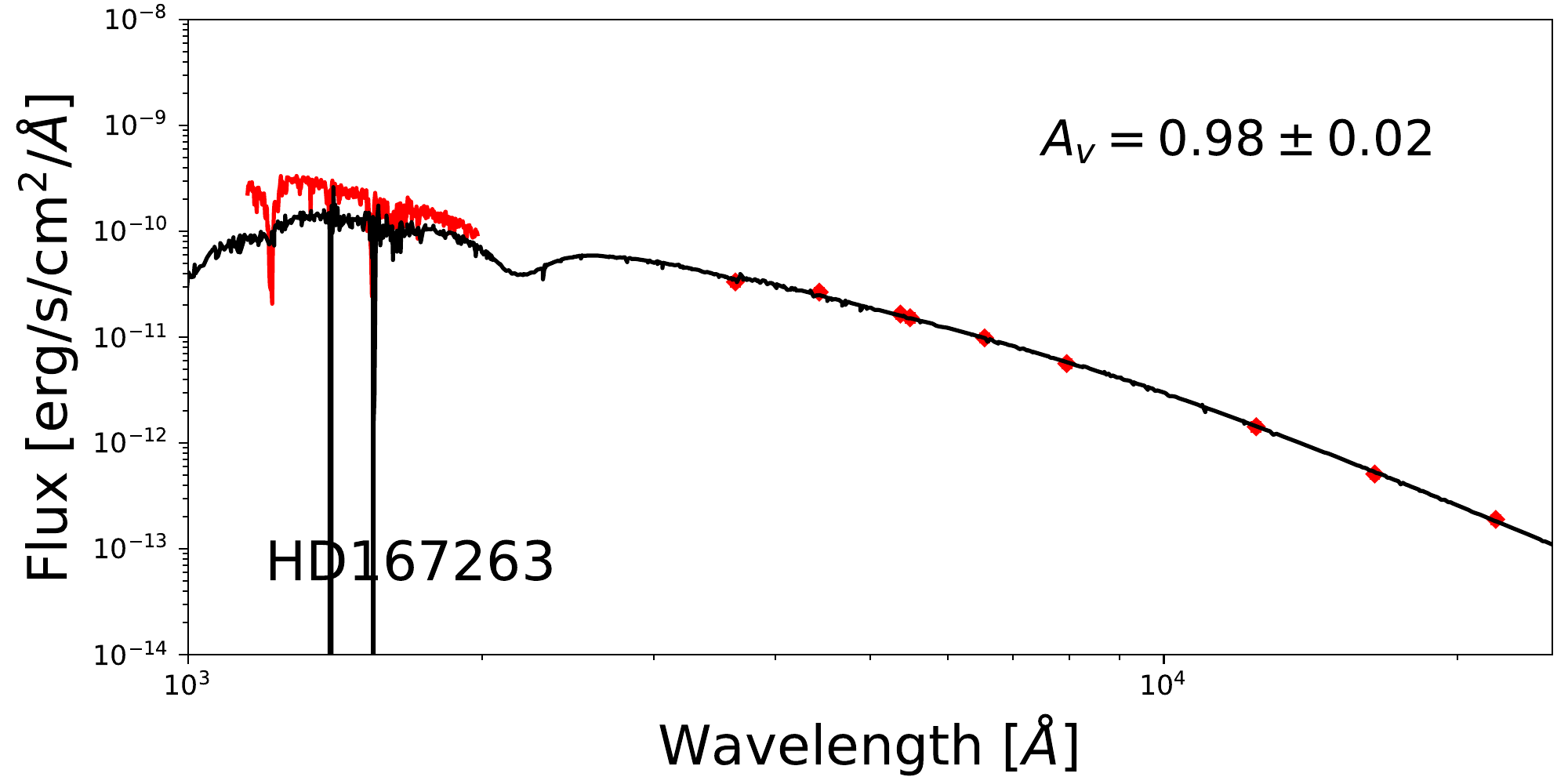}
    \caption{HD~167263}
    \end{subfigure}
    \hfill
    \begin{subfigure}{0.46\linewidth}
    \includegraphics[width = \textwidth, trim=0 0 0 0,clip]{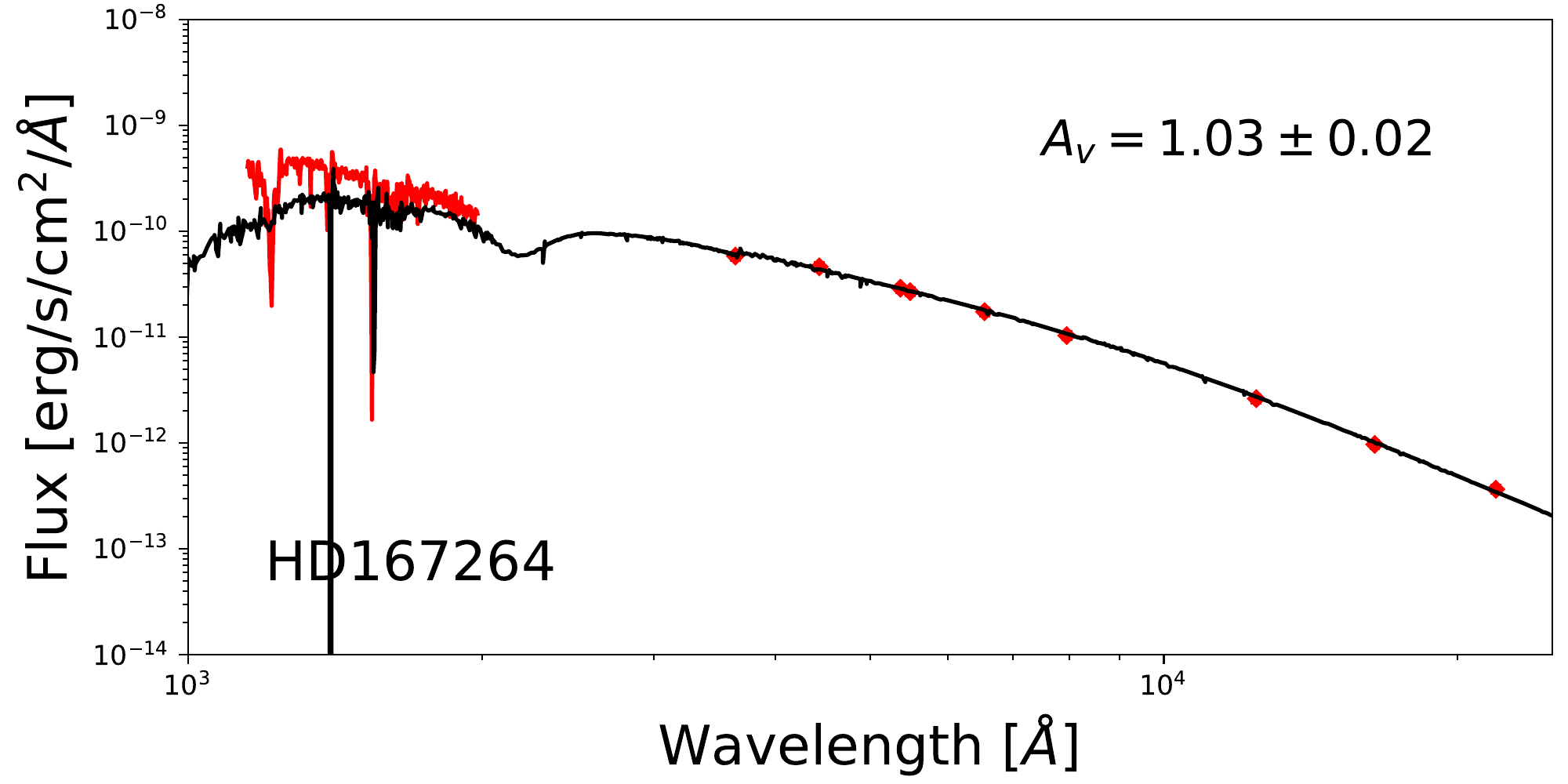} \caption{HD~167264}
    \end{subfigure}
    \hfill   
    \begin{subfigure}{0.46\linewidth}
    \includegraphics[width = \textwidth, trim=0 0 0 0,clip]{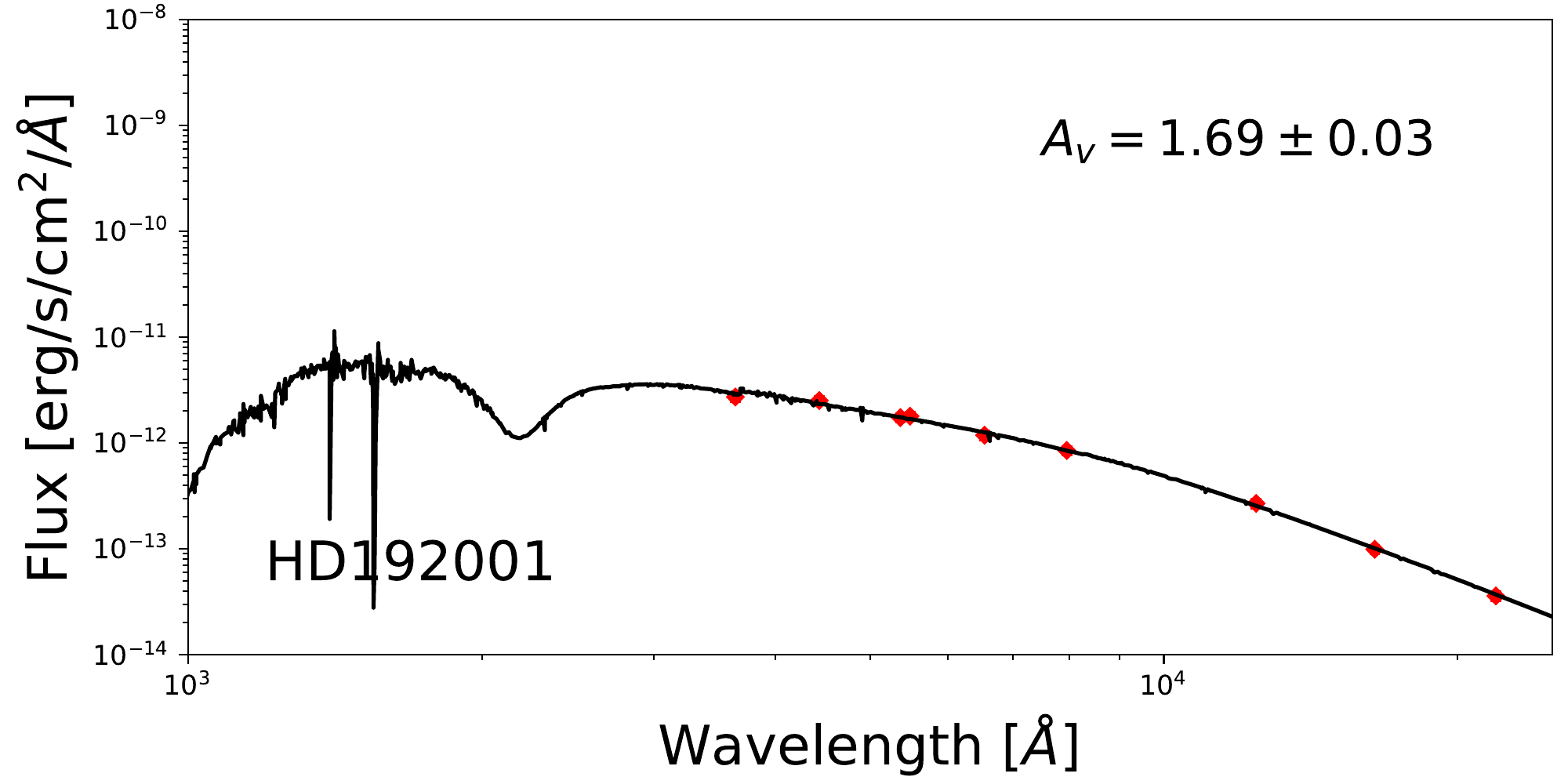} \caption{HD~192001}
    \end{subfigure}
    \hfill   
    \begin{subfigure}{0.46\linewidth}
    \includegraphics[width = \textwidth, trim=0 0 0 0,clip]{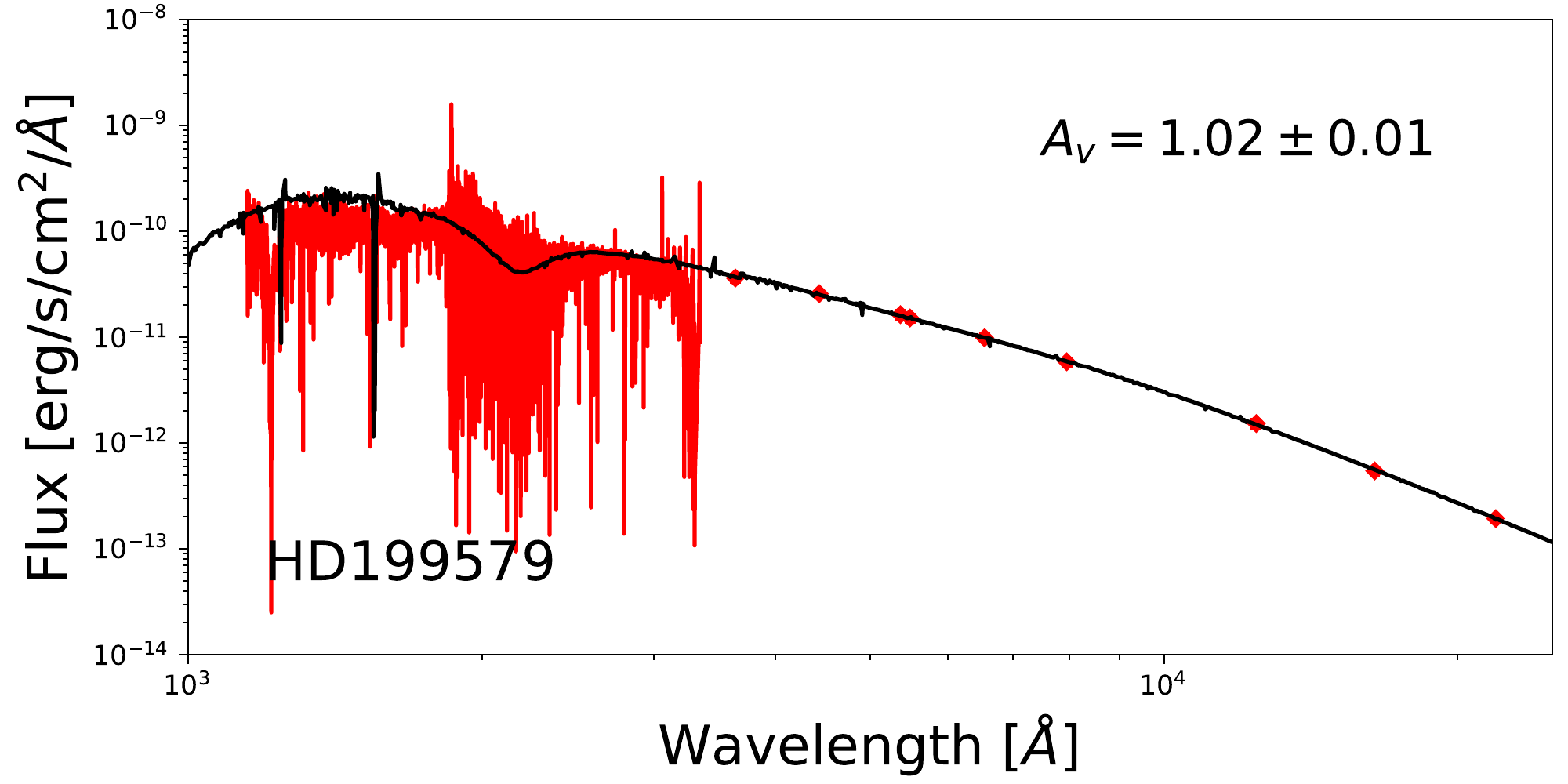} \caption{HD~199579}
    \end{subfigure}
    \hfill
    \begin{subfigure}{0.46\linewidth}
    \includegraphics[width = \textwidth, trim=0 0 0 0,clip]{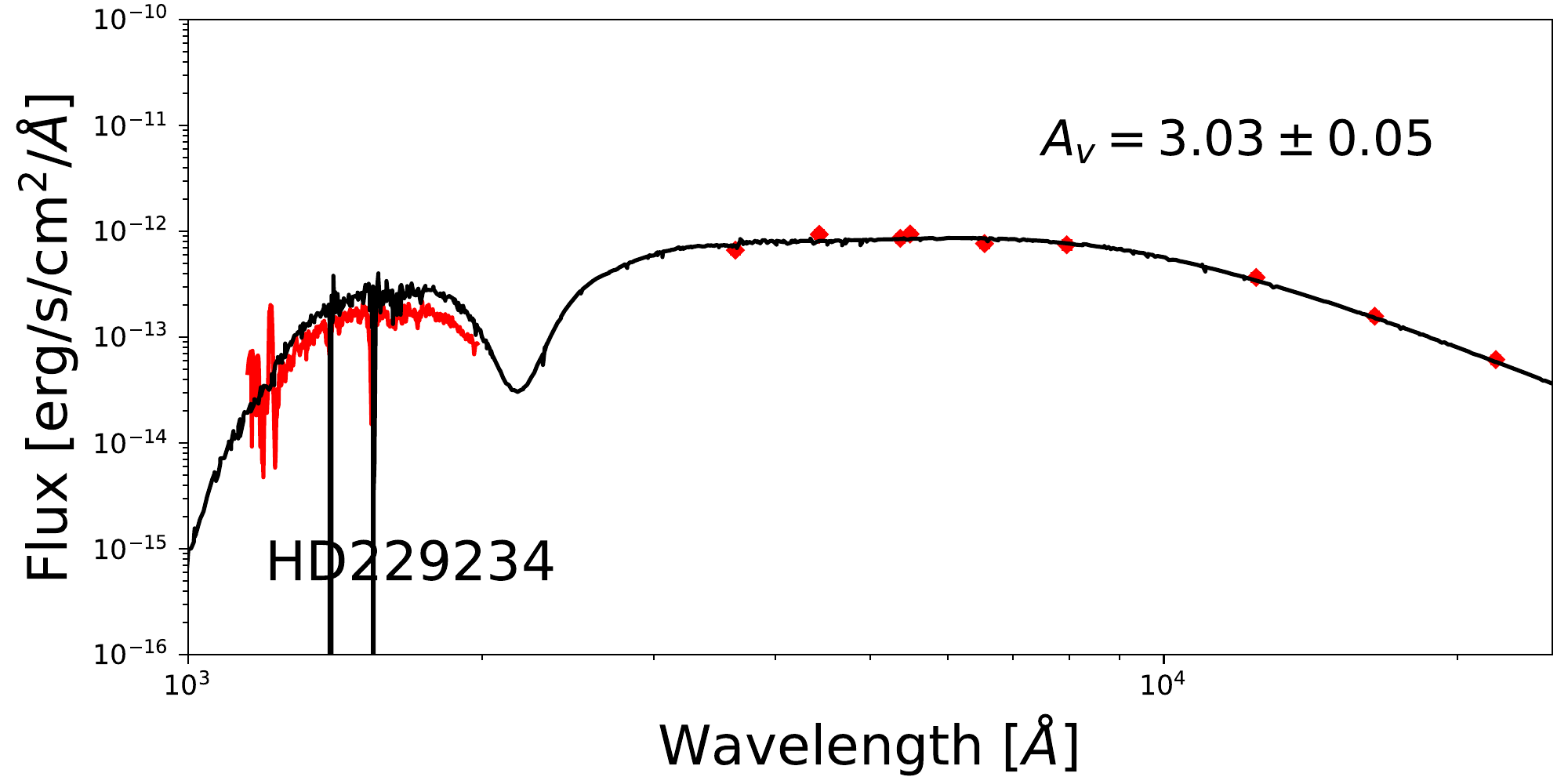} \caption{HD~229234}
    \end{subfigure}
    \hfill  
    \begin{subfigure}{0.46\linewidth}
    \includegraphics[width = \textwidth, trim=0 0 0 0,clip]{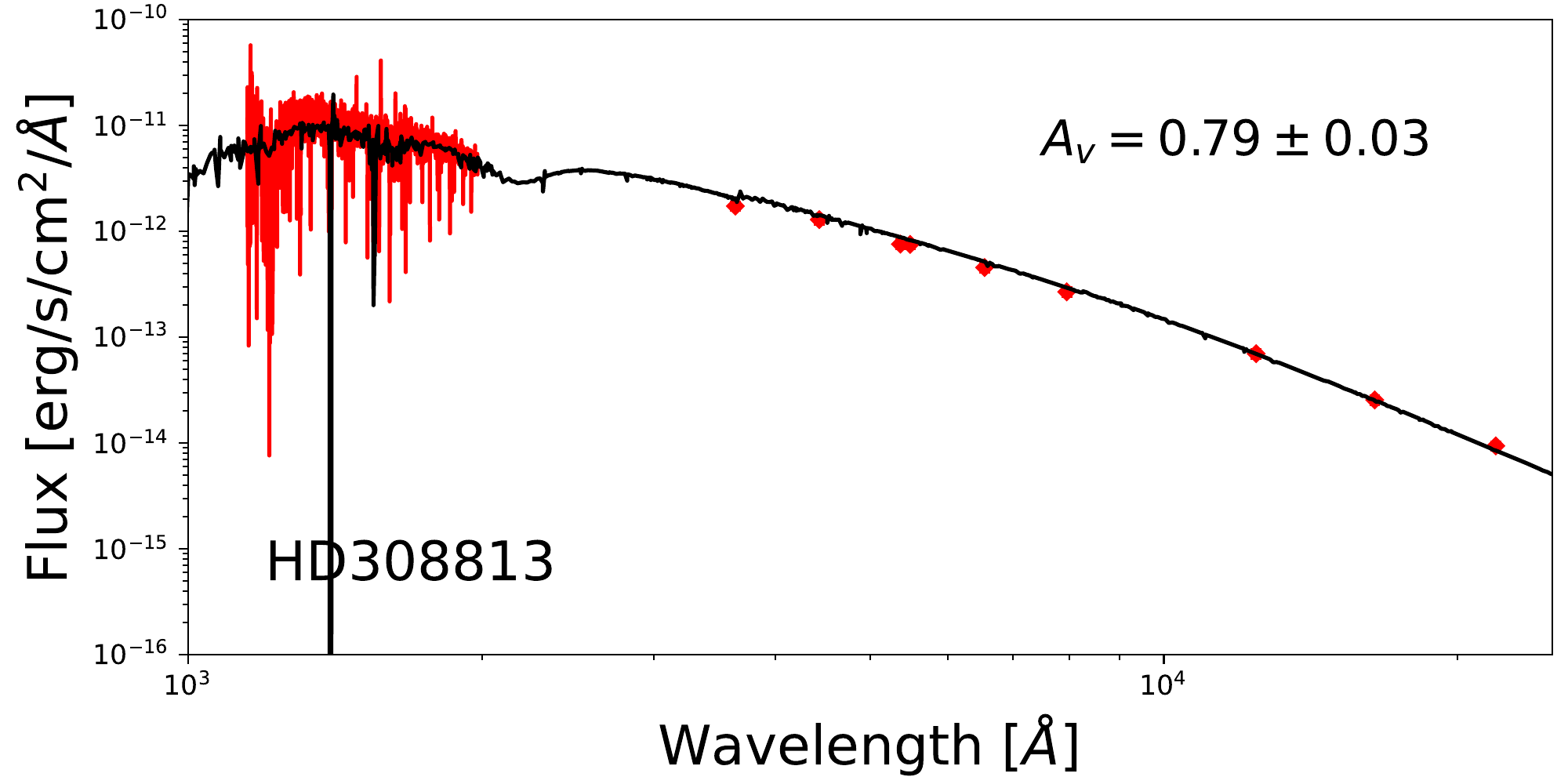} \caption{HD~308813}
    \end{subfigure}
    \hfill
    \begin{subfigure}{0.46\linewidth}
    \includegraphics[width = \textwidth, trim=0 0 0 0,clip]{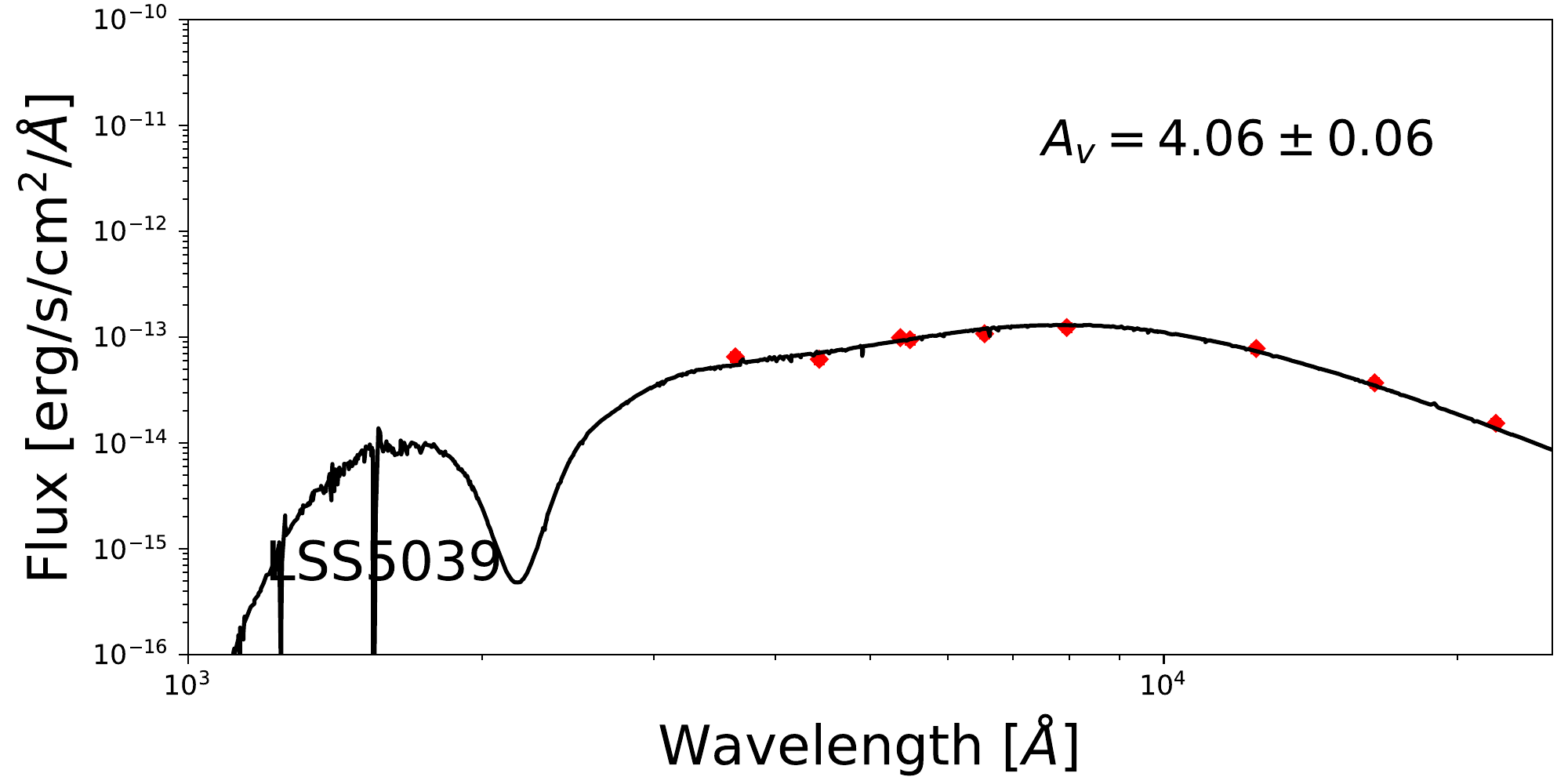}
    \caption{LS~5039}
    \end{subfigure}
    \caption{continued }
\end{figure*}
\begin{figure*}[h!]
\ContinuedFloat 
    \centering
    \begin{subfigure}{0.46\linewidth}
    \includegraphics[width = \textwidth, trim=0 0 0 0,clip]{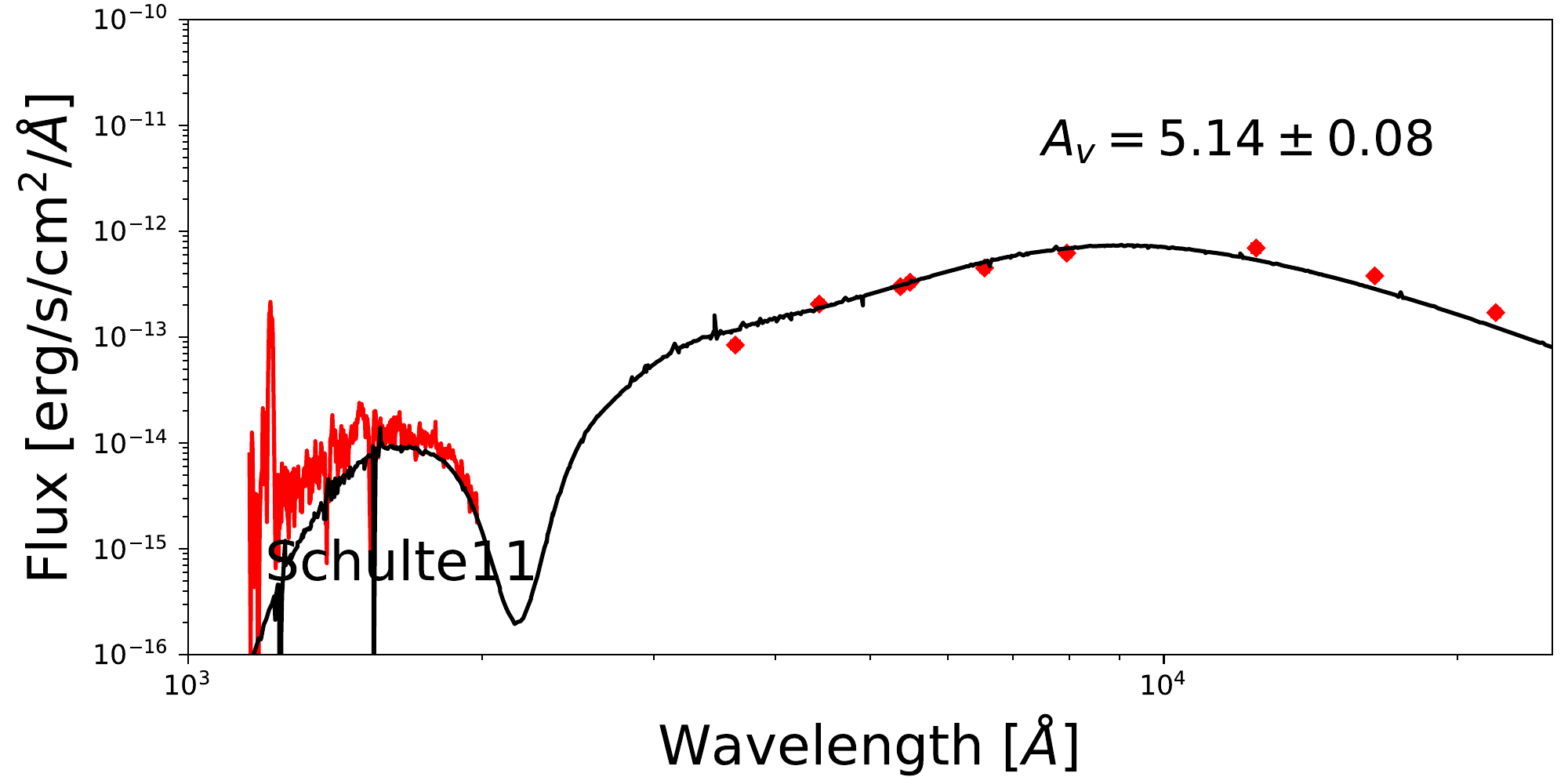} \caption{Schulte~11}
    \end{subfigure}
    \hfill
    \begin{subfigure}{0.46\linewidth}
    \includegraphics[width = \textwidth, trim=0 0 0 0,clip]{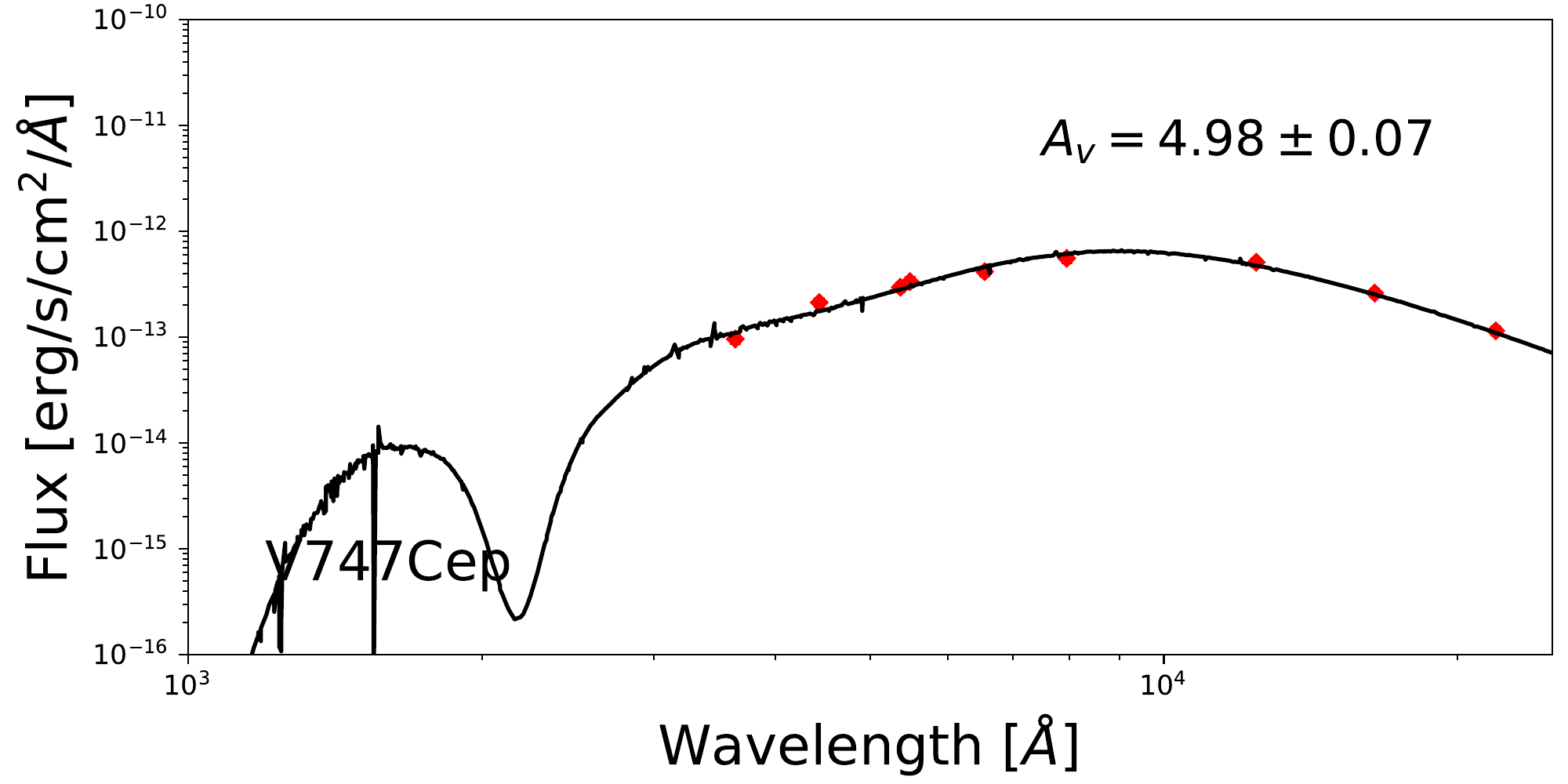} \caption{V747~Cep}
    \end{subfigure}
    \caption{continued }
  \end{figure*} 
\FloatBarrier

\section{Individual systems}
\label{sec:individual}
\subsection{SB1 systems}
\subsubsection{Cyg\,X-1}
Cyg~X-1 is already known as hosting an accreting BH, emitting X-rays. The orbital period is 5.6 days and the system has a low eccentricity of 0.02. The binary mass function is equal to $0.237 \pm 0.002\,\Msun$. For the visible star (classified as O9.7\,I), we derived an effective temperature of 29.8kK and a $\logg$ of 3.33 [cgs]. From the Gaia eDR3 distance and an extinction of $A_V = 3.2 \pm 0.1$, we compute a luminosity of $\log(L/\Lsun) = 5.48\pm0.06$, which gives a radius of $R = 20.7 \pm 1.2\,\Rsun$, resulting in a spectroscopic mass of $33.4_{-8.0}^{+8.0},\Msun$ for the visible star. Using BONNSAI, the predicted parameters for that star give an evolutionary masses of $30.0_{-3.1}^{+4.0}\,\Msun$, so that no mass discrepancy is observed within the error bars. These masses agree well with that measured by \citet{Miller-Jones21}. The spectral disentangling gave us a flat spectrum for the stellar-mass BH which is what is expected for that system. A $7~\Msun{}$ star or higher would have been detected from our data with spectral disentangling. From the TESS light curve (Fig.~\ref{fig:lightcurve}), we extracted the frequencies of the five highest peaks in the periodogram ($\nu_1 = 0.358(2)$, $\nu_1 = 0.150(5)$, $\nu_1 = 0.242(5)$, $\nu_1 = 0.424(5)$, and $\nu_1 = 0.109(6)$ d$^{-1}$). The first frequency corresponds to half the orbital period of the system. The light curve variation is similar to ellipsoidal variations, which is due to the deformation of the visible star (filling its Roche lobe and transferring its mass and angular momentum to the stellar-mass BH companion). The other signals, if one of them is due to the rotation of the visible star, provide us with possible inclinations between 13 and $40^{\circ}$. We note that an inclination of $27.5^{\circ}$ was reported by \citet{Miller-Jones21} but, based on our parameters, no peak related to that inclination is found in the periodogram. This range provides us with a mass estimate between 12 and $60~\Msun{}$ for the compact object.

\subsubsection{HD~12323}
HD~12323 is a short period ($P_{\rm orb}=1.9$ days) circular SB1 system. This system is a runaway \citep{maiz18}. The spectral disentangling does not allow us to extract the spectrum of the secondary component. From the orbital parameters, we measured a binary mass function equal to $0.0054 \pm 0.0008\,\Msun$. The visible star is nitrogen rich and was classified as an ON9.2V. We derived an effective temperature to be equal to 33.2kK and a surface gravity of 3.99 [cgs], once corrected for the centrifugal force. These stellar parameters agree very well with those given by \citet{martins15b}. Using the Gaia eDR3 and an extinction of $0.86 \pm 0.03$, we computed a luminosity of $\log(L/\Lsun) = 4.70 \pm 0.07$ and a radius of $R = 6.8 \pm 0.6\,\Rsun$. This gives a spectroscopic mass equal to $17.1 \pm 3.8\,\Msun$ and an evolutionary mass of $19.2_{-0.9}^{+1.0}\,\Msun$, showing no mass discrepancy. We estimated the mass of the secondary to be between 1.3 and $8\,\Msun$. The TESS light curve shows a clear oscillation with a frequency at $\nu= 1.039(1)$~d$^{-1}$, corresponding to a period of 0.962 days. This period corresponds to half the orbital period, and therefore suggests ellipsoidal variations, due to the deformation of the stars. If the star is in synchronous rotation with the system, the expected inclination is about $42^{\circ}$ which suggests a mass of about $2~\Msun{}$ for the secondary. In any cases, the spectral disentangling would have detected a MS companion down to a mass of $4~\Msun{}$. 

\subsubsection{HD\,14633}
HD\,14633 is a system with an 15.4-day orbital period and a high eccentricity of $0.698$. The orbital parameters derived in our analysis are similar to those from \citet{trigueros21}. The system is a runaway, and the presence of a bow shock was reported by \citet{peri12}. HD~14633 has been cited by \citet{mcswain07, mcswain10} as a potential system hosting a NS. The binary mass function is equal to $0.0041 \pm 0.0002\,\Msun$. The visible component is classified as an ON8.5\,V. The best-fit CMFGEN model gives an effective temperature of 33.9kK and a surface gravity of 3.93 [cgs]. From the Gaia distance and an extinction of $A_V = 0.32 \pm 0.04$, we computed a luminosity of $\log(L/\Lsun) = 4.60 \pm 0.10$ and a radius of $R = 5.8 \pm 0.7\,\Rsun$ for the visible star. Its spectroscopic mass is estimated to $10.6 \pm 3.6\,\Msun$ and the evolutionary mass to $19.0_{-1.1}^{+1.2}\,\Msun$. There is a clear mass discrepancy for this object, within the error bars. The companion is expected to have a mass between 1 and $7\,\Msun$, whether one considers the spectroscopic or the evolutionary mass estimate. The spectral disentangling does not allow us to extract the spectral signature of the secondary companion. From the simulations, we can, however, rule out the presence of a secondary earlier than B7~V. The TESS light curve shows stochastic variability and no significant frequency was detected. The companion therefore is probably an A or late-B-type star or a NS. 

\subsubsection{HD~15137}
HD~15137 was reported as a runaway SB1 system that might contain a NS or a low-mass BH by \citet{mcswain10}. The system is a runaway that was suggested by \citet{boyajian05} to have been ejected from NGC~654 through a supernova. We found an orbital period of 55.3 days and an eccentricity of	$e = 0.66$, confirming the orbital parameters derived by \citet{mcswain10}, and independently by \citet{trigueros21}. The spectral disentangling does not allow us to extract the secondary spectrum. From the orbital parameters, we calculated a binary mass function equals to $0.0092 \pm 0.0029\,\Msun$. The visible star is classified as O9.5~II-IIIn \citep{sota11}. The effective temperature is estimated to be 30.5kK and a surface gravity of 3.53 [cgs], once corrected for the centrifugal force. Using the Gaia eDR3 and an extinction of $1.08 \pm 0.02$, we computed a luminosity of $\log(L/\Lsun) = 4.97 \pm 0.08$ and a radius of $R = 10.9 \pm 1.0\,\Rsun$. This gives a spectroscopic mass equal to $14.9 \pm 3.3\,\Msun$ and an evolutionary mass of $22.2_{-1.8}^{+1.1}\,\Msun$, showing a clear mass discrepancy. We estimated the mass of the secondary to be between 1.5 and $3.0\,\Msun$, suggesting an F- or A-type star or a NS as spectral classification. The TESS light curve shows stochastic variability. However, \citet{trigueros21} suggested that the frequency at $\nu = 0.339$~d$^{-1}$ (i.e. giving a period of 2.95~days) might be due to rotation. Using this period with the stellar properties that we derived for the visible star of HD~15137, we calculated a possible inclination of $45^{\circ}$. Assuming that the rotational axis of the star is perpendicular to the orbital plane, we obtained a secondary mass of $2.5\,\Msun$. From the simulations, we would have detected the presence of a secondary more massive than $3~\Msun{}$. 

\subsubsection{HD\,46573}
HD\,46573 is a SB1 system detected for the first time by \citet{mahy09} and reported as runaway. This system has a period of 10.6 days and an eccentricity of 0.60. The binary mass function is computed to be $0.0008 \pm 0.0001\,\Msun$. The visible star is classified as O7\,V((f))z. It has an effective temperature of 35.3kK and a surface gravity of 3.85 [cgs]. These stellar parameters agree with those given by \citet{martins12} within the error bars. From the Gaia eDR3 and an extinction of $A_V = 1.88 \pm 0.03$, we compute a luminosity $\log(L/\Lsun) = 5.01 \pm 0.04$ and a radius of $R = 8.6 \pm 0.3\,\Rsun$. The spectroscopic mass is $18.9 \pm 4.0\,\Msun$ and the evolutionary mass is $24.0_{-1.1}^{+1.2}\,\Msun$. We do not detect any mass discrepancies. The spectral disentangling does not allow us to extract the spectral signature of the secondary from the composite spectra. We estimated the mass of the secondary to be between 0.7 and $7\,\Msun$, suggesting non-degenerate stars up to B5 on the MS or a compact object. The TESS light curve shows stochastic variability. No frequencies are found to be significant. The pseudo-synchronisation at the periastron is reached if the inclination is close to $30^{\circ}$, suggesting a secondary mass of about $1.7~\Msun{}$. The spectral disentangling would have detected a secondary star more massive than $3-5\Msun{}$.

\subsubsection{HD\,74194}
HD\,74194 is known as a supergiant fast X-ray transient (SFXT, \citealt{gamen15}). This system has a 9.5-day period and an eccentricity of 0.6. The secondary companion is a NS \citep{hainich20}. From the orbital parameters we compute a binary mass function of $0.0062 \pm 0.0019\,\Msun$. The visible star is classified as O8.5\,Ib-II(f)p. The effective temperature that we derived is 32.1kK and a surface gravity of 3.45. These parameters agree with those provided by \citet{hainich20}, within the error bars. From the Gaia eDR3 distance and an $A_V = 1.66 \pm 0.05$, we compute a luminosity $\log(L/\Lsun) = 5.41 \pm 0.04$ and we derive a radius of $R = 16.5 \pm 0.8\,\Rsun$. The resulting spectroscopic mass is estimated to $28.2 \pm 3.1\,\Msun$ and BONNSAI returns an evolutionary mass of $31.2_{-1.2}^{+1.4}\,\Msun$, which indicates no mass discrepancy. The secondary is expected to have a mass estimate between 1.8 and $6\,\Msun$. The TESS light curve of HD~74194 mainly shows SLF. Eight frequencies are, however, detected as significant from our criterion ($\nu_1 = 0.274(3)$, $\nu_2 = 0.188(3)$, $\nu_3 = 0.360(3)$, $\nu_4 = 0.411(2)$, $\nu_5 = 0.092(3)$, $\nu_6 = 0.150(3)$, $\nu_7 = 0.542(3)$, $\nu_8 = 0.596(3)$~d$^{-1}$). Assuming than one of these frequencies is originated from the rotation of the star, they provide a range of inclinations between 20 and $49^{\circ}$. Using this range and the binary mass function, the mass estimate for the unseen secondary star is between 2.5 and $5.7\,\Msun$. 

\subsubsection{HD\,75211}
HD\,75211 is an SB1 system \citep{sota14} with a period of 20.4 days and an eccentricity of 0.34. The binary mass function is $0.0162 \pm 0.0007\,\Msun$. The effective temperature of the visible star is estimated to 34.5kK and its surface gravity, corrected for the centrifugal force, to 3.59. These parameters agree very well with those provided by \citet{markova18}. From the Gaia eDR3 distance and an extinction of $A_V = 2.08 \pm 0.05$, the luminosity of HD\,75211 is $\log(L/\Lsun) = 5.36 \pm 0.03$ and we infer a radius of $R = 13.4 \pm 0.4\,\Rsun$. The spectroscopic mass is estimated to $25.3 \pm 2.7\,\Msun$ and the evolutionary mass to $31.0 \pm 1.0\,\Msun$. We observe a slight mass discrepancy for that object. The spectral disentangling does not allow us to extract the spectral lines of the secondary star. We estimated the mass of the secondary to be between 2.5 and $12\,\Msun$, whether we considered the spectroscopic or the evolutionary mass estimate. The TESS light curve shows stochastic variation. Its periodogram reveals five peaks higher than the threshold with frequencies $\nu_1= 0.356(2)$, $\nu_2= 0.423(4)$, $\nu_3= 0.512(3)$, $\nu_4= 0.061(2)$, $\nu_5= 0.239(2)$~d$^{-1}$. These frequencies correspond periods of 2.81, 2.36, 1.95, 16.39, and 4.18 days, respectively. None of these frequency is related to the orbital frequency ($\nu \sim 0.05$~d$^{-1}$). Speculating that this signal might come from rotation, we computed a range of inclinations between 23 and 58$^{\circ}$, suggesting a possible mass estimate for the secondary between 3.2 and $7.3\,\Msun$. This estimate suggests an early A/mid B spectral classification for the companion. From our simulations with the spectral disentangling, we can rule out the presence of a secondary star more massive than $5\,\Msun{}$.

\subsubsection{HD\,94024}
HD\,94024 is a short-period runaway system with an orbital period of 2.5 days and a circular orbit. The binary mass function is equal to $0.0068 \pm 0.0007\,\Msun$. We estimated an effective temperature for the visible star to be 33.7kK and a surface gravity $\logg = 3.75$ [cgs]. The luminosity is $\log(L/\Lsun) = 4.95 \pm 0.05$, computed from an extinction of $A_V = 1.22 \pm 0.01$, and the radius is $R = 8.7 \pm 0.4\,\Rsun$. The spectroscopic mass is calculated to be $15.6 \pm 2.8\,\Msun$ and the evolutionary one to $22.2_{-1.1}^{+1.0}\,\Msun$. There is a mass discrepancy between these two values. The spectral disentangling does not allow us to extract the spectral signature of the secondary star. We estimated the mass of the secondary to be between 1.4 and $6\,\Msun$ (i.e. of a spectral type between A and mid B, for a non-degenerate object). The TESS light curve (Fig.~\ref{fig:lightcurve}) shows two clear modulations with frequencies $\nu_1 = 0.811(1)$ and $\nu_2 = 0.070(4)$~d$^{-1}$. The period corresponding to the first frequency is half the orbital period, suggesting ellipsoidal variations due to the deformation of the visible star. If the star has a synchronous rotation with the orbit, the star must be seen under an inclination of $58^{\circ}$, which would suggest a mass of about $2~\Msun{}$ for the companion. The second significant frequency does not provide a physical value for the inclination (i.e. $\sin~i > 1$). From our simulations with the spectral disentangling, a secondary more massive than $3-5~\Msun{}$ would have been detected with our analysis.

\subsubsection{HD\,105627}
HD\,105627 is a system with a 4.3-day period and an eccentricity of 0.08. The binary mass function is equal to $0.0103 \pm 0.0007\,\Msun$. The visible component is classified as O9\,III. We derived an effective temperature of 32.5kK and a $\logg$, corrected for the centrifugal force, of 3.67 [cgs]. Our stellar parameters agree with those derived by \citet{dealmeida19}. From the Gaia eDR3 distance and an extinction of $A_V = 0.98 \pm 0.03$, we compute a luminosity of $\log(L/\Lsun) = 4.91 \pm 0.07$, giving a radius of $R = 9.0 \pm 0.7\,\Rsun$. We estimated a spectroscopic mass of $13.8 \pm 2.7\,\Msun$ for the visible star. From BONNSAI, we derived an evolutionary mass of $21.0_{-1.0}^{+1.3}\,\Msun$. There is a clear mass discrepancy for that object. The companion mass is expected to be between 1.4 and $6\,\Msun$, whether we consider the spectroscopic or evolutionary mass. There are two significant frequencies in the TESS light curve at $\nu_1 = 1.625(5)$~d$^{-1}$, and $\nu_2 = 0.379(4)$~d$^{-1}$. These frequencies are not related to the orbital motion. It is unlikely that the first frequency is due to rotation but rather from pulsations (it would indeed imply that the primary would rotate higher than critical). By assuming that the second frequency is  coming from a rotational modulation, we compute a possible inclination of about $70 \pm 36^{\circ}$. That would suggest a companion mass to be between $1.3$ and $2.5\,\Msun$. With this mass, the companion would either be an A- or F-type star or a NS. The spectral disentangling does not allow us to extract the secondary spectrum but that is justified due to the low number of observed spectra in our dataset. From our simulations, we would have detected a secondary companion down to $3-5~\Msun{}$.

\subsubsection{HD\,130298}
HD\,130298 is a highly eccentric runaway system ($e=0.468$) with an orbital period of 14.6 days and a bow shock was detected by \citet{peri12}. The calculated binary mass function is large ($0.3292 \pm 0.0073~\Msun{}$). The visible component is classified as O6.5\,III, with an effective temperature of 37.2kK and a $\logg$ of 3.82 [cgs]. We compute a luminosity of $\log(L/\Lsun) = 5.22 \pm 0.07$, giving a radius of $10.0 \pm 0.5\,\Rsun$. We compute a spectroscopic mass of $24.2 \pm 3.8\,\Msun$. The parameters predicted from BONNSAI give an evolutionary mass of $28.0_{-4.1}^{+5.2}\,\Msun$. There is no mass discrepancy for this object. The spectral disentangling does not allow us to extract the spectral signature of the secondary. With a minimum mass estimated to $7.7~\Msun{}$ for the secondary, its spectral lines should be visible in the disentangled and composite spectra. This could suggest that the secondary is candidate to be a quiet stellar-mass BH. The periodogram computed from the TESS light curve shows a clear peak at $\nu = 0.357(1)$~d$^{-1}$ (i.e. a period of 2.8 days, Fig.~\ref{fig:lightcurve}). The origin of this signal is not known but if it originates from a rotational modulation, it would correspond to an inclination of $54 \pm 16^{\circ}$ (based on the stellar parameters we derived). This inclination would suggest that the mass estimate for the companion would be equal to $8.8_{-1.5}^{+3.5}\,\Msun$, which corresponds to an early B-type star. From the simulations, we showed that a secondary star would have been detected down to a mass of $\sim 3-4~\Msun{}$, and therefore a $8~\Msun{}$ secondary would have been detected with the spectral disentangling.

\subsubsection{HD~165174}
HD~165174 is a SB1 system with a period of 23.9 days and an eccentricity of 0.16. The binary mass function is equal to $0.0313 \pm 0.0071\,\Msun$. The visible star is a fast rotator, classified as O9.7~IIn. The effective temperature is estimated to be 30.6kK and a surface gravity of 3.60 [cgs], after the correction for the centrifugal force. The spectral disentangling fails to extract the spectrum of the secondary star. Using the Gaia eDR3 and an extinction of $0.824 \pm 0.046$, we computed a luminosity of $\log(L/\Lsun) = 4.87 \pm 0.05$ and a radius of $R = 9.7 \pm 0.5\,\Rsun$. This gives a spectroscopic mass equal to $13.7 \pm 1.5\,\Msun$ and an evolutionary mass of $20.0_{-1.0}^{+1.1}\,\Msun$, showing a clear mass discrepancy. We estimated the mass of the secondary to be between 2.2 and $4.0\,\Msun$, depending on whether we consider the spectroscopic or the evolutionary mass for the primary, suggesting an A-type or late-B-type secondary if the component is not degenerate or a NS. The system was not observed with TESS. The analysis of the light curve of HD~165174 was done by \citet{handler12} from ground-based photometry. These authors found a significant frequency at 3.289~d$^{-1}$, corresponding to a period of 0.30 days. They ruled out the possibility that this signal might come from the rotation of the star but rather from pulsations. A secondary component, more massive than $3~\Msun{}$ would have been detected from spectral disentangling according to our simulations.

\subsubsection{HD~229234}
HD~229234 was reported as an SB1 system by \citet{mahy13}. The system has a period of 3.5 days and a circular orbit. The spectral disentangling does not allow us to extract the signature of the secondary spectrum. From the orbital parameters, we calculated a binary mass function of $0.0351 \pm 0.0057\,\Msun$. The visible star is classified as O9~III. The effective temperature is estimated to be 31.2kK and a surface gravity of 3.46 [cgs], after the correction for the centrifugal force. Using the Gaia eDR3 and an extinction of $3.03 \pm 0.05$, we computed a luminosity of $\log(L/\Lsun) = 5.12 \pm 0.03$ and a radius of $R = 12.4 \pm 0.4\,\Rsun$. This gives a spectroscopic mass equal to $16.1 \pm 1.4\,\Msun$ and an evolutionary mass of $23.2_{-0.4}^{+1.0}\,\Msun$, showing a clear mass discrepancy. We estimated the mass of the secondary to be between 2.6 and $20.0\,\Msun$, depending whether we consider the spectroscopic or the evolutionary mass for the primary. The secondary can thus be classified from an A-type star to an O-type star if it is a non-degenerate star. We stress, however, that an early-B or late-O-type star would have been detected with the spectral disentangling since, according to our simulations, the spectral disentangling would have detected a secondary object down to $3-4~\Msun{}$. Furthermore, we also stress that the systemic velocity of the 3.5-day period system varies as a function of time, suggesting a higher-order system. In this case, a $10~\Msun{}$ inner system would have been detected from our simulations. Similarities with HD~96670 \citep{Gomez21} can be assumed, but, so far, no clear evidence can be reported, and a more intensive monitoring of this object needs to be performed. The TESS light curve is dominated by two frequencies at $\nu_1 = 0.569(1)$ and $\nu_2 = 0.282(2)$~d$^{-1}$. These frequencies provide an inclination range between 15 and $31^{\circ}$, which combines with the binary mass function indicates a mass estimate between 6 and $14.4\,\Msun$ for the companion. By assuming that the primary star is rotating synchronously with the system, the inclination of the system would be calculated to be $31^{\circ}$, which would give a mass of $\sim 6~\Msun{}$ for the unseen secondary component.

\subsubsection{HD\,308813}
HD\,308813 is a 6.3-day period system with a highly eccentric orbit of 0.38. The binary mass function is equal to $0.0198 \pm 0.0030\,\Msun$. The visible star is classified as an O9.7\,IV(n) star. The effective temperature is estimated to be 30.3kK and a surface gravity of 3.81 [cgs], in agreement with the stellar parameters given by \citet{william13}. Using a distance of 2.38 kpc and an extinction of $0.79 \pm 0.03$, we computed a luminosity of $\log(L/\Lsun) = 4.77 \pm 0.20$ and a radius of $R = 10.7 \pm	2.3\,\Rsun$. This gives a spectroscopic mass equal to $10.7 \pm 4.3\,\Msun$ and an evolutionary mass of $17.8_{-1.5}^{+1.7}\,\Msun$, showing a small mass discrepancy. We estimated the mass of the secondary to be between 1.6 and $6.0\,\Msun$, depending whether we consider the spectroscopic or the evolutionary mass for the primary. This suggests a companion with a spectral classification between an A/F- and a late B-type star. The spectral disentangling prevents us from extracting the spectrum of the secondary but based on our simulations, we would have detected a secondary more massive than $2.5~\Msun{}$. In the TESS light curve, we detected 22 significant frequencies. The signal is dominated by a clear oscillation at a frequency $\nu = 0.158(1)$~d$^{-1}$, which corresponds to the orbital period. This signal is not due to ellipsoidal variations but might be related to the rotation of the star (if the star rotates synchronously with the system). We therefore estimated the inclination of the system to be $\sim 25^{\circ}$. That corresponds to a mass estimate for the secondary of $\sim 5~\Msun{}$ but an object with such a mass would have been detected from our methodology. In the TESS light curve, we also detected weak eclipses with a period of 3.85 days (i.e. $\nu = 0.521(1)$~d$^{-1}$). That period is not detected in spectroscopy, and could be induced by contamination from another object in the TESS field-of-view. 

\subsubsection{LS~5039}
LS~5039 is a short period ($P_{\rm orb}=3.9$ days) eccentric ($e = 0.25$) system. As mentioned by \citet{trigueros21}, LS~5039 is expected to host a compact object as secondary that could be a micro-quasar, a stellar-mass BH, or a NS \citep[][and references therein]{dubus13}. The binary mass function is equal to $0.0042 \pm 0.0008\,\Msun$. The visible star is classified as an ON6V((f))z by \citet{maiz16}. Its effective temperature is estimated to be 38.7kK and a surface gravity of 3.89 [cgs]. Using the Gaia eDR3 and an extinction of $4.06 \pm 0.06$, we computed a luminosity of $\log(L/\Lsun) = 4.90 \pm 0.04$ and a radius of $R = 6.3 \pm 0.3\,\Rsun$. This gives a spectroscopic mass equal to $11.1 \pm 1.5\,\Msun$ and an evolutionary mass of $32.2_{-3.7}^{+5.0}\,\Msun$, showing a clear mass discrepancy. We estimated the mass of the secondary to be between 1.3 and $3.0\,\Msun$ if we consider the spectroscopic mass or between 1.7 and 9 if we consider the evolutionary mass. This object was not observed by TESS. The spectral disentangling does not allow us to extract the spectrum of the secondary star. From our simulations, a secondary more massive than $6~\Msun{}$ would have, however, been detected. 

\subsection{SB2 systems}

\subsubsection{HD~29763}
HD\,29763 is a 3-day period system with circular orbit. The primary is a B3 star and the spectral disentangling allows us to characterise the secondary star. The RV semi-amplitudes are equal to $K_1 = 53.28$\,\kms\ and $K_2=138.53$\,\kms, giving a mass ratio equal to $M_2/M_1 \sim 0.38$. The secondary spectrum shows the \ion{Mg}{ii}~4481 line to be stronger than the \ion{He}{i}~4471 line, and the \ion{Si}{ii}~4128-30 doublet lines to be stronger than the \ion{He}{i}~4143 line. This suggests that the secondary is a B9 star or later. HD~29763 was not observed with TESS. From the minimum mass of the primary star and the masses estimated from the stellar parameters, we derived an inclination of about $40^{\circ}$ for the system. That gives a mass estimate for the secondary star between $1.8$ and $2.3~\Msun{}$. 

\subsubsection{HD\,30836}
HD\,30836 is 9.5-day system with a quasi-circular orbit ($e = 0.01$). The visible star is classified as B2\,III. The spectral disentangling reveals the spectral signature of the secondary star, and provided us with RV semi-amplitudes of $K_1 = 26.33$\,\kms\ and $K_2=87.21$\,\kms. We computed a mass ratio of $M_2/M_1 \sim 0.30$. The secondary shows \ion{Si}{ii}~4128-30 doublet (stronger than the \ion{He}{i}~4143), and the \ion{Mg}{ii}~4481 line stronger than the \ion{He}{i}~4471 line. This suggests a B9 secondary or even with a later type. By comparing the minimum mass of the primary with the estimated mass from its stellar parameters and its position in the Hertzsprung-Russell diagram (HRD), we computed an inclination of $31 \pm 2^{\circ}$. This inclination gives a mass for the secondary of about $3.0~\Msun{}$, in agreement with the derived spectral type. The TESS light curve is dominated by a signal with a frequency at $\nu= 0.254 \pm 0.003$~d$^{-1}$, corresponding to a period of 3.93 days. This period is not related to the orbital period, and could be produced by a rotational modulation of one component. Assuming that this signal is due to the rotation, we derived a possible inclination of $21 \pm 5^{\circ}$. 

\subsubsection{HD\,37737}
HD\,37737 is a 7.8-day period system with a highly eccentric orbit of 0.38. \citet{peri12} reported that this system is surrounded by a bow shock. The binary mass function is equal to $0.2224 \pm 0.0127\,\Msun$. The visible star is classified as an O9.5\,II-III(n) star. The effective temperature is estimated to be 29.2kK and a surface gravity of 3.49 [cgs]. Using the Gaia eDR3 and an extinction of $1.93 \pm 0.05$, we computed a luminosity of $\log(L/\Lsun) = 4.81 \pm 0.12$ and a radius of $R = 10.1 \pm 1.4\,\Rsun$. This gives a spectroscopic mass equal to $11.3 \pm 2.9\,\Msun$ and an evolutionary mass of $21.0_{-1.6}^{+1.2}\,\Msun$, showing a clear mass discrepancy. We estimated the mass of the secondary to be between 4 and $15.0\,\Msun$, depending on whether we consider the spectroscopic or the evolutionary mass for the primary. This suggests a B-type star companion. However, the spectral disentangling does not allow us to extract the spectrum of the secondary. Small RV semi-amplitude for the secondary and high rotation of the primary could be one reason, in addition to the S/N of the composite spectra, to explain why the spectral disentangling did not converge. The TESS light curve shows clear eclipses, which allows us to rule out the presence of a compact object. As mentioned by \citealt{trigueros21}), the two eclipses are really close from each other (see Fig.~\ref{fig:lightcurve}), with in between a pulse-like maximum. At the top of that signal, the light curve is also affected by a sinusoidal signal with a period of one-tenth the orbital period. We note that the periodogram shows a series of 15 harmonics of the orbital frequency. A fit of the light curve using PHOEBE (Fig.~\ref{Fig:phoebe}) indicates that the secondary is expected to have a mass between 3.9 and $5.5~M_{\odot}$. The inclination of the system is estimated to be equal to $76^{\circ}$. The light curve fit also provides us with a  characterisation of the physical parameters of the secondary. In addition to its mass, we derive a radius of $2.8~R_{\odot}$, and a $\logg{} \sim 4.2$. Such an object is at the limit of our detection technique. It is therefore not surprising that the secondary has not been detected in this work.

\subsubsection{HD~52533}
HD~52533 was reported as an SB1 system with a period of about 22 days and an eccentricity of 0.3 \citep{mcswain07,trigueros21}. We found similar orbital parameters ($P_{\rm orb} =21.95$ days and $e=0.39$).  The system was also reported to show eclipses, visible in the TESS light curve (see Fig.~\ref{fig:lightcurveSB2} and \citealt{trigueros21}). The spectral disentangling succeeded to extract for the first time the spectrum of the secondary component, providing us with RV semi-amplitudes equal to $K_1=88.42$~\kms\ and $K_2=208.98$~\kms. These values give a mass ratio of $M_2/M_1 \sim 0.40$. Both components appear to be fast rotators with projected rotational velocities of $\vsini \sim 300$~\,\kms for each component. The secondary spectrum does not show any \ion{He}{ii} lines, but we detect  \ion{Si}{iv} lines. We therefore classified the secondary as an B0-1 star. From the minimum mass of the primary and its estimated mass from its stellar parameters and its position in the HRD, we computed an inclination close to $90^{\circ}$. The facts that the separation between the two components is quite large and that the light curve shows eclipses also suggest that the inclination of the system is close to $90^{\circ}$. The fit of the light curve with PHOEBE (Fig.~\ref{Fig:phoebe}) confirms that inclination. We find that the primary has a mass between $24$ and $34~\Msun$, a radius of $R \sim 9.3~\Rsun$, and a $\logg{} = 4.03$ [cgs]. The secondary has a mass betweem $13$ and $16~\Msun$, a radius of $R \sim 5.5~\Rsun$, and a $\logg{} = 4.08$ [cgs].

\subsubsection{HD~57236}
HD~57236 is a long-period ($P_{\rm orb} = 212.5$ days) eccentric ($e = 0.58$) systems. The spectral disentangling allows us to characterise for the first time the spectral signature of the secondary component and provided us with RV semi-amplitudes of $K_1 = 59.81$\,\kms\ and $K_2=72.22$\,\kms. We computed a mass ratio of $M_2/M_1 \sim 0.83$. The secondary is a fast rotator with a projected rotational velocity of $\vsini \sim 200$~\kms, which might explain why the secondary has never been detected. The secondary spectrum shows the presence of \ion{He}{ii} and \ion{Si}{iv} lines, suggesting that late-O or early-B type star. From the minimum mass of the primary and its estimated masses, we computed an inclination of about $60^{\circ}$. This inclination gives a mass for the secondary between $18$ and $20~\Msun{}$. The TESS light curve shows clear oscillations with dominant frequencies at $\nu= 0.254\pm 0.003$, and $0.715 \pm 0.003$~d$^{-1}$, corresponding to periods of 3.94 and 1.40 days. These periods are not related to the orbital period, but could be linked with the rotations of both components. 

\subsubsection{HD\,91824}
HD\,91824 was reported as SB1 by \citet{sota14}. This object is a long-period system with a 112-day orbit and an eccentricity of 0.21. The spectral disentangling reveals for the first time the spectral signature of the secondary star, and provides us with RV semi-amplitudes of $K_1 = 36.19$\,\kms\ and $K_2=110.59$\,\kms, giving a mass ratio equal to $M_2/M_1 \sim 0.33$. In the disentangled spectrum of the secondary star, there are no \ion{He}{ii}, \ion{Si}{ii}, \ion{Si}{iv} lines and the \ion{Si}{iii} lines are stronger than the \ion{Mg}{ii} lines, suggesting a B2 spectral classification for the secondary (with an uncertainty of one subtype). From the minimum mass of the primary and its estimated masses, we computed an inclination close to $55^{\circ}$. This inclination gives a mass for the secondary betwee $10$ and $12~\Msun{}$, in agreement with the derived spectral type. The TESS light curve shows stochastic variability. We detected one main frequency at $\nu= 0.089 \pm 0.003$~d$^{-1}$, corresponding to a period of 11.28 days. 

\subsubsection{HD\,93028}
HD\,93028 has been reported to be an SB1 systems by \citet{sota11}. Its period is long with about 205 days and its eccentricity is equal to 0.13. The spectral disentangling succeeded to extract the signature of the secondary star. The full orbital solution provides us with RV semi-amplitudes of $K_1 = 35.58$\,\kms and $K_2 = 73.60$\,\kms, giving a mass ratio $M_2/M_1 \sim 0.48$. The primary rotates slowly with $\vsini \sim 30$~\kms\ while the secondary rotates faster with  with $\vsini \sim 150$~\kms. The higher projected rotational velocity of the secondary is probably the reason why this system was reported as SB1 in the literature. The secondary do not have \ion{He}{ii} lines, and is therefore classified as an early-B star. From the minimum mass of the primary and its estimated masses, we computed an inclination of $77^{\circ}$ (but with large error bars). This inclination gives a mass for the secondary between $8-11~\Msun{}$, in agreement with the derived spectral type. We note that the TESS light curve is heavily contaminated by other bright stars in the close neighbourhood.

\subsubsection{HD\,152405}
HD\,152405 is a 25.5-day period system with an eccentricity of 0.55. The system was reported as SB1 by \citet{sota14}. The spectral disentangling reveals for the first time the spectral signature of the secondary star, and provides us with RV semi-amplitudes of $K_1 = 30.18$\,\kms\ and $K_2=79.38$\,\kms, giving a mass ratio equal to $M_2/M_1 \sim 0.38$. The brightness ratio is low (i.e. less than 5\%). The secondary do not have \ion{He}{ii} line or \ion{Si}{ii} line, but we do detect the presence of \ion{Si}{iv} lines. We therefore classified the secondary to be an B1 star, but that classification is difficult because of the faintness of the star, and an uncertainty of two sub-groups must be mentioned. From the minimum mass of the primary and its estimated masses, we computed an inclination close to $25 \pm 5^{\circ}$. This inclination gives a mass for the secondary between 8 and $9~\Msun{}$, in agreement with the derived spectral type.

\subsubsection{HD\,152723}
HD~152723 was reported as SB1 system by \citet{sota14}. The system has a period of 18.9 days and an eccentricity of 0.51. The spectral disentangling reveals the contribution of the secondary companion in the composite spectra, even though this contribution is very weak with at least 5\% of the brightness. The secondary spectrum has no \ion{He}{ii} lines, and we note the presence of weak \ion{Si}{ii} lines. We therefore classified the secondary as an B5 star (with an uncertainty of two sub-types). The RV semi-amplitudes given by the spectral disentangling are $K_1 = 18.37$\,\kms\ and $K_2=89.37$\,\kms, giving a mass ratio equal to $M_2/M_1 \sim 0.21$. From the minimum mass of the primary and its estimated masses, we computed an inclination close to $17^{\circ}$. This inclination gives a mass for the secondary between 1 and $30~\Msun{}$. We did not retrieve the TESS light curve of HD~152723, because the star falls outside the field-of-view. 

\subsubsection{HD\,163892}
HD\,163892 is a 7.8-day period system that is almost circular ($e = 0.04$). The system was reported as SB1. The spectral disentangling reveals for the first time the spectral signature of the secondary star, and provides us with RV semi-amplitudes of $K_1 = 41.05$\,\kms\ and $K_2=232.46$\,\kms, giving a mass ratio equal to $M_2/M_1 \sim 0.18$. The secondary has the \ion{Mg}{ii}~4481 line with the same strength as the \ion{He}{i}~4471 line, suggesting a B5-B7 object. This classification is, however, difficult because of the faintness of the companion. From the minimum mass of the primary and its estimated masses, we computed an inclination close to $70^{\circ}$. This inclination gives a mass for the secondary $3 \pm 2~\Msun{}$, in agreement with the derived spectral type. The system has not been observed with TESS. 

\subsubsection{HD\,164438}
Reported as an SB1 by \citet{sota14}, HD\,164438 is a 10.2-day period system with an eccentricity of 0.28. The spectral disentangling reveals for the first time the spectral signature of the secondary star. The RV semi-amplitudes are equal to $K_1 = 28.68$\,\kms and $K_2 = 106.34$\,\kms, giving a mass ratio $M_2/M_1 \sim 0.27$. The secondary is faint with a brightness ratio lower than 0.1. We observed, in the disentangled spectrum of the secondary star, the \ion{Si}{ii}~4128-30 doublet stronger than the \ion{Si}{iii}~4552 line or even the \ion{He}{i}~4121 line. The \ion{He}{i}~4471 line is also with the same intensity as the \ion{Mg}{ii}~4481 line. We therefore classified the secondary as a B5 or later. From the minimum mass of the primary and its estimated masses, we computed an inclination close to $30^{\circ}$. This inclination gives a mass for the secondary $3 \pm 2~\Msun{}$, in agreement with the derived spectral type.

\subsubsection{HD\,164536}
HD\,164536 was reported as an SB1 system with a 13.4-day period by \citet{william13}. We found a slightly shorter orbital period of 11.7 days and an eccentricity of 0.07. The spectral disentangling succeeded to extract the signature of the secondary star. The full orbital solution provides us with RV semi-amplitudes of $K_1 = 22.95$\,\kms and $K_2 = 161.48$\,\kms, giving a mass ratio $M_2/M_1 \sim 0.14$. The secondary spectrum shows the presence of the \ion{Si}{ii}~4128-30 doublet slightly weaker than \ion{He}{i}~4143, and a ratio between \ion{He}{i}~4471 and \ion{Mg}{ii}~4481 close to unity. We classified the secondary as a B7 star or later. From the minimum mass of the primary and its estimated masses, we computed an inclination close to $40^{\circ}$. This inclination gives a mass for the secondary $5 \pm 2~\Msun{}$, in agreement with the derived spectral type. HD~164536 was not observed with TESS. 

\subsubsection{HD~167263}
HD~167263 was reported in the literature as being an SB1 system \citep{sota14} with an orbital period of 14.8 days \citep{stickland01} or 12.7 days \citep{mayer14}. We found a much longer period of 64.8 days for that system, and a very low eccentricity of $e=0.005$. The spectral disentangling reveals for the first time the SB2 nature of that system. It provided us with RV semi-amplitudes equal to $K_1=32.77$~\kms\ and $K_2=41.26$~\kms, giving a mass ratio of $M_2/M_1 \sim 0.79$. The secondary is a fast rotator with $\vsini \sim 220$~\kms. The secondary spectrum shows \ion{He}{ii} and \ion{Si}{iv} lines, indicating a late O-type star. From the minimum mass of the primary and its estimated masses, we computed an inclination close to $17 \pm 7^{\circ}$. This inclination gives a mass for the secondary $22 \pm 5~\Msun{}$, in agreement with the derived spectral type. HD~167263 was not observed with TESS. 

\subsubsection{HD~167264}
HD~167264 was reported as a long-period SB1 system by \citet{sota14}. We found a period of 674.4 days and an eccentricity of 0.23. The spectral disentangling succeeded to extract the spectrum of the secondary component. It provided us with RV semi-amplitudes equal to $K_1=26.28$~\kms\ and $K_2=34.36$~\kms, giving a mass ratio of $M_2/M_1 \sim 0.76$. The secondary spectrum shows weak \ion{He}{ii} lines and the \ion{Si}{iv} lines are clearly visible. That suggests an early-B or late-O as spectral classification for the secondary. From the minimum mass of the primary and its estimated mass from its stellar parameters and its position in the HRD, we computed an inclination close to $41 \pm 4^{\circ}$. This inclination gives a mass for the secondary $18 \pm 8~\Msun{}$, in agreement with the derived spectral type. HD~167264 was not observed with TESS. 

\subsubsection{HD\,192001}
HD\,192001 is a long period system with a period of 189 days on a very eccentric orbit ($e=0.83$). The spectral disentangling reveals the secondary star. The RV semi-amplitudes are equal to $K_1 = 71.64$\,\kms\ and $K_2=124.50$\,\kms, giving a mass ratio equal to $M_2/M_1 \sim 0.58$. The secondary spectrum does not show any \ion{He}{ii} lines, but \ion{Si}{iv} lines. The \ion{Si}{iv}~4089 line shows similar strength than the \ion{Si}{iii}~4552 line. We therefore classified the secondary of HD~192001 as an B0.7 star (with an uncertainty of one subgroup). From the minimum mass of the primary and its estimated mass from its stellar parameters and its position in the HRD, we computed an inclination close to $67 \pm 14^{\circ}$. This inclination gives a mass for the secondary $12 \pm 7~\Msun{}$, in agreement with the derived spectral type. The TESS light curve shows stochastic variation. No clear peak are detected, but the highest one reports a period of 3.09 days. 

\subsubsection{HD~199579}
HD\,199579 was reported as SB1 by \citet{sota11} and possible SB2 by \citet{william01}. The system has a period of 48.5 days and an quasi-circular eccentricity of 0.07, agreeing with the orbital parameters derived by \citet{william01}. The spectral disentangling confirms the SB2 nature of that system. It provided us with RV semi-amplitudes of $K_1 = 39.37$\,\kms\ and $K_2=119.48$\,\kms. We computed a mass ratio of $M_2/M_1 \sim 0.33$. The secondary does not show any \ion{He}{ii} lines. Given its fast rotation ($\sim 200$~\kms), we do not detect any \ion{Si}{ii} and \ion{Si}{iv} lines. We therefore classified the secondary as a B1-2 star. From the minimum mass of the primary and its estimated mass from its stellar parameters and its position in the HRD, we computed an inclination close to $58 \pm 6^{\circ}$. This inclination gives a mass for the secondary $8 \pm 2~\Msun{}$, in agreement with the derived spectral type. The TESS light curve shows stochastic variability dominated by signals with frequencies at $\nu= 0.217\pm 0.003$, $0.086 \pm 0.003$, $0.343\pm 0.003$, and $0.640 \pm 0.003$~d$^{-1}$, corresponding to periods of 4.61, 11.63, 2.92 and 1.56 days. These periods are not related to the orbital period.

\subsubsection{Schulte~11}
Schulte~11 was identified as an SB1 by \citet{kobulnicky12}. They found an orbital period of 72.4 days, and a large eccentricity ($e = 0.5$). We confirmed this period ($P_{\rm orb} = 72.6$ days) and we found a higher eccentricity of $e=0.61$. The eccentricity that we derived is higher than the eccentricity of $e=0.37$ presented by \citet{trigueros21}. The spectral disentangling allows us to extract the spectral signature of the secondary star. Given the O5.5~Ifc spectral classification of the primary, the spectral disentangling remains challenging and only the Balmer series could be extracted. The RV semi-amplitudes that we derived are equal to $K_1=29.91$~\kms\ and $K_2=134.92$~\kms, giving a mass ratio of $M_2/M_1 \sim 0.22$. From the minimum mass of the primary and its estimated masses, we computed an inclination of $31 \pm 9^{\circ}$. The TESS light curve shows stochastic variability but no clear peak is detected in the periodogram. 

\subsubsection{V747~Cep}
V747~Cep is an SB1 system with an orbital period of 5.3 days and an eccentric orbit \citep{majaess08}. The orbital parameters were confirmed by \citet{trigueros21} and through our analysis. We found a period of 5.3 days and an eccentricity of $e = 0.37$. The system was also reported to show eclipses in its TESS light curve \citep{trigueros21}. The spectral disentangling succeeded to extract for the first time the spectrum of the secondary component. While we can distinguish the spectral lines of the secondary, the disentangled spectrum is, however, very noisy and that prevents us from getting the physical parameters of the secondary. We found RV semi-amplitudes equal to $K_1=89.60$~\kms\ and $K_2=374.44$~\kms, giving a mass ratio of $M_2/M_1 \sim 0.24$. The secondary can be classified as B-type star. From the minimum mass of the primary and its estimated masses, we computed an inclination close to $75 \pm 6^{\circ}$. This inclination is in agreement with the fact that the light curve shows eclipses. We model the TESS light curve using PHOEBE (Fig.~\ref{Fig:phoebe}) to better constrain the fundamental properties of the secondary. From our fit, the primary has a mass of $33.8~M_{\odot}$, a radius of $8.9~R_{\odot}$, and a $\logg{} = 4.08$. We obtain for the secondary a mass of $7.3~M_{\odot}$, a radius of $3.5~R_{\odot}$, inferring a surface gravity of 4.2. The luminosities are computed to be equal to $\log(L/L_{\odot}) = 5.28$ for the primary and $3.56$ for the secondary.

\end{appendix}
\end{document}